\pdfoutput=1
\documentclass[12pt,twoside,doublespaced]{gatech-thesis}%
\usepackage{eurosym}
\usepackage{amsfonts}
\usepackage{amsmath}
\usepackage{amssymb}
\usepackage{dcolumn}
\usepackage{graphicx}
\usepackage{makeidx}
\usepackage{rotating}
\numberwithin{equation}{chapter}
\numberwithin{figure}{chapter}
\numberwithin{table}{chapter}
\titlepagetrue
\signaturepagetrue
\figurespagetrue
\tablespagetrue
\contentspagetrue
\bibpagetrue

\begin{document}

\title{Theory of Light-Matter Interactions in Cascade and Diamond Type Atomic Ensembles}
\author{Hsiang-Hua Jen}
\department{School of Physics}
\degree{Doctor of Philosophy}
\gradyear{2010}
\principaladvisor{Prof. T. A. Brian Kennedy}
\firstreader{Prof. Alex Kuzmich}
\secondreader{Prof. Michael S. Chapman}
\thirdreader{Prof. Carlos Sa de Melo}
\fourthreader{Prof. Ken Brown}
\submitdate{December 2010}%

\begin{preliminary}%

\begin{dedication}
\null\vfil

\medskip\ \ \ \ {\large To grandfather,}

\medskip{\large who supports and believes in me unconditionally throughout the
study,}

\medskip{\large and in memory of grandmother and father.}

\vfil\null

\end{dedication}

\begin{acknowledgements}
I am grateful to my thesis advisor, Professor Brian Kennedy, for his
instruction and support of my research. \ With his guidance and encouragement,
I learned and gained insights along the way of study. \ I am thankful to
Professor Alex Kuzmich for his direction in experimental perspective and to
Dr. S. D. Jenkins for his helpful discussions on theoretical background. \ I
am also thankful to the thesis committee, Professor Michael Chapman, Professor
Carlos Sa de Melo, and Professor Ken Brown.

Throughout this work, I had many useful discussions with quantum optics group
members, and I am thankful to them: Dr. D. N. Matsukevich, Dr. T.
Chaneli\`{e}re, Dr. S. Y. Lan, \ O. A. Collins, C. Campbell, Dr. R. Zhao, and
A. Radnaev. \ I am also appreciative to Professor Li You, Dr. P. Zhang, and
Dr. D. L. Zhou for the training in my early graduate studies. \ A special
thank you is due for the support of my friends, Shenshen Lin, Professor I-Tang
Yu, Dr. S. C. Lin, Dr. Yu Tsao, Dr. Pei Lin, Dr. T. Lee, and Dr. A. Liang.
\end{acknowledgements}

\contents

\begin{summary}
In this thesis, we investigate the quantum mechanical interaction of light
with matter in the form of a gas of ultracold atoms: the atomic ensemble. \ We
present a theoretical analysis of two problems, which involve the interaction
of quantized electromagnetic fields (called signal and idler) with the atomic
ensemble (i) cascade two-photon emission in an atomic ladder configuration,
and (ii) photon frequency conversion in an atomic diamond configuration. \ The
motivation of these studies comes from potential applications in long-distance
quantum communication where it is desirable to generate quantum correlations
between telecommunication wavelength light fields and ground level atomic
coherences. \ In the two systems of interest, the light field produced in the
upper arm of an atomic Rb level scheme is chosen to lie in the telecom window.
\ The other field, resonant on a ground level transition, is in the
near-infrared region of the spectrum. \ Telecom light is useful as it
minimizes losses in the optical fiber transmission links of any two
long-distance quantum communication device.

We develop a theory of correlated signal-idler pair correlation. \ The
analysis is complicated by the possible generation of multiple excitations in
the atomic ensemble. \ An analytical treatment is given in the limit of a
single excitation assuming adiabatic laser excitations. \ The analysis
predicts superradiant timescales in the idler emission in agreement with
experimental observation. \ To relax the restriction of a single excitation,
we develop a different theory of cascade emission, which is solved by
numerical simulation of classical stochastic differential equation using the
theory of open quantum systems. \ The simulations are in good qualitative
agreement with the analytical theory of superradiant timescales. \ We further
analyze the feasibility of this two-photon source to realize the DLCZ protocol
of the quantum repeater communication system.

We provide a quantum theory of near-infrared to telecom wavelength conversion
in the diamond configuration. \ The system provides a crucial part of a
quantum-repeater memory element, which enables a "stored" near-infrared photon
to be converted to a telecom wavelength for transmission without the
destruction of light-atom quantum correlation. \ We calculate the theoretical
conversion efficiency, analyzing the role of optical depth of the ensemble,
pulse length, and quantum fluctuations on the process.
\end{summary}%

\end{preliminary}%
%



\chapter{Introduction}

A quantum communication network based on the distribution and sharing of
entangled states is potentially secure to eavesdropping and is therefore of
great practical interest \cite{QI,cryp,QI2}. \ A protocol for the realization
of such a long distance system, known as the quantum repeater, was proposed by
Briegel \textit{et al}. \cite{repeater,Dur}. \ A quantum repeater based on the
use of atomic ensembles as memory elements, distributed over the network, was
subsequently suggested by Duan, Lukin, Cirac and Zoller \cite{dlcz}. \ The
storage of information in the atomic ensembles involves the Raman scattering
of an incident light beam from ground state atoms with the emission of a
signal photon. \ The photon is correlated with the creation of a phased,
ground-state, coherent excitation of the atomic ensemble. \ The information
may be retrieved by a reverse Raman scattering process, sending the excitation
back to the initial atomic ground state and generating an idler photon
directionally correlated with the signal photon
\cite{qubit,chou,vuletic,collective,store,collective2,single2,pan,kimble}.
\ In the alkali gases, the signal and the idler field wavelengths are in the
near-infrared spectral region. \ This presents a wavelength mismatch with
telecommunication wavelength optical fiber, which has a transmission window at
longer wavelengths\ (1.1-1.6 um). \ It is this mismatch that motivates the
search for alternative processes that can generate telecom wavelength photons
correlated with atomic spin waves \cite{telecom}. \ 

This motivates the research presented in this thesis where we study
multi-level atomic schemes in which the transition between the excited states
is resonant with a telecom wavelength light field \cite{telecom}. \ The basic
problem is to harness the absorption and the emission of telecom photons while
preserving quantum correlations between the atoms, which store information and
the photons that carry along the optical fiber channel of the network. \ In
this thesis, we theoretically study atomic cascade and diamond configurations
in this context.

\section{DLCZ Protocol for the Quantum Repeater}

A long-distance quantum repeater must overcome the exponential losses in the
optical fiber. \ To overcome this problem, the use of quantum memory was
proposed \cite{dlcz}. \ For a practical system, it is essential to maximize
quantum memory time, to preserve coherence during protocol operations, and
connect the memory elements by light signals in the low-loss window of the
optical fiber medium. \ The telecom wavelength range (1.1-1.6 $\mu$m) has a
loss rate as low as 0.2 dB/km. \ 

It is not common to have a telecom ground state transition in atomic gases
except for rare earth elements \cite{erbium,dysprosium} or in an erbium-doped
crystal \cite{solid}. \ However, a telecom wavelength (signal) can be
generated from transitions between excited levels in the alkali metals
\cite{telecom}. \ 

\subsection{Correlated cascade emission in quantum telecommunication}

The ladder configuration of atomic levels provides a source for telecom
photons (signal) from the upper atomic transition. \ For rubidium and cesium
atoms, the signal field has the range around 1.3-1.5 $\mu$m that can be
coupled to an optical fiber and transmitted to a remote location. \ Cascade
emission may result in pairs of photons, the signal entangled with the
subsequently emitted infrared photon (idler) from the lower atomic transition.
\ Entangled signal and idler photons were generated from a phase-matched
four-wave mixing configuration in a cold, optically thick $^{85}$Rb ensemble
\cite{telecom}. \ This correlated two-photon source is potentially useful as
the signal field has telecom wavelength.

The temporal emission characteristics of the idler field, generated on the
lower arm of the cascade transition, were observed in measurements of the
joint signal-idler correlation function. \ The idler decay time was shorter
than the natural atomic decay time and dependent on optical thickness in a way
reminiscent of superradiance \cite{Dicke,Stephen,Lehm,mu,OC:Mandel}.

We will develop an analytical theory of the cascade emission in an atomic
ensemble in Chapter 3. \ The influence of electromagnetic dipole-dipole
interactions between atoms is important to account for the idler field's
temporal profile. \ By developing the theory on the assumption of weak
adiabatic laser excitation, we are able to calculate the spectral
characteristics of the signal and idler fields, and make a connection with the
traditional theory of superradiance.

In Chapter 4, , we develop a more elaborate theory of the cascade emission
under similar physical conditions to Chapter 3, but without the assumption of
single atomic excitations. \ The theory is based on numerical solutions of
stochastic differential equations derived using open-systems methods of
quantum optics. \ We limit our analysis to the confirmation of the
superradiant emission of the idler field predicted in the simple theory and
observed experimentally.

In Chapter 5, we use this theory to discuss a potential application of the
cascade emission process in the DLCZ protocol, and discuss the role of
time-frequency entanglement.

\section{Quantum Memory with Light Frequency Conversion}

It is not sufficient to generate telecom wavelength light for quantum
communication. \ The light field must be quantum correlated with atomic
excitations stored in memory \cite{telecom}.

Recently there has been a breakthrough in this direction using a pair of cold,
non-degenerate rubidium gas samples \cite{radaev}. \ A correlated pair of
atomic spin wave and infrared fields are generated by conventional Raman
scattering in one ensemble. \ The light field is directed onto a second
ensemble where it is frequency converted to the telecom range by four-wave
mixing using a diamond configuration of atomic levels. \ The experiments were
designed to measure quantum correlations between the stored atomic excitation
and the telecom field.

The conversion scheme exploits an efficient low-noise parametric conversion
process that is facilitated by operating in the regime of high transparency
\cite{radaev}. \ This provides a basic quantum memory element for a scalable,
long distance quantum network. \ In Chapter 6, we investigate conditions
required to maximize the conversion efficiency as a function of optical
thickness of the atomic ensemble. \ The influence of the probe pulse duration
on the conversion efficiency is studied by numerical solution of the
Maxwell-Bloch equations.

\section{Outline}

The remainder of this thesis is organized as follows.

In Chapter 2, we review some theoretical methods to provide background for the
theories developed in Chapter 4 and 6. \ In particular we discuss the
derivation of quantum Heisenberg-Langevin equations for the interaction of a
group of atoms with a quantized propagating electromagnetic field. \ We
illustrate the connection of these operator equations with related classical
(c-number) stochastic Langevin equations. \ The latter have the useful
property that they may be numerically simulated, under certain conditions, and
we provide the Kubo oscillator as a numerical test case.

In Chapter 3, we present a theory of cascade two-photon emission in an atomic
ensemble. \ The radiative atomic dipole-dipole coupling is shown to influence
the emission of the idler photon, resulting in the appearance of superradiant
time scales. \ The theory is developed on the basis of Schr\"{o}dinger
probability amplitudes assuming single atomic excitations. This approach
allows a straightforward treatment of the spectral entanglement properties of
the signal-idler photons.

In Chapter 4, we relax the assumption of single atomic excitations and develop
a theory based on c-number stochastic partial differential equations, derived
using the methods reviewed in Chapter 2. \ Numerical solutions of the
equations are used to compare with the superradiant timescales derived in the
analytical theory.

In Chapter 5, the analysis of Chapter 3 is used to discuss the behavior of the
cascade emission on the DLCZ protocol for the quantum repeater. \ Entanglement
swapping and quantum teleportation are investigated, and the influence of
time-frequency entanglement is discussed.

In Chapter 6, the use of the diamond configuration in frequency up and down
conversion is analyzed using quantum-Heisenberg Langevin and Maxwell-Bloch
equation methods. \ We present results for the optimal conversion efficiency
as a function of optical thickness of the atomic ensemble. \ The role of pulse
length and quantum fluctuations are discussed.

In Chapter 7, we present some conclusions.

In Appendixes A-D, we present a great deal of supporting information on the
theoretical derivations that are quite lengthy on account of both the
multimode treatment of the light fields and the complicated atomic level
schemes and atomic dipole-dipole interactions.



\chapter{Review of Theoretical and Numerical methods}

In this Chapter, we review the derivations of quantum-Heisenberg equations and
c-number Langevin equations for light-atom interactions. \ The reason for
focusing on these methods is, in the first place that they are less familiar
than Schr\"{o}dinger picture methods (see Chapter 3 and 5) and that our
applications of these methods (Chapter 4 and 6) involve rather long
derivations that may obscure the basic ideas. \ 

We provide two methods of deriving the c-number Langevin equations and their
noise correlations. \ The equations may be found from the quantum
Heisenberg-Langevin equations using a "quantum-classical" correspondence
\cite{QS:Louisell}. \ Alternatively, c-number Langevin equations are deduced
by a Schr\"{o}dinger-picture approach that employs characteristic equation and
coherent state phase space methods. \ In the final step the Langevin equations
are deduced from a Fokker-Planck equation for a generalized statistical
distribution. \ Such methods were initially applied in quantum laser theory in
the 1960's by Haken \cite{LT:Haken}. \ The independent derivations will be
used to check the lengthy derivations involved in the case of cascade emission.

\section{Quantum and C-number Langevin Equations}%

\begin{figure}
[ptb]
\begin{center}
\includegraphics[
natheight=7.499600in,
natwidth=9.999800in,
height=4.7106in,
width=6.2716in
]%
{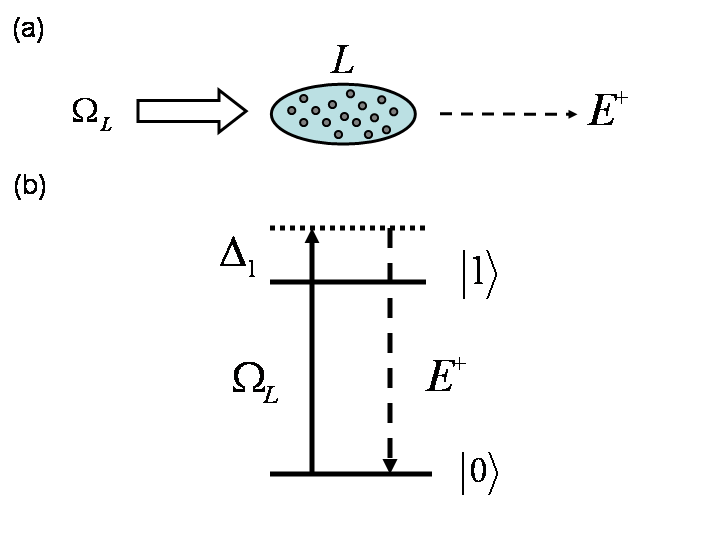}%
\caption{The two-level atomic ensemble interacts with a classical and quantum
field. \ (a) An elongated atomic ensemble of length $L$ is excited by a pump
field of Rabi frequency $\Omega_{a}$ and emits a propagating quantized field
denoted by annihilation operator $E^{+}.$ \ (b) Two-level structure for an
atomic ensemble with the ground ($|0\rangle$) and excited ($|1\rangle$) state.
\ The detuning of the pump field is $\Delta_{1}$.}%
\label{two_level}%
\end{center}
\end{figure}

Langevin equations were initially derived to describe Brownian motion
\cite{SM:Gardiner}. \ A fluctuating force is used to represent the random
impacts of the environment on the Brownian particle. \ A given realization of
the Langevin equation involves a trajectory perturbed by the random force.
\ Ensemble averaging such trajectories provides a natural and direct way to
investigate the dynamics of the stochastic variables. \ 

In this section, we review quantum and c-number Langevin equation approaches
for a two-level atomic ensemble interacting with a quantized electromagnetic
field. \ As shown in Figure \ref{two_level}, the atoms are excited by a pump
field of Rabi frequency $\Omega_{a}$, and a propagating quantized field
$E^{+}$ is considered to be emitted along the direction $\hat{z}$ of the
ensemble with length $L$. \ 

\subsection{Quantum Heisenberg-Langevin equations}

We consider the Hamiltonian of $N$ two-level (ground and excited states
$|0\rangle$, $|1\rangle$) atoms interacting with one pump field and a
multimode quantized fields with mode annihilation operators $\hat{a}_{l}$ that
satisfy the commutation relation $[\hat{a}_{l},\hat{a}_{l^{\prime}}^{\dag
}]=\delta_{ll^{\prime}}$ for the $l$th section along the propagation
direction. \ The propagation length $L$ is discretized into $2M+1 $ elements
\cite{quantization}. \ In the electric dipole approximation and rotating wave
approximation, the interaction is given by $-\vec{d}\cdot\vec{E}$, Appendix
B.1. \ The Hamiltonian $H$\ includes the free evolution ($H_{0} $) of atoms
with transition frequency $\omega_{1}$, the quantized field of central
frequency $\omega,$ and the dipole interaction ($H_{I}$),
\begin{align}
H  & =H_{0}+H_{I}\text{,}\\
H_{0}  & =\sum_{l=-M}^{M}\hbar\omega_{1}\hat{\sigma}_{11}^{l}(t)+\hbar
\omega\sum_{l=-M}^{M}\hat{a}_{l}^{\dag}(t)\hat{a}_{l}(t)+\hbar\sum
_{l,l^{\prime}}\omega_{ll^{\prime}}\hat{a}_{l}^{\dag}(t)\hat{a}_{l^{\prime}%
}(t)\text{ ,}\\
H_{I}  & =-\hbar\sum_{l=-M}^{M}\Big[\Omega_{L}\hat{\sigma}_{01}^{l\dagger
}(t)e^{ik_{L}z_{l}-i\omega_{L}t}+g\sqrt{2M+1}\hat{\sigma}_{01}^{l\dagger
}(t)\hat{a}_{l}(t)e^{ikz_{l}}+h.c.\Big]
\end{align}
where $\hat{\sigma}_{01}^{l}(t)\equiv\sum_{\mu}^{N_{z}}|0\rangle_{\mu}%
\langle1|\Big|_{r_{\mu}=z_{l}}$.$~\ $The Rabi frequency $\Omega_{L}%
=d_{10}\mathcal{E}(k_{L})/(2\hbar)$ is one-half the conventional definition.
\ The dipole matrix element $d_{10}\equiv$ $\langle1|\hat{d}|0\rangle$,
coupling strength $g\equiv d_{10}\mathcal{E}(k)/\hbar~$where $\mathcal{E}%
(k)=\sqrt{\hbar\omega/2\epsilon_{0}V}$ is the electric field per photon, and
$z_{p}=\frac{pL}{2M+1},$ $p=-M,...,M$. \ \ The matrix $\omega_{ll^{\prime}%
}\equiv\sum_{n=-M}^{M}k_{n}e^{ik_{n}(z_{l}-z_{l^{\prime}})}/(2M+1)$ accounts
for field propagation by coupling the local mode operators. \ 

The dynamical equations including dissipation due to spontaneous emission can
be treated by introducing the reservoir field that interacts with the system
\cite{QO:Scully}. \ After introducing the coupling to the reservoir, we may
write down by inspection the dissipation terms. \ We define $\gamma_{01}$ to
be the spontaneous emission rate from $|1\rangle\rightarrow|0\rangle.$ \ \ In
the co-moving frame coordinates $z$ and $\tau=t-z/c$, the quantum
Heisenberg-Langevin equations are%

\begin{align}
\frac{\partial}{\partial\tau}\tilde{\sigma}_{01}  & =(i\Delta_{1}-\frac
{\gamma_{01}}{2})\tilde{\sigma}_{01}+i\Omega_{a}(\tilde{\sigma}_{00}%
-\tilde{\sigma}_{11})+ig(\tilde{\sigma}_{00}-\tilde{\sigma}_{11})\tilde{E}%
^{+}+\mathcal{\tilde{F}}_{01},\\
\frac{\partial}{\partial\tau}\tilde{\sigma}_{11}  & =-\gamma_{01}\tilde
{\sigma}_{11}+i\Omega_{L}\tilde{\sigma}_{01}^{\dag}-i\Omega_{L}^{\ast}%
\tilde{\sigma}_{01}+ig\tilde{\sigma}_{01}^{\dag}\tilde{E}^{+}-ig^{\ast}%
\tilde{\sigma}_{01}\tilde{E}^{-}+\mathcal{\tilde{F}}_{11},\\
\frac{\partial}{\partial z}\tilde{E}^{+}  & =\frac{iNg^{\ast}}{c}\tilde
{\sigma}_{01}+\mathcal{\tilde{F}}_{E^{+}},\label{ch2_1}%
\end{align}
where various Langevin noises $\mathcal{\tilde{F}}$ associated with atomic
operators $\tilde{\sigma}_{01}$, $\tilde{\sigma}_{11}$ and field operator
${\tilde{E}}^{+}$ are necessary to preserve equal time commutation
relations.\ \ The detuning of the pump field is $\Delta_{1}=\omega_{L}%
-\omega_{1}$ and the slowly-varying operators are defined as $\tilde{\sigma
}_{01}\equiv\sigma_{01}^{l}e^{-ik_{a}z_{l}+i\omega_{a}t}/N_{z}$,
$\tilde{\sigma}_{11}\equiv\tilde{\sigma}_{11}^{l}/N_{z}$, and $\tilde{E}%
^{+}(z,t)\equiv\sqrt{2M+1}\hat{a}_{l}e^{i\omega_{a}t}$ where we let
$\omega=\omega_{L}$. \ \ The time evolution of atomic coherence ($\tilde
{\sigma}_{01}$) depends on the population difference ($\tilde{\sigma}%
_{00}-\tilde{\sigma}_{11}$), and in turn atomic population is influenced by
atomic coherence and the classical and quantized fields. \ The atomic
coherence couples to the quantized field along the propagation direction, $z$.

The noise operator correlations are related to the dissipation through the
fluctuation-dissipation theorem \cite{LP:Sargent, QO:Scully}. \ If we have a
quantum Langevin equation for variable $\hat{x}$
\begin{equation}
\dot{\hat{x}}(t)=\hat{A}_{x}(t)+\hat{F}_{x}(t)\label{ch2_ql}%
\end{equation}
where $\hat{A}_{x}$ is so-called the drift term for $\hat{x}$, and the
corresponding Langevin noise operator is $\hat{F}_{x}$,\ the quantum noise
correlation functions can be derived from the generalized Einstein relation,
\begin{equation}
\langle\hat{F}_{x}(t)\hat{F}_{y}(t)\rangle=-\langle\hat{x}(t)\hat{A}%
_{y}(t)\rangle-\langle\hat{A}_{x}(t)\hat{y}(t)\rangle+\frac{d}{dt}\langle
\hat{x}(t)\hat{y}(t)\rangle.\label{Einstein_1}%
\end{equation}
where the bracket denotes the quantum mechanical ensemble average. \ 

With the above recipe, we have the non-vanishing normally ordered quantum
noise correlation function from Eq. (\ref{ch2_1}),%

\begin{equation}
\hat{D}_{11,11}=\gamma_{01}\tilde{\sigma}_{11}.
\end{equation}
where $\left\langle \mathcal{\tilde{F}}_{11}^{\dag}(t,z)\mathcal{\tilde{F}%
}_{11}(t^{\prime},z^{\prime})\right\rangle =\frac{L}{N}\delta(t-t^{\prime
})\delta(z-z^{\prime})\left\langle \hat{D}_{11,11}\right\rangle $, and
$\hat{D}$ is also referred to as a diffusion matrix element by analogy with
classical diffusion processes.

Even for this relatively simple light-matter interaction, there is no
analytical solution possible. \ The c-number Langevin equation approach,
below, provides a possible way to attack the problem numerically by stochastic
simulation and to calculate normally-ordered quantities by ensemble averaging,
although we will not pursue such simulations here.

\subsection{C-number Langevin equation}

A c-number Langevin equation approach may be suitable for stochastic
simulation \cite{QN:Gardiner, SM:Gardiner}, and utilizes the methods developed
by Lax, Louisell, and Haken to describe the dynamics of the interaction
\cite{LT:Haken,QS:Louisell}. \ Their recipe involves a normal ordering
procedure and a so-called "quantum-classical correspondence" to derive the
c-number Langevin equations \cite{QS:Louisell,LP:Sargent,Fleischhauer94}.
\ The normal ordering chosen is $\tilde{\sigma}_{01}^{\dagger}$,
$\tilde{\sigma}_{11}$, $\tilde{\sigma}_{01}$, $\tilde{E}^{-}$, $\tilde{E}^{+}$
where\ the creation operators always appear to the left of the annihilation
operators. \ The population operator is put between the atomic coherence
operators since it is self conjugate.

The c-number Langevin equations are then derived from Eq. (\ref{ch2_1}) by
making the quantum-classical correspondence that we denote as%

\begin{equation}
\tilde{\sigma}_{01}^{\dagger}\rightarrow\alpha_{5},\text{ }\tilde{\sigma}%
_{11}\rightarrow\alpha_{4},\text{ }\tilde{\sigma}_{01}\rightarrow\alpha
_{3},\text{ }\tilde{E}^{-}\rightarrow E^{-},\text{ }\tilde{E}^{+}\rightarrow
E^{+}.
\end{equation}
Similarly for the Langevin noises,%

\begin{equation}
\mathcal{\tilde{F}}_{01}^{\dagger}\rightarrow\mathcal{F}_{5},\text{
}\mathcal{\tilde{F}}_{11}\rightarrow\mathcal{F}_{4},\text{ }\mathcal{\tilde
{F}}_{01}\rightarrow\mathcal{F}_{3},\text{ }\mathcal{\tilde{F}}_{E^{-}%
}\rightarrow\mathcal{F}_{2},\text{ }\mathcal{\tilde{F}}_{E^{+}}\rightarrow
\mathcal{F}_{1},
\end{equation}
where the notation is chosen to facilitate the comparison with an alternative
approach that we will discuss in the next Section. \ 

The classical noise correlation functions are also derived from an Einstein
relation. \ Consider the c-number Langevin equation for the variables $x$\ and
$y,$%
\begin{align}
\dot{x}(t)  & =A_{x}(t)+F_{x}(t),\\
\dot{y}(t)  & =A_{y}(t)+F_{y}(t).
\end{align}
\ From the requirement of equivalent time evolution of normally-ordered
operators and their c-number counterparts, we have for example%
\begin{equation}
\frac{d}{dt}\langle\hat{x}\hat{y}\rangle=\frac{d}{dt}\langle xy\rangle.
\end{equation}
\ Classical noise correlations can be derived from the quantum ones using
\begin{equation}
\langle F_{x}F_{y}\rangle=\langle\hat{F}_{x}\hat{F}_{y}\rangle+\langle\hat
{x}\hat{A}_{y}\rangle+\langle\hat{A}_{x}\hat{y}\rangle-\langle xA_{y}%
\rangle-\langle A_{x}y\rangle.\label{Einstein_2}%
\end{equation}
where the quantum and classical noise correlations are formally quite
different. \ For non-normally-ordered operators $\hat{x}\hat{z}$, we may use
the commutator to substitute that
\begin{equation}
\langle\hat{x}\hat{z}\rangle=\langle\hat{z}\hat{x}\rangle+\langle\lbrack
\hat{x},\hat{z}]\rangle.
\end{equation}

The drift term of the c-number Langevin equations are closely related to the
corresponding term in the quantum Heisenberg-Langevin equations. \ After the
quantum-classical correspondence is made, we derive the coupled equations with
c-number variables ($E^{+},$ $E^{-},$ $\alpha_{3},$ $\alpha_{4},$ $\alpha
_{5},$)\ and Langevin noises ($\mathcal{F}_{1,2,3,4,5}$) that satisfy%
\begin{align}
\frac{\partial}{\partial\tau}\alpha_{3}  & =(i\Delta_{1}-\frac{\gamma_{01}}%
{2})\alpha_{3}+i\Omega_{a}(\alpha_{0}-\alpha_{4})+ig(\alpha_{0}-\alpha
_{4})E^{+}+\mathcal{F}_{3},\\
\frac{\partial}{\partial\tau}\alpha_{4}  & =-\gamma_{01}\alpha_{4}+i\Omega
_{a}\alpha_{5}-i\Omega_{a}^{\ast}\alpha_{3}+ig\alpha_{5}E^{+}-ig^{\ast}%
\alpha_{3}E^{-}+\mathcal{F}_{4},\\
\frac{\partial}{\partial\tau}\alpha_{5}  & =(-i\Delta_{1}-\frac{\gamma_{01}%
}{2})\alpha_{5}-i\Omega_{a}^{\ast}(\alpha_{0}-\alpha_{4})-ig^{\ast}(\alpha
_{0}-\alpha_{4})E^{-}+\mathcal{F}_{5},\\
\frac{\partial}{\partial z}E^{+}  & =\frac{iNg^{\ast}}{c}\alpha_{3}%
+\mathcal{F}_{1},\text{ }\frac{\partial}{\partial z}E^{-}=-\frac{iNg}{c}%
\alpha_{5}+\mathcal{F}_{2},
\end{align}

The associated non-vanishing diffusion matrix elements, however look quite
different to their quantum counterparts%
\begin{align}
D_{3,3}  & =-i2\Omega_{a}\alpha_{3}-i2g\alpha_{3}E^{+},\nonumber\\
D_{4,4}  & =i\Omega_{a}\alpha_{5}-i\Omega_{a}^{\ast}\alpha_{3}+i\alpha
_{5}E^{+}-i\alpha_{3}E^{-}+\gamma_{01}\alpha_{4}.\label{dif1}%
\end{align}
The diffusion matrix elements are defined as $\left\langle \mathcal{F}%
_{i}(t,z)\mathcal{F}_{j}(t^{\prime},z^{\prime})\right\rangle =\frac{L}%
{N}\delta(t-t^{\prime})\delta(z-z^{\prime})\left\langle D_{ij}\right\rangle $
in the continuous limit. \ For the more complicated light-matter interactions
we will encounter in Chapter 4 involving four atomic levels interacting with
two propagating quantized light fields, the diffusion matrix calculation is
much more intricate. \ It is therefore important to have an independent check
of the c-number equations and the associated diffusion matrix. \ In the
following Section we review the Fokker-Planck equation approach based on a
Schrodinger picture treatment of the quantized light-atom interaction.

\section{Fokker-Planck Equations and Stochastic Differential Equations}

Here we review the alternative method, due to Haken \cite{LT:Haken}, to derive
the c-number Langevin equations or equivalently stochastic differential
equations via Fokker-Planck equations
\cite{LT:Haken,QO:Walls,QN:Gardiner,SM:Car}.

The Fokker-Planck equation is used to describe the fluctuations in Brownian
motion \cite{SM:Gardiner}, and its solution for probability distribution
$f(x,t)$ of Brownian particles in space $x$ and time $t$ is determined by the
drift and diffusion properties of the particles.

\subsection{Characteristic functions in P-representation}

The Characteristic function $\chi$ is convenient for the derivation of
Fokker-Planck equation, and it is the distribution function of the
Fokker-Planck equation in Fourier space. \ We follow the same procedure of
P-representation laser theory \cite{LT:Haken}. \ 

The relevant operators of our system are atomic coherences ($\tilde{\sigma
}_{01}^{l\dagger}$, $\tilde{\sigma}_{01}^{l}$), population ($\tilde{\sigma
}_{11}^{l}$) and field operators ($\hat{a}_{l}^{\dagger}$, $\hat{a}_{l}$).
\ The normally ordered exponential operator is chosen to be%

\begin{align}
E(\lambda)  & =\prod_{l}E^{l}(\lambda),\nonumber\\
E^{l}(\lambda)  & =e^{i\lambda_{5}^{l}\tilde{\sigma}_{01}^{l\dagger}%
}e^{i\lambda_{4}^{l}\tilde{\sigma}_{11}^{l}}e^{i\lambda_{3}^{l}\tilde{\sigma
}_{01}^{l}}e^{i\lambda_{2}^{l}\hat{a}_{l}^{\dagger}}e^{i\lambda_{1}^{l}\hat
{a}_{l}},
\end{align}
where $E(\lambda)$ the complete exponential operator and is decomposed into
products of $E^{l}(\lambda)$ for each section $l$ of the propagation
direction. \ We note that the ordering of operators is the same as we chose
for the quantum-classical correspondence in the previous Section. \ The
complex parameters $\lambda_{i}^{l}$ are classical counterparts of operators
in Fourier space, as will become clear when we derive the Fokker-Planck equation.

Then characteristic function $\chi$ can be calculated from a density matrix
$\rho,$%

\begin{align}
\chi & =\text{Tr}\left\{  E(\lambda)\rho\right\}  \text{,}\\
\frac{\partial\chi}{\partial t}  & =\text{Tr}\left\{  E(\lambda)\frac
{\partial\rho}{\partial t}\right\}  =\sum_{m}\left(  \frac{\partial\chi
}{\partial t}\right)  _{m},\text{ }m=A,L,I,sp
\end{align}
and time evolution of $\rho$\ is
\begin{align*}
\frac{\partial\rho}{\partial t}  & =\frac{1}{i\hbar}[H,\rho]+\left(
\frac{\partial\rho}{\partial t}\right)  _{sp},\text{ }H=H_{0}+H_{I},\\
\left(  \frac{\partial\rho}{\partial t}\right)  _{sp}  & =\sum_{l=-M}^{M}%
\sum_{\mu}^{N_{z}}\frac{\gamma_{01}}{2}\left[  2\hat{\sigma}_{01}^{\mu,l}%
\rho\hat{\sigma}_{01}^{\mu,l\dagger}-\hat{\sigma}_{01}^{\mu,l\dagger}%
\hat{\sigma}_{01}^{\mu,l}\rho-\rho\hat{\sigma}_{01}^{\mu,l\dagger}\hat{\sigma
}_{01}^{\mu,l}\right]  ,
\end{align*}
where $H_{0}=H_{A}+H_{L}$. $\ H_{A}$ is the Hamiltonian for atomic free
evolution, $H_{L}$ is the Hamiltonian for the pump field, and the dipole
interaction Hamiltonian is $H_{I}.$ \ The dissipation from spontaneous
emission is denoted as $sp$.

The contribution from $\left(  \frac{\partial\rho}{\partial t}\right)  _{sp}$
is calculated up to the second order in $\lambda_{i}$. \ The validity of
truncation to second order is due to the expansion in the small parameter
$1/N_{z}$. \ The dissipative contribution, identified by superscript (2),
takes the form,%

\begin{align}
& \gamma_{01}\text{Tr}\left\{  E(\lambda)\left[  \hat{\sigma}_{01}\rho
\hat{\sigma}_{01}^{\dagger}-\frac{1}{2}\hat{\sigma}_{11}\rho-\frac{1}{2}%
\rho\hat{\sigma}_{11}\right]  \right\}  ^{(2)}=\nonumber\\
& \gamma_{01}\left[  -\frac{i\lambda_{3}}{2}\frac{\partial}{\partial
(i\lambda_{3})}-\frac{i\lambda_{5}}{2}\frac{\partial}{\partial(i\lambda_{5}%
)}-i\lambda_{4}\frac{\partial}{\partial(i\lambda_{4})}+\frac{(i\lambda
_{4})^{2}}{2}\frac{\partial}{\partial(i\lambda_{4})}\right]  \chi.
\end{align}
where we drop the summation over spatial slices $l,$ which we will retrieve
later. \ Collecting together all contributions to the characteristic function,
we may proceed to write down a Fokker-Planck equation that leads to the
c-number Langevin equation.

\subsection{A Complimentary Derivation of C-number Langevin Equations}

The time derivative of the distribution function $f$ is found from the Fourier
transform of the characteristic function $\frac{\partial f}{\partial t}%
=\frac{1}{(2\pi)^{n}}\int...\int e^{-i\vec{\alpha}\cdot\vec{\lambda}}%
\frac{\partial\chi}{\partial t}d\lambda_{1}...d\lambda_{n}.$ \ Separating the
different contributions we may write%
\begin{equation}
\frac{\partial f}{\partial t}=\mathcal{L}f=\sum_{l,l^{\prime}}\left[
\mathcal{L}_{A}\delta_{ll^{\prime}}+\mathcal{L}_{L}+\mathcal{L}_{I}%
\delta_{ll^{\prime}}+\mathcal{L}_{sp}\delta_{ll^{\prime}}\right]  f.
\end{equation}

The details of the $\mathcal{L}$ operators can be found in Appendix B. \ Here
we show $\mathcal{L}_{I}$ as an example,%

\begin{align}
& \mathcal{L}_{I}=\nonumber\\
& i\Omega_{a}e^{ik_{a}z_{l}-i\omega_{a}t}\left[  -\frac{\partial^{2}}%
{\partial\alpha_{3}^{l}\partial\alpha_{3}^{l}}(\alpha_{3}^{l})-\frac{\partial
}{\partial\alpha_{3}^{l}}(-2\alpha_{4}^{l}+N_{z})+e^{-\frac{\partial}%
{\partial\alpha_{4}^{l}}}(\alpha_{5}^{l})\right]  -i\Omega_{a}e^{ik_{a}%
z_{l}-i\omega_{a}t}(\alpha_{5}^{l})\nonumber\\
& +ig\sqrt{2M+1}e^{ikz_{l}}\left[  -\frac{\partial^{2}}{\partial\alpha_{3}%
^{l}\partial\alpha_{3}^{l}}(\alpha_{3}^{l})-\frac{\partial}{\partial\alpha
_{3}^{l}}(-2\alpha_{4}^{l}+N_{z})+e^{-\frac{\partial}{\partial\alpha_{4}^{l}}%
}(\alpha_{5}^{l})\right]  \alpha_{1}^{l}\nonumber\\
& +ig^{\ast}\sqrt{2M+1}e^{-ikz_{l}}(\alpha_{3}^{l})\left(  \alpha_{2}%
^{l}-\frac{\partial}{\partial\alpha_{1}^{l}}\right)  +(C^{\prime})^{\ast},
\end{align}
where $C^{\prime}$ is the correspondence that $\alpha_{3}^{\ast}%
\leftrightarrow\alpha_{5,}$ $\alpha_{4}^{\ast}\leftrightarrow\alpha_{4,}$
$\alpha_{1}^{\ast}\leftrightarrow\alpha_{2},$ and $\ast$ denotes complex
conjugation. \ The results is a Fokker-Planck equation of the form
\begin{equation}
\frac{\partial f}{\partial t}=-\frac{\partial}{\partial\alpha}A_{\alpha
}f-\frac{\partial}{\partial\beta}A_{\beta}f+\frac{1}{2}\left(  \frac
{\partial^{2}}{\partial\alpha\partial\beta}+\frac{\partial^{2}}{\partial
\beta\partial\alpha}\right)  D_{\alpha\beta}f
\end{equation}
where $A_{\alpha,\beta}$ and $D_{\alpha\beta}$ are drift and diffusion terms.
\ The corresponding c-number Langevin equations may be derived rigorously when
$D$ is positive definite, and take the form%

\begin{equation}
\frac{\partial\alpha}{\partial t}=A_{\alpha}+\Gamma_{\alpha}\text{, }%
\frac{\partial\beta}{\partial t}=A_{\beta}+\Gamma_{\beta}%
\end{equation}
with a classical noise correlation $\langle\Gamma_{\alpha}(t)\Gamma_{\beta
}(t^{^{\prime}})\rangle=\delta(t-t^{\prime})D_{\alpha\beta}$. \ Higher order
derivatives (third order and higher, from the Taylor expansions of
$e^{-\frac{\partial}{\partial\alpha_{4}^{l}}}$) are ignored as they involve
the small parameter $1/N_{z}.$ \ The corresponding c-number Langevin, or
stochastic differential, equations are%

\begin{align}
\dot{\alpha}_{3}^{l}  & =(-i\omega_{1}-\frac{\gamma_{01}}{2})\alpha_{3}%
^{l}+i\Omega_{a}e^{ik_{a}z_{l}-i\omega_{a}t}(\alpha_{0}^{l}-\alpha_{4}%
^{l})\nonumber\\
& +ig\sqrt{2M+1}e^{ikz_{l}}(\alpha_{0}^{l}-\alpha_{4}^{l})\alpha_{1}%
^{l}+\Gamma_{3}^{l},\\
\dot{\alpha}_{4}^{l}  & =-\gamma_{01}\alpha_{4}^{l}+i\Omega_{a}e^{ik_{a}%
z_{l}-i\omega_{a}t}\alpha_{5}^{l}-i\Omega_{a}^{\ast}e^{-ik_{a}z_{l}%
+i\omega_{a}t}\alpha_{3}^{l}\nonumber\\
& +ig\sqrt{2M+1}e^{ikz_{l}}\alpha_{5}^{l}\alpha_{1}^{l}-ig^{\ast}\sqrt
{2M+1}e^{-ikz_{l}}\alpha_{3}^{l}\alpha_{2}^{l}+\Gamma_{4}^{l},\\
\dot{\alpha}_{1}^{l}  & =-i\omega\alpha_{1}^{l}-i\sum_{l^{\prime}}%
\omega_{ll^{\prime}}\alpha_{1}^{l^{\prime}}+ig^{\ast}\sqrt{2M+1}e^{-ikz_{l}%
}\alpha_{3}^{l}+\Gamma_{1}^{l}.
\end{align}

We can retrieve the continuous limit with the slowly varying variables,
$\alpha_{3}(z,t)\equiv\alpha_{3}^{l}e^{-ik_{a}z_{l}+i\omega_{a}t}/N_{z}$,
$\alpha_{4}(z,t)\equiv\alpha_{4}^{l}/N_{z}$, $E^{+}(z,t)\equiv\sqrt
{2M+1}\alpha_{1}^{l}e^{i\omega_{a}t}$,\ and note that

$-i\sum_{l^{\prime}}\omega_{ll^{\prime}}\alpha_{1}^{l^{\prime}}$ $=$
$-c\frac{\partial}{\partial z_{l}}\alpha_{1}^{l}$ and $\alpha_{0}^{l}%
=N_{z}-\alpha_{4}^{l}$. \ Define also the slowly-varying Langevin noises,%

\begin{align}
\mathcal{F}_{3}(z,t)  & =\frac{1}{N_{z}}\Gamma_{3}^{l}e^{-ik_{a}z_{l}%
+i\omega_{a}t},\mathcal{F}_{4}(z,t)=\frac{1}{N_{z}}\Gamma_{4}^{l},\nonumber\\
\mathcal{F}_{1}(z,t)  & =\sqrt{2M+1}e^{i\omega t}\Gamma_{1}^{l}.
\end{align}

Finally, in the co-moving frame coordinates $z$ and $\tau=t-z/c,$ the c-number
Langevin equation becomes%
\begin{align}
\frac{\partial}{\partial\tau}\alpha_{3}  & =(i\Delta_{1}-\frac{\gamma_{01}}%
{2})\alpha_{3}+i\Omega_{a}(\alpha_{0}-\alpha_{4})+ig(\alpha_{0}-\alpha
_{4})E^{+}+\mathcal{F}_{3},\\
\frac{\partial}{\partial\tau}\alpha_{4}  & =-\gamma_{01}\alpha_{4}+i\Omega
_{a}\alpha_{5}-i\Omega_{a}^{\ast}\alpha_{3}+ig\alpha_{5}E^{+}-ig^{\ast}%
\alpha_{3}E^{-}+\mathcal{F}_{4},\\
\frac{\partial}{\partial\tau}\alpha_{5}  & =(-i\Delta_{1}-\frac{\gamma_{01}%
}{2})\alpha_{5}-i\Omega_{a}^{\ast}(\alpha_{0}-\alpha_{4})-ig^{\ast}(\alpha
_{0}-\alpha_{4})E^{-}+\mathcal{F}_{5},\\
\frac{\partial}{\partial z}E^{+}  & =\frac{iNg^{\ast}}{c}\alpha_{3}%
+\mathcal{F}_{1},\text{ }\frac{\partial}{\partial z}E^{-}=-\frac{iNg}{c}%
\alpha_{5}+\mathcal{F}_{2},
\end{align}
where $\Delta_{1}=\omega_{a}-\omega_{1}$. \ The non-vanishing diffusion
coefficients extracted from the Fokker-Planck equation are
\begin{equation}
D_{3,3}=-i2\Omega_{a}\widetilde{\alpha}_{3}-i2\widetilde{\alpha}_{3}E_{i}%
^{+};\text{ }D_{4,4}=i\Omega_{a}\widetilde{\alpha}_{5}-i\Omega_{a}^{\ast
}\widetilde{\alpha}_{3}+i\widetilde{\alpha}_{5}E_{i}^{+}-i\widetilde{\alpha
}_{3}E_{i}^{-}+\gamma_{01}\widetilde{\alpha}_{4}.
\end{equation}

Comparing with the results in the previous Section and Eq. (\ref{dif1}), we
find complete agreement. \ As the c-number Langevin equations are derived from
a Fokker-Planck equation, they should be interpreted as Ito-type stochastic
differential equations (SDE), and this is important in the numerical solution
method \cite{QN:Gardiner}. \ In numerical simulation it is common to first
transform from the Ito equation to its corresponding Stratonovich form.

\section{Kubo Oscillator}

We present an example of the Kubo oscillator to illustrate numerical
simulation of a multiplicative noise stochastic differential equation. \ A
Kubo oscillator provides a good test case in the numerical solution of
stochastic differential equations. \ The Langevin equation of the
dimensionless Kubo oscillator with amplitude $z(t)$ is given by the
Stratonovich equation,%

\begin{equation}
\frac{d}{dt}z(t)=i\xi(t)z(t)\label{Kubo_one}%
\end{equation}
where $\xi(t)$ is a delta-correlated real Gaussian distributed noise with zero
mean, $\langle\xi(t)\rangle=0$ , and $\langle\xi(t)\xi(t^{\prime}%
)\rangle=\delta(t-t^{\prime}).$ \ The bracket denotes an ensemble average.
\ The exact analytical solution for the first moment is $\langle
z(t)\rangle=\langle z(t=0)\rangle e^{-t/2}$. \ To numerically simulate the
Stratonovich equation (\ref{Kubo_one}), we use the following discretization in
time \cite{Drummond91b, xmds}%

\begin{equation}
z(t_{m})=z_{n-1}+i\xi(t_{n-1})z(t_{m})\frac{\Delta t}{2}\label{kubo_two}%
\end{equation}
where $z(t_{m})$ is evaluated at the midpoint, $t_{m}=(t_{n}+t_{n-1})/2$ and
$\Delta t$ $=t_{n}-t_{n-1}$ is the time step. \ In this specific case where
the noise is linear in $z(t)$, we may solve Eq. (\ref{kubo_two}) to give
$z(t_{m})=z_{n-1}/(1-i\xi(t_{n-1})\Delta t/2)$. \ Setting $z(t_{n+1}%
)=2z(t_{m})-z(t_{n})$, we use $z(t_{n+1})$ for the next time step of the
integration.%
\begin{figure}
[ptb]
\begin{center}
\includegraphics[
natheight=15.499700in,
natwidth=18.000400in,
height=4.0365in,
width=5.3347in
]%
{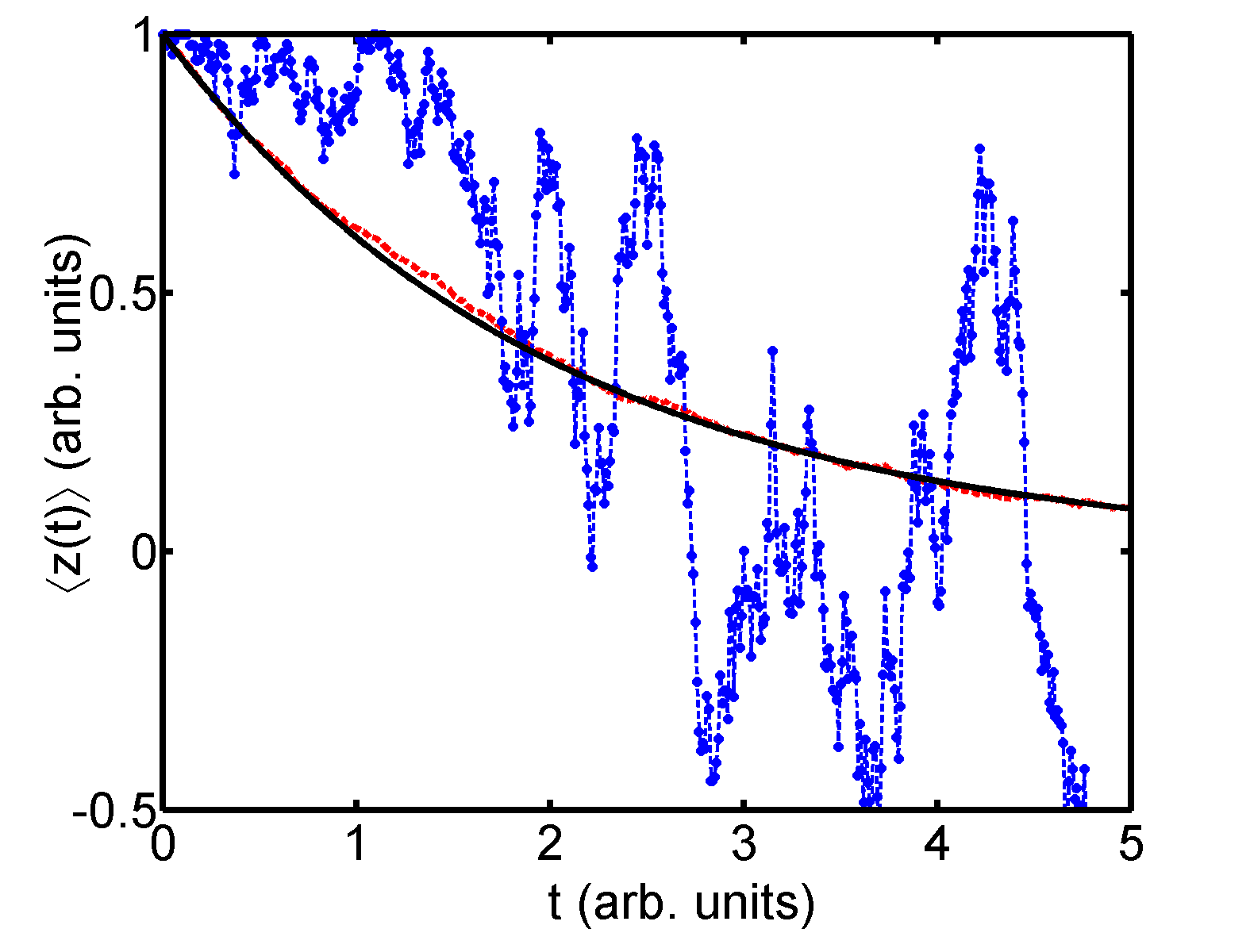}%
\caption{Kubo oscillator simulation. \ The time evolution of Re$\langle
z(t)\rangle$ (dashed-red) averaged from an ensemble of 1024 simulations.
$z(0)=1.$ \ We compare with the exact solution, $z(0)e^{-t/2}$ (solid-black),
and find good agreement. A demonstration of one stochastic realization
(dashed-circle blue) shows large fluctuation around the averaged and exact
results. \ Note that the imaginary part of the solution is almost vanishing as
it should be, and is not shown here.}%
\label{kubo}%
\end{center}
\end{figure}

The Langevin noise is numerically simulated as $\xi(t)=\operatorname{rand}%
$n$(t)/\sqrt{\Delta t}$, where $\operatorname{rand}$n$(t)$\ is a random number
generated from a Gaussian distribution with zero mean and unit variance. \ In
Figure \ref{kubo}, we compare the analytical and numerical results for the
Kubo oscillator. \ The initial condition is set as $z(0)=1$,\ and we use 1024
realizations for the converged numerical result $\langle z(t)\rangle$ with
$\Delta t=0.01$. \ The numerical result is in good agreement with the exact
solution, $z(0)e^{-t/2}$. \ The temporal evolution of one typical realization
of the stochastic process fluctuates significantly, as shown.

We will use the approach demonstrated here to simulate the more complicated
c-number Langevin noises in our investigation of cascade emission from an
atomic ensemble in Chapter 4 (see also Appendix B).



\chapter{Superradiant emission from a cascade atomic ensemble: Analytical
Method}

In this Chapter, we use Schr\"{o}dinger's equation to investigate cascade
emission from a four-level atomic ensemble. \ Quantum communication has opened
up the possibility to transmit quantum information over long distance. \ Due
to the transmission loss in long distance fiber-based quantum communication,
telecommunication (telecom) wavelength light is important to maximize the
transmission efficiency. \ The alkali atomic cascade transition shown in
Figure \ref{four} is able to generate telecom wavelength light, the signal,
from the upper transition and a near-infrared field, the idler, from the lower
one. \ The telecom light can travel through the fiber with minimal loss, while
the near-infrared field is suitable for storage and retrieval in an atomic
quantum memory element. \ Their use in a quantum information system requires
quantum correlations between stored excitations and the telecom field. \ \ 

\ We develop a quantum theory to characterize the properties of the correlated
signal and idler photons and study how the laser excitation pulse modifies
their spectral profile. \ The wave packets of this entangled source are found,
and Schmidt decomposition provides the basis for engineering a pure photon
source that is crucial in quantum information processing.

\section{Introduction}

The spontaneous emission from an optically dense atomic ensemble is a
many-body problem due to the radiative coupling between atoms. \ This coupling
is responsible for the phenomenon of superradiance firstly discussed by Dicke
\cite{Dicke} in 1954.

Since then, this collective emission has been extensively studied in two atom
systems indicating a dipole-dipole interaction \cite{Stephen,Lehm}, in the
totally inverted N atom systems \cite{stehle,Tallet}, and in the extended
atomic ensemble \cite{mu}. \ The emission intensity has been investigated
using the master equation approach \cite{master,Bon,Bon1} and with
Maxwell-Bloch equations \cite{Bon2,Feld}. \ A useful summary and review of
superradiance can be found in the reference \cite{Gross,phase}. \ Recent
approaches to superradiance include the quantum trajectory method
\cite{trajectory,eig1} and the quantum correction method \cite{Fleischhauer99}%
. \ 

In the limit of single atomic excitation, superradiant emission
characteristics have been discussed in the reference \cite{Eberly} and
\cite{Scully}. \ For a singly excited system, the basis set reduces to N
rather than $2^{N}$ states. \ Radiative phenomena have been investigated using
dynamical methods \cite{kurizki,Scully2,eig2} and by the numerical solution of
an eigenvalue problem \cite{Friedberg08a,
Svidzinsky08,Friedberg08b,Friedberg08c}. \ A collective frequency shift
\cite{Arecchi, Morawitz} can be significant at a high atomic density
\cite{Scully09} and has been observed recently in an experiment where atoms
are resonant with a planar cavity \cite{supershift}.

\section{The example of two-state atoms interacting with a pump field}

The atomic dynamics of $N$ two-state atoms interacting with a pump field
generally requires a basis of $2^{N}$ orthogonal states. \ In this Section we
investigate multiple excitations by a laser by solving numerically the master
equation for few atom systems ($N=2,3,4$), using the quantum optics toolbox
\cite{QObox}. \ The complete orthogonal states may be chosen as 1 symmetric
state and $(C_{n}^{N}-1)$ non-symmetric states for any excitation number n
where $C_{n}^{N}$ is the combination coefficient. \ It is natural to construct
the complete orthogonal states using this decomposition because the
interaction Hamiltonian of the pump field, $H_{I}=[-\hbar\frac{\Omega_{a}}%
{2}\sum_{\mu}^{N}|1\rangle\langle0|e^{i\vec{k}\cdot\vec{r}_{\mu}}%
+c.c.]-\hbar\Delta_{1}\sum_{\mu}^{N}|1\rangle\langle1|$, has the same form for
each atom.\ 

For the example of two two-state atoms, there are 4 orthogonal basis states:
the ground state $|00\rangle,$ the symmetric state of a single excitation
$(e^{i\vec{k}\cdot\vec{r}_{1}}|10\rangle+e^{i\vec{k}\cdot\vec{r}_{2}%
}|01\rangle)/\sqrt{2}$, the associated anti-symmetric state $(e^{i\vec{k}%
\cdot\vec{r}_{1}}|10\rangle-e^{i\vec{k}\cdot\vec{r}_{2}}|01\rangle)/\sqrt{2}$,
and the state of two excitations $e^{i\vec{k}\cdot(\vec{r}_{1}+\vec{r}_{2}%
)}|11\rangle.$ \ Note that the spatial phase factor for different atomic
position $\vec{r}$ is included due to the pump field of the wavevector
$\vec{k}$ that is directed along the $\hat{z}$ axis. \ If more atoms are
involved, the complete states of multiple excitations can be derived by
extending the results of reference \cite{kurizki}, and here we list the states
of four atoms ($N=4$),%

\begin{align}
n  & =0,\text{ }|\phi_{1}\rangle=|0,0,0,0\rangle,\nonumber\\
n  & =1\left\{
\begin{array}
[c]{l}%
|\phi_{2}\rangle=\frac{1}{\sqrt{N}}\sum_{\mu=1}^{N}e^{i\vec{k}\cdot\vec
{r}_{\mu}}|1\rangle_{\mu}|0\rangle_{\lambda\neq\mu};\\
|\phi_{l+2}\rangle=\sum_{j=1}^{N-1}\left(  \frac{1+1/\sqrt{N}}{N-1}%
-\delta_{jl}\right)  e^{i\vec{k}\cdot\vec{r}_{j}}|1\rangle_{j}|0\rangle
_{\lambda\neq j}-\frac{e^{i\vec{k}\cdot\vec{r}_{N}}}{\sqrt{N}}|1\rangle
_{N}|0\rangle_{\lambda\neq N},\\
l=1,2,...N-1;
\end{array}
\right. \nonumber\\
n  & =2\left\{
\begin{array}
[c]{l}%
|\phi_{N+2}\rangle=\frac{1}{\sqrt{N(N-1)/2}}\sum_{\mu>\nu}^{N}\sum_{\nu=1}%
^{N}e^{i\vec{k}\cdot(\vec{r}_{\mu}+\vec{r}_{\nu})}|1\rangle_{\mu}%
|1\rangle_{\nu}|0\rangle_{\lambda\neq\mu,\nu};\\
|\phi_{m+N+2}\rangle=\sum_{l>j}^{N}\sum_{j=1}^{N-2}\left(  \frac
{1+1/\sqrt{N_{2}}}{N_{2}-1}-\delta_{m,(j,l)}\right)  e^{i\vec{k}\cdot(\vec
{r}_{j}+\vec{r}_{l})}|1\rangle_{j}|1\rangle_{l}|0\rangle_{\lambda\neq j,l}\\
-\frac{e^{i\vec{k}\cdot(\vec{r}_{N-1}+\vec{r}_{N})}}{\sqrt{N_{2}}}%
|1\rangle_{N-1}|1\rangle_{N}|0\rangle_{\lambda\neq N-1,N},\text{
}m=1,2,...N_{2}-1\text{, }\\
\text{where }N_{2}\equiv N(N-1)/2;
\end{array}
\right. \nonumber\\
& \bullet\nonumber\\
& \bullet\nonumber\\
& \bullet\nonumber\\
n  & =N,\text{ }|\phi_{2^{N}}\rangle=|1\rangle^{\otimes N}\prod_{j=1}%
^{N}e^{i\vec{k}\cdot\vec{r}_{j}},
\end{align}
where $(j,l)$ in the subscript of the Kronecker delta function of two
excitation states is defined so that $(1,2)=1,(1,3)=2,(1,4)=3,(2,3)=4,$ and
$(2,4)=5.$ \ Note that the $n=0,$ $|\phi_{1}\rangle,$ and $n=N,$ $|\phi
_{2^{N}}\rangle,$ states are symmetric. \ For $n\neq0,N$ excitations, the
states are constructed from one symmetric state and $C_{n}^{N}$ non-symmetric states.%

\begin{figure}
[ptb]
\begin{center}
\includegraphics[
natheight=20.000400in,
natwidth=25.499700in,
height=5.5449in,
width=7.0891in
]%
{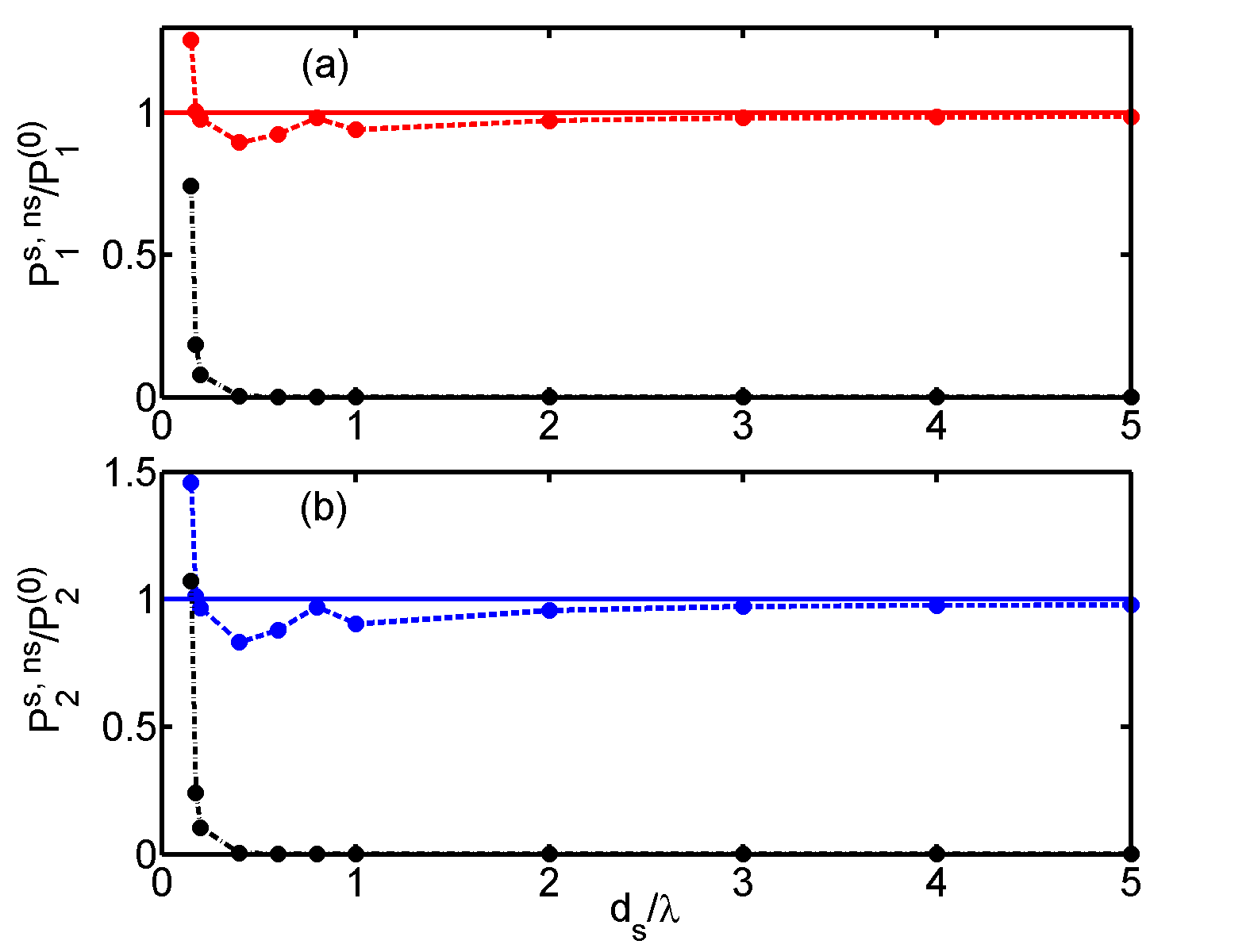}%
\caption{Single- and double-excitation populations as a function of distance
$d_{s}.\ \ $(a) The populations of the symmetric state for a single excitation
$P_{1}^{\text{s}}$ (dashed-red) and the sum of non-symmetric single-excitation
states $P_{1}^{\text{ns}}$ (dashed-dotted black). \ (b) The populations of the
symmetric state for double excitations $P_{2}^{\text{s}}$ (dashed-blue) and
the sum of non-symmetric double-excitation states $P_{2}^{\text{ns}}$
(dashed-dotted black). \ $P_{1}^{\text{s, ns}}$ and $P_{2}^{\text{s, ns}}$ are
normalized respectively by the solutions of non-interacting atoms
$P_{1}^{\text{(0)}}$(solid-red) and $P_{2}^{\text{(0)}}$\ (solid-blue).
\ $P_{1}^{\text{(0)}}=1.58\times10^{-3}$ and $P_{2}^{\text{(0)}}%
=9.4\times10^{-7},$ \ are the single- and double-excitation probabilities for
independent atoms. }%
\label{excitation_density}%
\end{center}
\end{figure}

To investigate the probability of multiple atomic excitations in conditions of
weak off-resonant excitation, we choose a configuration of four atoms that sit
on the vertices of a square with side $d_{s}.$ \ The atomic density matrix
includes a laser excitation term in addition to the one and two-atom
dissipation terms; these arise from spontaneous emission and radiative
coupling due to dipole-dipole interaction \cite{Lehm}; see Eq. (\ref{dd}) in
Appendix A. \ We numerically solve for the time evolution of the density
matrix. \ The result of steady state single- and double-excitation populations
are shown in Figure \ref{excitation_density} as a function of $d_{s}.$ \ We
have assumed a continuous laser field with peak Rabi frequency $\Omega
_{a}=0.2\gamma$ and detuning $\Delta_{1}=5\gamma,$ where $\gamma$ is the
single-atom spontaneous decay rate for the excited state. \ The populations of
the symmetric states are $P_{1}^{\text{s}}\equiv$Tr$(\hat{\rho}|\phi
_{2}\rangle\langle\phi_{2}|)$ for a single excitation and $P_{2}^{\text{s}%
}\equiv$Tr$(\hat{\rho}|\phi_{6}\rangle\langle\phi_{6}|)$ for double
excitations where $\hat{\rho}$ is the density operator of the atomic system.
\ The total populations of the non-symmetric excitation states are
$P_{1}^{\text{ns}}\equiv$Tr$\left(  \hat{\rho}\left(  \sum_{x=2}^{5}|\phi
_{x}\rangle\langle\phi_{x}|\right)  \right)  ,$ for a single excitation, and
$P_{2}^{\text{ns}}\equiv$Tr$\left(  \hat{\rho}\left(  \sum_{x=7}^{11}|\phi
_{x}\rangle\langle\phi_{x}|\right)  \right)  ,$ for double excitations,
respectively. \ \ The probabilities of three and four excitations are
negligible under the weak excitation conditions we consider.

As $d_{s}$ approaches and exceeds $\lambda$ (the transition wavelength)$,$ the
populations tend to the independent atom limit when dipole-dipole coupling is
omitted. \ In this limit, the probability of exciting any non-symmetric states
goes to zero. \ The single and double excitation probabilities, $P_{1}%
^{\text{s}}$ and $P_{2}^{\text{s}}$, are normalized to their independent atom
values, $P_{1}^{\text{(0)}}=P_{e}P_{g}^{3}C_{1}^{4}$ and $P_{2}^{\text{(0)}%
}=P_{e}^{2}P_{g}^{2}C_{2}^{4}$ where $P_{e}=\Omega_{a}^{2}/(4\Delta_{1}%
^{2}+\gamma^{2})$ \cite{QL:Loudon}$,$ $P_{g}=1-P_{e},$ and $C_{i}^{n}$ is the
combination coefficient. \ For $d_{s}\ll\lambda$, the populations of the
non-symmetric states are comparable to the symmetric ones, indicating the
importance of dipole-dipole interactions. \ We see no evidence of a dipole
blockade effect in this limit for four atoms, but we have observed it in the
case of two atoms. \ Dipole blockade refers to the predominance of single
excitations as dipole shifts detune double and higher excitation states.

\begin{figure}
[ptb]
\begin{center}
\includegraphics[
natheight=20.499700in,
natwidth=25.000400in,
height=4.6224in,
width=5.9637in
]%
{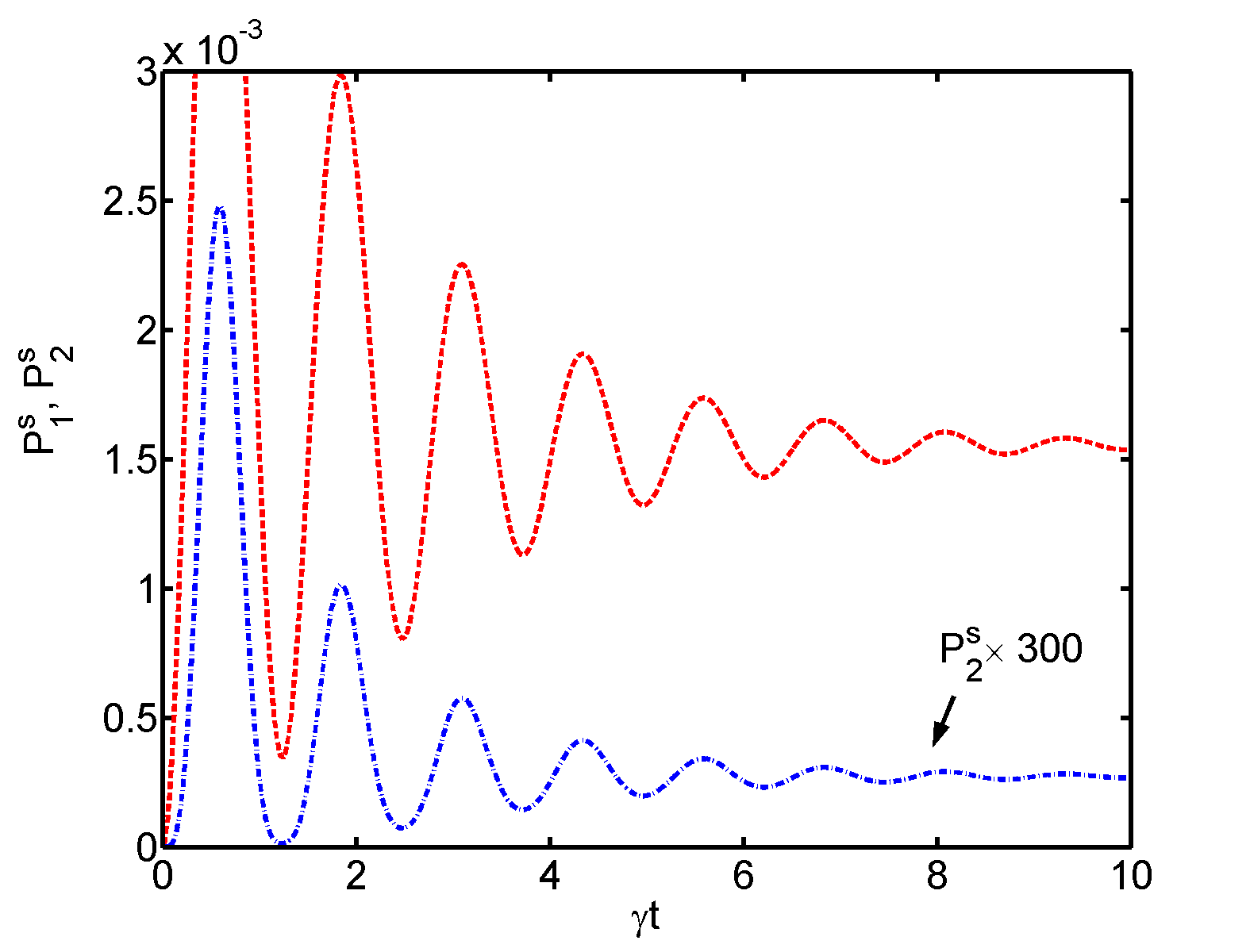}%
\caption{Time evolution of populations for symmetric states $P_{1}^{\text{s}}$
and $P_{2}^{\text{s}}$. \ The population of the symmetric state for a single
excitation is $P_{1}^{\text{s}}$ (dashed-red), and that for the symmetric
state for double excitations is $P_{2}^{\text{s}}$ (dashed--dotted blue).
\ The pump condition is the same as in Figure \ref{excitation_density} for
$d_{s}=3\lambda.$}%
\label{weak_field}%
\end{center}
\end{figure}

In Figure \ref{weak_field}, we show the time evolution of $P_{1}^{\text{s}%
}(t)$ and $P_{2}^{\text{s}}(t)$ for $d_{s}=3\lambda$ (this corresponds to an
atomic density $8\times10^{10}$ cm$^{-3}$). \ The period of the Rabi
oscillation is determined by $2\pi/\Delta_{1}$, and the asymptotic steady
state value for $P_{2}^{\text{s}}$ is about $1.6\times10^{-3}.$ \ This
coincides with the approximate result $|\sqrt{N}\Omega_{a}/(2\Delta_{1})|^{2}$
that is found when we truncate the basis to the ground state and the
orthogonal states of a single atomic excitation.

We also numerically solve a line of atoms ($N=2,3,$ or $4$) with an equal
separation from$\ d_{s}=1$ to $5\lambda,$ and the results of steady state
populations indicate the condition for truncation of the basis set at a single
atomic excitation is valid when $\left\vert \Delta_{1}\right\vert \gg\sqrt
{N}\Omega_{a}/2.$ \ If the condition of a single atomic excitation $\Delta
_{1}\gg\sqrt{N}\Omega_{a}/2$ is relaxed, we will also have dynamical couplings
between symmetric and non-symmetric states (at least for $d_{s}\lesssim
3\lambda$). \ It is the dipole-dipole interaction that couples the
non-symmetric and symmetric states in the presence of the pump laser.%

\section{Theory of Cascade Emission}

\begin{figure}
[ptb]
\begin{center}
\includegraphics[
natheight=7.499600in,
natwidth=9.999800in,
height=3.2837in,
width=4.3682in
]%
{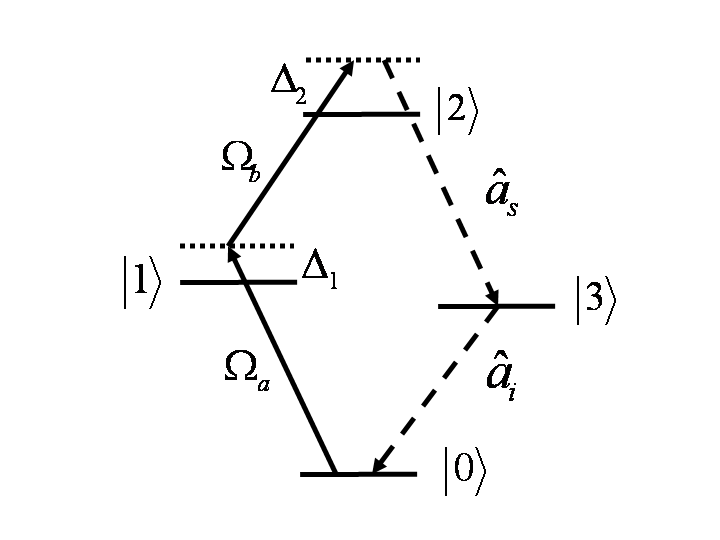}%
\caption{Four-level atomic ensemble interacting with two driving lasers
(solid) with Rabi frequencies $\Omega_{a}$ and $\Omega_{b}.$ \ Signal and
idler fields are labelled by $\hat{a}_{s}$ and $\hat{a}_{i},$ respectively and
$\Delta_{1}$ and $\Delta_{2}$ are one and two-photon laser detunings. }%
\label{four}%
\end{center}
\end{figure}

We consider $N$ cold atoms that are initially prepared in the ground state
interacting with four independent electromagnetic fields.\ \ As shown in
Figure \ref{four}, two driving lasers (of Rabi frequencies $\Omega_{a}$ and
$\Omega_{b}$) excite a ladder configuration $|0\rangle\rightarrow
|1\rangle\rightarrow|2\rangle.$ \ Two quantum fields, signal $\hat{a}_{s}$ and
idler $\hat{a}_{i},$ are generated spontaneously. \ The atoms adiabatically
follow the two excitation pulses and decay through the cascade emission of
signal and idler photons.\ \ Based on the discussion in the previous Section,
we permit only single atomic excitations under the condition of large
detuning, $\Delta_{1}\gg\sqrt{N}\Omega_{a}/2$. \ The Hamiltonian and the
coupled equations of the atomic dynamics are detailed in Appendix A. \ %

To correctly describe the frequency shifts arising from dipole-dipole
interactions, we do not make the rotating wave approximation on the electric
dipole interaction Hamiltonian. \ The frequency shift has contributions from
the single atom Lamb shift and a collective frequency shift. \ The Lamb shift
is assumed to be renormalized into the single atom transition frequency
distinguishing it from the collective shift due to the atom-atom interaction.

\subsection{Probability amplitudes for signal and signal-idler emissions}

Writing the state-vector $|\psi(t)\rangle$ in a basis restricted to single
atomic excitations, and single pairs of signal and idler photons, we can
introduce the probability amplitudes,%

\begin{equation}
C_{s,k_{i}}(t)=\sum_{\mu=1}^{N}e^{-i\vec{k}_{i}\cdot\vec{r}_{\mu}}%
\langle3_{\mu},1_{k_{s},\lambda_{s}}|\psi(t)\rangle
\end{equation}
and%

\begin{equation}
D_{s,i}(t)=\langle0,1_{k_{s},\lambda_{s}},1_{k_{i},\lambda_{i}}|\psi(t)\rangle
\end{equation}
defined in Appendix A. \ Note that $C_{s,k_{i}}(t)$ is an amplitude for a
phased excitation of the ensemble of atoms subsequent to signal photon emission.

After adiabatically eliminating the laser excited levels in the equations of
motion, we are able to simplify and derive the amplitude $C_{s,k_{i}}$\ and
the signal-idler (two-photon) state amplitude $D_{s,i}$ as shown in Appendix A,%

\begin{equation}
C_{s,k_{i}}(t)=g_{s}^{\ast}(\epsilon_{k_{s},\lambda_{s}}^{\ast}\cdot\hat
{d}_{s})\sum_{\mu}e^{i\Delta\vec{k}\cdot\vec{r}_{\mu}}\int_{0}^{t}dt^{\prime
}e^{i(\omega_{s}-\omega_{23}-\Delta_{2})t^{\prime}}e^{(-\frac{\Gamma_{3}^{N}%
}{2}+i\delta\omega_{i})(t-t^{\prime})}b(t^{\prime})
\end{equation}

\begin{align}
D_{s,i}(t)  & =g_{i}^{\ast}g_{s}^{\ast}(\epsilon_{k_{i},\lambda_{i}}^{\ast
}\cdot\hat{d}_{i})(\epsilon_{k_{s},\lambda_{s}}^{\ast}\cdot\hat{d}_{s}%
)\sum_{\mu}e^{i\Delta\vec{k}\cdot\vec{r}_{\mu}}\int_{0}^{t}\int_{0}%
^{t^{\prime}}dt^{\prime\prime}dt^{\prime}e^{(-\frac{\Gamma_{3}^{N}}{2}%
+i\delta\omega_{i})(t^{\prime}-t^{\prime\prime})}\nonumber\\
& e^{i(\omega_{i}-\omega_{3})t^{\prime}}e^{i(\omega_{s}-\omega_{23}-\Delta
_{2})t^{\prime\prime}}b(t^{\prime\prime}).
\end{align}

The factor $\sum_{\mu}e^{i\Delta\vec{k}\cdot\vec{r}_{\mu}}$ reflects
phase-matching of the interaction under conditions of four-wave mixing when
the wavevector mismatch $\Delta\vec{k}=\vec{k}_{a}+\vec{k}_{b}-\vec{k}%
_{s}-\vec{k}_{i}\rightarrow0.$ \ The radiative coupling between atoms results
in the appearance of the superradiant decay constant%

\begin{equation}
\Gamma_{3}^{N}=(N\bar{\mu}+1)\Gamma_{3}%
\end{equation}
where $\Gamma_{3}$ is the natural decay rate of the $|3\rangle\rightarrow
|0\rangle$ transition, and $\bar{\mu}$ is a geometrical constant depending on
the shape of the atomic ensemble. \ An expression for the collective frequency
shift $\delta\omega_{i}$\ is given in the Appendix A. \ As shown in Figure
\ref{muplot}, we numerically calculate the geometrical factor $\bar{\mu},$
Eq.(\ref{mu}), to demonstrate how the decay factor $N\bar{\mu}+1$ depends on
the height and radius of a cylindrical ensemble. \ The arrows in the figure
point out the contour lines (yellow and green) of $N\bar{\mu}+1\approx4$ and
$6$ which are comparable to the operating conditions of the experiment
\cite{telecom}.
\begin{figure}
[ptb]
\begin{center}
\includegraphics[
natheight=20.000400in,
natwidth=25.499700in,
height=5.1546in,
width=6.7795in
]%
{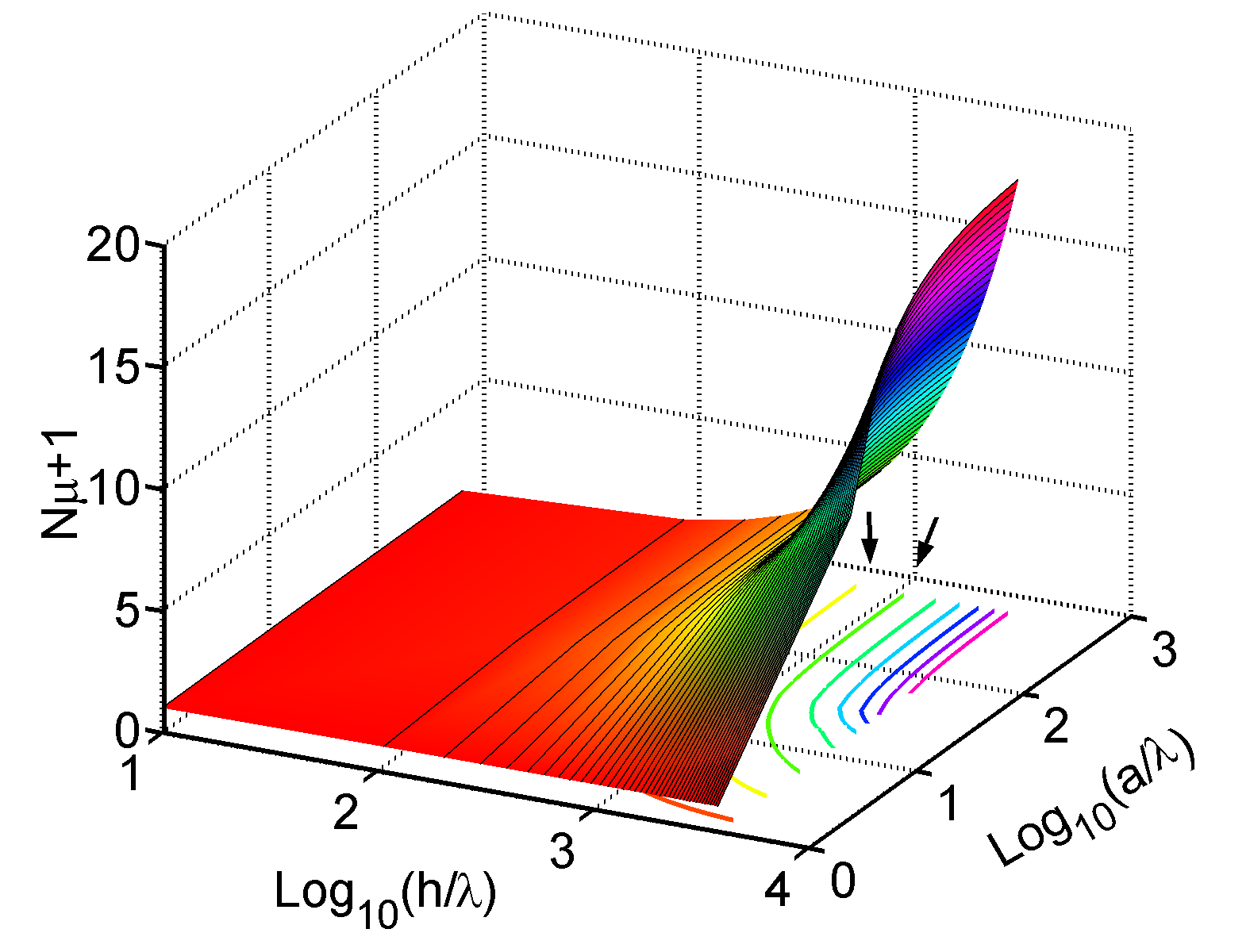}%
\caption{The superradiance decay factor $N\mu+1$ ($\mu=\bar{\mu}$) for a
cylindrical ensemble of length $h$ and radius $a$ in unit of transition
wavelength $\lambda$. \ The atomic density is $8\times10^{10}$ cm$^{-3}$ and
$\lambda=795$ nm corresponding to the D1 line of $^{85}$Rb. \ See the text for
the explanation of the arrows.}%
\label{muplot}%
\end{center}
\end{figure}

In the above expressions for the probability amplitudes, $b(t)=\frac
{\Omega_{a}(t)\Omega_{b}(t)}{4\Delta_{1}\Delta_{2}}$ is proportional to the
product of the Rabi frequencies. \ We use normalized Gaussian pulses as an
example where $\Omega_{a}(t)=\frac{1}{\sqrt{\pi}\tau}\tilde{\Omega}%
_{a}e^{-t^{2}/\tau^{2}}$,$~\Omega_{b}(t)=\frac{1}{\sqrt{\pi}\tau}\tilde
{\Omega}_{b}e^{-t^{2}/\tau^{2}}$, so that the two pulses are overlapped with
the same pulse width. $\ \tilde{\Omega}_{a,b}$ is the pulse area, and let
$\Delta\omega_{s}\equiv\omega_{s}-\omega_{23}-\Delta_{2}-\delta\omega
_{i},~\Delta\omega_{i}\equiv\omega_{i}-\omega_{3}+\delta\omega_{i}$. \ We have
the probability amplitude for signal photon emission and atoms in a phased state,%

\begin{align}
& C_{s,k_{i}}(t,\Delta\omega_{s})\nonumber\\
& =\frac{\tilde{\Omega}_{a}\tilde{\Omega}_{b}g_{s}^{\ast}(\epsilon
_{_{ks,\lambda_{s}}}^{\ast}\cdot\hat{d}_{s})}{4\Delta_{1}\Delta_{2}}\sum_{\mu
}e^{i\Delta\vec{k}\cdot\vec{r}_{\mu}}\frac{1}{\pi\tau^{2}}e^{(-\frac
{\Gamma_{3}^{N}}{2}+i\delta\omega_{i})t}\int_{-\infty}^{t}dt^{\prime}%
e^{\frac{\Gamma_{3}^{N}}{2}t^{\prime}}e^{i\Delta\omega_{s}t^{\prime}%
}e^{-2t^{\prime}{}^{2}/\tau^{2}}\nonumber\\
& =\frac{\tilde{\Omega}_{a}\tilde{\Omega}_{b}g_{s}^{\ast}(\epsilon
_{_{ks,\lambda_{s}}}^{\ast}\cdot\hat{d}_{s})}{4\Delta_{1}\Delta_{2}}\sum_{\mu
}e^{i\Delta\vec{k}\cdot\vec{r}_{\mu}}\frac{1}{\pi\tau^{2}}\frac{\tau}{2}%
\sqrt{\frac{\pi}{2}}e^{(-\frac{\Gamma_{3}^{N}}{2}+i\delta\omega_{i}%
)t}e^{(\frac{\Gamma_{3}^{N}}{2}+i\Delta\omega_{s})^{2}\tau^{2}/8}%
\times\nonumber\\
& \Big(1+\text{erf}(\frac{4t-(\frac{\Gamma_{3}^{N}}{2}+i\Delta\omega_{s}%
)\tau^{2}}{2\sqrt{2}\tau})\Big),
\end{align}
and the two-photon probability amplitude is
\begin{align}
& D_{si}(t,\Delta\omega_{s},\Delta\omega_{i})\nonumber\\
& =\frac{\tilde{\Omega}_{a}\tilde{\Omega}_{b}g_{i}^{\ast}g_{s}^{\ast}%
(\epsilon_{k_{i},\lambda_{i}}^{\ast}\cdot\hat{d}_{i})(\epsilon_{_{ks,\lambda
_{s}}}^{\ast}\cdot\hat{d}_{s})}{4\Delta_{1}\Delta_{2}}\sum_{\mu}e^{i\Delta
\vec{k}\cdot\vec{r}_{\mu}}\frac{1}{\pi\tau^{2}}\sqrt{\frac{\pi}{2}}\frac{\tau
e^{-\frac{\Gamma_{3}^{N}}{2}t}}{2(\frac{\Gamma_{3}^{N}}{2}-i\Delta\omega_{i}%
)}\Big\{\nonumber\\
& -e^{i\Delta\omega_{i}t+(\frac{\Gamma_{3}^{N}}{2}+i\Delta\omega_{s})^{2}%
\tau^{2}/8}\Big(1+\text{erf}(\frac{4t-(\frac{\Gamma_{3}^{N}}{2}+i\Delta
\omega_{s})\tau^{2}}{2\sqrt{2}\tau})\Big)\nonumber\\
& +e^{-(\Delta\omega_{s}+\Delta\omega_{i})^{2}\tau^{2}}e^{\frac{\Gamma_{3}%
^{N}}{2}t}\Big(1+\text{erf}(\frac{4t-i(\Delta\omega_{s}+\Delta\omega_{i}%
)\tau^{2}}{2\sqrt{2}\tau})\Big)\Big\},
\end{align}
where erf is the error function%

\begin{equation}
\text{erf}(x)=\frac{2}{\sqrt{\pi}}\int_{0}^{x}e^{-t^{2}}dt.
\end{equation}

Asymptotically $D_{si}$ approaches the value,%

\begin{equation}
D_{si}(\Delta\omega_{s},\Delta\omega_{i})=\frac{\tilde{\Omega}_{a}%
\tilde{\Omega}_{b}g_{i}^{\ast}g_{s}^{\ast}(\epsilon_{k_{i},\lambda_{i}}^{\ast
}\cdot\hat{d}_{i})(\epsilon_{_{s}}^{\ast}\cdot\hat{d}_{s})}{4\Delta_{1}%
\Delta_{2}}\frac{\sum_{\mu}e^{i\Delta\vec{k}\cdot\vec{r}_{\mu}}}{\sqrt{2\pi
}\tau}\frac{e^{-(\Delta\omega_{s}+\Delta\omega_{i})^{2}\tau^{2}/8}}%
{\frac{\Gamma_{3}^{N}}{2}-i\Delta\omega_{i}},\label{longtwo}%
\end{equation}
indicating a spectral width $\Gamma_{3}^{N}/2$ for idler photon in a
Lorentzian distribution modulating a Gaussian profile with a spectral width
$2\sqrt{2}/\tau$ for signal and idler. \ Energy conservation of signal and
idler photons with driving fields at their central frequencies corresponds to
$\omega_{s}+\omega_{i}=\omega_{a}+\omega_{b}$, which makes $\Delta\omega
_{s}+\Delta\omega_{i}=0$; the collective frequency shifts cancel.

\section{A Correlated Two-photon State}

Using the asymptotic form of the two-photon state given in Eq. (\ref{longtwo}%
), the second-order correlation function $G_{s,i}^{(2)}$ is calculated as
\cite{QO:Scully}%

\begin{align}
&  G_{s,i}^{(2)}=\langle\psi(t\rightarrow\infty)|\hat{E}_{s}^{-}(\vec{r}%
_{1},t_{1})\hat{E}_{i}^{-}(\vec{r}_{2},t_{2})\hat{E}_{i}^{+}(\vec{r}_{2}%
,t_{2})\hat{E}_{s}^{+}(\vec{r}_{1},t_{1})|\psi(t\rightarrow\infty
)\rangle=|\Phi_{s,i}|^{2}\\
&  \Phi_{s,i}=\langle0|\hat{E}_{i}^{+}(\vec{r}_{2},t_{2})\hat{E}_{s}^{+}%
(\vec{r}_{1},t_{1})|\psi(t\rightarrow\infty)\rangle\\
&  \hat{E}_{s}^{+}(\vec{r}_{1},t_{1})=\sum_{k_{s},\lambda}\sqrt{\frac
{\hbar\omega_{s}}{2\epsilon_{0}V}}\hat{a}_{k_{s},\lambda}\vec{\epsilon}%
_{k_{s},\lambda_{s}}e^{i\vec{k}_{s}\cdot\vec{r}_{1}-i\omega_{s}t_{1}}\\
&  \hat{E}_{i}^{+}(\vec{r}_{2},t_{2})=\sum_{k_{i},\lambda}\sqrt{\frac
{\hbar\omega_{i}}{2\epsilon_{0}V}}\hat{a}_{k_{i},\lambda}\vec{\epsilon}%
_{k_{i},\lambda_{i}}e^{i\vec{k}_{i}\cdot\vec{r}_{2}-i\omega_{i}t_{2}}%
\end{align}
where $|\psi(t\rightarrow\infty)\rangle$ denotes the state vector in the long
time limit that involves the ground state and two-photon state vectors. \ Free
electromagnetic fields, signal and idler photons, at space ($\vec{r}_{1}%
,\vec{r}_{2}$) and time ($t_{1},t_{2}$) are $\hat{E}_{s}^{+}$ and $\hat{E}%
_{i}^{+}$ where ($+$) denotes their positive frequency part. \ For second
order correlation function, only $D_{si},$ derived in the previous Section
contributes to it, then we have,%

\begin{align}
&  \Phi_{s,i}=\sum_{k_{s},\lambda_{s}}\sum_{k_{i},\lambda_{i}}\frac{\omega
_{s}}{2\epsilon_{0}V}\frac{\omega_{i}}{2\epsilon_{0}V}(\vec{d}_{s}\cdot
\vec{\epsilon}_{k_{s},\lambda_{s}})\vec{\epsilon}_{k_{s},\lambda_{s}}%
\times\nonumber\\
&  (\vec{d}_{i}\cdot\vec{\epsilon}_{k_{i},\lambda_{i}})\vec{\epsilon}%
_{k_{i},\lambda_{i}}\frac{\tilde{\Omega}_{a}\tilde{\Omega}_{b}}{4\Delta
_{1}\Delta_{2}}\frac{\sum_{\mu}e^{i\Delta\vec{k}\cdot\vec{r}_{\mu}}}%
{\sqrt{2\pi}\tau}\frac{e^{-(\Delta\omega_{s}+\Delta\omega_{i})^{2}\tau^{2}/8}%
}{\frac{\Gamma_{3}^{N}}{2}-i\Delta\omega_{i}}~e^{i\vec{k}_{s}\cdot\vec{r}%
_{1}-i\omega_{s}t_{1}}e^{i\vec{k}_{i}\cdot\vec{r}_{2}-i\omega_{i}t_{2}%
}\nonumber\\
&  =\frac{\tilde{\Omega}_{a}\tilde{\Omega}_{b}}{4\Delta_{1}\Delta_{2}}%
\frac{\sum_{\mu}e^{i\Delta\vec{k}\cdot\vec{r}_{\mu}}}{\sqrt{2\pi}\tau}%
\frac{|\vec{d}_{s}||\vec{d}_{i}|}{4\epsilon_{0}^{2}c^{6}(2\pi)^{6}}\int
d\Omega_{s}[\hat{d}_{s}-\hat{k}_{s}(\hat{k}_{s}\cdot\hat{d}_{s})]\nonumber\\
&  \int d\Omega_{i}[\hat{d}_{i}-\hat{k}_{i}(\hat{k}_{i}\cdot\hat{d}_{i})]\int
d\omega_{i}\omega_{i}^{3}e^{-i\omega_{i}(t_{2}-\frac{\vec{r}_{2}\cdot\hat
{k}_{i}}{c})}\frac{(\omega_{23}-\Delta\omega_{i})^{3}}{\frac{\Gamma_{3}^{N}%
}{2}-i\Delta\omega_{i}}e^{-i(\omega_{23}+\Delta_{2})(t_{1}-\frac{\vec{r}%
_{1}\cdot\hat{k}_{s}}{c})}\nonumber\\
&  e^{i(\Delta\omega_{i}-\delta\omega_{i})(t_{1}-\frac{\vec{r}_{1}\cdot\hat
{k}_{s}}{c})}\int d\Delta\omega_{s}e^{-i\Delta\omega_{s}(t_{1}-\frac{\vec
{r}_{1}\cdot\hat{k}_{s}}{c})}e^{-\Delta\omega_{s}^{2}\tau^{2}/8}%
\end{align}
where we have used the change of variables in the first step, replaced
$\omega_{s}=\omega_{23}+\Delta_{2}+\Delta\omega_{s}+\delta\omega_{i}$, and
changed the variable $\Delta\omega_{s}\rightarrow\Delta\omega_{s}-\Delta
\omega_{i}$. \ Solid angle integration is denoted as $d\Omega_{s,i}$ for
signal (idler) photon. \ The divergent part of $\omega_{s}^{3}$ (which varies
relatively slowly) has been moved out from the integral of $d\Delta\omega_{s}%
$, and we replace $\omega_{s}$ with the signal transition frequency
$\omega_{23}.$ \ We then have
\begin{align}
&  \Phi_{s,i}\nonumber\\
&  =\frac{\tilde{\Omega}_{a}\tilde{\Omega}_{b}}{4\Delta_{1}\Delta_{2}}%
\frac{\sum_{\mu}e^{i\Delta\vec{k}\cdot\vec{r}_{\mu}}}{\sqrt{2\pi}\tau}%
\frac{|\vec{d}_{s}||\vec{d}_{i}|\omega_{3}^{3}\omega_{23}^{3}}{4\epsilon
_{0}^{2}c^{6}(2\pi)^{6}}\int d\Omega_{s}d\Omega_{i}[\hat{d}_{s}-\hat{k}%
_{s}(\hat{k}_{s}\cdot\hat{d}_{s})][\hat{d}_{i}-\hat{k}_{i}(\hat{k}_{i}%
\cdot\hat{d}_{i})]\nonumber\\
&  \frac{2\sqrt{2\pi}}{\tau}\int d\Delta\omega_{i}\frac{e^{-i\Delta\omega
_{i}(t_{2}-\frac{\vec{r}_{2}\cdot\hat{k}_{i}}{c}-t_{1}+\frac{\vec{r}_{1}%
\cdot\hat{k}_{s}}{c})}}{(\frac{\Gamma_{3}^{N}}{2}-i\Delta\omega_{i}%
-i\delta\omega_{i})}e^{-i(\omega_{23}+\Delta_{2})(t_{1}-\frac{\vec{r}_{1}%
\cdot\hat{k}_{s}}{c})}e^{-i\omega_{3}(t_{2}-\frac{\vec{r}_{2}\cdot\hat{k}_{i}%
}{c})}e^{-2(t_{1}-\frac{\vec{r}_{1}\cdot\hat{k}_{s}}{c})^{2}/\tau^{2}%
}\nonumber\\
&
\end{align}
where we replace $\omega_{i}=\omega_{3}+\Delta\omega_{i}-\delta\omega_{i}$ and
change the variable $\Delta\omega_{i}\rightarrow\Delta\omega_{i}+\delta
\omega_{i}$. \ The divergent part of $\omega_{i}^{3}$ is again moved out from
the integral of $d\Delta\omega_{i}$ and replace $\omega_{i} $ with the signal
transition frequency $\omega_{3}.$ \ Finally we have%

\begin{align}
&  \Phi_{s,i}\nonumber\\
&  =\frac{\tilde{\Omega}_{a}\tilde{\Omega}_{b}}{4\Delta_{1}\Delta_{2}}%
\frac{|\vec{d}_{s}||\vec{d}_{i}|\omega_{3}^{3}\omega_{23}^{3}}{2\epsilon
_{0}^{2}c^{6}\tau^{2}(2\pi)^{6}}\sum_{\mu}e^{i\Delta\vec{k}\cdot\vec{r}_{\mu}%
}\int d\Omega_{s}d\Omega_{i}[\hat{d}_{s}-\hat{k}_{s}(\hat{k}_{s}\cdot\hat
{d}_{s})][\hat{d}_{i}-\hat{k}_{i}(\hat{k}_{i}\cdot\hat{d}_{i})]\nonumber\\
&  e^{-2(t_{1}-\frac{\vec{r}_{1}\cdot\hat{k}_{s}}{c})^{2}/\tau^{2}%
}e^{-i(\omega_{23}+\Delta_{2})(t_{1}-\frac{\vec{r}_{1}\cdot\hat{k}_{s}}{c}%
)}e^{-i\omega_{3}(t_{2}-\frac{\vec{r}_{2}\cdot\hat{k}_{i}}{c})}e^{(-\frac
{\Gamma_{3}^{N}}{2}+i\delta\omega_{i})(t_{2}-\frac{\vec{r}_{2}\cdot\hat{k}%
_{i}}{c}-t_{1}+\frac{\vec{r}_{1}\cdot\hat{k}_{s}}{c})}\nonumber\\
&  \Theta(t_{2}-\frac{\vec{r}_{2}\cdot\hat{k}_{i}}{c}-t_{1}+\frac{\vec{r}%
_{1}\cdot\hat{k}_{s}}{c})
\end{align}
where the complex integral with the pole at $\Delta\omega_{i}=-i\frac
{\Gamma_{3}^{N}}{2}-\delta\omega_{i}$ in the lower half plane leads to a step
function $\Theta$ that shows the causal connection between signal and idler
emission. \ The emission time for the signal field ($t_{1}-\frac{\vec{r}%
_{1}\cdot\hat{k}_{s}}{c}$) is within the pulse envelope of width $\tau$, and
the idler photon decays with a superradiant constant $\Gamma_{3}^{N}/2$.
\ Note that the collective frequency shift $\delta\omega_{i}$ appears in the
signal ($\omega_{23}+\Delta_{2}+\delta\omega_{i}$) and idler ($\omega
_{3}-\delta\omega_{i}$) frequency consistent with energy conservation. \ Let
$\Delta t_{s}\equiv t_{1}-\frac{\vec{r}_{1}\cdot\hat{k}_{s}}{c}$ and $\Delta
t_{i}\equiv t_{2}-\frac{\vec{r}_{2}\cdot\hat{k}_{i}}{c}$, we then \ have%

\begin{align}
| &  \Phi_{s,i}(\Delta t_{s},\Delta t_{i})|\nonumber\\
&  =\frac{\tilde{\Omega}_{a}\tilde{\Omega}_{b}}{4\Delta_{1}\Delta_{2}}%
\frac{|\vec{d}_{s}||\vec{d}_{i}|\omega_{3}^{3}\omega_{23}^{3}}{2\epsilon
_{0}^{2}c^{6}\tau^{2}(2\pi)^{6}}\sum_{\mu}e^{i\Delta\vec{k}\cdot\vec{r}_{\mu}%
}\int d\Omega_{s}d\Omega_{i}[\hat{d}_{s}-\hat{k}_{s}(\hat{k}_{s}\cdot\hat
{d}_{s})][\hat{d}_{i}-\hat{k}_{i}(\hat{k}_{i}\cdot\hat{d}_{i})]\nonumber\\
&  e^{-2(\Delta t_{s})^{2}/\tau^{2}}e^{-\frac{\Gamma_{3}^{N}}{2}(\Delta
t_{i}-\Delta t_{s})}\Theta(\Delta t_{i}-\Delta t_{s}).\label{g2}%
\end{align}

If we let $\Delta t\equiv\Delta t_{i}-\Delta t_{s}$ and choose $\Delta
t_{s}=0$ as the origin in time (idler gating time), then we have the
second-order correlation function%

\begin{equation}
G_{s,i}^{(2)}(\Delta t)=|\Phi_{s,i}(\Delta t)|^{2}\propto e^{-\Gamma_{3}%
^{N}\Delta t}\text{ where }\Delta t\geq0.
\end{equation}
\ It resembles the result for the second-order correlation function in the
case of single atom, whereas here we have an enhanced decay rate due to the
atomic dipole-dipole interaction.

\begin{figure}
[ptb]
\begin{center}
\includegraphics[
natheight=20.000400in,
natwidth=25.499700in,
height=6.2399in,
width=6.8245in
]%
{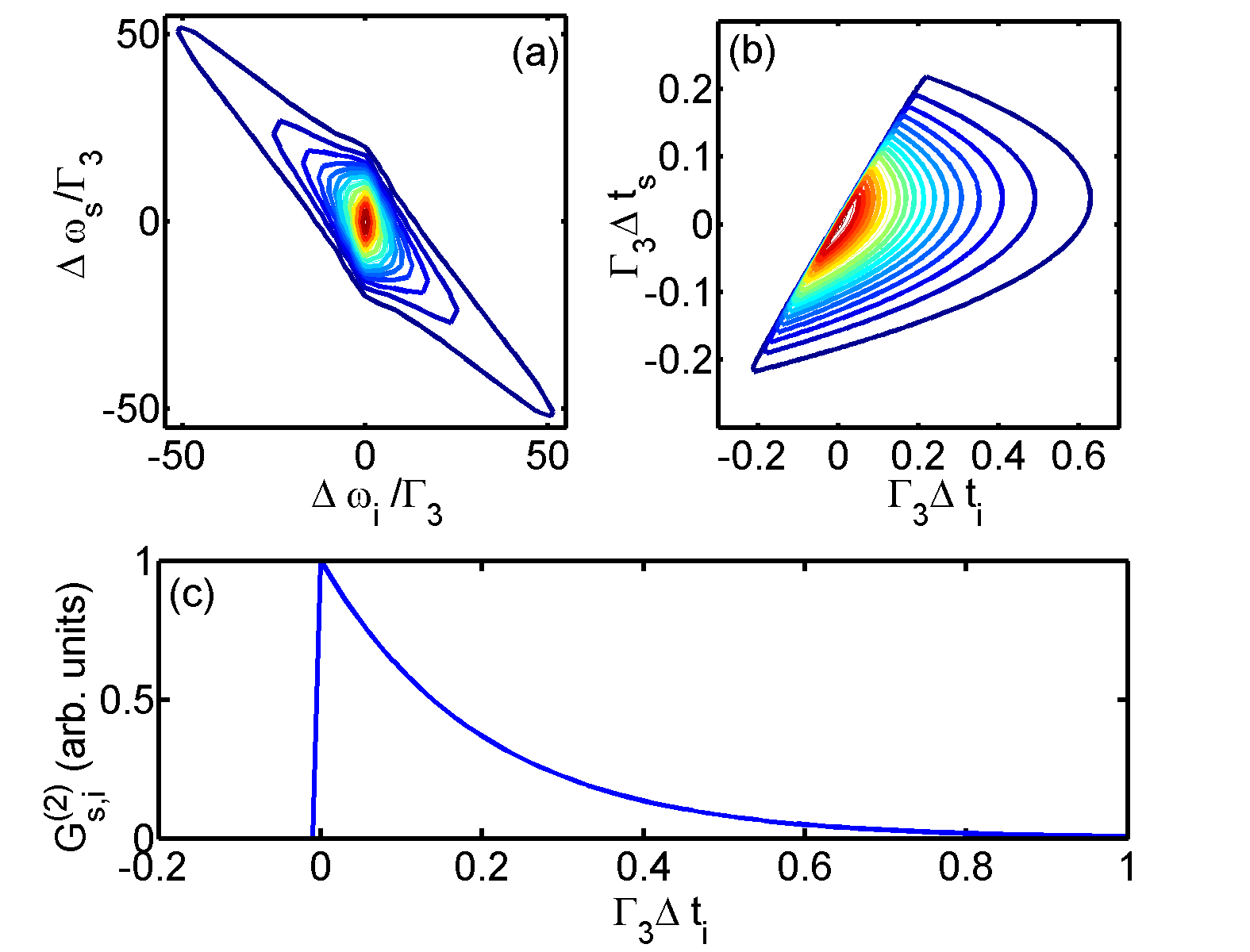}%
\caption{(a) Absolute value of the spectrum for two-photon state probability
amplitude $D_{s,i}$ and (b) the second-order correlation function
$G_{s,i}^{(2)}(\Delta t_{s},\Delta t_{i}).$ \ (c) A normalized $G_{s,i}%
^{(2)}(\Delta t_{s}=0,\Delta t_{i})$ with $\Gamma_{3}\tau=0.2$. \ The
exponential decay corresponds to the superradiant decay factor $N\bar{\mu
}+1=5.$}%
\label{spec_fft}%
\end{center}
\end{figure}

In Figure \ref{spec_fft}, we plot out the absolute value of spectrum
$D_{si}(\Delta\omega_{s},\Delta\omega_{i})$ and the second-order correlation
function $G_{s,i}^{(2)}(\Delta t_{s},\Delta t_{i})$. \ In (c), we show for
$\Gamma_{3}\Delta t_{i}=0.2.$ The width of $1/\Delta t_{i}=5\Gamma_{3}%
$\ corresponds to $\Gamma_{3}^{N}=(N\bar{\mu}+1)\Gamma_{3}=5\Gamma_{3}$.%

\section{Schmidt Decomposition}

Correlated photon pairs may be generated by parametric down conversion (PDC)
\cite{clock,pulsepump,branning}. \ The degree of entanglement can be
quantified by Schmidt mode decomposition \cite{law,parker}, allowing the
influence of group-velocity matching \cite{eliminate} to be assessed. \ A pure
single photon source is a basis element for quantum computation by linear
optics (LOQC) \cite{LOQC}, and it can be conditionally generated by
measurement \cite{single}. \ A similar approach can be applied to the study of
the transverse degrees of freedom in type-II PDC \cite{transverse} and PDC in
a distributed microcavity \cite{micro}. \ In photonic-crystal fiber (PCF), a
factorizable photon pair can be generated by spectral engineering
\cite{fiber}. The spectral effect has been discussed in relation to a quantum
teleportation protocol \cite{spectral} as a first step toward quantum communication.

We would like to perform an analysis of entanglement properties of our cascade
emission source. \ In addition to polarization entanglement, a
characterization of frequency space entanglement is required to clarify its
suitability in, for example, the DLCZ protocol \cite{dlcz}.

In the long time limit, the state function is given by, Eq. (\ref{longtwo}),%

\begin{equation}
|\psi\rangle=|0,\text{vac}\rangle+\sum_{s,i}D_{s,i}|0,1_{\vec{k}_{s}%
,\lambda_{s}},1_{\vec{k}_{i},\lambda_{i}}\rangle
\end{equation}
where $s=(k_{s},\lambda_{s})$, $i=(k_{i},\lambda_{i}),$ and $|0,$vac$\rangle$
is the joint atomic ground and photon vacuum state.

The spatial correlation of two-photon state in FWM condition can be eliminated
by pinholes or by coupling to single mode fiber so we consider only the
continuous frequency space. \ For some specific polarizations $\lambda_{s}$
and $\lambda_{i}$, we have the state vector $|\Psi\rangle$,%

\begin{equation}
|\Psi\rangle=\int f(\omega_{s},\omega_{i})\hat{a}_{\lambda_{s}}^{\dag}%
(\omega_{s})\hat{a}_{\lambda_{i}}^{\dag}(\omega_{i})|0\rangle d\omega
_{s}d\omega_{i},
\end{equation}
where
\begin{equation}
f(\omega_{s},\omega_{i})=\frac{e^{-(\Delta\omega_{s}+\Delta\omega_{i})^{2}%
\tau^{2}/8}}{\frac{\Gamma_{3}^{N}}{2}-i\Delta\omega_{i}}.\label{longtwo2}%
\end{equation}

The quantification of entanglement can be determined in the Schmidt basis
where the state vector is expressed as%

\begin{align}
& |\Psi\rangle=\sum_{n}\sqrt{\lambda_{n}}\hat{b}_{n}^{\dag}\hat{c}_{n}^{\dag
}|0\rangle,\\
& \hat{b}_{n}^{\dag}\equiv\int\psi_{n}(\omega_{s})\hat{a}_{\lambda_{s}}^{\dag
}(\omega_{s})d\omega_{s},\\
& \hat{c}_{n}^{\dag}\equiv\int\phi_{n}(\omega_{i})\hat{a}_{\lambda_{i}}^{\dag
}(\omega_{i})d\omega_{i},
\end{align}
where $\hat{b}_{n}^{\dag},$ $\hat{c}_{n}^{\dag}$ are effective creation
operators. \ Eigenvalues $\lambda_{n}$, and eigenfunctions $\psi_{n}$ and
$\phi_{n},$ are the solutions of the eigenvalue equations,%

\begin{align}
\int K_{1}(\omega,\omega^{\prime})\psi_{n}(\omega^{\prime})d\omega^{\prime}  &
=\lambda_{n}\psi_{n}(\omega),\\
\int K_{2}(\omega,\omega^{\prime})\phi_{n}(\omega^{\prime})d\omega^{\prime}  &
=\lambda_{n}\phi_{n}(\omega),
\end{align}
where $K_{1}(\omega,\omega^{\prime})\equiv\int f(\omega,\omega_{1})f^{\ast
}(\omega^{\prime},\omega_{1})d\omega_{1}$ and $K_{2}(\omega,\omega^{\prime
})\equiv\int f(\omega_{2},\omega)f^{\ast}(\omega_{2},\omega^{\prime}%
)d\omega_{2}$ are the kernels for the one-photon spectral correlations
\cite{law,parker}. \ Orthogonality of eigenfunctions is $\int\psi_{i}%
(\omega)\psi_{j}(\omega)d\omega=\delta_{ij}$, $\int\phi_{i}(\omega)\phi
_{j}(\omega)d\omega=\delta_{ij},$ and the normalization of quantum state
requires $\sum_{n}\lambda_{n}=1$.

In the Schmidt basis, the von Neumann entropy may be written%

\begin{equation}
S=-\sum_{n=1}^{\infty}\lambda_{n}\text{ln}\lambda_{n}.
\end{equation}

If there is only one non-zero Schmidt number $\lambda_{1}=1$, the entropy is
zero, which means no entanglement and a factorizable state. \ For more than
one non-zero Schmidt number, the entropy is larger than zero and bipartite
entanglement is present.

The kernel in Eq. (\ref{longtwo2}) has all the frequency entanglement
information, entanglement means $f(\omega_{s},\omega_{i})$ cannot be
factorized in the form $g(\omega_{s})h(\omega_{i}),$ a multiplication of two
separate spectral functions. \ By inspection the Gaussian profile of signal
and idler emission is a source of correlation. \ The joint spectrum
$\Delta\omega_{s}+\Delta\omega_{i}$ is confined within the width of order of
$1/\tau$. \ The Lorentzian factor associated with the idler emission has a
width governed by the superradiant decay rate.%

\begin{figure}
[ptb]
\begin{center}
\includegraphics[
natheight=20.000400in,
natwidth=25.499700in,
height=5.425in,
width=6.4601in
]%
{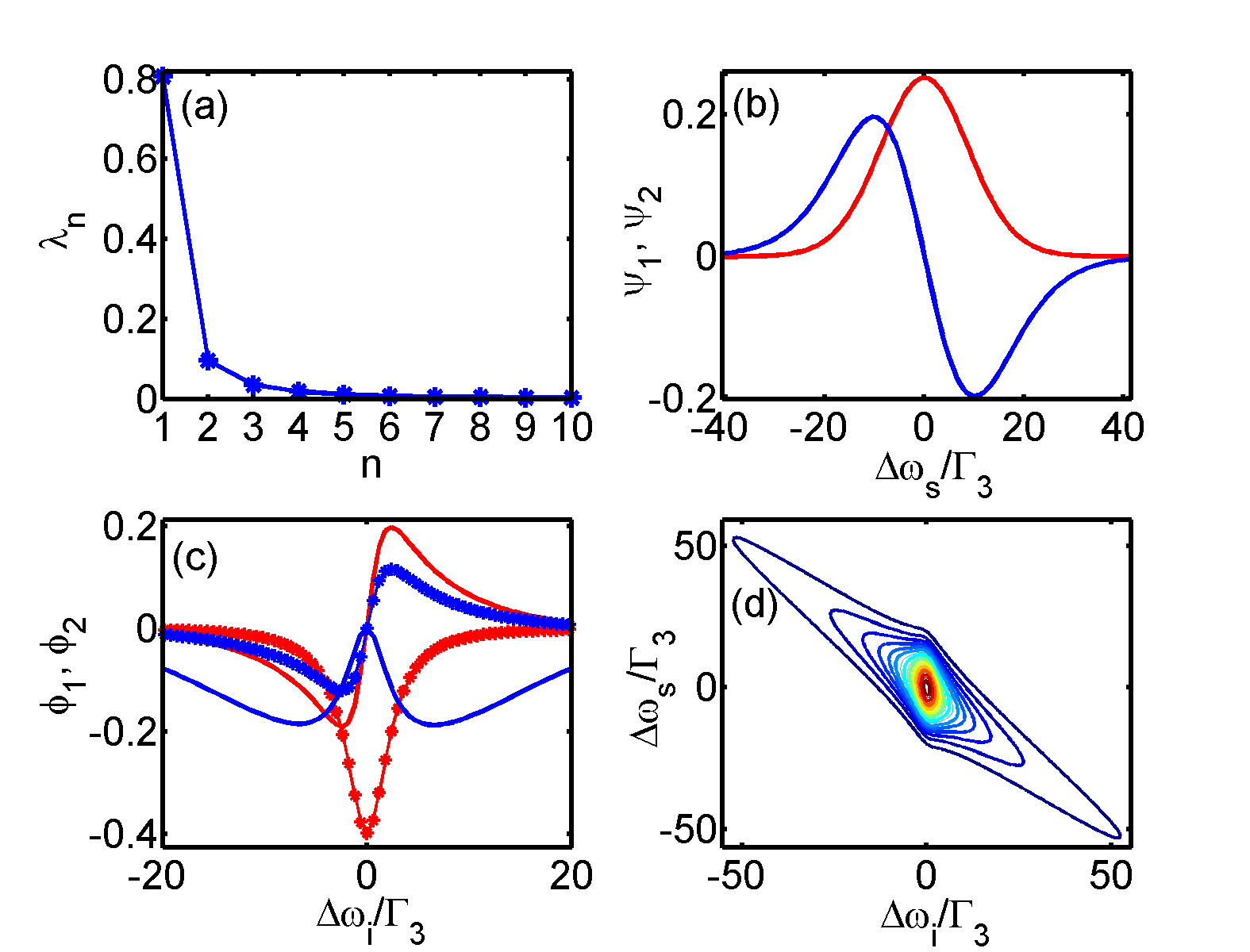}%
\caption{Schmidt mode analysis with pulse width $\tau=0.25$ and superradiance
decay factor $N\bar{\mu}+1=5.$ \ (a) Schmidt number and (b) signal mode
functions: Re[$\psi_{1}$] (solid-red) and Re[$\psi_{2}$] (solid-blue).
Imaginary parts are not shown, then are zero. (c) Real (solid) and imaginery
(dotted) parts of first (red) and second (blue) idler mode functions,
$\phi_{1}$ and $\phi_{2}$. \ (d) The absolute spectrum $|f(\Delta\omega
_{s},\Delta\omega_{i})|$.}%
\label{spec_mode}%
\end{center}
\end{figure}

In Figure \ref{spec_mode}, we show the Schmidt decomposition of the spectrum.
\ We use a moderate superradiant decay constant $N\bar{\mu}+1=5,$ comparable
to the reference \cite{telecom}, and a nanosecond pulse duration
$\tau=0.25~(\approx26/4\text{ ns})$, and $\Gamma_{3}/2\pi=6$ MHz. \ Due to
slow convergence associated with the Lorentzian profile, we use a frequency
range up to $\pm1200$ (in unit of $\Gamma_{3}$) with $2000\times2000$ grid.
\ The numerical error in the eigenvalue calculation is estimated to be about
$1\%$ error. \ In this case, the largest Schmidt number is $0.8$ and
corresponding signal mode function has a FWHM Gaussian profile $4\sqrt{2\text{
ln}(2)}/\tau\approx19\Gamma_{3}$. \ The idler mode function $\phi_{1}%
$\ reflects the Lorentzian profile in the spectrum at the signal peak
frequency ($\Delta\omega_{s}=0$),%

\begin{equation}
f(\Delta\omega_{s}=0,\Delta\omega_{i})=\frac{e^{-\Delta\omega_{i}^{2}\tau
^{2}/8}}{(N\mu+1)\Gamma_{3}/2-i\Delta\omega_{i}}%
\end{equation}
where a relatively broad Gaussian distribution is overlapped with a narrow
spread of superradiant decay rate [FWHM $>(N\bar{\mu}+1)\Gamma_{3}/2$].

\begin{figure}
[ptb]
\begin{center}
\includegraphics[
natheight=20.000400in,
natwidth=25.499700in,
height=6.3745in,
width=6.3036in
]%
{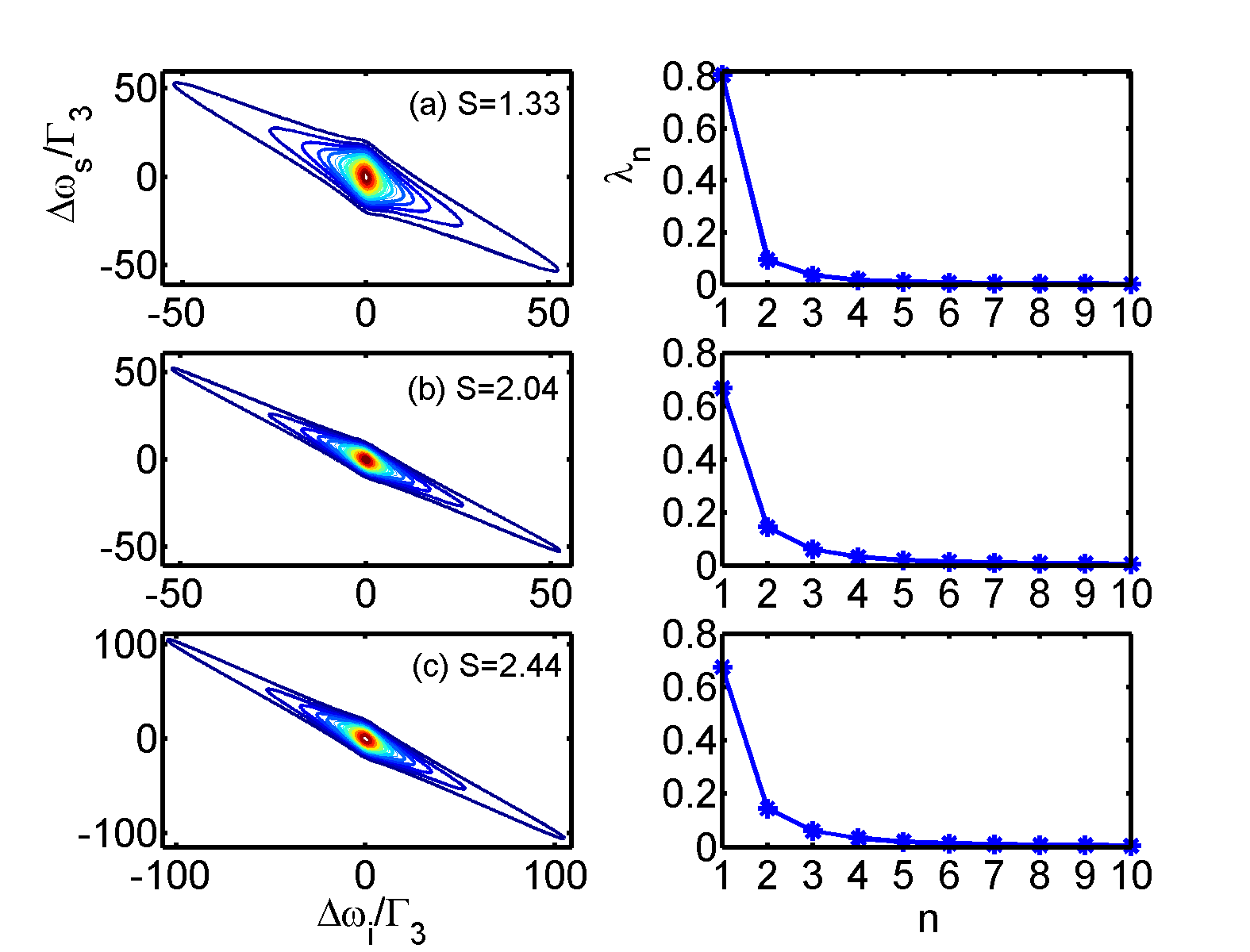}%
\caption{Absolute spectrum of two-photon state and the eigenvalues of Schmidt
decomposition. $N\bar{\mu}+1=5$ for both (a) $\tau=0.25$ (b) $\tau=0.5$.
$N\bar{\mu}+1=10$ for (c) $\tau=0.25$. \ The von Neumann entropy (S) is
indicated in the plots.}%
\label{spec_schmidt}%
\end{center}
\end{figure}

Figure \ref{spec_schmidt} shows that the cascade emission source is more
entangled if the superradiant decay constant, or the pulse duration increases.
\ We note that the Gaussian profile aligns the spectrum along the axis
$\Delta\omega_{s}=-\Delta\omega_{i}$ and the spectral width for signal photon
at the center of the idler frequency distribution ($\Delta\omega_{i}=0$) is
determined by pulse duration $\tau$. \ For a shorter pulse $\tau^{-1}%
>(N\bar{\mu}+1)\Gamma_{3}/2$, the joint Gaussian profile has a larger width,
and the spectrum is cut off by the Lorentzian idler distribution. \ A larger
width leads to a less entangled source and distributes the spectral weight
mainly along the crossed axes $\Delta\omega_{s}=0$ and $\Delta\omega_{i}=0$.
\ A narrow Lorentzian profile cuts off the entanglement source term
$e^{-(\Delta\omega_{s}+\Delta\omega_{i})^{2}\tau^{2}/8}$ tilting the spectrum
along the line $\Delta\omega_{s}+\Delta\omega_{i}=0.$ In the opposite limit,
$\tau^{-1}<(N\bar{\mu}+1)\Gamma_{3}/2$, the spectrum is highly entangled
corresponding to tight alignment along the axis $\Delta\omega_{s}%
=-\Delta\omega_{i}$ (Figure \ref{spec_schmidt} (c)).%

Note that the short pulse duration ($\tau\geq0.25$ ($6.5$ ns)) should not
violate the assumption of adiabaticity $\tau\gtrsim1/\Delta_{1}\text{ or
}1/\Delta_{2}$.

The Schmidt analysis and calculation of von Neumann entropy shows that
signal-idler fields are more entangled if the ensemble is more optically
dense, corresponding to stronger superradiance. \ For the DLCZ protocol, we
wish to avoid frequency entanglement. The superradiance may be reduced with
smaller atomic densities but good qubit storage and retrieval efficiency
require a moderate optical thickness \cite{telecom}. \ A better approach
involves using short pulse excitation $\tau^{-1}>(N\bar{\mu}+1)\Gamma_{3}$. We
will investigate the spectral properties in more details for the DLCZ scheme
in Chapter 5.



\chapter{Superradiant emission from a cascade atomic ensemble: Numerical
Approach}

In this Chapter, we investigate the cascade emission (signal and idler) from
an atomic ensemble using a numerical approach. \ In Chapter 3, we studied the
correlated emission using Schr\"{o}dinger's equation assuming single atomic
excitations. \ To relax the assumption of single atomic excitations, we derive
a set of c-number stochastic differential equations derived using the quantum
statistical methods reviewed in Chapter 2. \ We solve numerically for the
dynamics of the atoms and counter-propagating signal and idler fields. \ The
signal and idler field intensities are calculated, and the signal-idler
correlation function is studied for different optical depths of the atomic
ensemble, and compared with the analytical results of Chapter 3.

\section{Introduction}

To account for multiple atomic excitations in the signal-idler emission from a
cascade atomic ensemble, the Schr\"{o}dinger's equation approach becomes
cumbersome. \ An alternative theory based on c-number Langevin equations as
discussed in Chapter 2, is suitable for solution by stochastic simulations.
\ An essential element in the stochastic simulations is a proper
characterization of the Langevin noises. \ These represent the quantum
fluctuations responsible for the initiation of the spontaneous emission from
the inverted \cite{Feld,Haake1,Haake2,Polder79}, or pumped atomic system
\cite{Chiao88,Chiao95} as in our case.\ \ 

The positive-P phase space method \cite{QN:Gardiner,
quantization,Smith88,Smith0,Smith1,Boyd89,Drummond91} is employed to derive
the Fokker-Planck equations that lead directly to the c-number Langevin
equations. \ The classical noise correlation functions, equivalently diffusion
coefficients, are alternatively\ confirmed by use of the Einstein relations
reviewed in Chapter 2. \ The c-number Langevin equations correspond to
Ito-type stochastic differential equations that may be simulated numerically.
\ The noise correlations can be represented either by using a square
\cite{Carmichael86} or a non-square "square root" diffusion matrix
\cite{Smith1}. \ The approach enables us to calculate normally-ordered
quantities, signal-idler field intensities, and the second-order correlation
function. \ The numerical approach involves a semi-implicit difference
algorithm and shooting method \cite{numerical} to integrate the stochastic
"Maxwell-Bloch" equations.

Recently a new positive-P phase space method involving a stochastic gauge
function \cite{Drummond02} has been developed. \ This approach has an improved
treatment of sampling errors and boundary errors in the treatment of quantum
anharmonic oscillators \cite{Drummond01,Collett01}. \ It has also been applied
to a many-body system of bosons \cite{Drummond03} and fermions
\cite{Drummond06}. \ In this Chapter, we follow the traditional positive-P
representation method \cite{drummond80}.\ 

\section{Theory of Cascade emission}

The complete derivation of the c-number Langevin equations for cascade
emission from the four-level atomic ensemble is described in detail in
Appendix B. \ After setting up the Hamiltonian, we follow the standard
procedure to construct the characteristic functions \cite{LT:Haken} in
Appendix B.2 using the positive-P representation \cite{QN:Gardiner}. \ In
Appendix B.3.1, the Fokker-Planck equation is found by directly Fourier
transforming the characteristic functions, and making a $1/N_{z}$ expansion.\ 

\ Finally the Ito stochastic differential equations are written down from
inspection of the first-order derivative (drift term) and second-order
derivative (diffusion term) in the Fokker-Planck equation. \ The equations are
then written in dimensionless form by introducing the Arecchi-Courtens
cooperation units \cite{scale} in Appendix B.3.2. \ From Eq. (\ref{bloch2})
and the field equations that follow, these c-number Langevin equations in a
co-moving frame are,%

\begin{align}
\frac{\partial}{\partial\tau}\pi_{01}  & =(i\Delta_{1}-\frac{\gamma_{01}}%
{2})\pi_{01}+i\Omega_{a}(\pi_{00}-\pi_{11})+i\Omega_{b}^{\ast}\pi_{02}%
-i\pi_{13}^{\dag}E_{i}^{+}+\mathcal{F}_{01}\text{ (I),}\nonumber\\
\frac{\partial}{\partial\tau}\pi_{12}  & =i(\Delta_{2}-\Delta_{1}%
+i\frac{\gamma_{01}+\gamma_{2}}{2})\pi_{12}-i\Omega_{a}^{\ast}\pi_{02}%
+i\Omega_{b}(\pi_{11}-\pi_{22})+i\pi_{13}E_{s}^{+}e^{-i\Delta kz}\nonumber\\
& +\mathcal{F}_{12},\nonumber\\
\frac{\partial}{\partial\tau}\pi_{02}  & =(i\Delta_{2}-\frac{\gamma_{2}}%
{2})\pi_{02}-i\Omega_{a}\pi_{12}+i\Omega_{b}\pi_{01}+i\pi_{03}E_{s}%
^{+}e^{-i\Delta kz}-i\pi_{32}E_{i}^{+}+\mathcal{F}_{02},\nonumber\\
\frac{\partial}{\partial\tau}\pi_{11}  & =-\gamma_{01}\pi_{11}+\gamma_{12}%
\pi_{22}+i\Omega_{a}\pi_{01}^{\dag}-i\Omega_{a}^{\ast}\pi_{01}-i\Omega_{b}%
\pi_{12}^{\dag}+i\Omega_{b}^{\ast}\pi_{12}+\mathcal{F}_{11},\nonumber\\
\frac{\partial}{\partial\tau}\pi_{22}  & =-\gamma_{2}\pi_{22}+i\Omega_{b}%
\pi_{12}^{\dag}-i\Omega_{b}^{\ast}\pi_{12}+i\pi_{32}^{\dag}E_{s}%
^{+}e^{-i\Delta kz}-i\pi_{32}E_{s}^{-}e^{i\Delta kz}+\mathcal{F}%
_{22},\nonumber\\
\frac{\partial}{\partial\tau}\pi_{33}  & =-\gamma_{03}\pi_{33}+\gamma_{32}%
\pi_{22}-i\pi_{32}^{\dag}E_{s}^{+}e^{-i\Delta kz}+i\pi_{32}E_{s}^{-}e^{i\Delta
kz}+i\pi_{03}^{\dag}E_{i}^{+}-i\pi_{03}E_{i}^{-}\nonumber\\
& +\mathcal{F}_{33},\nonumber\\
\frac{\partial}{\partial\tau}\pi_{13}  & =-(i\Delta_{1}+\frac{\gamma
_{01}+\gamma_{03}}{2})\pi_{13}-i\Omega_{a}^{\ast}\pi_{03}-i\Omega_{b}\pi
_{32}^{\dag}+i\pi_{12}E_{s}^{-}e^{i\Delta kz}+i\pi_{01}^{\dag}E_{i}%
^{+}\nonumber\\
& +\mathcal{F}_{13},\nonumber\\
\frac{\partial}{\partial\tau}\pi_{03}  & =-\frac{\gamma_{03}}{2}\pi
_{03}-i\Omega_{a}\pi_{13}+i\pi_{02}E_{s}^{-}e^{i\Delta kz}+i(\pi_{00}-\pi
_{33})E_{i}^{+}+\mathcal{F}_{03},\nonumber\\
\frac{\partial}{\partial\tau}\pi_{32}  & =i\Delta_{2}-\frac{\gamma_{03}%
+\gamma_{2}}{2}\pi_{32}+i\Omega_{b}\pi_{13}^{\dag}-i(\pi_{22}-\pi_{33}%
)E_{s}^{+}e^{-i\Delta kz}-i\pi_{02}E_{i}^{-}+\mathcal{F}_{32},\nonumber\\
\frac{\partial}{\partial z}E_{s}^{+}  & =-i\pi_{32}e^{i\Delta kz}\frac
{|g_{s}|^{2}}{|g_{i}|^{2}}-\mathcal{F}_{s},\text{ }\frac{\partial}{\partial
z}E_{i}^{+}=i\pi_{03}+\mathcal{F}_{i},\nonumber\\
& \label{bloch3}%
\end{align}
where (I) stands for Ito type SDE. \ $\pi_{ij}$ is the stochastic variable
that corresponds to the atomic populations of state $|i\rangle$ when $i=j$ and
to atomic coherence when $i\neq j$, and $\mathcal{F}_{ij}$ are c-number
Langevin noises. \ The remaining equations of motion, which close the set, can
be found by replacing the above classical variables, $\pi_{jk}^{\ast
}\rightarrow\pi_{jk}^{\dag},$ $(\pi_{jk}^{\dag})^{\ast}\rightarrow\pi_{jk},$
$(E_{s,i}^{+})^{\ast}\rightarrow E_{s,i}^{-},$ $(E_{s,i}^{-})^{\ast
}\rightarrow E_{s,i}^{+}$ , and $\mathcal{F}_{jk}^{\ast}\rightarrow
\mathcal{F}_{jk}^{\dag}$.\ \ Note that the atomic populations satisfy
$\pi_{jj}^{\ast}=\pi_{jj}.$ \ The superscripts, dagger ($\dag$) for atomic
variables and ($-$) for field variables, denote the independent variables,
which is a feature of the positive-P representation: there are double
dimension spaces for each variable. \ These variables are complex conjugate to
each other when ensemble averages are taken, for example $\left\langle
\pi_{jk}\right\rangle =\left\langle \pi_{jk}^{\dag}\right\rangle ^{\ast}$ and
$\left\langle E_{s,i}^{+}\right\rangle =\left\langle E_{s,i}^{-}\right\rangle
^{\ast}.$ \ The doubled spaces allow the variables to explore trajectories
outside the classical phase space.

Before going further to discuss the numerical solution of the SDE, we point
out that the diffusion matrix elements have been computed using Fokker-Planck
equations and by the Einstein relations described in Appendix B.3.3. \ This
provides the important check on the lengthy derivations of the diffusion
matrix elements we need for the simulations.

The next step is to find expressions for the Langevin noises, and the details
are given in Appendix B.3.4 in terms of a non-square matrix $B$
\cite{QO:Walls,Smith1}. \ The matrix $B$ is used to construct the symmetric
diffusion matrix $D(\alpha)=B(\alpha)B^{T}(\alpha)$ for a Ito SDE,%

\begin{equation}
dx_{t}^{i}=A_{i}(t,\overrightarrow{x_{t}})dt+\sum\limits_{j}B_{ij}%
(t,\overrightarrow{x_{t}})dW_{t}^{j}(t)\text{ \ (I)}\label{Ito}%
\end{equation}
where $\xi_{i}dt=dW_{t}^{i}(t)$ (Wiener process) and $\left\langle \xi
_{i}(t)\xi_{j}(t^{\prime})\right\rangle =\delta_{ij}\delta(t-t^{\prime}).$
\ Note that $B\rightarrow BS,$ where $S$ is an orthogonal matrix ($SS^{T}=I$),
leaves $D$\ unchanged, so $B$ is not unique. \ We could also construct a
square matrix representation $B$ \cite{QN:Gardiner,SM:Gardiner,Carmichael86}.
\ This involves a procedure of matrix decomposition into a product of lower
and upper triangular matrix factors. \ A Cholesky decomposition can be used to
determine the $B$ matrix elements successively row by row. \ The downside of
this procedure is that the $B$ matrix elements must be differentiated in
converting the Ito SDE to its equivalent Stratonovich form for numerical solution.

The Stratonovich SDE is necessary for the stability and the convergence of
semi-implicit methods. \ Because of the analytic difficulties in transforming
to the Stratonovich form, we use instead the non-square form of $B$
\cite{Smith1} that is shown explicitly in Appendix B.3.4. \ 

In this case a typical $B$ matrix element is a sum of terms, each one of which
is a product of the square root of a diffusion matrix element with a unit
strength real (if the diffusion matrix element is diagonal) or complex (if the
diffusion matrix element is off-diagonal) Gaussian unit white noise. \ It is
straightforward to check that a $B$ matrix constructed in this way reproduces
the required diffusion matrix $D=BB^{T}$.

As pointed out in the reference \cite{Drummond91}, the transverse
dipole-dipole interaction can be neglected and nonparaxial spontaneous decay
rate can be accounted for by a single atom decay rate one if the atomic
density is not too high. \ We are interested here in conditions where the
ensemble length $L$ is significant and propagation effects are non-negligible,
and the average distance between atoms $d=\sqrt[3]{V/N}$ is larger than the
transition wavelength $\lambda.$ \ The length scales satisfy $\lambda\lesssim
d\ll L,$ and we consider a pencil-like cylindrical atomic ensemble. \ The
paraxial or one-dimensional assumption for field propagation is then valid,
and the transverse dipole-dipole interaction is not important for the atomic
density we focus here.

\section{Numerical Simulation}

In this Section, we discuss the numerical integration of the atomic and field
equations derived given in the last Section.

There are several possible ways to integrate the differential equation
numerically. \ Three main categories of algorithm used are forward (explicit),
backward (implicit), and mid-point (semi-implicit) methods \cite{numerical}.
\ The\ midpoint method is in a sense between the explicit and implicit
methods, and we will use an algorithm of this type in the following. \ Let
$t_{m}=t_{n}+\frac{\Delta t}{2}$ for $n$th segment and iterate (m denotes mid point)%

\begin{equation}
x(t_{m})=x(t_{n})+f[t_{m},x(t_{m})]\frac{\Delta t}{2}%
\end{equation}
until convergence is reached. \ Then step forward with $x(t_{n+1}%
)=2x(t_{m})-x(t_{n}).$

The forward difference method, which Euler or Runge-Kutta methods utilizes, is
not guaranteed to\ converge in stochastic integrations \cite{xmds}. \ There it
is shown that the semi-implicit method \cite{semi} is more robust in
Stratonovich type SDE simulations \cite{Drummond91b}. \ More extensive studies
of the stability and convergence of SDE can be found in the reference
\cite{SDE:Kloeden}. \ The Stratonovich type SDE equivalent to the Ito type
equation (\ref{Ito}), is%
\begin{align}
dx_{t}^{i}  & =[A_{i}(t,\overrightarrow{x_{t}})-\frac{1}{2}\sum\limits_{j}%
\sum\limits_{k}B_{jk}(t,\overrightarrow{x_{t}})\frac{\partial}{\partial x^{j}%
}B_{ik}(t,\overrightarrow{x_{t}})]dt\nonumber\\
& +\sum\limits_{j}B_{ij}(t,\overrightarrow{x_{t}})dW_{t}^{j}\text{
\ (Stratonovich),}%
\end{align}
which has the same diffusion terms $B_{ij},$ but with modified drift terms.
\ This "correction" term arises from the different definitions of stochastic
integral in the Ito and Stratonovich calculus.

At the end of Appendix B 3.3, we derive the Stratonovich SDE with the
(underlined) "correction" terms noted above. \ We then have 19 classical
variables including atomic populations, coherences, and two
counter-propagating cascade fields. \ With 64 diffusion matrix elements and an
associated 117 random numbers required to represent the instantaneous Langevin
noises, we are ready to solve the equations numerically using the robust
midpoint difference method.

\subsection{Shooting and secant method}

The problem we encounter here involves counter-propagating field equations in
the space dimension and initial value type atomic equations in the time
dimension. \ The initial value problem is addressed by the difference method
discussed in the previous Section. \
\begin{figure}
[ptb]
\begin{center}
\includegraphics[
natheight=7.499600in,
natwidth=9.999800in,
height=4.4053in,
width=5.8634in
]%
{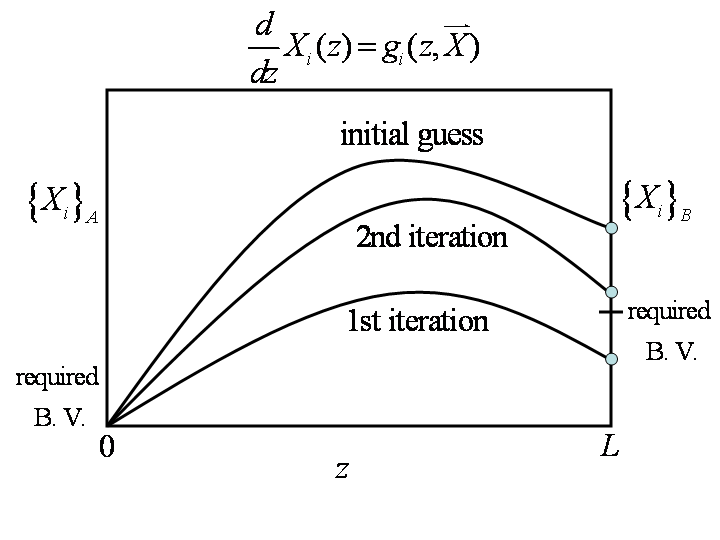}%
\caption{Schematic illustration of the principle of the shooting method for
two-point boundary value problems.}%
\label{shooting2}%
\end{center}
\end{figure}

The counter-propagating field equations have a boundary condition specified at
each end of the medium. \ This is a two-point boundary value problem, and a
numerical approach to its solution, the shooting method \cite{numerical}, is
illustrated in Figure \ref{shooting2}.

Consider the set of differential equations $dX_{i}(z)/dz=g_{i}(z,\vec{X})$.
\ A subset $A$ of \{$X_{i}$\} satisfy boundary conditions at $z=0$, and the
complementary subset $B$ satisfy boundary conditions at $z=L.$

The shooting method augments the set $A$ with a set of "guesses" $A^{\prime},$
so that $A\cup A^{\prime}$ enable the differential equations to be integrated
as an initial value problem (from $z=0$ to $z=L$). \ The idea is that
$A^{\prime}$ is the correct choice when the integrated values at $z=L$
reproduce the true boundary conditions, set $B,$ within a permissible
tolerance. \ The set $A^{\prime}$ is updated to enable convergence of the
output at $z=L$ to the set $B.$

\begin{figure}
[ptb]
\begin{center}
\includegraphics[
natheight=7.499600in,
natwidth=9.999800in,
height=4.4053in,
width=5.8634in
]%
{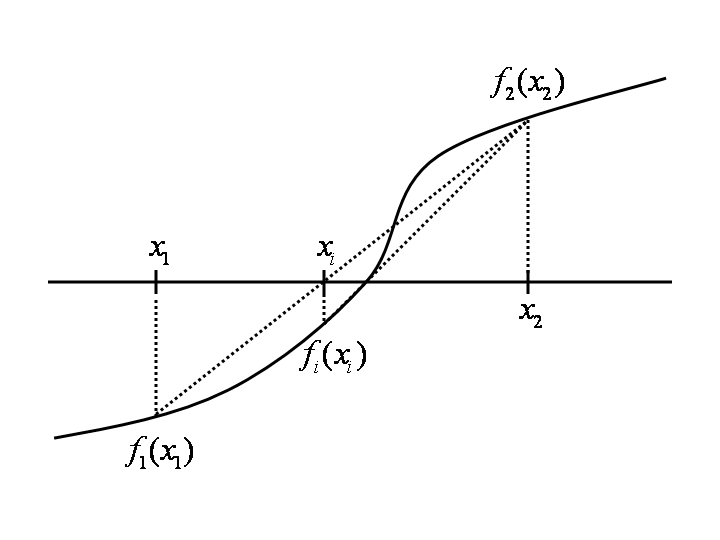}%
\caption{Secant method. \ The root is bracketed by two initial guesses of
$x_{1}$ and $x_{2}$ and an updated guess $x_{i}$ is located at the
intersection of two straight lines.}%
\label{secant}%
\end{center}
\end{figure}

The secant method that is used to update each element of $A^{\prime}$ takes
two guesses $x_{1}$ and $x_{2}$ for each variable of $A^{\prime}$ and returns
an updated value $x_{i},$%

\begin{equation}
x_{i}=x_{2}-f_{2}\frac{f_{2}-f_{1}}{x_{2}-x_{1}}.
\end{equation}
where $f_{1}$ and $f_{2}$ are the differences between the required values of
that variable in set $B$ and the numerically computed values assuming $x_{1}$
and $x_{2}$ values at $z=0.$ This method is iterated until convergence to all
values in $B$ is obtained. \ The secant method is illustrated in Figure
\ref{secant}.%

\subsection{Outline of the numerical solution}

We use Matlab to perform the numerical integrations. \ For simplicity, we
label the atomic and field variables as $a_{i}$ and $e_{i}$. \ \ The
counter-propagating field ($-z$ direction) variables are $e_{1}$ and $e_{2}$
(signal fields) and $e_{3}$ and $e_{4}$ (idler fields) propagate in the $+z$
direction.\ \ We set the local time $\tau\rightarrow t$ in the following
description of the algorithm.

We initialize 15 $a_{i}(z,t)$, 4 $e_{i}(z,t)$ in time $t\in(0,T)$ and space
$z\in(0,L),$ and select 19 Gaussian random numbers $n_{i}(z,t).$ \ Set time
and space grids with spacings $\Delta t,\Delta z$ respectively.\ \ For each
realization among $R$ statistical ensemble averages, we update the variables
governed by the symbolic equations of motion,%

\begin{align}
\frac{\partial}{\partial z}e_{i}  & =P_{i}(\vec{e},\vec{a},n_{e_{i}}),\\
\frac{\partial}{\partial t}a_{i}  & =A_{i}(\vec{e},\vec{a},n_{a_{i}}),
\end{align}
where $P_{i}$ and $A_{i}$ are in general the functions of variables that are
denoted as vectors $\vec{e}$ and $\vec{a}$. \ Each variable has its own
stochastic source term as $n_{e_{i}}$ or $n_{a_{i}}.$ \ 

The algorithm proceeds by using the midpoint difference method for the
evolutions in space and time and the shooting method for $e_{i},$%

\begin{align*}
e_{i}(z_{m},t)  & =e_{i}(z,t)+\frac{\Delta z}{2}P_{i}[\vec{e}(z_{m},t),\vec
{a}(z,t),n_{e_{i}}(z,t)],\\
a_{i}(z,t_{m})  & =a_{i}(z,t)+\frac{\Delta t}{2}A_{i}[\vec{e}(z,t),\vec
{a}(z,t_{m}),n_{a_{i}}(z,t)],
\end{align*}
where $z_{m}=z+\Delta z/2$ and $t_{m}=t+\Delta t/2.$ \ The two guesses
required in the secant method used in the shooting method are chosen as
$x_{1}=\{e_{1}(0,t),e_{2}(0,t)\}$ and $x_{2}=\{e_{3}(L,t),e_{4}(L,t)\}.$

Any normally-ordered quantity $\left\langle Q\right\rangle $ can be derived by
ensemble averages that $\left\langle Q\right\rangle =\sum_{i=1}^{R}Q_{i}/R $
where $Q_{i}$ is the result for each realization.\ \ Note that the update for
field variables in space precedes the update for atomic variables, which takes
into account that field variables evolve faster than atomic variables. \ The
order should not matter when finer grids are used.

\subsection{Results for signal, idler intensities, and the second-order
correlation function}

In this subsection, we present the second-order correlation function of
signal-idler fields, and their intensity profiles. $\ $We define the
intensities of signal and idler fields by \ %

\begin{equation}
I_{s}(t)=\left\langle E_{s}^{-}(t)E_{s}^{+}(t)\right\rangle ,\text{ }%
I_{i}(t)=\left\langle E_{i}^{-}(t)E_{i}^{+}(t)\right\rangle ,
\end{equation}
respectively, and the second-order signal-idler correlation function%

\begin{equation}
G_{s,i}(t,\tau)=\left\langle E_{s}^{-}(t)E_{i}^{-}(t+\tau)E_{i}^{+}%
(t+\tau)E_{s}^{+}(t)\right\rangle
\end{equation}
where $\tau$ is the delay time of the idler field with respect a reference
time $t$\ of the signal field. \ Since the correlation function is not
stationary \cite{QL:Loudon}, we choose $t$ as the time when $G_{s,i}$ is at
its maximum.

We consider a cigar shaped $^{85}$Rb ensemble of radius $0.25$ mm and $L=3$
mm. \ The operating conditions of the pump lasers are ($\Omega_{a},$
$\Omega_{b},$ $\Delta_{1},$ $\Delta_{2}$) $=$ ($0.4,$ $1,$ $1,$ $0$%
)$\gamma_{03}$ where $\Omega_{a}$ is the peak value of a $50$ ns square pulse,
and $\Omega_{b}$ is the Rabi frequency of a continuous wave laser. \ The four
atomic levels are chosen as ($|0\rangle,$ $|1\rangle,$ $|2\rangle,$
$|3\rangle$) $=$ ($|$5S$_{1/2},$F=3$\rangle,$ $|5$P$_{3/2},$F=4$\rangle,$
$|5$P$_{3/2},$F=4$\rangle,$ $|4$D$_{5/2},$F=5$\rangle$). \ The natural decay
rate for atomic transition $|1\rangle\rightarrow|0\rangle$ or $|3\rangle
\rightarrow|0\rangle$ is $\gamma_{01}=\gamma_{03}=1/26$ ns and they have a
wavelength 780 nm. \ For atomic transition $|2\rangle\rightarrow|1\rangle$ or
$|2\rangle\rightarrow|3\rangle$ is $\gamma_{12}=\gamma_{32}=0.156\gamma_{03}$
\cite{gsgi} with a telecom wavelength 1.53$\mu$m. \ The scale factor of the
coupling constants for signal and idler transitions is $g_{s}/g_{i}=0.775.$

\begin{figure}
[ptb]
\begin{center}
\includegraphics[
natheight=15.104400in,
natwidth=20.333700in,
height=4.3762in,
width=7.3175in
]%
{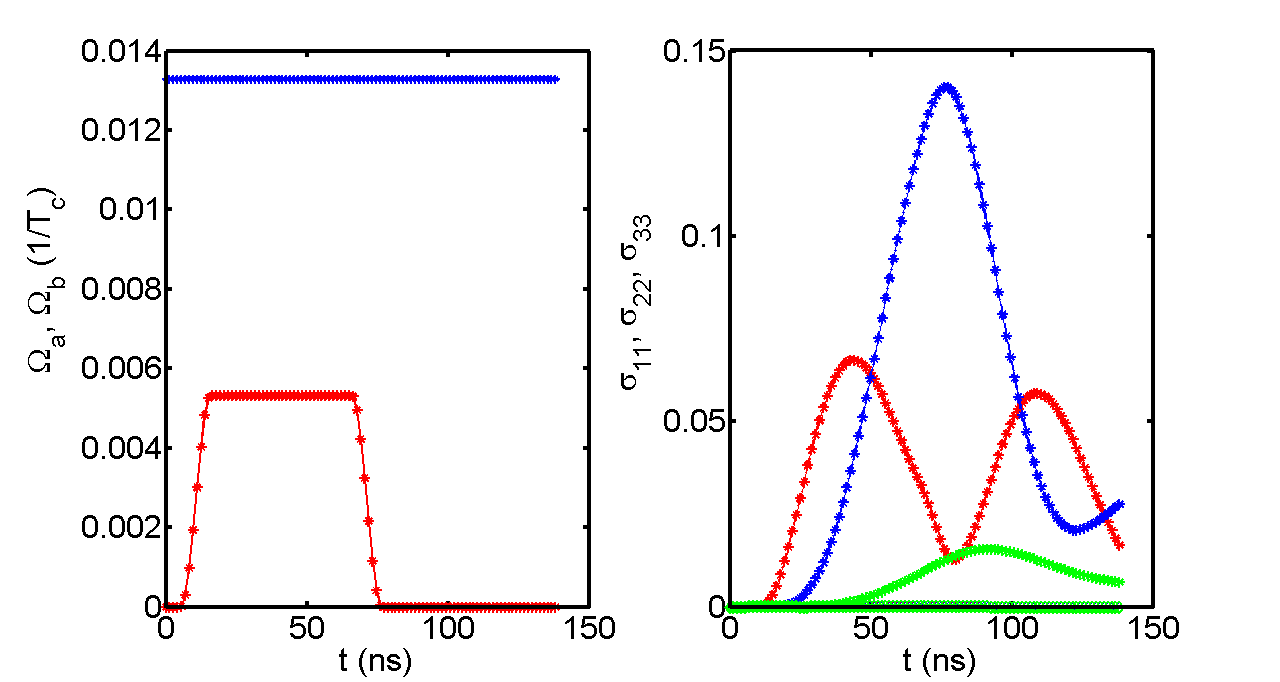}%
5natwidth=6.317500in,
\caption{Time-varying pump fields and time evolution of atomic populations.
(Left) The first pump field $\Omega_{a}$ (dotted-red) is a square pulse of
duration 50 ns and $\Omega_{b}$ is continuous wave (dotted-blue). \ (Right)
The time evolution of the real part of populations for three atomic levels
$\sigma_{11}=\left\langle \tilde{\alpha}_{13}\right\rangle $ (dotted-red),
$\sigma_{22}=\left\langle \tilde{\alpha}_{12}\right\rangle $ (dotted-blue),
$\sigma_{33}=\left\langle \tilde{\alpha}_{11}\right\rangle $ (dotted-green) at
$z=0,L$, and almost vanishing imaginary parts for all three of them. indicate
convergence of the ensemble averages.\ Note that these atomic populations are
uniform as a function of $z.$}%
\label{omega_population}%
\end{center}
\end{figure}

\begin{figure}
[ptb]
\begin{center}
\includegraphics[
natheight=15.885600in,
natwidth=20.333700in,
height=5.8493in,
width=6.9525in
]%
{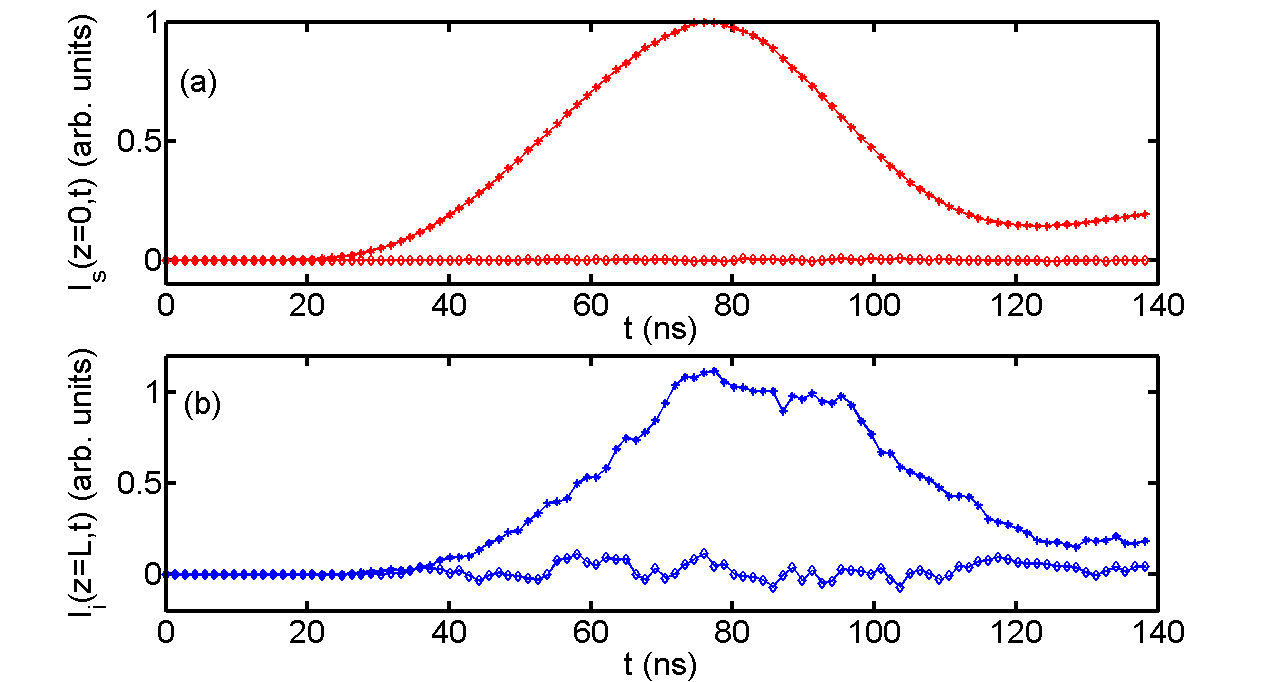}%
\caption{Temporal intensity profiles of counter-propagating signal and idler
fields. \ (a) At $z=0,$ real (dotted-red) and imaginary (diamond-red) parts of
signal intensity. \ (b) At $z=L,$ real (dotted-blue) and imaginary
(diamond-blue) parts of idler intensity. \ Both intensities are normalized by
the peak value of signal intensity that is $7.56\times10^{-12}$ $E_{c}^{2}$.
\ Note that the idler fluctuations and its non-vanishing imaginary part
indicate a relatively slower convergence compared with the signal intensity.
\ The ensemble size was 8$\times10^{5},$ and the atomic density $\rho=10^{10}%
$cm$^{-3}$.}%
\label{intensity}%
\end{center}
\end{figure}

\begin{figure}
[ptb]
\begin{center}
\includegraphics[
natheight=15.104400in,
natwidth=20.333700in,
height=6.6579in,
width=7.5416in
]%
{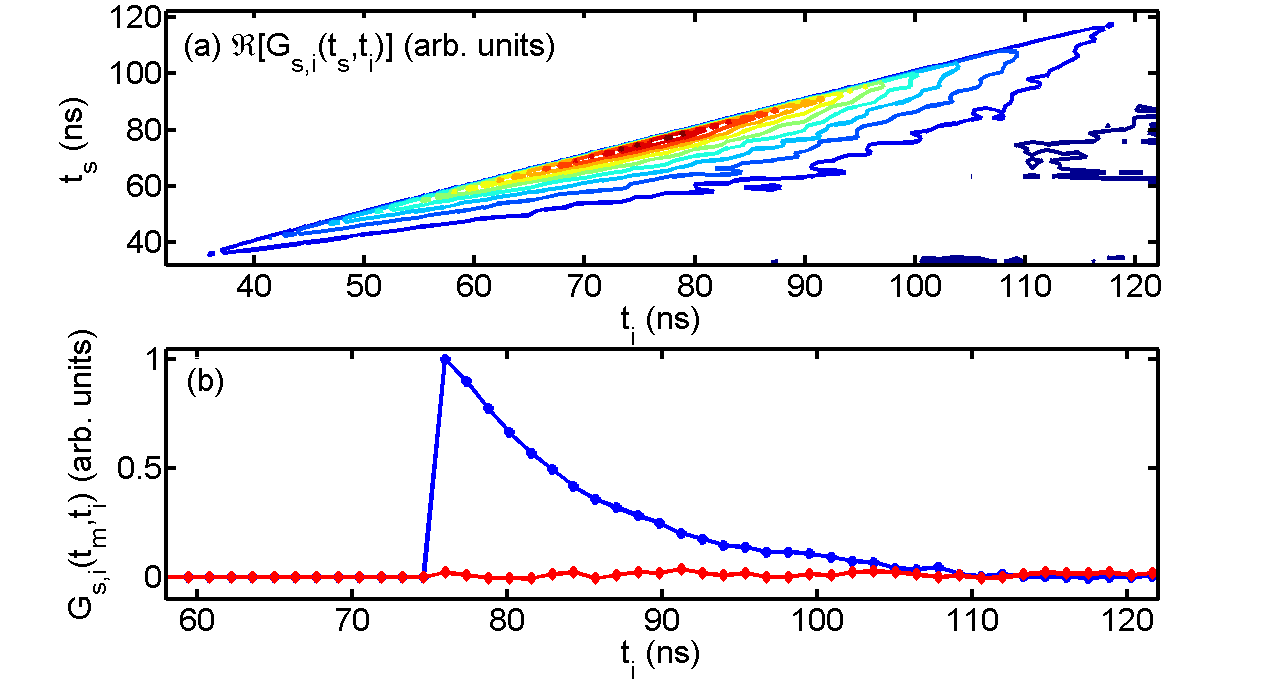}%
\caption{Second-order correlation function $G_{s,i}(t_{s},t_{i}).$ The 2-D
contour plot of the real part of $G_{s,i}$ with a causal cut-off at
$t_{s}=t_{i}$ is shown in (a). \ The plot (b) gives a cross-section at
$t_{s}=t_{m}\approx75$ ns, which is normalized to the maximum of the real part
(dotted-blue) of $G_{s,i}.$ \ The imaginary part (diamond-red) of $G_{s,i}$ is
nearly vanishing, and the number of realizations is 8$\times10^{5}$ for
$\rho=10^{10}$cm$^{-3}.$}%
\label{gsi_t}%
\end{center}
\end{figure}

We have investigated four different atomic densities from a dilute ensemble
with an optical density (opd) of 0.11 to a opd = 4.35. \ In Figure
\ref{omega_population}, \ref{intensity}, and \ref{gsi_t}, we take the atomic
density $\rho=10^{10}$ cm$^{-3}$ (opd = 2.18) for example, and the grid sizes
for dimensionless time $\Delta t=4$ and space $\Delta z=0.0007$ are chosen.
\ The convergence of the grid spacings is fixed in practice by convergence to
the signal intensity profile with an estimated relative error less than 0.5\%.%

The temporal profiles of the exciting lasers are shown in the left panel of
Figure \ref{omega_population}. \ The atomic density is chosen as $\rho
=10^{10}$ cm$^{-3},$ and the cooperation time $T_{c}$ is 0.35 ns. \ The right
panel shows time evolution of atomic populations for levels $|1\rangle$,
$|2\rangle,$ and $|3\rangle$\ at $z=0,L,$ that are spatially uniform. \ The
populations are found by ensemble averaging the complex stochastic population
variables. \ The imaginary parts of the ensemble averages tend to zero as the
ensemble size is increased, and this is a useful indicator of convergence, see
Appendix B.2 for a discussion. \ In this example, the ensemble size was
8$\times10^{5}.$ \ The small rise after the pump pulse $\Omega_{a}$ is turned
off is due to the modulation caused by the pump pulse $\Omega_{b},$ which has
a generalized Rabi frequency $\sqrt{\Delta_{2}^{2}+4\Omega_{b}^{2}}$. This
influences also the intensity profiles and the correlation functions.%

In Figure \ref{intensity}, we show that counter-propagating signal ($-\hat{z}
$) and idler ($+\hat{z}$) fields at the respective ends of the atomic
ensemble. \ The plots show the real and imaginary parts of the observables,
and both are normalized to the peak value of signal intensity. \ Note that the
characteristic field strength in terms of natural decay rate of the idler
transition ($\gamma_{03}$) and dipole moment ($d_{i}$) is $(d_{i}/\hbar
)E_{c}\approx36.3\gamma_{03}$. \ The fluctuation in the real idler field
intensity at $z=L$ and non-vanishing imaginary part indicates a slower
convergence compared to the signal field that has an almost vanishing
imaginary part. \ The slow convergence is a practical limitation of the
method. \ 

In Figure \ref{gsi_t} (a), we show a contour plot of the second-order
correlation function $G_{s,i}(t_{s},t_{i})$ where $t_{i}\geq t_{s}.$ \ In
Figure \ref{gsi_t} (b), a section is shown through $t_{s}\approx75$ ns where
$G_{s,i}$ is at its maximum. \ The approximately exponential decay of
$G_{s,i}$ is clearly superradiant consistent with the theory of Chapter 3 and
the reference \cite{telecom}. \ The non-vanishing imaginary part of $G_{s,i}$
calculated by ensemble averaging is also shown in (b) and indicates a
reasonable convergence after 8$\times10^{5}$ realizations. \ In Table
\ref{table1}, we display numerical parameters of our simulations for four
different atomic densities. \ The number of dimensions in space and time is
$M_{t}\times M_{z}$ with grid sizes ($\Delta t,\Delta z$) in terms of
cooperation time ($T_{c}$), length ($L_{c}$). \ The superradiant time scale
($T_{f}$) is found by fitting $G_{s,i}$ to an exponential function
($e^{-t/T_{f}}$), with $95\%$ confidence range.%

\begin{figure}
[ptb]
\begin{center}
\includegraphics[
natheight=15.104400in,
natwidth=20.333700in,
height=4.0179in,
width=6.2604in
]%
{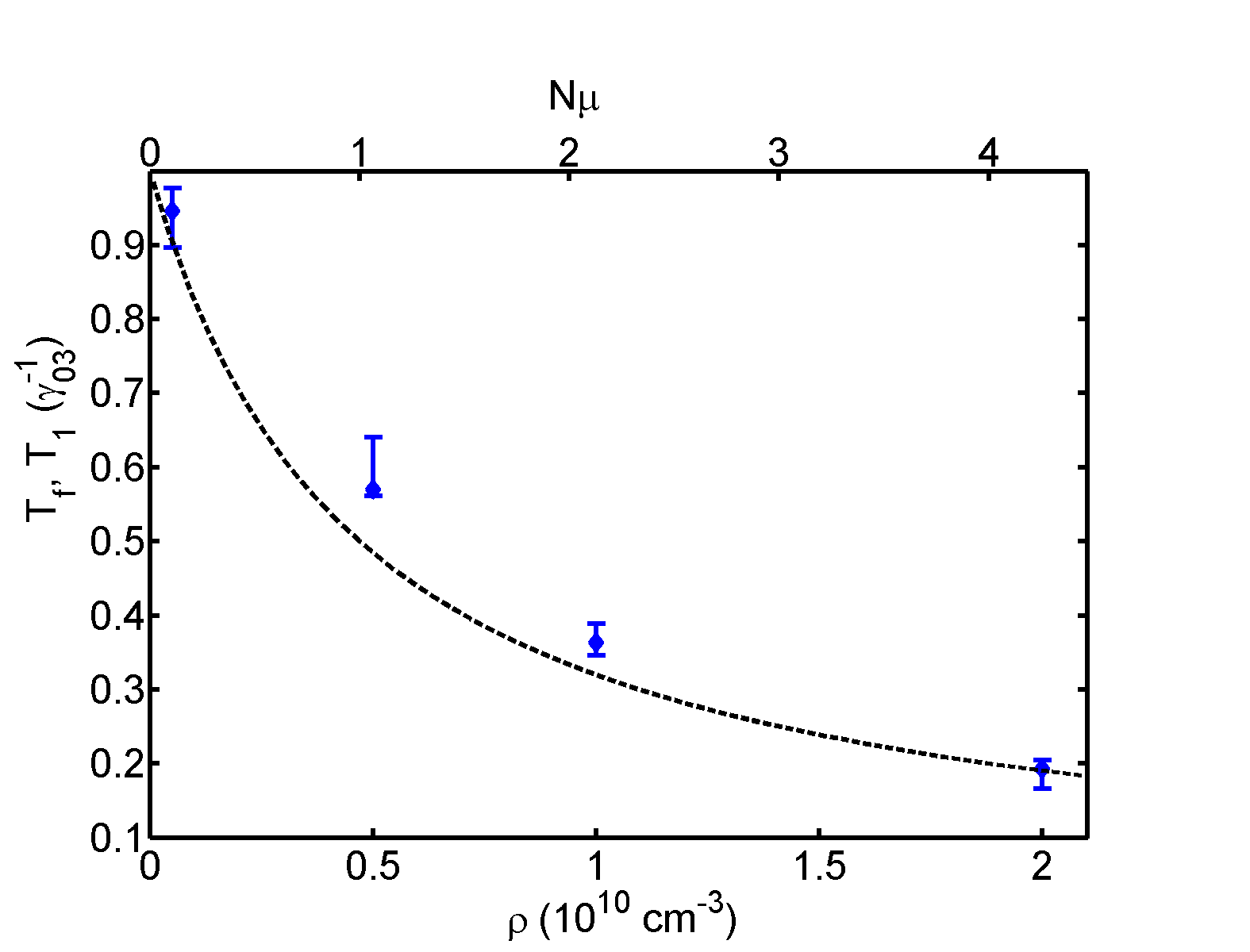}%
\caption{Characteristic timescales, $T_{f}$ and $T_{1}$ vs atomic density
$\rho$ and the superradiant enhancement factor $N\mu$ ($\mu=\bar{\mu}$).
$\ T_{f}$ (dotted-blue) is the fitted characteristic timescale for
$G_{s,i}(t_{s}=t_{m},t_{i}=t_{m}+\tau)$ where $t_{m}$ is chosen at its
maximum, as in Figure \ref{gsi_t}. \ The errorbars indicate the fitting
uncertainties. \ As a comparison, $T_{1}$=$\gamma_{03}^{-1}/$($N\mu+1$)
(dashed-black) is plotted where $\gamma_{03}^{-1}=26$ ns is the natural decay
time of D1 line of $^{87}$Rb atom, and $\mu$ is the geometrical constant for a
cylindrical atomic ensemble, as discussed in Chapter 3. \ The number of
realizations is 4$\times10^{5}$ for $\rho=5\times10^{8}$, $5\times10^{9}$
cm$^{-3}$ and 8$\times10^{5}$ for $\rho=10^{10}$, $2\times10^{10}$ cm$^{-3}.$}%
\label{gsi}%
\end{center}
\end{figure}

In Figure \ref{gsi}, the characteristic time scale is plotted as a function of
atomic density and the factor $N\bar{\mu}$, and shows faster decay for
optically denser atomic ensembles. \ We also plot the timescale $T_{1}%
=\gamma_{03}^{-1}/(N\mu+1)$ (ns) that is derived from the theory of Chapter 3,
in which $\bar{\mu}$ is the geometrical constant for a cylindrical ensemble,
Eq. (\ref{mu}). $\ $The natural decay time $\gamma_{03}^{-1}=26$ ns
corresponds to the D2 line of $^{87}$Rb. \ The error bar indicates the
deviation due to the fitting range from the peak of $G_{s,i}$ to approximately
25\% and 5\% of the peak value. \ The theory and simulations are in good
qualitative agreement, approaching independent atom behavior at lower
densities. \ For larger opd atomic ensembles, larger statistical ensembles are
necessary for numerical simulations to converge. \ The integration of
8$\times10^{5}$ realizations used in the case of $\rho=10^{10}$ cm$^{-3}$
consumes about 14 days with Matlab's parallel computing toolbox (function
"\textit{parfor"}) with a Dell precision workstation T7400 (64-bit Quad-Core
Intel Xeon processors).%

\begin{table}[t] \centering
\caption{Numerical simulation parameters for different atomic densities $\rho$.  Corresponding optical depth (opd),
time and space grids ($M_{t}\times M_{z}$) with grid sizes ($\Delta t,\Delta z$) in terms of cooperation time ($T_c$) and
length ($L_c$), and the fitted characteristic time $T_{f}$ for $G_{s,i}$ (see text).}%
\begin{tabular}
[c]{|c|c|c|c|c|c|}\cline{1-5}\cline{4-4}\cline{6-6}%
$\rho($cm$^{-3})$ & opd & $M_{t}\times M_{z}$ &
\begin{tabular}
[c]{c}%
$\Delta t(T_{c}),$\\
$\Delta z(L_{c})$%
\end{tabular}
&
\begin{tabular}
[c]{c}%
$T_{c}($ns$),$\\
$L_{c}($m$)$%
\end{tabular}
&
\begin{tabular}
[c]{c}%
fitted $T_{f}$(ns)\\
\lbrack$95\%$ confidence range]
\end{tabular}
\\\cline{1-5}\cline{4-4}\cline{6-6}%
5$\times10^{8}$ & $0.11$ & $101\times44$ & 0.9, 1.5$\times10^{-4}$ & 1.55,
0.46 & $24.6$ $[24.2,$ $25.0]$\\\hline
5$\times10^{9}$ & $1.09$ & $101\times42$ & 2.8, 4.5$\times10^{-4}$ & 0.49,
0.15 & $14.8$ $[14.4,$ $15.3]$\\\hline
1$\times10^{10}$ & $2.18$ & $101\times42$ & 4, 7$\times10^{-4}$ & 0.35, 0.10 &
$9.4$ $[9.2,$ $9.7]$\\\hline
2$\times10^{10}$ & $4.35$ & $101\times42$ & 5.5, 1$\times10^{-3}$ & 0.24,
0.07 & $5.0$ $[4.6,$ $5.5]$\\\hline
\end{tabular}
\label{table1}%
\end{table}%

\section{Conclusion}

We have derived c-number Langevin equations in the positive-P representation
for the cascade signal-idler emission process in an atomic ensemble. \ The
complete c-number Langevin noise correlations are derived and confirmed by an
alternative theoretical method. \ The equations are solved numerically by a
stable and convergent semi-implicit difference method, while the
counter-propagating spatial evolution is solved by implementing the shooting
method. \ 

We investigate four different atomic densities readily obtainable in a
magneto-optical trap experiment. \ Signal and idler field intensities and
their correlation function are calculated by ensemble averages. \ Vanishing of
the unphysical imaginary parts within some tolerance is used as a guide to
convergence. \ We find an enhanced characteristic time scale for idler
emission in the second-order correlation functions from a dense atomic
ensemble, consistent with the superradiance timescales predicted by the
analytical method in Chapter 3, and observed experimentally \cite{telecom}.



\chapter{Spectral analysis for cascade-emission-based quantum communication}

Cascade emission in alkali atoms is a source of telecommunication photons.
\ In this Chapter, we investigate the DLCZ \cite{dlcz} scheme using the
cascade emission from an atomic ensemble.

\section{Introduction}

Long distance quantum communication based on atomic ensembles was proposed by
Duan, Lukin, Cirac, and Zoller \cite{dlcz}. \ This scheme involves Raman
scattering of light by the atoms. \ The cascade transitions investigated in
Chapter 3 and 4 provide a source of telecommunication wavelength photons. \ It
is interesting to assess the cascade scheme in the DLCZ protocol given that it
could potentially reduce transmission losses in a quantum telecommunication
system. \ The DLCZ scheme is based on entanglement generation and swapping and
quantum state transfer.

In this Chapter, we first discuss entanglement generation and then investigate
how frequency entanglement of the cascade photon pair influences entanglement swapping.

\section{DLCZ Scheme with Cascade Emission}

In the DLCZ protocol, a weak pump laser Raman scatters a single photon
generating a quantum correlated spin excitation in the ensemble. \ By
interfering the Raman photons generated from two separate atomic ensembles on
a beam splitter (B.S.), the DLCZ entangled state $(|01\rangle+|10\rangle
)/\sqrt{2}$ \cite{0110} is prepared conditioned on one and only one click of
the detectors after the B.S. \ Hence $|0\rangle$ and $|1\rangle$ represent the
state of zero or one collective spin excitations stored in the hyperfine
ground state coherences. \ This state originates from an indistinguishable
photon paths. \ The error from multiple excitations can be made negligible if
the pump laser is weak enough. \ 

\begin{figure}
[ptb]
\begin{center}
\includegraphics[
natheight=7.499600in,
natwidth=9.999800in,
height=4.9208in,
width=6.3192in
]%
{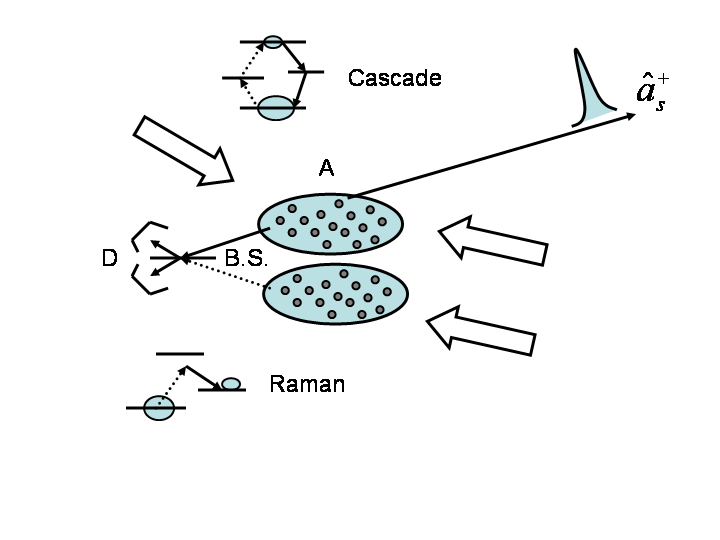}%
\caption{Entanglement generation in the DLCZ scheme using the cascade and
Raman transitions in two different atomic ensembles. \ Large white arrows
represent laser pump excitations corresponding to the dashed lines in either
cascade or Raman level structures. \ Here $\hat{a}_{s}^{\dagger}$ represents
the emitted telecom photon. \ B.S. means beam splitter that is used to
interfere the incoming photons measured by the photon detector D. \ The label
A refers to the pair of ensembles for later reference.}%
\label{ent_gen}%
\end{center}
\end{figure}

As shown in Figure \ref{ent_gen}, we consider instead that one of the
ensembles employ cascade emission. \ The idea is for cascade emission to
generate a telecom photon ($\hat{a}_{s}^{\dag}$) for transmission in the
optical fiber, and an infrared photon that interferes locally with the Raman
photon generated in the $\Lambda$-type atomic ensemble. \ In this way
interference of the infrared photons generate the entangled state,%

\begin{equation}
|\Psi\rangle=\frac{1}{\sqrt{2}}(|01\rangle_{a,s}+|10\rangle_{a,s}),
\end{equation}
similar to the conventional DLCZ entanglement generation scheme. \ Now,
however, instead of a stored spin excitation, we generate a telecom photon.%

The entanglement swapping with the cascade emission may be implemented as
shown in Figure \ref{ent_swap}, and will be discussed in detail in the next
Section. \ The initial state is a tensor product of two state vectors
generated locally at the sites A and B.%
\begin{align}
|\Psi\rangle &  =(\sqrt{1-\eta_{1A}}|0\rangle+\sqrt{\eta_{1A}}|1\rangle
_{i}^{A}|1\rangle_{s}^{A})\otimes(\sqrt{1-\eta_{2A}}|0\rangle+\sqrt{\eta_{2A}%
}|1\rangle_{r}^{A}|1\rangle_{a}^{A})\otimes\nonumber\\
&  (\sqrt{1-\eta_{1B}}|0\rangle+\sqrt{\eta_{1B}}|1\rangle_{i}^{B}|1\rangle
_{s}^{B})\otimes(\sqrt{1-\eta_{2B}}|0\rangle+\sqrt{\eta_{2B}}|1\rangle_{r}%
^{B}|1\rangle_{a}^{B}),\label{joint}%
\end{align}
where (s,~i) represent the signal and idler photons from the cascade emission,
and (r, a) are Raman scattered photon and the collective spin excitation.
\ Here$\ \eta_{1}$ and $\eta_{2}$ are efficiencies to generate cascade and
Raman emission. \ Since $\eta_{1}$ and $\eta_{2}\ll1,$ multiple atomic
excitations or multi-photon generation can be excluded.

\section{Entanglement Swapping}%

\begin{figure}
[ptb]
\begin{center}
\includegraphics[
natheight=7.499600in,
natwidth=9.999800in,
height=5.0436in,
width=5.7856in
]%
{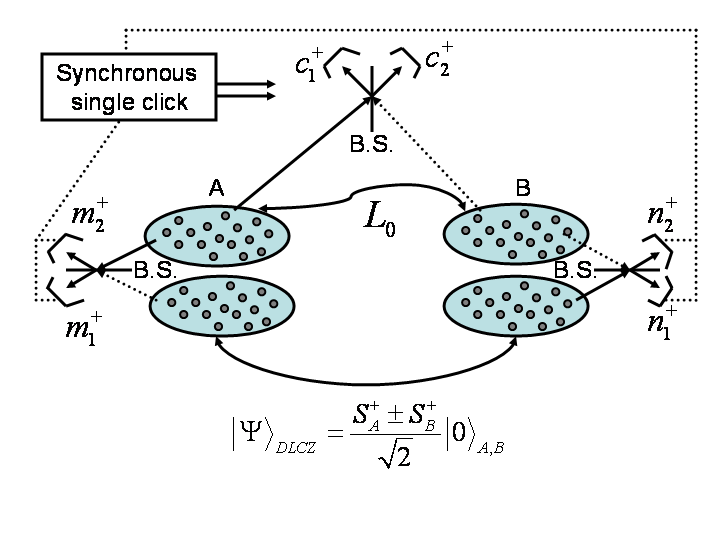}%
\caption{Entanglement swapping of DLCZ scheme using the cascade transition.
The site A is described in detail in Figure \ref{ent_gen} and equivalently for
the site B. \ The telecom signal photons are sent from both sites and
interfere by B.S. midway between with detectors represented by $c_{1}%
^{\dagger}$ and $c_{2}^{\dagger}$. \ Synchronous single clicks of the
detectors from both sites ($m_{1,2}^{\dagger}$, $n_{1,2}^{\dagger}$) and the
midway detector ($c_{1,2}^{\dagger}$) generate the entangled state between
lower atomic ensembles at sites A and B. \ The locally generated entanglement
is swapped to distantly separated sites in this cascade-emission-based DLCZ
protocol.}%
\label{ent_swap}%
\end{center}
\end{figure}

Consider the product state generated from A and B, Figure \ref{ent_swap},%

\begin{align}
|\Psi\rangle & =(\frac{|10\rangle_{as}+|01\rangle_{as}}{\sqrt{2}})_{A}%
\otimes(\frac{|10\rangle_{as}+|01\rangle_{as}}{\sqrt{2}})_{B}\nonumber\\
& =\frac{1}{2}(|1010\rangle_{asas}+|1001\rangle_{asas}+|0110\rangle
_{asas}+|0101\rangle_{asas}),
\end{align}
where the subscript (a) represents a stored local atomic excitation, and (s)
means a telecom photon propagating toward the B.S. in the middle. \ We can
tell from this effective state that the first component ($|1010\rangle_{asas}%
$) contributes no telecom photons at all (two local excitations) and can be
ruled out by measuring a "click" at one of the middle detectors. \ The second
and the third components have components of the entangled state of quantum
swapping, and the fourth one is the source of error if the photodetector
cannot resolve one from two photons. \ The error could be corrected by using a
photon number resolving detector (PNRD) if other drawbacks like dark counts,
photon losses during propagation, and detector inefficiency are not considered.

Now we will formulate the entanglement swapping including the spectral effects
discussed in Chapter 3. \ We ignore pump-phase offsets, assuming $50/50$ B.S.
and a symmetric set-up ($\eta_{1A}=\eta_{1B}=\eta_{1},~\eta_{2A}=\eta
_{2B}=\eta_{2}$) for simplicity. \ Expand the previous joint state, Eq.
(\ref{joint}) and keep the terms up to the second order of $\eta_{1,2}$ that
can contribute to detection events ($\hat{m}_{1,2}^{\dag},\hat{n}_{1,2}^{\dag
}$),%

\begin{align}
&  |\Psi\rangle_{eff}\nonumber\\
&  =\eta_{1}(1-\eta_{2})|1\rangle_{i}^{A}|1\rangle_{s}^{A}|1\rangle_{i}%
^{B}|1\rangle_{s}^{B}+\eta_{2}(1-\eta_{1})|1\rangle_{r}^{A}|1\rangle_{cs}%
^{A}|1\rangle_{r}^{B}|1\rangle_{cs}^{B}+\nonumber\\
&  \sqrt{\eta_{1}\eta_{2}(1-\eta_{1})(1-\eta_{2})}|1\rangle_{i}^{A}%
|1\rangle_{s}^{A}|1\rangle_{r}^{B}|1\rangle_{cs}^{B}+\sqrt{\eta_{1}\eta
_{2}(1-\eta_{1})(1-\eta_{2})}|1\rangle_{r}^{A}|1\rangle_{cs}^{A}|1\rangle
_{i}^{B}|1\rangle_{s}^{B},\label{mode}%
\end{align}
where the cascade emission state $|1\rangle_{s}|1\rangle_{i}\equiv\int
f(\omega_{s},\omega_{i})\hat{a}_{\lambda_{s}}^{\dag}(\omega_{s})\hat
{a}_{\lambda_{i}}^{\dag}(\omega_{i})|0\rangle d\omega_{s}d\omega_{i}$ has the
spectral distribution $f(\omega_{s},\omega_{i})$\ as derived in Chapter 3.

As shown in Figure \ref{ent_swap}, entanglement swapping protocol is fulfilled
by measuring three clicks from the three pairs of the detectors respectively
($\hat{m}_{1,2}^{\dag},\hat{n}_{1,2}^{\dag},\hat{c}_{1,2}^{\dag}$). \ The
quantum efficiency of the detector is considered in the protocol, and we
describe a model for quantum efficiency in Appendix C.1. \ We then use this
model to describe photodetection events registered by non-resolving photon
detectors (NRPD). \ Starting with the input density operator $\hat{\rho}%
_{in}=|\Psi\rangle_{eff}\langle\Psi|,$ we derive the projected density
operator, Eq. (\ref{out}), conditioned on the three clicks of $\hat{m}%
_{1}^{\dag},\hat{n}_{1}^{\dag},$ and $\hat{c}_{1}^{\dag}$ in Appendix C.2. We
use the Schmidt decomposition of the projected density operator and assume a
single mode for the Raman scattered photon. \ We find the un-normalized
density operator $\hat{\rho}_{out}^{(2)}$\ given in Eq. (\ref{out}),%

\begin{align}
&  \hat{\rho}_{out}^{(2)}=\nonumber\\
&  \frac{\eta_{1}^{2}(1-\eta_{2})^{2}}{16}(2-\eta_{t})\eta_{t}\eta_{eff}%
^{2}\Big(1+\sum_{j}\lambda_{j}^{2}\Big)|0\rangle\langle0|+\frac{\eta_{1}%
\eta_{2}(1-\eta_{1})(1-\eta_{2})}{8}\eta_{t}\eta_{eff}^{2}\nonumber\\
&  \bigg\{\Big(\hat{S}_{B}^{\dag}|0\rangle\langle0|\hat{S}_{B}+\hat{S}%
_{A}^{\dag}|0\rangle\langle0|\hat{S}_{A}\Big)+\sum_{j}\lambda_{j}\int\phi
_{j}(\omega_{i})\phi_{j}^{\ast}(\omega_{i}^{\prime})\Phi^{\ast}(\omega
_{i})\Phi^{\ast}(\omega_{i}^{\prime})d\omega_{i}d\omega_{i}^{\prime
}\nonumber\\
&  \Big(\hat{S}_{B}^{\dag}|0\rangle\langle0|\hat{S}_{A}+\hat{S}_{A}^{\dag
}|0\rangle\langle0|\hat{S}_{B}\Big)\bigg\},\label{out3}%
\end{align}
where $\eta_{t}$ and $\eta_{eff}$ are quantum efficiencies of the detectors at
the telecom and infrared wavelengths respectively. \ The first term in Eq.
(\ref{out3}) is the atomic vacuum state at sites A and B and contributes an
error to the output density operator. \ The second term contains the
components of the DLCZ entangled state.

We can define the fidelity $F$, the success probability $P_{S}$ of
entanglement swapping of the entangled state $|\Psi\rangle_{DLCZ}=(S_{A}%
^{\dag}+S_{B}^{\dag})|0\rangle/\sqrt{2},$ and the heralding probability
$P_{H}$ for the third click as%

\begin{align}
F  & \equiv\frac{\text{Tr}(\hat{\rho}_{out}^{(2)}|\Psi\rangle_{DLCZ}%
\langle\Psi|)}{\text{Tr}(\hat{\rho}_{out}^{(2)})},\\
P_{H}  & =P_{1}+P_{2},~P_{1}=P_{2}=\frac{\text{Tr}(\hat{\rho}_{out}^{(2)}%
)}{\mathcal{N}},\\
P_{S}  & =P_{1}\times F_{1}+P_{2}\times F_{2},~F_{1}=F_{2}=F,
\end{align}
where $P_{1,2}$ is the heralding probability of the single click from the
midway detector ($\hat{c}_{1,2}^{\dag})$ as shown in Figure \ref{ent_swap},
and a trace (Tr) is taken over atomic degrees of freedom. \ The normalization
factor $\mathcal{N}$\ is calculated in Eq. (\ref{normalization}) and is given
by
\begin{equation}
\mathcal{N}=\frac{\eta_{1}^{2}(1-\eta_{2})^{2}}{4}\eta_{eff}^{2}+\frac
{\eta_{1}\eta_{2}(1-\eta_{1})(1-\eta_{2})}{2}\eta_{eff}^{2}+\frac{\eta_{2}%
^{2}(1-\eta_{1})^{2}}{4}\eta_{eff}^{2}.
\end{equation}

We have used the following properties for the calculation of $\hat{\rho}%
_{out}^{(2)}$ and $\mathcal{N}$,%

\begin{equation}
\int d\omega_{s}d\omega_{i}|f(\omega_{s},\omega_{i})|^{2}=1,
\end{equation}
where orthonormal relations in the mode functions are used, and%
\begin{equation}
\int d\omega_{s}d\omega_{s}^{\prime}d\omega_{i}d\omega_{i}^{\prime}%
f(\omega_{s}^{\prime},\omega_{i}^{\prime})f^{\ast}(\omega_{s}^{\prime}%
,\omega_{i})f(\omega_{s},\omega_{i})f^{\ast}(\omega_{s},\omega_{i}^{\prime
})=\sum_{j}\lambda_{j}^{2}.
\end{equation}
Note that the single mode spectral function for the Raman photon satisfies
$\int d\omega|\Phi(\omega)|^{2}=1.$

The fidelity, heralding, and success probability become
\begin{align}
& F=\frac{1+\sum_{j}\lambda_{j}\int\phi_{j}(\omega_{i})\phi_{j}^{\ast}%
(\omega_{i}^{\prime})\Phi^{\ast}(\omega_{i})\Phi^{\ast}(\omega_{i}^{\prime
})d\omega_{i}d\omega_{i}^{\prime}}{\eta_{r}(2-\eta_{t})(1+\sum_{j}\lambda
_{j}^{2})/2+2},\\
& P_{H}=\frac{\eta_{r}\eta_{t}(2-\eta_{t})(1+\sum_{j}\lambda_{j}^{2}%
)/2+2\eta_{t}}{(\sqrt{\eta_{r}}+1/\sqrt{\eta_{r}})^{2}},\\
& P_{S}=\eta_{t}\frac{1+\sum_{j}\lambda_{j}\int\phi_{j}(\omega_{i})\phi
_{j}^{\ast}(\omega_{i}^{\prime})\Phi^{\ast}(\omega_{i})\Phi^{\ast}(\omega
_{i}^{\prime})d\omega_{i}d\omega_{i}^{\prime}}{(\sqrt{\eta_{r}}+1/\sqrt
{\eta_{r}})^{2}},
\end{align}
where $\frac{1-\eta_{2}}{1-\eta_{1}}\approx1$ and $\eta_{r}=\eta_{1}/\eta_{2}$.

The fidelity depends on a sum of square of Schmidt numbers in the denominator
and the mode mismatch between the idler and Raman photons in the numerator.
\ Let us assume that the Raman photon mode is engineered to be matched with
the idler photon mode of the largest Schmidt number ($\phi_{1}(\omega_{i})$ in
our case), which is required to have a larger fidelity (so is the success
probability) compared to other modes. \ We may also compare the NRPD with the
performance of PNRD in the midway detectors, then we have the fidelity,
heralding, and success probability,%

\begin{align}
F  & =\left\{
\begin{array}
[c]{c}%
\frac{1+\lambda_{1}}{\eta_{r}(2-\eta_{t})(1+\sum_{j}\lambda_{j}^{2}%
)/2+2},~\text{NRPD}\\
\frac{1+\lambda_{1}}{\eta_{r}(1-\eta_{t})(1+\sum_{j}\lambda_{j}^{2}%
)+2},~\text{PRND}%
\end{array}
\right. \label{fidelity}\\
P_{H}  & =\left\{
\begin{array}
[c]{c}%
\frac{\eta_{r}\eta_{t}(2-\eta_{t})(1+\sum_{j}\lambda_{j}^{2})/2+2\eta_{t}%
}{(\sqrt{\eta_{r}}+1/\sqrt{\eta_{r}})^{2}},~\text{NRPD}\\
\frac{\eta_{r}\eta_{t}(1-\eta_{t})(1+\sum_{j}\lambda_{j}^{2})+2\eta_{t}%
}{(\sqrt{\eta_{r}}+1/\sqrt{\eta_{r}})^{2}},~\text{PRND}%
\end{array}
\right. \label{herald}\\
P_{S}  & =\left\{
\begin{array}
[c]{c}%
\frac{\eta_{t}(1+\lambda_{1})}{(\sqrt{\eta_{r}}+1/\sqrt{\eta_{r}})^{2}%
},~\text{NRPD}\\
\frac{\eta_{t}(1+\lambda_{1})}{(\sqrt{\eta_{r}}+1/\sqrt{\eta_{r}})^{2}%
},~\text{PRND}%
\end{array}
\right.  .\label{success}%
\end{align}

When the relative efficiency is made arbitrarily small, the fidelity
approaches $(1+\lambda_{1})/2$ for both types of detectors. \ It reaches one
if a pure cascade emission source is generated (von Neumann entropy $E=0$ and
$\lambda_{1}=1$). \ When $\eta_{r}=1$ with a pure source using NRPD with a
perfect quantum efficiency, $F=2/3,~P_{H}=3/4,~P_{S}=1/2,$ which coincide with
the results of the reference \cite{shapiro} (with perfect quantum efficiency).

We discuss the frequency entanglement for various pulse widths and
superradiant decay rates in Chapter 3.4. \ We find that for shorter driving
pulses and smaller superradiant decay rates, the cascade emission source is
less spectrally entangled. \ That means when $\eta_{r}$ is fixed, a shorter
driving pulse heralds a higher fidelity DLCZ entangled state.%

\begin{figure}
[ptb]
\begin{center}
\includegraphics[
natheight=20.000400in,
natwidth=25.499700in,
height=5.6386in,
width=6.4636in
]%
{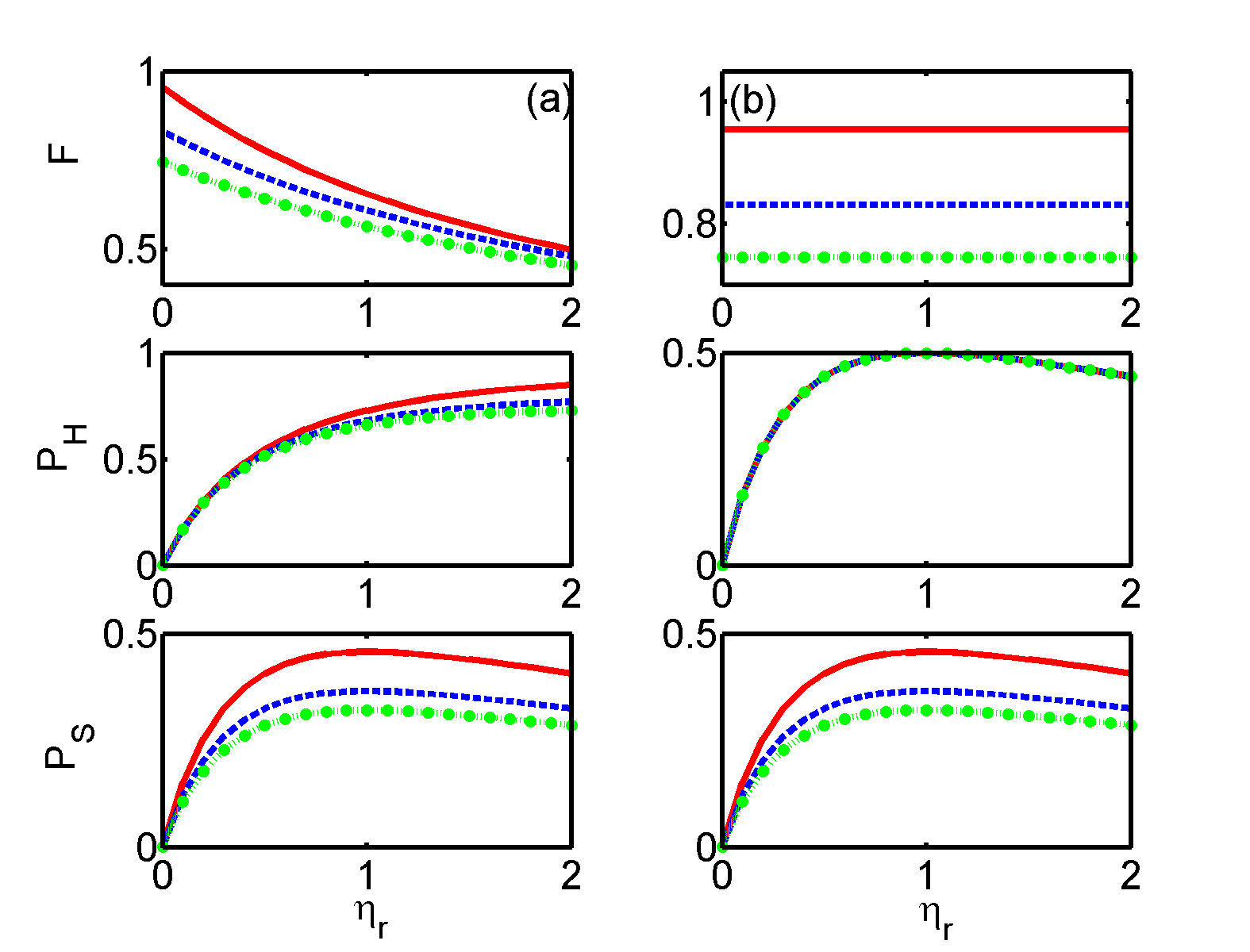}%
\caption{Fidelity $F$, heralding $P_{H}$, and success $P_{S}$ probabilities of
entanglement swapping versus relative efficiency $\eta_{r}$\ with perfect
detection efficiency $\eta_{t}=1.$ \ Column (a) NRPD and (b) PNRD.
\ Solid-red, dashed-blue, and dotted-green curves correspond to the pulse
width parameters $\tau=(0.1,0.5,0.5)$ and superradiant factor $N\bar{\mu
}+1=(5,5,10)$ (see Chapter 3 and Appendix A)$.$ \ The von Neumann entropy is
$S=(0.684,2.041,2.886),$ respectively.}%
\label{F_etar}%
\end{center}
\end{figure}

In Figure \ref{F_etar}, we numerically calculate the entropy and plot out the
fidelity from Eq. (\ref{fidelity}), the heralding probability from Eq.
(\ref{herald}), and the success probability from Eq. ( \ref{success}) as a
function of the relative efficiency $\eta_{r}.$ \ With a perfect detection
efficiency ($\eta=1$), we find that at a smaller $\eta_{r},$ the less
entangled source gives us a higher fidelity DLCZ entangled state but with a
smaller success probability. \ Small generation probability for cascade
emission ($\eta_{r}<1$) reduces the error of NRPD from two telecom photons
interference, but it reduces the successful entanglement swapping at the same time.

The optimal success probability occurs by using the same excitation efficiency
for both cascade and Raman configurations. \ For PNRD, the fidelity is higher
than NRPD, and the heralding probability is the same independent of the degree
of frequency space entanglement. \ The success probabilities for both types of
detectors are equal. \ The advantage of PNRD shows up in the fidelity of
quantum swapping.

\begin{figure}
[ptb]
\begin{center}
\includegraphics[
natheight=20.000400in,
natwidth=25.499700in,
height=5.6239in,
width=6.4809in
]%
{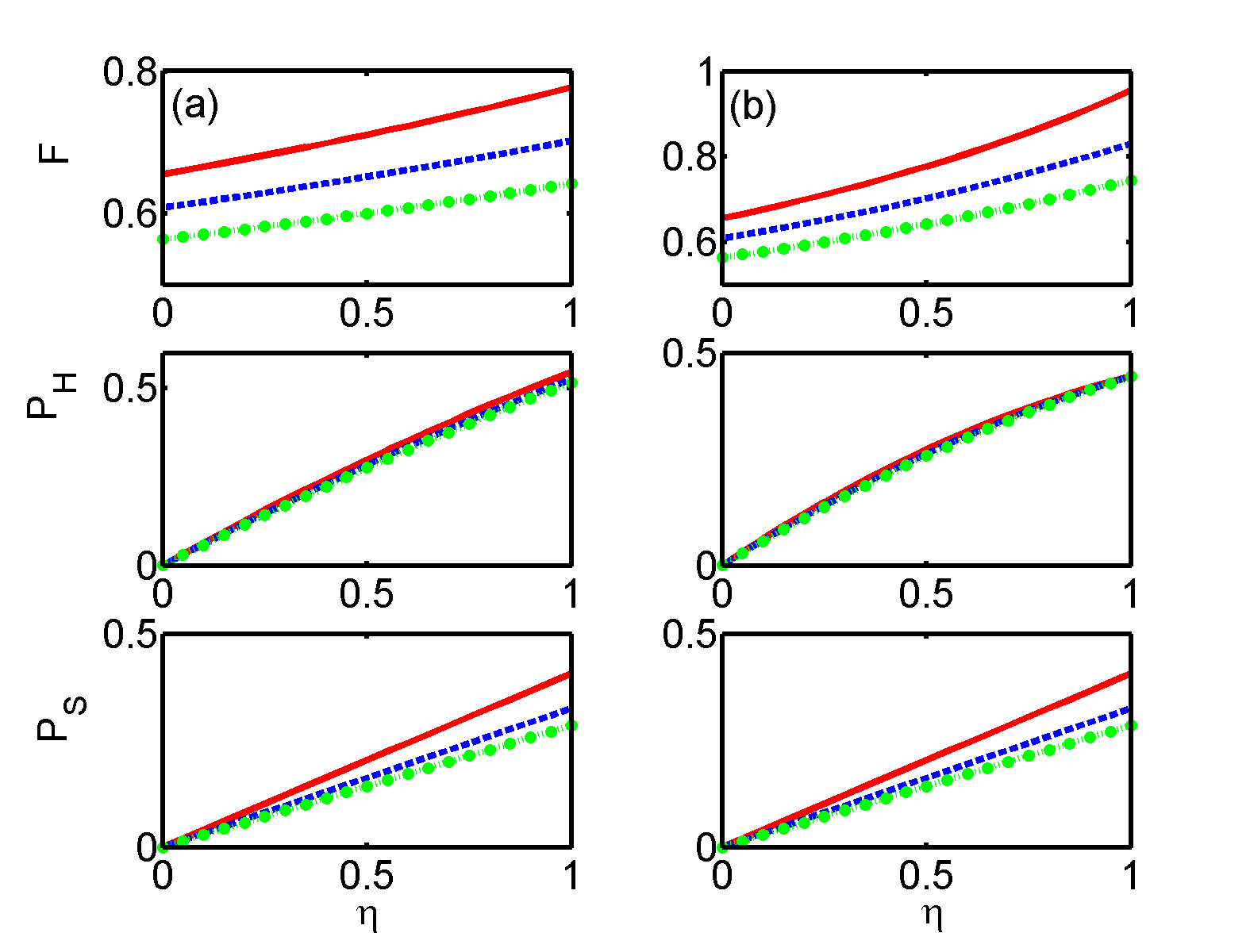}%
\caption{Fidelity $F$, heralding $P_{H}$, and success \ $P_{S}$ probabilities
of entanglement swapping versus telecom detector quantum efficiency $\eta$ for
the case of (a) NRPD and (b) PNRD. \ Solid-red, dashed-blue, and dotted-green
curves correspond to the same parameters used in Figure \ref{F_etar}.}%
\label{F_eta}%
\end{center}
\end{figure}

In Figure \ref{F_eta}, we show that the measures improve monotonically with
the quantum efficiency ($\eta=\eta_{t}$) of the detector at telecom
wavelength, with $\eta_{r}=0.5$. \ The success probabilities for both types of
detectors are the same and again the advantage of PNRD shows up in the fidelity.%

\section{Polarization Maximally Entangled State (PME State) and Quantum
Teleportation}

\begin{figure}
[ptb]
\begin{center}
\includegraphics[
natheight=7.499600in,
natwidth=9.999800in,
height=5.7086in,
width=5.9421in
]%
{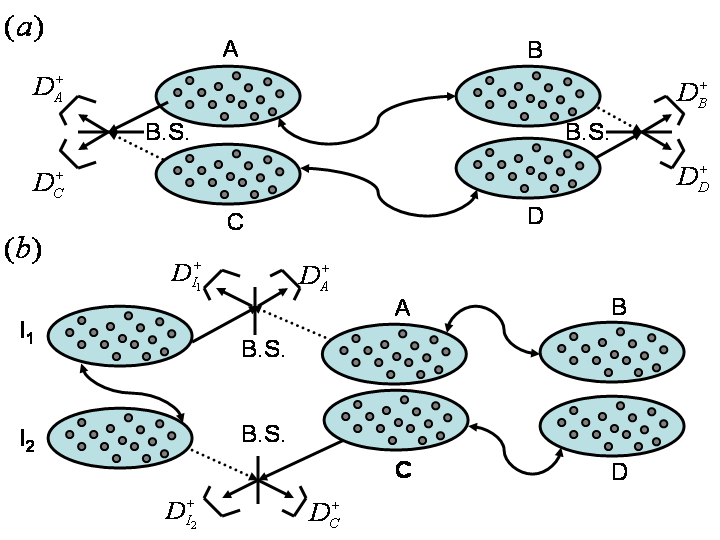}%
\caption{PME projection (a) and quantum teleportation (b) in the DLCZ scheme.
\ Four atomic ensembles\ (A,B,C,D) are used to generate two DLCZ entangled
states at (A,B) and (C,D). \ PME state is projected probabilistically
conditioned on four possible detection events of ($D_{A}^{\dagger}$ or
$D_{C}^{\dagger}$) and ($D_{B}^{\dagger}$ or $D_{D}^{\dagger}$) in (a). \ In
the quantum teleportation protocol (b), another two ensembles (I$_{1},$I$_{2}%
$) are used to prepare a quantum state that is teleported to atomic ensembles
B and D conditioned on four possible detection events of ($\hat{D}_{I_{1}}$ or
$\hat{D}_{A}$) and ($\hat{D}_{I_{2}}$ or $\hat{D}_{C}$).}%
\label{pme_qt}%
\end{center}
\end{figure}

In Figure \ref{pme_qt}, we illustrate schematically a scheme for probabilistic
PME state preparation and quantum teleportation. \ Four ensembles (ABCD) are
used to generate two entangled pairs of DLCZ entangled states, and another two
ensembles ($I_{1},~I_{2}$) are used to prepare a quantum state to be teleported.%

With the conditional output density matrix from Eq. (\ref{out}), we proceed to
construct the PME state $|\Psi\rangle_{PME}=\frac{1}{\sqrt{2}}(\hat{S}%
_{A}^{\dag}\hat{S}_{D}^{\dag}+\hat{S}_{B}^{\dag}\hat{S}_{C}^{\dag})|0\rangle$
where $(C,D)$ represents another parallel entanglement connection setup,
Figure \ref{pme_qt} (a). \ This PME state is useful in entanglement-based
communication schemes \cite{dlcz}, and we will here calculate its success
probability. \ The normalized density matrix for the AB\ system is from Eq.
(\ref{out3}) (let $\eta_{t}=\eta$),%

\begin{align}
\hat{\rho}_{out,n}^{(2),AB} &  =\frac{a}{a+4}|0\rangle\langle0|+\frac{2}%
{a+4}\Big(\hat{S}_{B}^{\dag}|0\rangle\langle0|\hat{S}_{B}+\hat{S}_{A}^{\dag
}|0\rangle\langle0|\hat{S}_{A}\nonumber\\
&  +\lambda_{1}\hat{S}_{B}^{\dag}|0\rangle\langle0|\hat{S}_{A}+\lambda_{1}%
\hat{S}_{A}^{\dag}|0\rangle\langle0|\hat{S}_{B}\Big),
\end{align}
where the largest Schmidt number ($\lambda_{1}$) of mode overlap is chosen and
$a\equiv\eta_{r}(2-\eta)\left(  1+\sum_{j}\lambda_{j}^{2}\right)  $.

A parallel pair of entangled ensembles (C,D) is introduced, and the joint
density operator is $\hat{\rho}_{out,n}^{(2),AB}\otimes\hat{\rho}%
_{out,n}^{(2),CD}.$ \ The latter expression is developed mathematically in
Appendix C.3.

With projection of the PME state, we have the post measurement success
probability [a click from each side; the side of (A or C) and (B or D)],%

\begin{align}
P_{S,PME}  & =\langle\Psi|\hat{\rho}_{out,n}^{(2),AB}\otimes\hat{\rho}%
_{out,n}^{(2),CD}|\Psi\rangle_{PME},\nonumber\\
& =\frac{4(1+\lambda_{1}^{2})}{[\eta_{r}(2-\eta_{t})(1+\sum_{j}\lambda_{j}%
^{2})+4]^{2}}.
\end{align}
\ For $\eta_{r}\ll1$, $P_{S,PME}$ reaches the maximum of $1/2$ when a pure
source ($\lambda_{1}=1$) is used.

For an arbitrary quantum state transfer to long distance, quantum
teleportation scheme may be used. \ Another two ensembles ($I_{1},I_{2}$) are
introduced \cite{dlcz}, and the quantum state can be described by
$|\Psi\rangle=(d_{0}\hat{S}_{I_{1}}^{\dag}+d_{1}\hat{S}_{I_{2}}^{\dag
})|0\rangle$ with $|d_{0}|^{2}+|d_{1}|^{2}=1$. \ The joint density matrix for
quantum teleportation is%

\begin{equation}
\hat{\rho}_{QT}=(d_{0}\hat{S}_{I_{1}}^{\dag}+d_{1}\hat{S}_{I_{2}}^{\dag
})|0\rangle\langle0|(d_{0}^{\ast}\hat{S}_{I_{1}}+d_{1}^{\ast}\hat{S}_{I_{2}%
})\otimes\hat{\rho}_{out,n}^{(2),AB}\otimes\hat{\rho}_{out,n}^{(2),CD}.
\end{equation}

Atomic ensembles (A,B) in parallel with (C,D) provide a scheme for PME state
preparation. \ Retrieve the quantum state [ensemble ($I_{1},I_{2}$)] into
photons and interfere them at B.S., respectively, with photons from A and C.
We have the teleported quantum state at B and D conditioned on the single
click of ($\hat{D}_{I_{1}}$ or $\hat{D}_{A}$) and ($\hat{D}_{I_{2}}$ or
$\hat{D}_{C}$).

Consider single detection events at $\hat{D}_{I_{1}}$ and $\hat{D}_{I_{2}}$ as
an example. \ With the NRPD measurement operators $\hat{M}_{I_{1},I_{2}}%
\equiv(\hat{I}_{D1}^{\dag}-|0\rangle_{D1}\langle0|)\otimes|0\rangle_{D_{A}%
}\langle0|\otimes(\hat{I}_{D2}^{\dag}-|0\rangle_{D2}\langle0|)\otimes
|0\rangle_{D_{C}}\langle0|$ (we use $D_{1},D_{2}$ for $D_{I_{1}},D_{I_{2}}$),
the density matrix after the measurement becomes%

\begin{align}
&  \hat{\rho}_{1}\equiv\text{Tr}(\hat{\rho}_{QT,eff}\hat{M}_{I_{1},I_{2}%
})=\nonumber\\
&  \frac{a+2}{2(a+4)^{2}}|0\rangle_{ABCD}\langle0|+\frac{4}{(a+4)^{2}%
}\Big(\frac{|d_{0}|^{2}}{4}\hat{S}_{B}^{\dag}|0\rangle\langle0|\hat{S}%
_{B}+\frac{|d_{1}|^{2}}{4}\hat{S}_{D}^{\dag}|0\rangle\langle0|\hat{S}%
_{D}+\nonumber\\
&  \frac{\lambda_{1}^{2}d_{0}d_{1}^{\ast}}{4}\hat{S}_{B}^{\dag}|0\rangle
\langle0|\hat{S}_{D}+\frac{\lambda_{1}^{2}d_{0}^{\ast}d_{1}}{4}\hat{S}%
_{D}^{\dag}|0\rangle\langle0|\hat{S}_{B}\Big),
\end{align}
where $\hat{\rho}_{QT,eff}$ is calculated in Eq. (\ref{QT}), and the trace is
taken over the electromagnetic field degrees of freedom.

For a successful transfer of the quantum state $|\Phi\rangle=(d_{0}\hat{S}%
_{B}^{\dag}+d_{1}\hat{S}_{D}^{\dag})|0\rangle$, the fidelity $F_{1}%
=\langle\Phi|\hat{\rho}_{1}|\Phi\rangle/$Tr$(\hat{\rho}_{1}),$ and the
heralding probability is$~P_{1}=\text{Tr}(\hat{\rho}_{1})$, with the trace
over all atomic degrees of freedom. \ Except for the detection event we
consider here, there are three other detection events including ($D_{A}%
,~D_{C}$), ($D_{I_{1}},~D_{C}$) and ($D_{A},$ $D_{I_{2}}$). \ The teleported
state from the detection events\ ($D_{I_{1}},~D_{C}$) and ($D_{I_{2}},~D_{A}$)
requires a $\pi$ rotation correction\ on the relative phase ($d_{0}\rightarrow
d_{0},$ $d_{1}\rightarrow-d_{1}$).

The fidelity and heralding probabilities conditioned on the other three pairs
of clicks are the same as $F_{1}$ and $P_{1}$ respectively, so the success
probability is
\begin{align}
P_{S,QT}  & =\sum_{i}^{4}P_{i}F_{i}=4P_{1}F_{1},\nonumber\\
& =\frac{F^{2}}{(1+\lambda_{1})^{2}}[1+(2\lambda_{1}^{2}-2)|d_{0}|^{2}%
|d_{1}|^{2}],
\end{align}
where $F$ is the fidelity of entanglement swapping for NRPD, Eq.
(\ref{fidelity}). \ For PNRD, the success probability for quantum
teleportation is unchanged.

\begin{figure}
[ptb]
\begin{center}
\includegraphics[
natheight=18.000400in,
natwidth=24.499700in,
height=5.0462in,
width=6.3105in
]%
{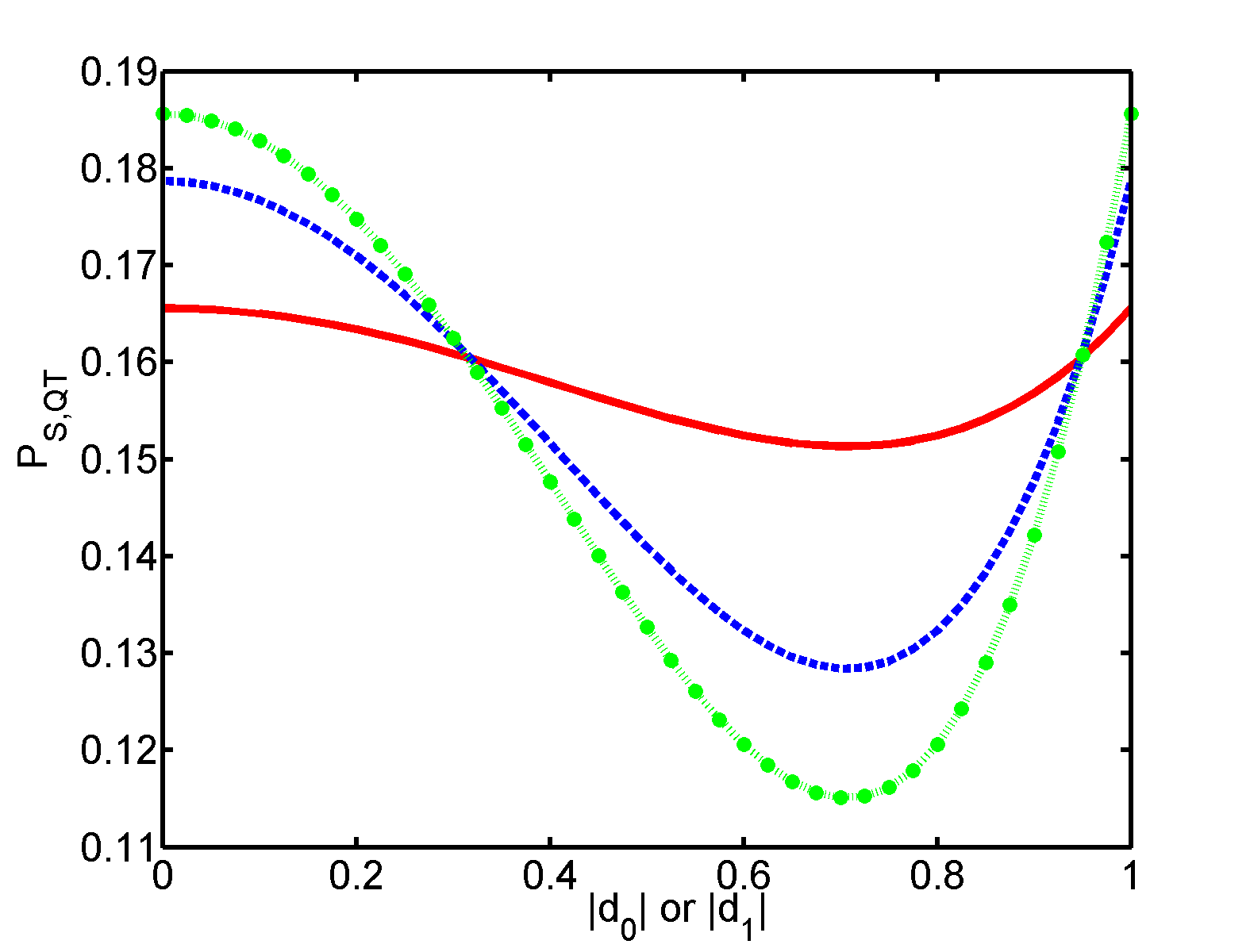}%
\caption{Success probability of quantum teleportation as a function of the
probability amplitude of teleported quantum state with $\eta_{r}=0.5$ and a
perfect detector efficiency $\eta_{t}=1.$ Solid-red, dashed-blue, and
dotted-green curves correspond to the same parameters used in Figure
\ref{F_etar}.}%
\label{P_QT}%
\end{center}
\end{figure}

The success probability for quantum teleportation depends on the probability
amplitude of the quantum state and the fidelity $F$ of the entanglement
swapping. \ In Figure \ref{P_QT}, for $\eta_{r}=0.5$ and $\eta_{t}=1$, we can
see in the region $|d_{0}|\approx0.3\sim0.9$, higher success probability
requires a less entangled cascade emission source. \ Outside this region, it
prefers a more entangled source. \ When a pure source is used ($\lambda_{1}%
=1$) and let $\eta_{r}\ll1,$ $\eta_{t}=1$, we can achieve the maximum of the
success probability $P_{S,QT}=\frac{1}{4}$ when $F=1$, which is also achieved
in the traditional DLCZ scheme with perfect quantum efficiencies
\cite{shapiro}.%

\section{Conclusion}

We have described probabilistic protocols for the DLCZ scheme implementing the
cascade emission source.\ \ We characterize the spectral properties of the
cascade emission by Schmidt mode analysis and investigate the fidelity and
success probability of the protocols using photon resolving and non-resolving
photon detectors. \ The success probability is independent of the detector
type, but photon number resolving \ detection improves the fidelity. \ 

The performance of the protocol also depends on the ratio of efficiencies in
generating the cascade and Raman photons. \ The success probability is
optimized for equal efficiencies while the fidelity is higher when the ratio
is smaller than one for non-resolving photon detectors.

The frequency space entanglement of telecom photons produced in cascade
emission deteriorates the performance of DLCZ protocols. \ The harmful effect
can be diminished by using shorter pump pulses to generate the cascade
emission. \ A state dependent success probability of quantum teleportation was
calculated, and in some cases a more highly frequency entangled cascade
emission source teleports more successfully. \ An improved performance could
be achieved if the error source (vacuum part) were removed. \ This could be
done by entanglement purification \cite{QI2} at the stage of entanglement
swapping and then using the purified source to teleport the quantum state.



\chapter{Efficiency of light-frequency conversion in an atomic ensemble}

In this Chapter \footnote{This Chapter is based on reference \cite{conversion}%
.}, the efficiency of frequency up and down conversion of light in an atomic
ensemble, with a diamond level configuration, is analyzed theoretically. The
conditions of pump field intensities and detunings required to maximize the
conversion as a function of optical thickness of the ensemble are determined.
The influence of the probe pulse duration on the conversion efficiency is
investigated by the numeric solution of the Maxwell-Bloch equations.\ \ The
set of equations are similar to those in Chapter 4, but a c-number version of
the interaction is considered here. \ The properties of absorption and
dispersion of fields are extracted from the steady state solutions to
demonstrate the parametric coupling between the fields. \ We will show that,
in calculating conversion efficiency, a quantum version of the equation
including Langevin noises is equivalent to the c-number one. \ Frequency
conversion provides the bridge for transmitted qubit (telecommunication
wavelength) and local quantum memory (near-infrared light), in which a large
scale quantum communication can be fulfilled.

In Section II, we discuss the four-wave mixing process and present solutions
for the up- and down- converted fields. \ The dressed state picture is used as
a guide to understand the characteristic features of the absorption and
signal-idler field coupling. \ In Section III, we present the results of an
optimization in conversion efficiency as a function of the optical depth of
the atomic ensemble. \ In Section IV, we investigate the effects of a finite
pulse duration by numerically integrating the Maxwell-Bloch equations.
\ Section V demonstrates the results of Langevin noise correlations and we
conclude in Section VI. The derivations of the Maxwell-Bloch and parametric
equations are relegated to the Appendix D.

\section{Introduction}

The frequency conversion of light fields has been an important theme in
optical physics for around half a century. In quantum information physics the
conversion of single photons to and from the telecom wavelength band is a
topic of more recent vintage, and is motivated by the desire to minimize
optical fiber transmission losses when distributing entangled states over
distant quantum memory elements in a quantum repeater \cite{repeater}.

\begin{figure}
[ptb]
\begin{center}
\includegraphics[
natheight=7.499600in,
natwidth=9.999800in,
height=3.2292in,
width=4.5463in
]%
{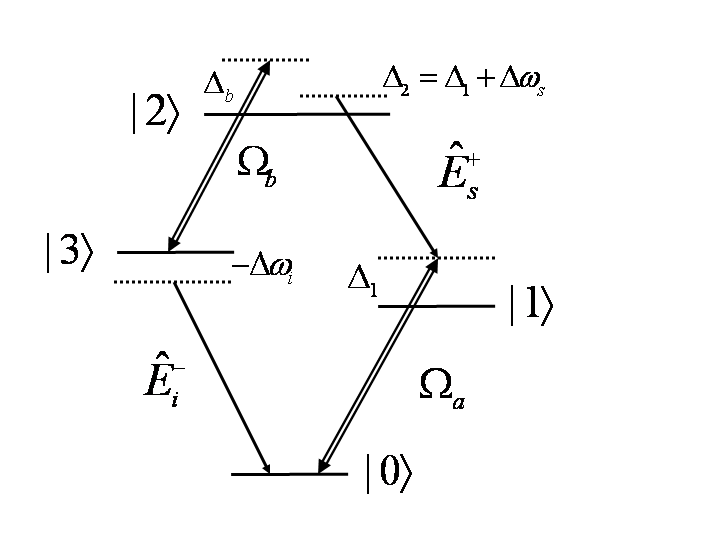}%
\caption{The diamond configuration of atomic system for conversion scheme. Two
pump lasers (double line) with Rabi frequencies $\Omega_{a},\Omega_{b}$ and
propagated probe fields (single line) $E_{s}^{+},E_{i}^{+}$ interact with the
atomic medium. Various detunings are defined in the Appendix D, and the atomic
levels used in the experiment \cite{radaev} are $(|0\rangle,|1\rangle
,|2\rangle,|3\rangle)=(|5\text{S}_{1/2},\text{F}=1\rangle,|5\text{P}%
_{3/2},\text{F}=2\rangle,|6\text{S}_{1/2},\text{F}=1\rangle,|5\text{P}%
_{1/2},\text{F}=2\rangle).$}%
\label{conversion}%
\end{center}
\end{figure}

An associated technical problem is that telecom light is not readily stored in
ground level atomic memory coherences. Retrieval processes in atomic
ensembles, for example using electromagnetically induced transparency
\cite{eit}, or more specifically the dark-polariton mechanism
\cite{Lukin,polariton}, generate shorter wavelength radiation correlated to
the stored atomic excitation by Raman scattering. Such radiation, optically
resonant to the ground level of typical atoms and ions, has been retrieved in
numerous experiments
\cite{qubit,chou,collective,vuletic,store,single2,pan,kimble,ionent,ion}. An
important advance would involve generation of atomic memory coherences
quantum-correlated with telecom wavelength radiation, thereby minimizing
transmission losses over long distances. Recently there has been a
breakthrough in this direction using a pair of cold, non-degenerate rubidium
gas samples \cite{radaev}. The stored excitation is correlated with an
infra-red field (idler) in one gas sample, and the idler is then frequency
converted to a telecom wavelength signal field in the other ensemble. The
frequency conversion mechanism involves the diamond configuration of atomic
levels shown in Figure \ref{conversion}.

In a probabilistic protocol it is important to maximize all efficiencies,
e.g., fiber transmission, single-photon detection, and quantum memory lifetime
\cite{ran}. In the present work we investigate the efficiency of frequency up-
and down- conversion in the diamond atomic configuration \cite{telecom,orozco}%
, as a function of the ensemble's optical thickness, and the intensity and
detuning of the pump fields involved in the near-resonant, four-wave mixing process.%

\section{Theory}

We consider a cold and cigar-shaped $^{87}$Rb atomic ensemble with
co-propagating light fields similar to the experimental setup in the reference
\cite{radaev}.

The conversion scheme shown in Figure \ref{conversion} involves two pump
lasers with frequencies $\omega_{a}$ and $\omega_{b}$, respectively; their
Rabi frequencies are given by $\Omega_{a}$ and $\Omega_{b}$. Two weak probe
fields, signal and idler, with frequency $\omega_{s}$ and $\omega_{i} $,
respectively, propagate through the optically thick atomic medium. Unlike the
cascade driving scheme, where two-photon excitation generates a photon pair
spontaneously \cite{telecom}, pump laser b experiences a transparent medium if
both the signal and idler fields are in the vacuum state. With an incident
signal field, four-wave mixing with the pumps generates an up-converted idler
field, while an incident idler field generates a down-converted signal.

The Maxwell-Bloch equations for the interacting system of light and four light
fields is derived in the Appendix D. \ By linearizing the equations with
respect to the signal and idler field amplitudes, and adiabatically
eliminating the atoms, one arrives at coupled parametric equations for the
signal and idler fields. We discuss their solution in this section, and leave
numerical solutions of the Maxwell Bloch equations to Section IV.

The calculation of conversion efficiencies can also be carried out with the
quantized Heisenberg-Langevin version of the coupled parametric equations,
which we will show in Section V. \ The resulting conversion efficiencies are
identical to the semiclassical treatment; the additional quantum noise
contributions vanish as the $|2\rangle\rightarrow|3\rangle$ transition driven
by pump laser b has vanishing populations and atomic coherence. A similar
simplification occurs in the calculation of the storage efficiency of spin
waves in a system of atoms in the $\Lambda$ configuration
\cite{gorshkov,quantum_interface}.

The co-moving propagation equation for c-number signal and idler fields
(respectively, ${E}_{s}^{+}$ and ${E}_{i}^{+}$) under energy conservation
($\Delta\omega=\omega_{a}+\omega_{s}-\omega_{b}-\omega_{i}=0$) and four-wave
mixing conditions ($\Delta k=k_{a}-k_{s}+k_{b}-k_{i}=0$) are%

\begin{align}
\frac{\partial}{\partial z}{E}_{s}^{+} &  =\beta_{s}{E}_{s}^{+}+\kappa_{s}%
{E}_{i}^{+}\nonumber\\
\frac{\partial}{\partial z}{E}_{i}^{+} &  =\kappa_{i}{E}_{s}^{+}+\alpha_{i}%
{E}_{i}^{+}.
\end{align}

The coupled equations are similar to those found for the double $\Lambda$
system \cite{braje,couple}. The self-coupling coefficients $\beta_{s},$
$\alpha_{i}$ and parametric coefficients $\kappa_{s}$, $\kappa_{i}$ are
defined in Appendix D. \ The set of equations can be simplified as%

\begin{equation}
\frac{\partial}{\partial z}x(z)=Ax
\end{equation}
where%

\begin{equation}
x=%
\begin{pmatrix}
E_{s}^{+}\\
E_{i}^{+}%
\end{pmatrix}
,A=\left(
\begin{array}
[c]{cc}%
\beta_{s} & \kappa_{s}\\
\kappa_{i} & \alpha_{i}%
\end{array}
\right)  .
\end{equation}

\bigskip The equations are solved by considering a similarity transformation
$S$ that $\Lambda=S^{-1}AS$ is diagonalized and $y=S^{-1}x$ such that
\begin{align}
\frac{\partial}{\partial z}y  & =\Lambda y\\
y(z)  & =e^{\Lambda(z-z_{0})}y(z_{0})
\end{align}
where $y(z_{0})$ is the boundary condition. \ With the known boundary
condition $x_{1}(0)$ and $x_{2}(0)$ where we choose the input face of
propagation as $z_{0}=0$, we have
\begin{equation}
x(z)=Se^{\Lambda z}S^{-1}x(0).
\end{equation}

And the diagonalized and transformation matrix are
\begin{align}
\Lambda & =\left(
\begin{array}
[c]{cc}%
(\alpha_{i}+\beta_{s})/2+w & 0\\
0 & (\alpha_{i}+\beta_{s})/2-w
\end{array}
\right)  ,\\
S  & =\left(
\begin{array}
[c]{cc}%
q+w & \kappa_{s}\\
\kappa_{i} & -q-w
\end{array}
\right)  ,~\\
S^{-1}  & =\frac{1}{2w(w+q)}\left(
\begin{array}
[c]{cc}%
q+w & \kappa_{s}\\
\kappa_{i} & -q-w
\end{array}
\right)
\end{align}
where $w\equiv\sqrt{q^{2}+\kappa_{s}\kappa_{i}}$, and $q\equiv(-\alpha
_{i}+\beta_{s})/2$. \ 

The solution of fields from down conversion is%

\begin{equation}%
\begin{bmatrix}
E_{s}^{+}(L)\\
E_{i}^{+}(L)
\end{bmatrix}
=Se^{\Lambda L}S^{-1}%
\begin{bmatrix}
0\\
E_{i}^{+}(0)
\end{bmatrix}
=\frac{E_{i}^{+}(0)e^{(\alpha_{i}+\beta_{s})L/2}}{2w}%
\begin{bmatrix}
\kappa_{s}(e^{wL}-e^{-wL})\\
\frac{1}{(w+q)}[\kappa_{s}\kappa_{i}e^{wL}+(q+w)^{2}e^{-wL}]
\end{bmatrix}
.
\end{equation}

Similarly, the solution of fields from up conversion is%

\begin{equation}%
\begin{bmatrix}
E_{s}^{+}(L)\\
E_{i}^{+}(L)
\end{bmatrix}
=Se^{\Lambda L}S^{-1}%
\begin{bmatrix}
E_{s}^{+}(0)\\
0
\end{bmatrix}
=\frac{E_{s}^{+}(0)e^{(\alpha_{i}+\beta_{s})L/2}}{2w}%
\begin{bmatrix}
\frac{1}{(w+q)}[(q+w)^{2}e^{wL}+\kappa_{s}\kappa_{i}e^{-wL}]\\
\kappa_{i}(e^{wL}-e^{-wL})
\end{bmatrix}
.
\end{equation}

We define the down conversion efficiency $\eta_{\text{d}}$ and transmission of
input idler field $T_{\text{d}}$ as%

\begin{align}
\eta_{\text{d}}  & =\left\vert \frac{E_{s}^{+}(L)}{E_{i}^{+}(0)}\right\vert
^{2}=\left\vert \frac{\kappa_{s}}{2w}e^{(\alpha_{i}+\beta_{s})L/2}%
(e^{wL}-e^{-wL})\right\vert ^{2}\label{down}\\
T_{\text{d}}  & =\left\vert \frac{E_{i}^{+}(L)}{E_{i}^{+}(0)}\right\vert
^{2}=\left\vert \frac{e^{(\alpha_{i}+\beta_{s})L/2}}{2w(w+q)}[\kappa_{s}%
\kappa_{i}e^{wL}+(q+w)^{2}e^{-wL}]\right\vert ^{2}.
\end{align}

Similarly the up conversion efficiency $\eta_{\text{u}}$ and transmission of
input signal field $T_{\text{u}}$ is%

\begin{align}
\eta_{\text{u}}  & =\left\vert \frac{E_{i}^{+}(L)}{E_{s}^{+}(0)}\right\vert
^{2}=\left\vert \frac{\kappa_{i}}{2w}e^{(\alpha_{i}+\beta_{s})L/2}%
(e^{wL}-e^{-wL})\right\vert ^{2}\label{up}\\
T_{\text{u}}  & =\left\vert \frac{E_{s}^{+}(L)}{E_{s}^{+}(0)}\right\vert
^{2}=\left\vert \frac{e^{(\alpha_{i}+\beta_{s})L/2}}{2w(w+q)}[(q+w)^{2}%
e^{wL}+\kappa_{s}\kappa_{i}e^{-wL}]\right\vert ^{2}.
\end{align}
%

\begin{figure}
[ptb]
\begin{center}
\includegraphics[
natheight=7.499600in,
natwidth=9.999800in,
height=3.4169in,
width=4.5463in
]%
{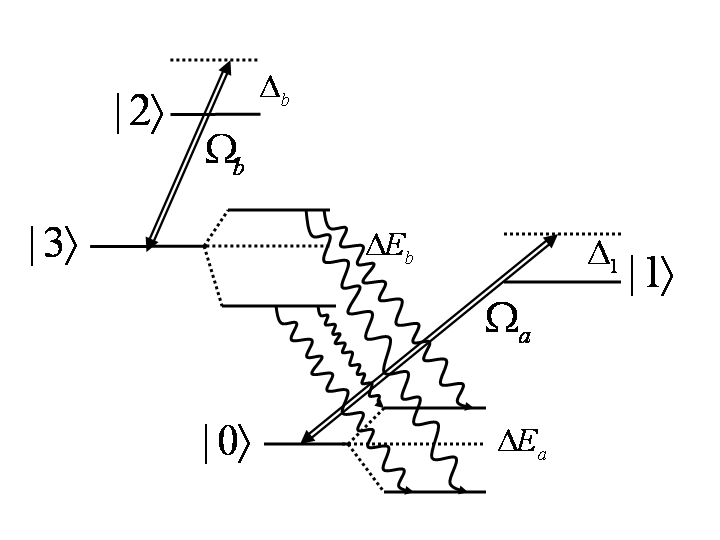}%
\caption{Dressed-state picture from the perspective of the probe idler
transition between atomic levels $|0\rangle$ and $|3\rangle.$ \ Two strong
fields $\Omega_{a},\Omega_{b}$ shift the levels with energy $\Delta E_{a,b}$
and wavy lines represent the idler field resonances.}%
\label{dress}%
\end{center}
\end{figure}

The above is the central result of this section. \ The up and down conversion
efficiencies differ only in the parametric coupling coefficients $\kappa_{i}%
$\ and $\kappa_{s}.$ \ In the strong parametric coupling regime where
$|\kappa_{i}|,|\kappa_{s}|\gg|\alpha_{i}|,|\beta_{s}|$, the coefficients can
be simplified to $\eta_{\text{u}}\simeq\sqrt{\frac{\kappa_{i}}{\kappa_{s}}%
}\sinh(\sqrt{\kappa_{s}\kappa_{i}}L),\eta_{\text{d}}\simeq\sqrt{\frac
{\kappa_{s}}{\kappa_{i}}}\sinh(\sqrt{\kappa_{s}\kappa_{i}}L)$ and
$T_{\text{u}}=T_{\text{d}}\simeq\cosh(\sqrt{\kappa_{s}\kappa_{i}}L).$ \ Under
the further assumptions $\alpha_{i}=\beta_{s}=0$ and $\kappa_{i},\kappa_{s}$
are pure imaginary, we find $\eta_{\text{u}}=\eta_{\text{d}}$ $=\sin
^{2}[\operatorname{Im}(\kappa_{s}L)]$ and $T_{\text{u}}=T_{\text{d}}=\cos
^{2}[\operatorname{Im}(\kappa_{s}L)]$. This result was recently derived by
Gogyan using a dressed state approach \cite{gogyan08}, in the case of resonant
pump fields $\Delta_{1}=\Delta_{b}=0$ \cite{gogyan}. In this ideal limit there
is a conservation condition $\eta_{\text{u}}+T_{\text{u}}=\eta_{\text{d}%
}+T_{\text{d}}=1.$ \ The parametric coupling coefficients are not identical,
but in the regime of strong coupling they approach each other. As noted by
Gogyan, when the pump-a intensity is large ($\Omega_{a}>>|\Delta_{1}%
|,\gamma_{03}$) the $|0\rangle\leftrightarrow|1\rangle$ is saturated, the
atomic coherence is negligible and $\kappa_{s}\approx\kappa_{i}\propto
\tilde{\sigma}_{00,s}(\frac{\Omega_{a}^{\ast}\Omega_{b}}{T_{02}}+\frac
{\Omega_{a}^{\ast}\Omega_{b}}{T_{13}})$. Alternatively, in the limit
$\Omega_{b}>>\Omega_{a},\gamma_{32}$ and $|\Delta_{1}|>>\gamma_{01}$ the
atomic coherence of $|0\rangle\leftrightarrow|1\rangle$ dominates and once
again $\kappa_{s}\approx\kappa_{i}\propto\frac{i\Omega_{b}|\Omega_{b}%
|^{2}\tilde{\sigma}_{01,s}^{\dag}}{T_{13}T_{02}}$. \ Note that this scheme is
also similar to the frequency conversion in nonlinear materials \cite{Kumar}.

The ac-Stark splitting induced by the pump lasers shifts the resonant
absorption condition for the idler and signal fields. The idler and signal
experience resonant absorption at the transition frequency of the dressed
atom. The corresponding transitions for the idler are shown in Figure
\ref{dress}. The bare states are shifted by $\Delta E_{a}=\left\vert
\Delta_{1}\pm\sqrt{\Delta_{1}^{2}+4\Omega_{a}^{2}}\right\vert /2$ and $\Delta
E_{b}=\left\vert \Delta_{b}\pm\sqrt{\Delta_{b}^{2}+4\Omega_{b}^{2}}\right\vert
/2,$ respectively. Note that our Rabi frequencies are smaller by a factor 2
than the standard definitions to avoid a plethora of prefactors in the
equations of the Appendix.

For resonant pump fields, $\Delta E_{a,b}=\pm\Omega_{a,b}$ . The idler
transition resonances are at $\Delta\omega_{i}=-(\Omega_{a}+\Omega_{b}),$
$-\left\vert \Omega_{a}-\Omega_{b}\right\vert ,$ $\left\vert \Omega_{a}%
-\Omega_{b}\right\vert ,$ $(\Omega_{a}+\Omega_{b})$ and these delineate three
windows separated by these four absorption peaks. For $\Omega_{a}>\Omega_{b}$
the centers of these windows are at $-\Omega_{a},$ $0,$ and $\Omega_{a}$,
respectively. Choosing the idler detuning $\Delta\omega_{i}=\pm\Omega_{a}$ as
in Ref. \cite{gogyan}, the idler interacts with the atomic medium at the
center of the left or right window.%

\begin{figure}
[ptb]
\begin{center}
\includegraphics[
natheight=18.000400in,
natwidth=24.499700in,
height=4.7132in,
width=6.1791in
]%
{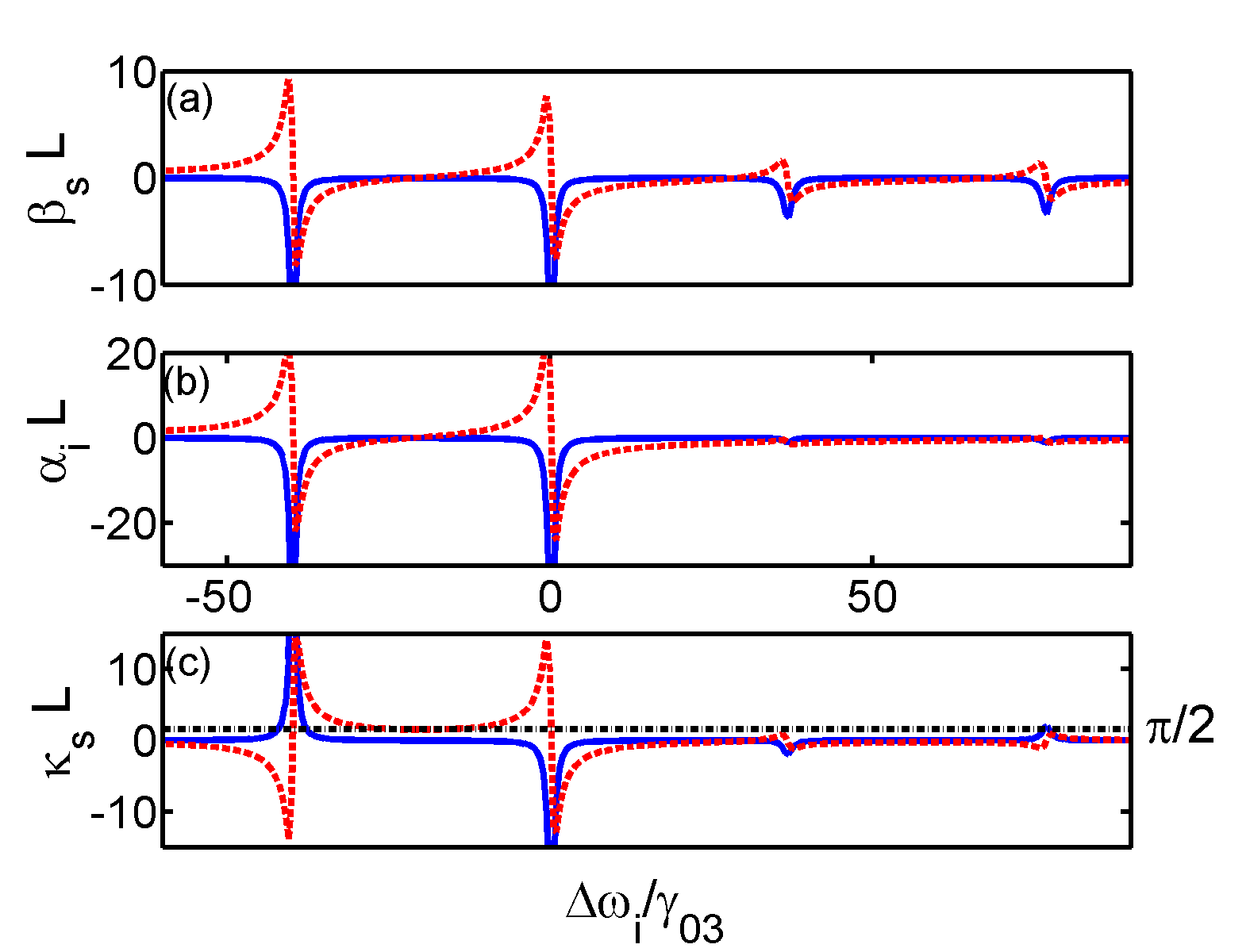}%
\caption{Self-coupling coefficients $\beta_{s},\alpha_{i}$ and cross-coupling
coefficient $\kappa_{s}$. Dimensionless quantities (a) $\beta_{s}L$, (b)
$\alpha_{i}L$ and (c) $\kappa_{s}L$ with real (solid blue) and imaginary
(dashed red) parts are plotted as a dependence of idler detuning $\Delta
\omega_{i}$ [same label in (b)] showing four absorption peaks to construct
three parametric coupling windows. A black dashed-dot line of the constant
$\pi/2$ is added in (c) to demonstrate the crossover with $\Im(\kappa_{s}L)$
indicating the ideal conversion efficiency condition in the left window. The
parameters we use are ($\Omega_{a}$, $\Omega_{b}$, $\Delta_{1}$, $\Delta_{b}$,
$\Delta\omega_{i}$) $=(33$, $20$, $39$, $2$, $-21)\gamma_{03}$ for optical
depth $\rho\sigma L=150$ with $L=6$mm. Various natural decay rates are
$\gamma_{03}=1/27.7\text{ns}$, $\gamma_{01}=1/26.24\text{ns}$, $\gamma
_{12}=\gamma_{03}/2.76,$ and $\gamma_{32}=\gamma_{03}/5.38$ \cite{gsgi}.}%
\label{coefficient}%
\end{center}
\end{figure}

As an example of the strong coupling windows created by intense pump lasers,
we show in Figure \ref{coefficient} the self and cross coupling coefficients
for the signal and idler fields as a function of the idler frequency. Note
that the corresponding frequency of signal field is determined by
$\Delta\omega_{s}=\Delta\omega_{i}-\Delta_{1}+\Delta_{b}$. \ The dimensionless
quantities $\alpha_{i}L$, $\beta_{s}L$ and $\kappa_{s}L$ are shown under the
conditions of maximum conversion efficiency to be discussed in the next
section. \ We choose the optical depth (opd) $\rho\sigma L=150$ where $\rho$
is the number density, $\sigma\equiv3\lambda^{2}/(4\pi)$ the resonant
absorption cross-section, and $L$ the atomic ensemble length in the
propagation direction. \ Three parametric coupling windows are separated by
two strong absorption peaks on the left and two relatively weak ones on the
right. The imaginary part of the self-coupling coefficients are seen to vanish
in each window at a certain point, while the real parts are small away from
resonances. At the same time the cross-coupling coefficients have a large
imaginary part.\ The positive gradient of $\Im(\beta_{s}L)$ and $\Im
(\alpha_{i}L)$ inside the windows is indicative of normal dispersion. \ 

\section{Optimal Conversion Efficiency}

It is important to ascertain the parameters that allow maximum efficiency of
conversion due its potential in practical quantum information processing. \ In
principle we need to search the three parametric coupling windows to find the
optimum conditions for an atomic ensemble of a given optical thickness.%

\begin{figure}
[ptb]
\begin{center}
\includegraphics[
natheight=17.333600in,
natwidth=24.239400in,
height=4.3232in,
width=5.6403in
]%
{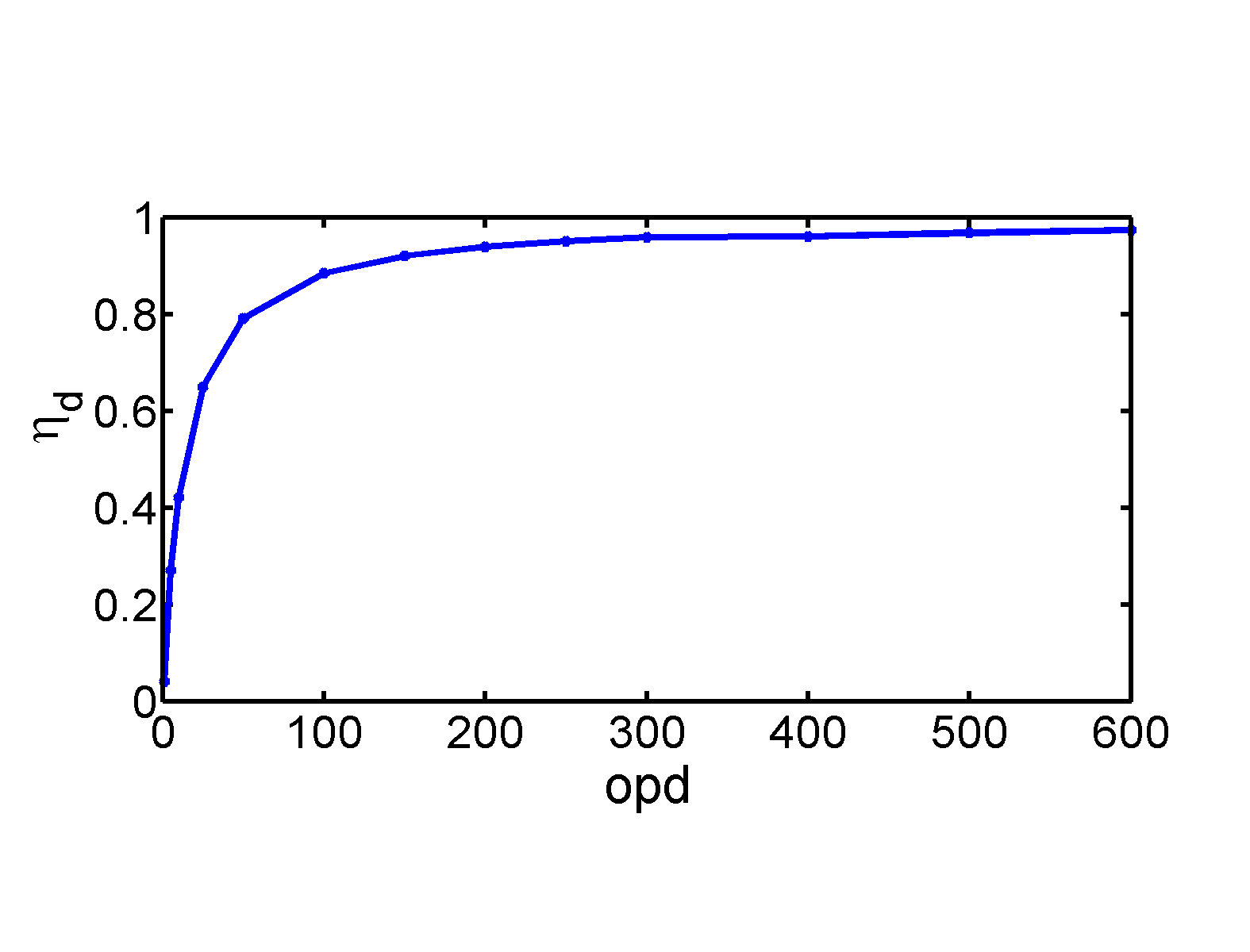}%
\caption{Down conversion efficiency $\eta_{\text{d}}$ vs optical depth (opd)
from $1$ to $600.$ \ Each dotted point is the maximum for five variational
parameters $\Omega_{a}$, $\Omega_{b}$, $\Delta_{1}$, $\Delta_{b}$, and
$\Delta\omega_{i}$.}%
\label{opd}%
\end{center}
\end{figure}

In the previous section we have discussed how three parametric coupling
windows appear for some particular values of pump laser parameters.\ In the
search for the maximal conversion efficiency, five parameters $\Omega_{a}$,
$\Omega_{b}$, $\Delta_{1}$, $\Delta_{b}$, and $\Delta\omega_{i}$ are varied to
maximize the conversion efficiency for a fixed optical depth of atomic
ensemble, using functional optimization.

The optical depth $\rho\sigma L$ appears through the dependence on atomic
number $N$ in the Arecchi-Courtens cooperation time $T_{c}$ \cite{scale}
\[
T_{c}^{-2}\equiv N|g_{i}|^{2}=\frac{\gamma_{03}c}{2L}\ \ \rho\sigma L.
\]
\ In Figure \ref{opd}, we show the maximum of down conversion efficiency using
Eq. (\ref{down}) for different optical depths from $1$ to $600$. \ The maximum
is found by varying five parameters mentioned above and the conversion
efficiency reaches $100\%$ asymptotically when the optical depth becomes
larger. \ In the strong parametric coupling regime as we discussed in the
previous section, $\eta_{\text{d}}\simeq\sin^{2}[\operatorname{Im}(\kappa
_{s}L)]$ and it has a maximum when $\operatorname{Im}(\kappa_{s}L)=\frac{\pi
}{2},$ see Figure \ref{coefficient}.\ Since $\operatorname{Im}(\kappa_{s}L)$
is proportional to optical depth and inversely proportional to the Rabi
frequencies of the driving lasers, an order of magnitude estimate of the
optical depth necessary for near unit conversion efficiency is opd$\simeq
\frac{\pi}{2}\Omega_{a,b}/\gamma_{03}>>1$.

\begin{figure}
[ptb]
\begin{center}
\includegraphics[
natheight=17.413200in,
natwidth=24.320700in,
height=4.5576in,
width=5.847in
]%
{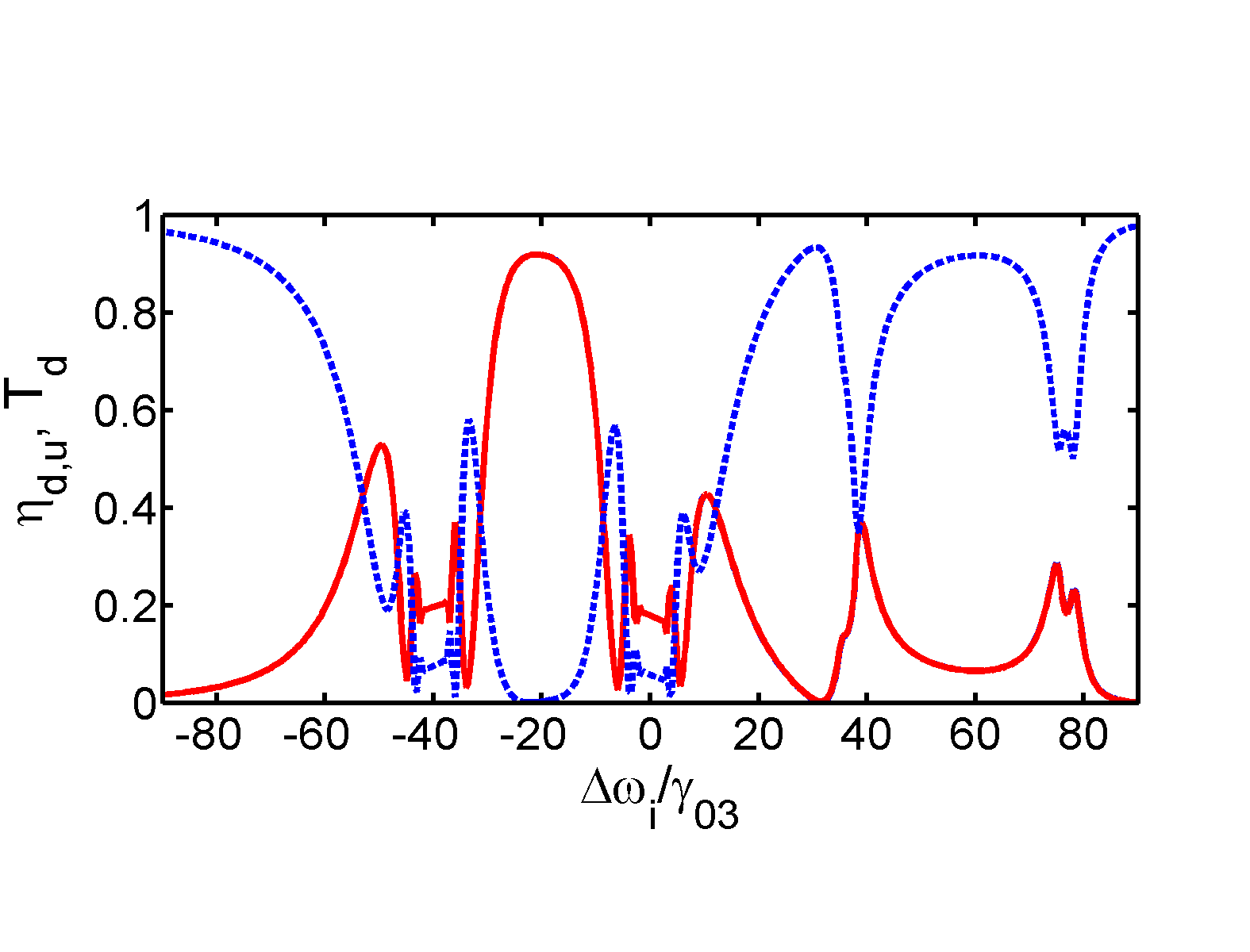}%
\caption{Conversion efficiency $\eta_{\text{d}}$, $\eta_{\text{u}}$ and
transmission $T_{\text{d}}$ vs $\Delta\omega_{i}$ for opd=150. \ $\eta
_{\text{d}}$ and $\eta_{\text{u}}$ are indistinguishable and shown in solid
red line, and $T_{\text{d}}$ is in dashed blue line. \ High transmission
efficiency corresponds to low conversion efficiency indicating the approximate
conservation condition within each parametric coupling window. The maximum
conversion efficiency is found in the left window at around $\Delta\omega
_{i}=-20\gamma_{03}$ and other relevant parameters are the same as in Figure
\ref{coefficient}.}%
\label{trans}%
\end{center}
\end{figure}

The behavior of the cross-coupling coefficient $\operatorname{Im}(\kappa
_{s}L)$ as a function of idler detuning indicates where large conversion is to
be found, as a comparison with Figure \ref{trans} shows. \ The maximum
efficiency of about $0.92$ is located in the left parametric coupling window
at the intersection of $\operatorname{Im}(\kappa_{s}L)$ and $\frac{\pi}{2}$.
\ Inside the windows the trade-off between conversion and transmission is
clear. \ In the region where absorption is large, on the sides of the window
(especially for the left window), the efficiency and the transmission are both
low although the valley in conversion efficiency corresponds to a peak in
transmission as expected in parametric coupling. The transmission approaches
unity when the incident idler field is far off-resonance.

We note that the symmetry $(\Delta_{1},\Delta_{b},\Delta\omega_{i}%
)\rightarrow-(\Delta_{1},\Delta_{b},\Delta\omega_{i})$ gives degenerate
optimal conversion conditions.%

Moreover, for the region where absorption is large on the sides of the window
(especially for the left window), the efficiency and the transmission are both
low but the valley of efficiency corresponds to a peak for transmission
indicating the feature of parametric coupling. \ The plateau for efficiency in
absorption region can be estimated as $\eta_{\text{d}}\approx\left\vert
\frac{\kappa_{s}}{2w}\right\vert ^{2}$ where $\alpha_{i}\beta_{s}\approx
\kappa_{s}\kappa_{i}$. \ Based on the study of finding the maximum efficiency
for frequency conversion, we will investigate the situation when the input is
a pulse and numerical integration of full equation of motion is required in
the next section.

\section{Pulse Conversion: Solution of the Maxwell-Bloch Equations}

The effect of finite-duration input probe pulses, which are often employed in
practice, can be assessed by numerically solving the Maxwell-Bloch equations
for the coupled atoms-fields system. The characteristic scales of time and
length are given by the Arecchi-Courtens time $T_{c}$ and $L_{c}=cT_{c}$,
respectively, which are inversely proportional to the square root of the opd.
The cooperative electric field is the product of the atomic number and the
idler electric field per photon, i.e., $E_{c}=\sqrt{\rho\hbar\omega
_{i}/(2\epsilon_{0})}.$

Scaling the space, time, electric field amplitude, various detunings, and
natural decay rates accordingly, indicated by tildes, the Maxwell-Bloch
equations of Eqs. (\ref{bloch_conv},\ref{maxwell1},\ref{maxwell2}) under
energy conservation ($\Delta\omega=\omega_{a}+\omega_{s}-\omega_{b}-\omega
_{i}=0$) and four-wave mixing conditions ($\Delta k=k_{a}-k_{s}+k_{b}-k_{i}%
=0$) become
\begin{align}
\frac{\partial}{\partial\tilde{\tau}}\tilde{\sigma}_{01}  & =(i\tilde{\Delta
}_{1}-\frac{\tilde{\gamma}_{01}}{2})\tilde{\sigma}_{01}+i\tilde{\Omega}%
_{a}(\tilde{\sigma}_{00}-\tilde{\sigma}_{11})+i\tilde{\sigma}_{02}\tilde
{E}_{s}^{-}-i\tilde{\sigma}_{13}^{\dag}\tilde{E}_{i}^{+},\nonumber\\
\frac{\partial}{\partial\tilde{\tau}}\tilde{\sigma}_{12}  & =(i\Delta
\tilde{\omega}_{s}-\frac{\tilde{\gamma}_{01}+\tilde{\gamma}_{2}}{2}%
)\tilde{\sigma}_{12}-i\tilde{\Omega}_{a}^{\ast}\tilde{\sigma}_{02}%
+i(\tilde{\sigma}_{11}-\tilde{\sigma}_{22})\tilde{E}_{s}^{+}+i\tilde{\Omega
}_{b}\tilde{\sigma}_{13},\nonumber\\
\frac{\partial}{\partial\tilde{\tau}}\tilde{\sigma}_{02}  & =(i\tilde{\Delta
}_{2}-\frac{\tilde{\gamma}_{2}}{2})\tilde{\sigma}_{02}-i\tilde{\sigma}%
_{12}\tilde{\Omega}_{a}+i\tilde{\sigma}_{01}\tilde{E}_{s}^{+}+i\tilde{\sigma
}_{03}\tilde{\Omega}_{b}-i\tilde{\sigma}_{32}\tilde{E}_{i}^{+},\nonumber\\
\frac{\partial}{\partial\tilde{\tau}}\tilde{\sigma}_{11}  & =-\tilde{\gamma
}_{01}\tilde{\sigma}_{11}+\tilde{\gamma}_{12}\tilde{\sigma}_{22}%
+i\tilde{\Omega}_{a}\tilde{\sigma}_{01}^{\dag}-i\tilde{\Omega}_{a}^{\ast
}\tilde{\sigma}_{01}-i\tilde{\sigma}_{12}^{\dag}\tilde{E}_{s}^{+}%
+i\tilde{\sigma}_{12}\tilde{E}_{s}^{-},\nonumber\\
\frac{\partial}{\partial\tilde{\tau}}\tilde{\sigma}_{22}  & =-\tilde{\gamma
}_{2}\tilde{\sigma}_{22}+i\tilde{\sigma}_{12}^{\dag}\tilde{E}_{s}^{+}%
-i\tilde{\sigma}_{12}\tilde{E}_{s}^{-}+i\tilde{\Omega}_{b}\tilde{\sigma}%
_{32}^{\dag}-i\tilde{\Omega}_{b}^{\ast}\tilde{\sigma}_{32},\nonumber\\
\frac{\partial}{\partial\tilde{\tau}}\tilde{\sigma}_{33}  & =-\tilde{\gamma
}_{03}\tilde{\sigma}_{33}+\tilde{\gamma}_{32}\tilde{\sigma}_{22}%
-i\tilde{\Omega}_{b}\tilde{\sigma}_{32}^{\dag}+i\tilde{\Omega}_{b}^{\ast
}\tilde{\sigma}_{32}+i\tilde{\sigma}_{03}^{\dag}\tilde{E}_{i}^{+}%
-i\tilde{\sigma}_{03}\tilde{E}_{i}^{-},\nonumber\\
\frac{\partial}{\partial\tilde{\tau}}\tilde{\sigma}_{13}  & =(i\Delta
\tilde{\omega}_{i}-i\tilde{\Delta}_{1}-\frac{\tilde{\gamma}_{01}+\tilde
{\gamma}_{03}}{2})\tilde{\sigma}_{13}-i\tilde{\Omega}_{a}^{\ast}\tilde{\sigma
}_{03}-i\tilde{\sigma}_{32}^{\dag}\tilde{E}_{s}^{+}+i\tilde{\Omega}_{b}^{\ast
}\tilde{\sigma}_{12}+i\tilde{\sigma}_{01}^{\dag}\tilde{E}_{i}^{+},\nonumber\\
\frac{\partial}{\partial\tilde{\tau}}\tilde{\sigma}_{03}  & =(i\Delta
\tilde{\omega}_{i}-\frac{\tilde{\gamma}_{03}}{2})\tilde{\sigma}_{03}%
-i\tilde{\Omega}_{a}\tilde{\sigma}_{13}+i\tilde{\Omega}_{b}^{\ast}%
\tilde{\sigma}_{02}+i(\tilde{\sigma}_{00}-\tilde{\sigma}_{33})\tilde{E}%
_{i}^{+},\nonumber\\
\frac{\partial}{\partial\tilde{\tau}}\tilde{\sigma}_{32}^{\dag}  &
=(-i\tilde{\Delta}_{b}-\frac{\tilde{\gamma}_{03}+\tilde{\gamma}_{2}}{2}%
)\tilde{\sigma}_{32}^{\dag}-i\tilde{\sigma}_{13}\tilde{E}_{s}^{-}%
+i\tilde{\Omega}_{b}^{\ast}(\tilde{\sigma}_{22}-\tilde{\sigma}_{33}%
)+i\tilde{\sigma}_{02}^{\dag}\tilde{E}_{i}^{+},
\end{align}
and%
\begin{equation}
\frac{\partial}{\partial\tilde{z}}\tilde{E}_{s}^{+}=i\tilde{\sigma}_{12}%
\frac{|g_{s}|^{2}}{|g_{i}|^{2}}\text{, \ }\frac{\partial}{\partial\tilde{z}%
}\tilde{E}_{i}^{+}=i\tilde{\sigma}_{03}%
\end{equation}
where $\tilde{z}$ $=z/L_{c}$, $\tilde{\tau}$ $=\tau/T_{c}$, $\tilde{\Omega
}_{a,b}=\Omega_{a,b}T_{c}$, $\tilde{E}_{s,i}^{+}$ $=E_{s,i}^{+}/E_{c},$ and
$|g_{s}|^{2}/|g_{i}|^{2}$ is a factor of unit transformation from signal to
idler field strength. \ Natural life time \cite{gsgi} for signal and idler
transitions is used to calculate the ratio of coupling strength $g_{s}%
/g_{i}=1.035.$ \ The above equations were integrated with a semi-implicit
finite difference method \cite{semi}. \ The midpoint integration method is
stable and has high accuracy without sacrificing memory for finer grids
\cite{numerical}. \ The algorithm has been tested by comparing with the
parametric equations' solutions in appropriate limits, and these solutions are
recovered when fine enough grids are employed.

To illustrate the influence of finite pump pulse duration, we compute the down
conversion efficiency
\begin{equation}
\eta_{\text{d}}=\frac{\int|{E}_{s}^{+}(z=L,\tau)|^{2}d\tau}{\int|{E}_{i}%
^{+}(z=0,\tau)|^{2}d\tau}.
\end{equation}
%

\begin{figure}
[ptb]
\begin{center}
\includegraphics[
natheight=17.333600in,
natwidth=24.239400in,
height=4.6224in,
width=5.8349in
]%
{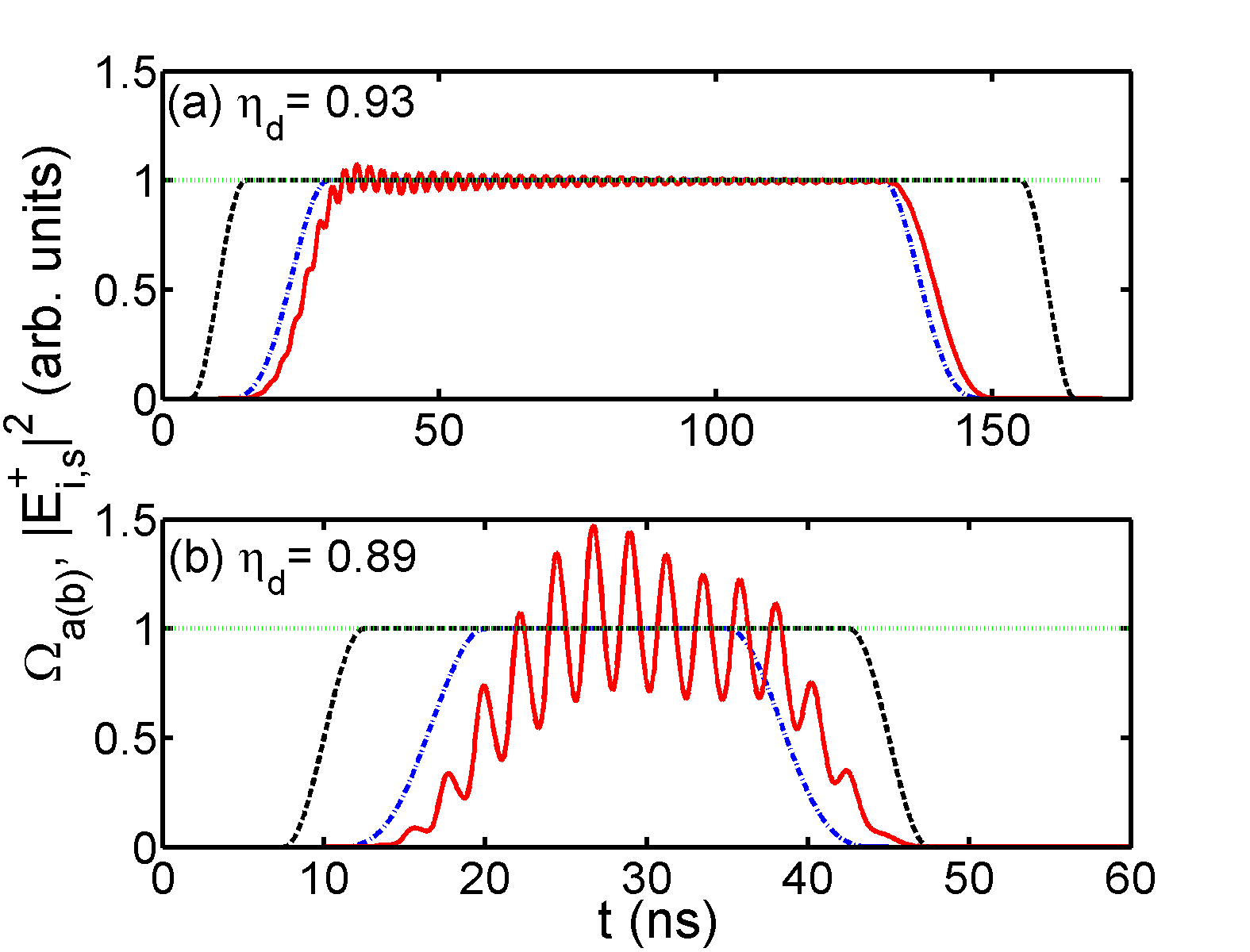}%
\caption{Time-varying pump fields of Rabi frequencies $\Omega_{a,b}(t)$ and
down-converted signal intensity ($|E_{s}^{+}(t,z=L)|^{2})$ from an input idler
pulse ($|E_{i}^{+}(t,z=0)|^{2})$. \ Here we let $t=\tau$ that is the delayed
time in co-moving frame. \ Pump-b (dotted green) is a continuous wave and
pump-a (dashed black) is a square pulse long enough to enclose input idler
pulse with (a) 100 ns and (b) 15 ns (dashed-dot blue). \ Output signal
intensity (solid red) at the end of atomic ensemble $z=L$ is oscillatory due
to the pump fields. \ The square pulse in rising region ($t_{r}-\frac{t_{s}%
}{2}<t<t_{r}+\frac{t_{s}}{2}$) has the form of $\frac{1}{2}[1+\sin(\frac
{\pi(t-t_{r})}{t_{s}})]$\ that in (a) ($t_{r},t_{s}$)=(10,10)ns for pump-a and
($t_{r},t_{s}$)=(20,20)ns for input idler; (b) ($t_{r},t_{s}$)=(10,5)ns for
pump-a and ($t_{r},t_{s}$)=(15,10)ns for input idler where $t_{r}$ is the
rising time indicating the center of rising period $t_{s}.$ \ Note that the
falling region of square pulse is symmetric to the rising one.}%
\label{pulse}%
\end{center}
\end{figure}

In Figure \ref{pulse}, we show the computed values of $\eta_{\text{d}}$ for
two different input idler pulse durations. \ We fix the opd=$150$ and use the
near optimum parameters ($\Omega_{a}$, $\Omega_{b}$, $\Delta_{1}$, $\Delta
_{b}$, $\Delta\omega_{i}$) $=(33$, $20$, $39$, $2$, $-21)\gamma_{03}$
determined from the coupled parametric equations. The temporal shape of the
pump laser intensities is also shown. \ Pump laser b is taken to be continuous
wave, while pump a is a square pulse with duration large enough to completely
overlap the input idler pulse. \ To compare with the steady state solutions,
we choose the Rabi frequency of idler as $0.1\gamma_{03},$ which is small
compared to those of the pumps. \ We find that the conversion efficiency is
reduced for shorter idler pulse inputs.\ \ A $100$ ns idler pulse is long
enough that it has \textit{\textbf{a }}almost the same maximum conversion
efficiency of \ $0.92$ as in Figure \ref{opd} for opd=150. \ While for the
shorter idler pulse of $15$ ns, the signal develops significant temporal
modulation, and this reduces the conversion efficiency, although it is still
quite appreciable. The modulation frequency is at the generalized Rabi
frequency of pump-a $\sqrt{\Delta_{1}^{2}+4\Omega_{a}^{2}} $. We note the
characteristic time and space scales of the calculations are $T_{c}=0.086$ ns
and $L_{c}=26$ mm for a moderate atomic density $\rho=1.7\times10^{11}%
$cm$^{-3}$ and $L=6$ mm. \ The grid size for dimensionless time $\Delta
\tilde{t}=0.5$ and space $\Delta\tilde{z}=0.001$ were chosen for both $100$
and $15$ ns idler pulse durations, and the convergence is reached with an
estimated relative error less than $1\%$.

\begin{figure}
[ptb]
\begin{center}
\includegraphics[
natheight=18.499700in,
natwidth=24.000400in,
height=5.2667in,
width=6.026in
]%
{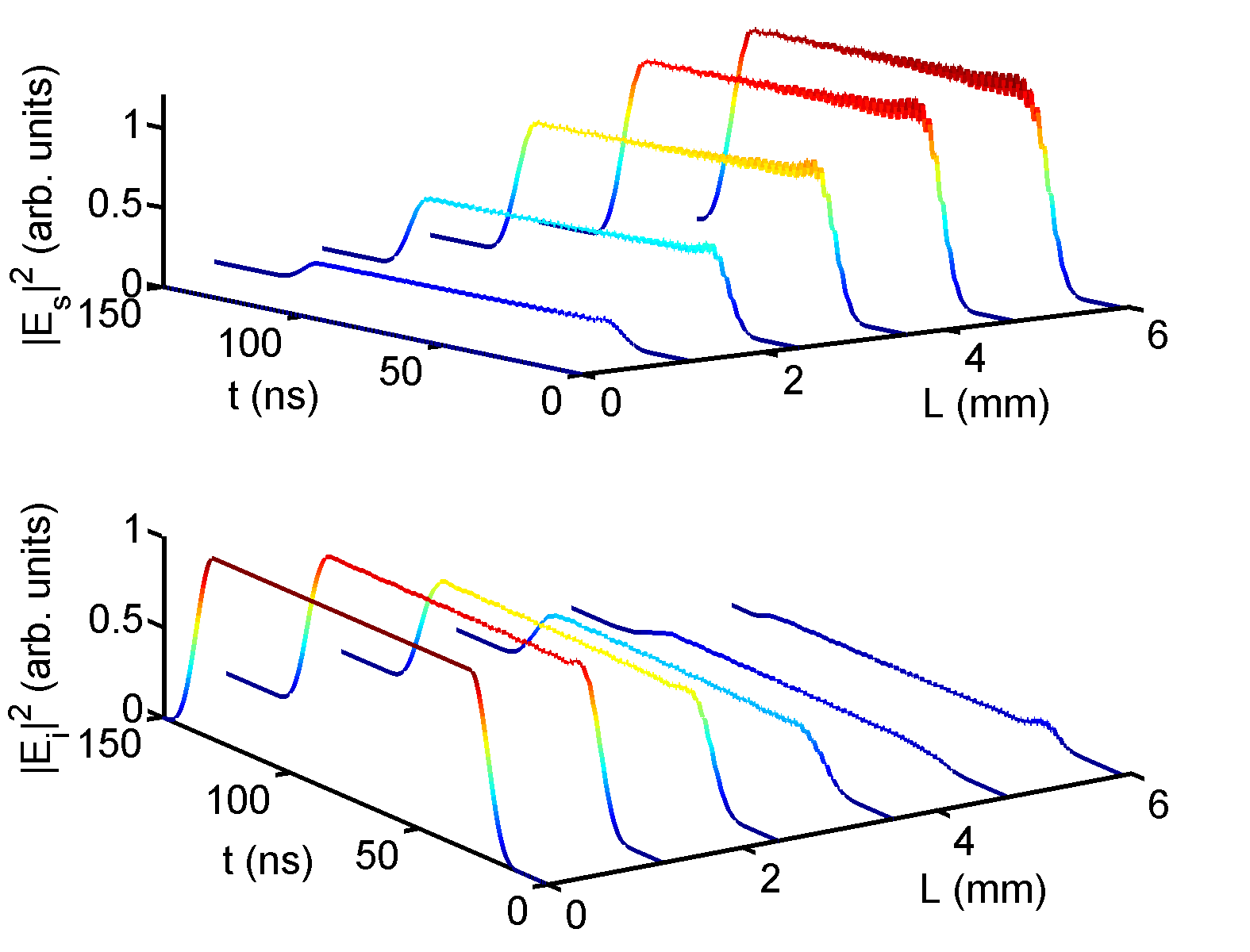}%
\caption{Three-dimensional line plots of converted signal and input idler
intensities in $t$ (ns) and $L$ (mm). \ Here we let $t=\tau$ that is the
delayed time in co-moving frame, and the parameters are the same as in Figure
\ref{pulse} (a).}%
\label{conv_3d}%
\end{center}
\end{figure}

Moreover, we show in Figure \ref{conv_3d} of three dimensional plots of signal
$|{E}_{s}^{+}(z,t)|^{2}$ and idler intensities $|{E}_{i}^{+}(z,t)|^{2} $. \ A
$100$ ns input idler pulse is demonstrated in time-space propagation which is
converted to signal pulse at the output surface of the ensemble.%

\section{Discussion of Quantum Fluctuations}

In this Section, we derive quantized Heisenberg-Langevin equations by adding
corresponding Langevin noises to the coupled equations. \ Similar to the
results of Appendix D, we have
\begin{align}
\frac{\partial}{\partial z}E_{s}^{+}  & =\beta_{s}E_{s}^{+}+\kappa_{s}%
E_{i}^{+}+\hat{f}_{s},\\
\frac{\partial}{\partial z}E_{i}^{+}  & =\kappa_{i}E_{s}^{+}+\alpha_{i}%
E_{i}^{+}+\hat{f}_{i}%
\end{align}
where signal and idler fields (respectively, ${E}_{s}^{+}$ and ${E}_{i}^{+}$)
are now quantized, and Langevin noises ($\hat{f}_{s}$ and $\hat{f}_{i}$) in
linearized equations are%
\begin{align}
\hat{f}_{s}  & =\frac{iNg_{s}^{\ast}}{cD}[(T_{03}+\frac{|\Omega_{a}|^{2}%
}{T_{13}}+\frac{|\Omega_{b}|^{2}}{T_{02}})\mathcal{\tilde{F}}_{12}%
+(\frac{\Omega_{a}^{\ast}\Omega_{b}}{T_{02}}+\frac{\Omega_{a}^{\ast}\Omega
_{b}}{T_{13}})\mathcal{\tilde{F}}_{03}\nonumber\\
& +\frac{i\Omega_{b}}{T_{13}}(T_{03}+\frac{|\Omega_{b}|^{2}-|\Omega_{a}|^{2}%
}{T_{02}})\mathcal{\tilde{F}}_{13}+\frac{i\Omega_{a}^{\ast}}{T_{02}}%
(-T_{03}+\frac{|\Omega_{b}|^{2}-|\Omega_{a}|^{2}}{T_{13}})\mathcal{\tilde{F}%
}_{02}]+\mathcal{\tilde{F}}_{s},\nonumber\\
& \\
\hat{f}_{i}  & =\frac{iNg_{i}^{\ast}}{cD}[(\frac{\Omega_{a}\Omega_{b}^{\ast}%
}{T_{02}}+\frac{\Omega_{a}\Omega_{b}^{\ast}}{T_{13}})\mathcal{\tilde{F}}%
_{12}+(T_{12}+\frac{|\Omega_{a}|^{2}}{T_{02}}+\frac{|\Omega_{b}|^{2}}{T_{13}%
})\mathcal{\tilde{F}}_{03}\nonumber\\
& +\frac{i\Omega_{a}}{T_{13}}(-T_{12}+\frac{|\Omega_{b}|^{2}-|\Omega_{a}|^{2}%
}{T_{02}})\mathcal{\tilde{F}}_{13}+\frac{i\Omega_{a}^{\ast}}{T_{02}}%
(T_{12}+\frac{|\Omega_{b}|^{2}-|\Omega_{a}|^{2}}{T_{13}})\mathcal{\tilde{F}%
}_{02}]+\mathcal{\tilde{F}}_{i}\nonumber\\
&
\end{align}
where various atomic and field Langevin noises $\mathcal{\tilde{F}}_{y}$ are
associated with coupled equations of atomic operators $\hat{\sigma}_{y}$\ and
field operators ${E}_{s,i}^{+}$ when $y=s,i$.

Note that various normal correlation functions of quantum Langevin noises have
$\left\langle \mathcal{\tilde{F}}_{i}^{\dag}(t,z)\mathcal{\tilde{F}}%
_{j}(t^{\prime},z^{\prime})\right\rangle =\frac{L}{N}\delta(t-t^{\prime
})\delta(z-z^{\prime})\hat{D}_{ij}$ in continuous limit. \ If ensemble average
is taken over the above field equations, with the property of Langevin noises
that $\left\langle \mathcal{\tilde{F}}_{i}^{\dag}(t,z)\right\rangle =$
$\left\langle \mathcal{\tilde{F}}_{i}(t^{\prime},z^{\prime})\right\rangle =0,$
the field equations are reduced to c-number ones where fluctuations due to
Langevin noises do not matter. \ For calculation of normally-ordered
operators, semi-classical approximation is valid even in quantum regime. \ We
demonstrate in the following and include Langevin noises in derivation of
solutions of field operators.

The set of equations can be written as
\begin{equation}
\frac{\partial}{\partial z}x(z)=Ax+f
\end{equation}
where $x(z)=%
\begin{bmatrix}
E_{s}^{+}(z)\\
E_{i}^{+}(z)
\end{bmatrix}
$ and $f=%
\begin{bmatrix}
\hat{f}_{s}(z)\\
\hat{f}_{i}(z)
\end{bmatrix}
.$

Consider a similarity transformation $S$ that $S^{-1}AS=\Lambda$ and
$y=S^{-1}x$, then we have
\begin{align}
\frac{\partial}{\partial z}y  & =\Lambda y+S^{-1}f,\\
y(z)  & =e^{\Lambda(z-z_{0})}y(z_{0})+\int_{z_{0}}^{z}dz^{\prime}%
e^{\Lambda(z-z^{\prime})}S^{-1}f(z^{\prime}).
\end{align}

With the boundary condition $x_{1}(0)$ and $x_{2}(0)$, we have
\begin{equation}
x(z)=Se^{\Lambda z}S^{-1}x(0)+\int_{0}^{z}dz^{\prime}Se^{\Lambda(z-z^{\prime
})}S^{-1}f(z^{\prime}).
\end{equation}

So we have correspondingly
\begin{align}
& A=\left(
\begin{array}
[c]{cc}%
\beta_{s} & \kappa_{s}\\
\kappa_{i} & \alpha_{i}%
\end{array}
\right)  ,~\Lambda=\left(
\begin{array}
[c]{cc}%
(\alpha_{i}+\beta_{s})/2+w & 0\\
0 & (\alpha_{i}+\beta_{s})/2-w
\end{array}
\right) \nonumber\\
& S=\left(
\begin{array}
[c]{cc}%
q+w & \kappa_{s}\\
\kappa_{i} & -q-w
\end{array}
\right)  ,~S^{-1}=\frac{1}{2w(w+q)}\left(
\begin{array}
[c]{cc}%
q+w & \kappa_{s}\\
\kappa_{i} & -q-w
\end{array}
\right)
\end{align}
where $w\equiv\sqrt{q^{2}+\kappa_{s}\kappa_{i}}$, and $q\equiv(-\alpha
_{i}+\beta_{s})/2$. \ 

The solutions of fields from down conversion for example are%

\begin{align}%
\begin{bmatrix}
E_{s}^{+}(L)\\
E_{i}^{+}(L)
\end{bmatrix}
& =%
\begin{bmatrix}
q+w & \kappa_{s}\\
\kappa_{i} & -q-w
\end{bmatrix}%
\begin{bmatrix}
e^{[(\alpha_{i}+\beta_{s})/2+w]L} & 0\\
0 & e^{[(\alpha_{i}+\beta_{s})/2-w]L}%
\end{bmatrix}
\frac{1}{2w(w+q)}\times\nonumber\\
&
\begin{bmatrix}
q+w & \kappa_{s}\\
\kappa_{i} & -q-w
\end{bmatrix}%
\begin{bmatrix}
0\\
E_{i}^{+}(0)
\end{bmatrix}
\nonumber\\
& +\int_{0}^{L}dz^{\prime}S%
\begin{bmatrix}
e^{[(\alpha_{i}+\beta_{s})/2+w](L-z^{\prime})} & 0\\
0 & e^{[(\alpha_{i}+\beta_{s})/2-w](L-z^{\prime})}%
\end{bmatrix}
S^{-1}f(z^{\prime}),\nonumber\\
& =%
\begin{bmatrix}
\frac{\kappa_{s}}{2w}(e^{[(\alpha_{i}+\beta_{s})/2+w]L}-e^{[(\alpha_{i}%
+\beta_{s})/2-w]L})\\
\frac{1}{2w(w+q)}[\kappa_{s}\kappa_{i}e^{[(\alpha_{i}+\beta_{s})/2+w]L}%
+(q+w)^{2}e^{[(\alpha_{i}+\beta_{s})/2-w]L}]
\end{bmatrix}
E_{i}^{+}(0)\nonumber\\
& +\int_{0}^{L}dz^{\prime}S%
\begin{bmatrix}
e^{[(\alpha_{i}+\beta_{s})/2+w](L-z^{\prime})} & 0\\
0 & e^{[(\alpha_{i}+\beta_{s})/2-w](L-z^{\prime})}%
\end{bmatrix}
S^{-1}f(z^{\prime}),\nonumber\\
& =%
\begin{bmatrix}
\sqrt{\eta_{d}}e^{i\arg(\sqrt{\eta_{d}})}\\
\sqrt{T_{d}}e^{i\arg(\sqrt{T_{d}})}%
\end{bmatrix}
E_{i}^{+}(0)+\int_{0}^{L}dz^{\prime}%
\begin{bmatrix}
\xi_{s}(z^{\prime})\hat{f}_{s}(z^{\prime})+\xi_{i}(z^{\prime})\hat{f}%
_{i}(z^{\prime})\\
\zeta_{s}(z^{\prime})\hat{f}_{s}(z^{\prime})+\zeta_{i}(z^{\prime})\hat{f}%
_{i}(z^{\prime})
\end{bmatrix}
\end{align}
where $\eta_{d}$ and $T_{d}$ are conversion efficiency and transmission for
down conversion as derived in Section II, arg represents the argument for
complex numbers, and we see that extra terms involve Langevin noises.
\ $\xi_{s,i}$ and $\zeta_{s,i}$ are defined in the below,%

\begin{align}
\xi_{s}  & =\frac{1}{2w(w+q)}[(w+q)^{2}e^{[(\alpha_{i}+\beta_{s}%
)/2+w](L-z^{\prime})}+\kappa_{s}\kappa_{i}e^{[(\alpha_{i}+\beta_{s}%
)/2-w](L-z^{\prime})}],\\
\xi_{i}  & =\frac{\kappa_{s}}{2w}[e^{[(\alpha_{i}+\beta_{s})/2+w](L-z^{\prime
})}-e^{[(\alpha_{i}+\beta_{s})/2-w](L-z^{\prime})}],\\
\zeta_{s}  & =\frac{\kappa_{i}}{2w}[e^{[(\alpha_{i}+\beta_{s}%
)/2+w](L-z^{\prime})}-e^{[(\alpha_{i}+\beta_{s})/2-w](L-z^{\prime})}],\\
\zeta_{i}  & =\frac{1}{2w(w+q)}[\kappa_{s}\kappa_{i}e^{[(\alpha_{i}+\beta
_{s})/2+w](L-z^{\prime})}+(w+q)^{2}e^{[(\alpha_{i}+\beta_{s})/2-w](L-z^{\prime
})}].
\end{align}
The down conversion efficiency $\eta_{\text{down}}$ is defined as%

\begin{align}
\eta_{\text{down}}  & =\frac{\left\langle E_{s}^{-}(L)E_{s}^{+}%
(L)\right\rangle }{\left\langle E_{i}^{-}(0)E_{i}^{+}(0)\right\rangle
}\nonumber\\
& =\eta_{d}+\frac{1}{\left\langle E_{i}^{-}(0)E_{i}^{+}(0)\right\rangle
}\times\nonumber\\
& \left\langle \int_{0}^{L}dz^{\prime}[\xi_{s}^{\ast}(z^{\prime})\hat{f}%
_{s}^{\dag}(z^{\prime})+\xi_{i}(z^{\prime})\hat{f}_{i}^{\dag}(z^{\prime}%
)]\int_{0}^{L}dz^{\prime\prime}[\xi_{s}(z^{\prime\prime})\hat{f}_{s}%
(z^{\prime\prime})+\xi_{i}(z^{\prime\prime})\hat{f}_{i}(z^{\prime\prime
})]\right\rangle \nonumber\\
&
\end{align}
in which expectation value of normally-ordered operators involves
contributions from semi-classical treatment and normally-ordered noise
correlation functions. \ The relevant normally-ordered quantum diffusion
coefficients $\hat{D}_{ij}$ from Einstein's relation are (note that $\hat
{D}_{ij}=\hat{D}_{ji}^{\dag}$)%

\begin{align}
(\text{i})\text{ }\hat{D}_{12,12}  & =\gamma_{01}\left\langle \tilde{\sigma
}_{22}\right\rangle \approx\gamma_{01}\tilde{\sigma}_{22,s}^{{}}=0;\text{
}\nonumber\\
\hat{D}_{12,03}  & =\hat{D}_{12,02}=0;\text{ }\nonumber\\
\hat{D}_{12,13}  & =\gamma_{01}\left\langle \tilde{\sigma}_{23}\right\rangle
\approx\gamma_{01}\tilde{\sigma}_{23,s}^{{}}=0;\nonumber\\
\hat{D}_{12,s}  & =\hat{D}_{12,i}=0;\\
(\text{ii})\text{ }\hat{D}_{03,03}  & =\gamma_{32}\left\langle \tilde{\sigma
}_{22}\right\rangle \approx\gamma_{32}\tilde{\sigma}_{22,s}^{{}}=0;\text{
}\nonumber\\
\hat{D}_{03,02}  & =\hat{D}_{03,13}=0;\nonumber\\
\hat{D}_{03,s}  & =\hat{D}_{03,i}=0;\\
(\text{iii})\text{ }\hat{D}_{02,02}  & =\hat{D}_{02,13}=0;\nonumber\\
\hat{D}_{02,s}  & =\hat{D}_{02,i}=0;\\
(\text{iV})\text{ }\hat{D}_{13,13}  & =\gamma_{01}\left\langle \tilde{\sigma
}_{33}\right\rangle +\gamma_{32}\left\langle \tilde{\sigma}_{22}\right\rangle
\approx\gamma_{01}\tilde{\sigma}_{33,s}^{{}}+\gamma_{32}\tilde{\sigma}%
_{22,s}^{{}}=0;\text{ }\nonumber\\
\hat{D}_{13,s}  & =\hat{D}_{13,i}=0;\\
(\text{V})\text{ }\hat{D}_{s,s}  & =\hat{D}_{s,i}=0;\\
(\text{Vi})\text{ }\hat{D}_{i,i}  & =0;
\end{align}
where we have approximated various nonvanishing quantum diffusion coefficients
by zeroth order properties of atomic operators (the steady state solutions).
\ The above normally-ordered correlation functions give zero contributions in
the linearized equations of motion, so c-number Langevin equation is
sufficient to derive the conversion efficiency. \ The normally-ordered noise
correlations are zero because the population ($\tilde{\sigma}_{22,s}^{{}%
},\tilde{\sigma}_{33,s}^{{}}$) and coherence ($\tilde{\sigma}_{23,s}^{{}}$)
properties are zero for the atomic level driven by pump-b. \ The linearized
field equations in diamond structure have similar noise properties to
$\Lambda$\ system in which most atoms are on the ground state, and Langevin
noise can be neglected if normally-ordered quantities, say storage efficiency,
are considered \cite{gorshkov,quantum_interface}.

\section{Conclusion}

We have studied light frequency conversion in an atomic ensemble with a
diamond configuration of atomic levels such as $^{87}$Rb. The motivation stems
from the need to efficiently convert light resonant with ground state
transitions (storable in the sense of quantum memories) to and from the
telecom wavelength band for low-loss quantum network communication. The
optically thick atomic sample is driven by two strong co-propagating pump
fields, and a probe idler or signal field depending on whether we consider
down- or up-conversion. \ Parametric equations for the probe fields are
derived and used to compute conversion efficiencies. \ They can be understood
by dressed-state picture where we can visualize four absorption lines due to
two strong pump lasers and thus three parametric coupling windows are created.
\ There are two major contributions to the conversion efficiency, which are
related to atomic populations and coherences in the lower arm of the diamond
level driven by laser pump-a.\ \ When this transition is saturated by a large
pump Rabi frequency or when the coherence dominates due to a large pump-b Rabi
frequency in the upper transition, the cross-coupling coefficients and hence
the conversion efficiencies are equal.\ 

By performing a global parameter search we find conditions of pump Rabi
frequencies, detunings, and signal/idler input frequency to maximize the
conversion efficiency as a function of optical depth of the ensemble. Only in
the limit of very large optical depth does the maximum efficiency approach the
ideal strong coupling result \cite{gogyan}. \ Under conditions routinely
obtained in cold, non-degenerate rubidium gas, with opd $\simeq100-200$,
optimal conversion efficiencies of the order $80\%$ to $90\%$ are predicted.
Numerical solution of the Maxwell-Bloch equations confirms the solution of the
parametric equations in the limit of long pulse duration, and indicates that
for shorter pulses, pump pulse induced modulation may reduce the conversion efficiency.



\chapter{Conclusion}

We provide a theoretical study of light-matter interactions in cascade and
diamond type atomic ensembles. \ A\ correlated two-photon (telecom
signal-infrared idler) state vector is derived in the long time limit within
the adiabatic approximation. \ The second-order correlation function is
calculated, and shows a superradiant time scale in the infrared idler
emission. \ The entanglement in frequency space for such a two-photon state is
analyzed by Schmidt decomposition. \ We are able to derive the mode functions
and investigate the influence of pump pulse duration and superradiant decay
rate that depends on optical density and ensemble geometry.

To investigate multiple atomic excitations on the correlated emission from the
atomic cascade transitions, we use the coherent state positive-P
representation and derive an equivalent Ito type stochastic differential
equation (SDE). \ The equations are solved numerically by a stable and
convergent semi-implicit difference method, while the counter-propagating
spatial evolution is solved by implementing the shooting method. \ We find an
enhanced characteristic time scale for idler emission in the second-order
correlation functions, consistent with the superradiance timescales predicted
by the analytical method in Chapter 3, and observed experimentally.

In Chapter 5, the correlated two-photon state derived in Chapter 3 is used to
investigate the spectral effects on DLCZ protocols involving entanglement
generation, swapping, and quantum teleportation. \ We analyze the performance
of the protocol using, photon-number resolving and non-resolving photon
detectors. We find that a more genuine and high fidelity protocol requires a
source with reduced frequency space entanglement.

In Chapter 6, we present the analytical results on the efficiency of
light-frequency conversion in a diamond atomic configuration. \ We find the
optimum efficiency as a function of optical density. \ We find the maximum
conversion efficiency by studying parametric coupling windows that are created
by strong pump fields, and provide numerical solutions for the pulse conversion.

\appendix
%



\chapter{Derivation of a Schr\"{o}dinger wave equation for spontaneous
emissions from a cascade type atomic ensemble}

In this appendix, we derive the Hamiltonian for the cascade emission
(signal-idler) from a four-level atomic ensemble. \ We use Schr\"{o}dinger's
equation to study the correlated two-photon state from a two-photon laser
excitation. \ Apart from the rotating wave approximation, non-rotating wave
probability amplitudes are introduced to take into account the proper
frequency shift. \ The adiabatic approximation on laser-excited states is used
to simplify the atomic dynamics and solve for the signal-idler probability
amplitude. \ 

\section{Hamiltonian and Equation of Motion}

Consider an ensemble of N four-level atoms interacting with two classical
fields and spontaneously emitted signal and idler photons as shown in Figure
\ref{four}. \ These identical atoms distribute randomly with a uniform
density. \ Use dipole approximation of light-matter interactions, $-\vec
{d}\cdot\vec{E}$ where $\vec{E}$ is classical or quantum electric field, and
include non-rotating wave approximation (RWA) terms in the interaction of
quantum fields, the Hamiltonian in interaction picture is%

\begin{align}
& V_{I}(t)=\nonumber\\
& -\hbar\Delta_{1}\sum_{\mu=1}^{N}|1\rangle_{\mu}\langle1|-\hbar\Delta_{2}%
\sum_{\mu=1}^{N}|2\rangle_{\mu}\langle2|-\frac{\hbar}{2}\sum_{\mu=1}%
^{N}\Big[\Omega_{a}|1\rangle_{\mu}\langle0|e^{i\vec{k}_{a}\cdot\vec{r}_{\mu}%
}+\Omega_{b}|2\rangle_{\mu}\langle1|e^{i\vec{k}_{b}\cdot\vec{r}_{\mu}%
}\nonumber\\
& +h.c.\Big]-i\hbar\sum_{k_{s},\lambda_{s}}g_{k_{s}}\left[  \vec{\epsilon
}_{k_{s},\lambda_{s}}\hat{a}_{k_{s},\lambda_{s}}e^{-i\omega_{ks}t+i\vec{k}%
_{s}\cdot\vec{r}_{\mu}}-\vec{\epsilon}_{k_{s},\lambda_{s}}^{\ast}\hat
{a}_{k_{s},\lambda_{s}}^{\dagger}e^{i\omega_{ks}t-i\vec{k}_{s}\cdot\vec
{r}_{\mu}}\right]  \cdot\hat{d}_{s}\sum_{\mu=1}^{N}\nonumber\\
& \Big[|2\rangle_{\mu}\langle3|e^{i(\omega_{23}+\Delta_{2})t}+|3\rangle_{\mu
}\langle2|e^{-i(\omega_{23}+\Delta_{2})t}\Big]-i\hbar\sum_{k_{i},\lambda_{i}%
}g_{k_{i}}\Big[\vec{\epsilon}_{k_{i},\lambda_{i}}\hat{a}_{k_{i},\lambda_{i}%
}e^{-i\omega_{ki}t+i\vec{k}_{i}\cdot\vec{r}_{\mu}}-\nonumber\\
& \vec{\epsilon}_{k_{i},\lambda_{i}}^{\ast}\hat{a}_{k_{i},\lambda_{i}%
}^{\dagger}e^{i\omega_{ki}t-i\vec{k}_{i}\cdot\vec{r}_{\mu}}\Big]\cdot\hat
{d}_{i}\sum_{\mu=1}^{N}\left[  |3\rangle_{\mu}\langle0|e^{i\omega_{3}%
t}+|0\rangle_{\mu}\langle3|e^{-i\omega_{3}t}\right]  ,\label{H1}%
\end{align}
where the time dependence of laser frequency is absorbed into interaction
terms of signal and idler fields. \ Single photon detuning $\Delta_{1}%
=\omega_{a}-\omega_{1}$, two-photon detuning $\Delta_{2}=\omega_{a}+\omega
_{b}-\omega_{2},$ and $\omega_{23}=\omega_{2}-\omega_{3}$. \ Rabi frequencies
are $\Omega_{a}\equiv(1||\hat{d}||0)\mathcal{E}(k_{a})/\hbar$, $\Omega
_{b}\equiv(2||\hat{d}||1)\mathcal{E}(k_{b})/\hbar, $ and coupling coefficients
are $g_{ks}\equiv(3||\hat{d}||2)\mathcal{E}(k_{s})/\hbar$, $g_{ki}%
\equiv(0||\hat{d}||3)\mathcal{E}(k_{i})/\hbar$. \ The double matrix element of
the dipole moment is independent of the hyperfine structure, and
$\mathcal{E}(k)=\sqrt{\frac{\hbar kc}{2\epsilon_{0}V}}$. \ Polarizations of
signal and idler fields are $\vec{\epsilon}_{k_{s},\lambda_{s}}$,
$\vec{\epsilon}_{k_{i},\lambda_{i}}$, and the unit direction of dipole
operators are $\hat{d}_{s}$, $\hat{d}_{i}$.

In the limit of large detuned and weak driving fields, $\Delta_{1}\gg
\frac{\sqrt{N}\Omega_{a}}{2},$ that is discussed in Chapter 3.2, we consider
only single excitations and ignore the spontaneous decay during the excitation
process. \ The state function can be written as%

\begin{align}
&  |\psi(t)\rangle=\nonumber\\
&  \mathcal{E}(t)|0,\text{vac}\rangle+\sum_{\mu=1}^{N}A_{\mu}(t)|1_{\mu
},\text{vac}\rangle+\sum_{\mu=1}^{N}B_{\mu}(t)|2_{\mu},\text{vac}\rangle
+\sum_{\mu=1}^{N}\sum_{k_{s},\lambda_{s}}C_{s}^{\mu}(t)|3_{\mu},1_{\vec{k}%
_{s},\lambda_{s}}\rangle\nonumber\\
&  +\sum_{\substack{k_{s},\lambda_{s} \\k_{i},\lambda_{i}}}D_{s,i}%
(t)|0,1_{\vec{k}_{s},\lambda_{s}},1_{\vec{k}_{i},\lambda_{i}}\rangle
+\underbrace{\sum_{\mu=1}^{N}\sum_{k_{i},\lambda_{i}}C_{i}^{\mu}(t)|3_{\mu
},1_{\vec{k}_{i},\lambda_{i}}\rangle+\sum_{\mu=1}^{N}C^{\mu}(t)|3_{\mu}%
\rangle}\nonumber\\
&  +\underbrace{\sum_{\mu=1}^{N}\sum_{k_{s},\lambda_{s}}B_{s}^{\mu}(t)|2_{\mu
},1_{\vec{k}_{s},\lambda_{s}}\rangle+\sum_{\nu<\mu}\sum_{\mu=1}^{N}%
\sum_{\substack{k_{s},\lambda_{s} \\k_{i},\lambda_{i}}}C_{s,i}^{\mu\nu
}(t)|3_{\mu},3_{\nu},1_{\vec{k}_{s},\lambda_{s}},1_{\vec{k}_{i},\lambda_{i}%
}\rangle}%
\end{align}
where $|$vac$\rangle$ is the photon vacuum state, $s\equiv(k_{s},\lambda_{s}%
)$, $i\equiv(k_{i},\lambda_{i})$, $|m_{\mu}\rangle\equiv|m_{\mu}%
\rangle|0\rangle_{\nu\neq\mu}^{\otimes N-1}$, $m=1,2,3$ and $|3_{\mu},3_{\nu
}\rangle\equiv|3_{\mu}\rangle|3_{\nu}\rangle|0\rangle_{\lambda\neq\mu,\nu
}^{\otimes N-2}$. \ The probability amplitudes coupled from rotating wave
terms in the Hamiltonian are $\mathcal{E}(t),$ $A_{\mu}(t),$ $B_{\mu}(t),$
$C_{s}^{\mu}(t),$ $D_{s,i}(t),$\ which indicate the complete cycle of single
excitation process from the ground state, intermediate, upper excited state,
intermediate excited state with emission of a signal photon, and the ground
state with the signal-idler emission. \ Note that the states underlined are
coupled through non-RWA terms that describe a transition from upper excited
state to intermediate one by absorbing a photon for $B_{s}^{\mu}(t)$ and
$C^{\mu}(t)$ and a transition from the ground state to the intermediate one by
emitting a photon for $C_{i}^{\mu}(t)$ and $\mathcal{E}(t)$ or $C_{s}^{\mu
}(t)$ and $C_{s,i}^{\mu\nu}(t)$.\ \ Apply the Schr\"{o}dinger equation
$i\hbar\frac{\partial}{\partial t}|\psi(t)\rangle=V_{I}(t)|\psi(t)\rangle$,
and we have the coupled equations of motion,%

\begin{align}
i\dot{\mathcal{E}} &  =-\frac{\Omega_{a}^{\ast}}{2}\sum_{\mu}e^{-i\vec{k}%
_{a}\cdot\vec{r}_{\mu}}A_{\mu}-\underbrace{i\sum_{i,\mu}g_{i}(\vec{\epsilon
}_{i}\cdot\hat{d}_{i}^{\ast})e^{i\vec{k}_{i}\cdot\vec{r}_{\mu}}e^{-i(\omega
_{ki}+\omega_{3})t}C_{i}^{\mu}},\nonumber\\
i\dot{C}_{i}^{\mu} &  =\underbrace{ig_{i}^{\ast}(\vec{\epsilon}_{i}^{\ast
}\cdot\hat{d}_{i})e^{-i\vec{k}_{i}\cdot\vec{r}_{\mu}}e^{i(\omega_{i}%
+\omega_{3})t}\mathcal{E}},\nonumber\\
i\dot{A}_{\mu} &  =-\Delta_{1}A_{\mu}-\frac{\Omega_{a}}{2}e^{i\vec{k}_{a}%
\cdot\vec{r}_{\mu}}\mathcal{E}-\frac{\Omega_{b}^{\ast}}{2}e^{-i\vec{k}%
_{b}\cdot\vec{r}_{\mu}}B_{\mu},\nonumber\\
i\dot{B}_{\mu} &  =-\Delta_{2}B_{\mu}-\frac{\Omega_{b}}{2}e^{i\vec{k}_{b}%
\cdot\vec{r}_{\mu}}A_{\mu}-i\sum_{_{s}}g_{s}(\vec{\epsilon}_{_{s}}\cdot\hat
{d}_{s}^{\ast})e^{i\vec{k}_{s}\cdot\vec{r}_{\mu}}e^{-i(\omega_{s}-\omega
_{23}-\Delta_{2})t}C_{s}^{\mu},\nonumber\\
i\dot{C}_{s}^{\mu} &  =ig_{s}^{\ast}(\epsilon_{_{s}}^{\ast}\cdot\hat{d}%
_{s})e^{-i\vec{k}_{s}\cdot\vec{r}_{\mu}}e^{i(\omega_{ks}-\omega_{23}%
-\Delta_{2})t}B_{\mu}-i\sum_{i}g_{i}(\vec{\epsilon}_{_{i}}\cdot\hat{d}%
_{i}^{\ast})e^{i\vec{k}_{i}\cdot\vec{r}_{\mu}}e^{-i(\omega_{i}-\omega_{3}%
)t}D_{s,i}\nonumber\\
&  -\underbrace{i\sum_{i}g_{i}(\vec{\epsilon}_{_{i}}\cdot\hat{d}_{i}^{\ast
})e^{i(\omega_{i}+\omega_{3})t}\Big[\sum_{\nu<\mu}C_{s,i}^{\mu\nu}e^{i\vec
{k}_{i}\cdot\vec{r}_{\nu}}+\sum_{\nu>\mu}e^{i\vec{k}_{i}\cdot\vec{r}_{\nu}%
}C_{s,i}^{\nu\mu}\Big]},\nonumber\\
i\dot{C}_{s,i}^{\mu\nu} &  =\underbrace{ig_{i}^{\ast}(\epsilon_{_{i}}^{\ast
}\cdot\hat{d}_{i})e^{i(\omega_{i}+\omega_{3})t}\Big[e^{-i\vec{k}_{i}\cdot
\vec{r}_{\nu}}C_{s}^{\mu}+C_{s}^{\nu}e^{-i\vec{k}_{i}\cdot\vec{r}_{\mu}}%
\Big]}\Big|_{\nu<\mu},\nonumber\\
i\dot{D}_{s,i} &  =ig_{i}^{\ast}(\epsilon_{_{i}}^{\ast}\cdot\hat{d}_{i}%
)\sum_{\mu}e^{-i\vec{k}_{i}\cdot\vec{r}_{\mu}}e^{i(\omega_{ki}-\omega_{3}%
)t}C_{s}^{\mu},\nonumber\\
i\dot{C}_{\mu} &  =-\underbrace{i\sum_{_{s}}g_{s}(\vec{\epsilon}_{_{s}}%
\cdot\hat{d}_{s}^{\ast})e^{i\vec{k}_{s}\cdot\vec{r}_{\mu}}e^{-i(\omega
_{s}+\omega_{23}+\Delta_{2})t}B_{s}^{\mu}},\nonumber\\
i\dot{B}_{s}^{\mu} &  =\underbrace{ig_{s}^{\ast}(\epsilon_{_{s}}^{\ast}%
\cdot\hat{d}_{s})e^{-i\vec{k}_{s}\cdot\vec{r}_{\mu}}e^{i(\omega_{ks}%
+\omega_{23}+\Delta_{2})t}C_{\mu}}.
\end{align}

The Lamb shift for the atomic transition $|3\rangle\rightarrow|0\rangle$\ with
the optical frequency $\omega_{3}$ and spontaneous decay rate $\Gamma$ is
$\int_{0}^{\infty}d\omega\frac{\Gamma}{2\pi}[$P.V.$(\omega-\omega_{3})^{-1}%
-$P.V.$(\omega+\omega_{3})^{-1}]$ that can be identified partly within the
substitution of these non-RWA terms. \ We substitute $C_{i}^{\mu} $ into
$\mathcal{E}$, $C_{s,i}^{\mu\nu}$ into $C_{s}^{\mu}$, and $B_{s}^{\mu}$ into
$C_{\mu}$, and they are%

\begin{align}
\dot{\mathcal{E}}  & =\frac{i\Omega_{a}^{\ast}}{2}\sum_{\mu}e^{-i\vec{k}%
_{a}\cdot\vec{r}_{\mu}}A_{\mu}-N\sum_{i}|g_{i}|^{2}|(\vec{\epsilon}_{i}%
\cdot\hat{d}_{i}^{\ast})|^{2}\int_{0}^{t}dt^{\prime}e^{i(\omega_{i}+\omega
_{3})(t^{\prime}-t)}\mathcal{E(}t^{\prime})\nonumber\\
& =\frac{i\Omega_{a}^{\ast}}{2}\sum_{\mu}e^{-i\vec{k}_{a}\cdot\vec{r}_{\mu}%
}A_{\mu}-N\mathcal{E}\oint d\Omega_{i}[1-(\hat{k}_{i}\cdot\hat{d}_{i}%
)^{2}]\frac{V}{(2\pi)^{3}}\int_{0}^{\infty}dk_{i}k_{i}^{2}\frac{\hbar
\omega_{i}}{2\epsilon_{0}V}\frac{|d_{i}|^{2}}{\hbar^{2}}\nonumber\\
& \times\lbrack\pi\delta(\omega_{i}+\omega_{3})-i\text{P.V.}(\omega_{i}%
+\omega_{3})^{-1}]\nonumber\\
& =\frac{i\Omega_{a}^{\ast}}{2}\sum_{\mu}e^{-i\vec{k}_{a}\cdot\vec{r}_{\mu}%
}A_{\mu}+iN\mathcal{E}\int_{0}^{\infty}d\omega_{i}\frac{\Gamma_{i}}{2\pi
}\text{P.V.}(\omega_{i}+\omega_{3})^{-1},
\end{align}

\begin{align}
\dot{C}_{s}^{\mu}  & =g_{s}^{\ast}(\epsilon_{_{s}}^{\ast}\cdot\hat{d}%
_{s})e^{-i\vec{k}_{s}\cdot\vec{r}_{\mu}}e^{i(\omega_{ks}-\omega_{23}%
-\Delta_{2})t}B_{\mu}-\sum_{i}g_{i}(\vec{\epsilon}_{_{i}}\cdot\hat{d}%
_{i}^{\ast})e^{i\vec{k}_{i}\cdot\vec{r}_{\mu}}e^{-i(\omega_{i}-\omega_{3}%
)t}D_{s,i}\nonumber\\
& -\sum_{i}|g_{i}|^{2}|(\vec{\epsilon}_{i}\cdot\hat{d}_{i}^{\ast})|^{2}%
\int_{0}^{t}dt^{\prime}e^{i(\omega_{i}+\omega_{3})(t^{\prime}-t)}%
\mathcal{E(}t^{\prime})\Big\{\sum_{\nu<\mu}\Big[C_{s}^{\mu}(t^{\prime
})+e^{i\vec{k}_{i}\cdot(\vec{r}_{\nu}-\vec{r}_{\mu})}C_{s}^{\nu}(t^{\prime
})\Big]\nonumber\\
& {\Large +}\sum_{\nu>\mu}\Big[e^{i\vec{k}_{i}\cdot(\vec{r}_{\nu}-\vec{r}%
_{\mu})}C_{s}^{\nu}(t^{\prime})+C_{s}^{\mu}(t^{\prime})\Big]\Big\}\nonumber\\
& =g_{s}^{\ast}(\epsilon_{_{s}}^{\ast}\cdot\hat{d}_{s})e^{-i\vec{k}_{s}%
\cdot\vec{r}_{\mu}}e^{i(\omega_{ks}-\omega_{23}-\Delta_{2})t}B_{\mu}-\sum
_{i}g_{i}(\vec{\epsilon}_{_{i}}\cdot\hat{d}_{i}^{\ast})e^{i\vec{k}_{i}%
\cdot\vec{r}_{\mu}}e^{-i(\omega_{i}-\omega_{3})t}D_{s,i}\nonumber\\
& +i(N-1)C_{s}^{\mu}\int_{0}^{\infty}d\omega_{i}\frac{\Gamma_{i}}{2\pi
}\text{P.V.}(\omega_{i}+\omega_{3})^{-1}-\sum_{i}|g_{i}|^{2}|(\vec{\epsilon
}_{i}\cdot\hat{d}_{i}^{\ast})|^{2}\times\nonumber\\
& \int_{0}^{t}dt^{\prime}e^{i(\omega_{i}+\omega_{3})(t^{\prime}-t)}\sum
_{\nu\neq\mu}e^{i\vec{k}_{i}\cdot(\vec{r}_{\mu}-\vec{r}_{\nu})}C_{s}^{\nu
}(t^{\prime}),
\end{align}
where we have used the symmetric property of $\Omega_{\nu\mu}^{+}(\xi
)=\Omega_{\mu\nu}^{+}(\xi)\equiv$ $\sum_{i}|g_{i}|^{2}|(\vec{\epsilon}%
_{i}\cdot\hat{d}_{i}^{\ast})|^{2}\int_{0}^{t}dt^{\prime}e^{i(\omega_{i}%
+\omega_{3})(t^{\prime}-t)}e^{i\vec{k}_{i}\cdot(\vec{r}_{\nu}-\vec{r}_{\mu})}$
\cite{Lehm}. \ The spontaneous decay rate for the idler transition is
$\Gamma_{i}\equiv\frac{|d_{i}|^{2}\omega_{i}^{3}}{3\pi\hbar\epsilon_{0}c^{3}}$
\cite{Weisskopf, QO:Scully}, and\ the same thing for signal transition
$\Gamma_{s}\equiv\frac{|d_{s}|^{2}\omega_{s}^{3}}{3\pi\hbar\epsilon_{0}c^{3}}$ that%

\begin{equation}
\dot{C}_{\mu}=iC_{\mu}\int_{0}^{\infty}d\omega_{s}\frac{\Gamma_{s}}{2\pi
}\text{P.V.}(\omega_{s}+\omega_{23}+\Delta_{2})^{-1}.
\end{equation}

It is now clear the contribution from non-RWA terms to the Lamb shift of the
idler transition resides in $\dot{\mathcal{E}}$ and $\dot{C}_{s}^{\mu}$, which
are proportional to $N$ and $N-1$. \ The difference of the level shifts then
gives rise to $-\int_{0}^{\infty}d\omega\frac{\Gamma}{2\pi}$P.V.$(\omega
+\omega_{3})^{-1}$, and the other part can be derived from substitutions of
RWA terms. \ The signal transition has the same effect as shown in $\dot
{C}_{\mu}$. \ The frequency shift due to dipole-dipole interaction also
appeared in $\dot{C}_{s}^{\mu}$ that has the contribution of interactions from
other atoms. \ There also will be contributions from RWA terms, and we will
show the complete expression for collective decay rate and frequency shift. \ 

Define $C_{s,q_{i}}=\sum_{\mu}C_{s}^{\mu}e^{-i\vec{q}_{i}\cdot\vec{r}_{\mu}}$,
substitute $D_{s,i}$ into $C_{s}^{\mu},$ and we have%

\begin{align}
\dot{C}_{s,q_{i}}  & =g_{s}^{\ast}(\epsilon_{_{s}}^{\ast}\cdot\hat{d}_{s}%
)\sum_{\mu}e^{-i(\vec{k}_{s}+\vec{q}_{i})\cdot\vec{r}_{\mu}}e^{i(\omega
_{ks}-\omega_{23}-\Delta_{2})t}B_{\mu}-\sum_{i}|g_{i}|^{2}|(\vec{\epsilon}%
_{i}\cdot\hat{d}_{i}^{\ast})|^{2}\times\nonumber\\
& \sum_{\mu}e^{i(\vec{k}_{i}-\vec{q}_{i})\cdot\vec{r}_{\mu}}\int_{0}%
^{t}dt^{\prime}e^{i(\omega_{i}-\omega_{3})(t^{\prime}-t)}C_{s,ki}(t^{\prime
})-\sum_{i}|g_{i}|^{2}|(\vec{\epsilon}_{i}\cdot\hat{d}_{i}^{\ast})|^{2}%
\times\nonumber\\
& \int_{0}^{t}dt^{\prime}e^{i(\omega_{i}+\omega_{3})(t^{\prime}-t)}%
\Big[\sum_{\mu}e^{i(\vec{k}_{i}-\vec{q}_{i})\cdot\vec{r}_{\mu}}C_{s,ki}%
-C_{s,qi}\Big]\nonumber\\
& +i(N-1)C_{s,q_{i}}\int_{0}^{\infty}d\omega_{i}\frac{\Gamma_{i}}{2\pi
}\text{P.V.}(\omega_{i}+\omega_{3})^{-1}\nonumber\\
& =g_{s}^{\ast}(\epsilon_{_{s}}^{\ast}\cdot\hat{d}_{s})\sum_{\mu}e^{-i(\vec
{k}_{s}+\vec{q}_{i})\cdot\vec{r}_{\mu}}e^{i(\omega_{ks}-\omega_{23}-\Delta
_{2})t}B_{\mu}-\frac{3}{8\pi}\oint d\Omega_{i}[1-(\hat{k}_{i}\cdot\hat{d}%
_{i})^{2}]\frac{\Gamma_{3}}{2}\nonumber\\
& \times\sum_{\mu}e^{i(\vec{k}_{i}-\vec{q}_{i})\cdot\vec{r}_{\mu}}%
\Big|_{|\vec{k}_{i}|=k_{3}}C_{s,k_{3}\hat{k}_{i}}+i\frac{3}{8\pi}\oint
d\Omega_{i}[1-(\hat{k}_{i}\cdot\hat{d}_{i})^{2}]\int_{0}^{\infty}d\omega
_{i}\frac{\Gamma_{i}}{2\pi}\nonumber\\
& \Big[\text{P.V.}(\omega_{i}-\omega_{3})^{-1}+\text{P.V.}(\omega_{i}%
+\omega_{3})^{-1}\Big]\Big[\sum_{\mu}e^{i(\vec{k}_{i}-\vec{q}_{i})\cdot\vec
{r}_{\mu}}C_{s,ki}-C_{s,qi}\Big]\nonumber\\
& +iC_{s,q_{i}}\int_{0}^{\infty}d\omega_{i}\frac{\Gamma_{i}}{2\pi}%
\text{P.V.}(\omega_{i}-\omega_{3})^{-1}\nonumber\\
& +i(N-1)C_{s,q_{i}}\int_{0}^{\infty}d\omega_{i}\frac{\Gamma_{i}}{2\pi
}\text{P.V.}(\omega_{i}+\omega_{3})^{-1}.
\end{align}
Renormalize the Lamb shift (last two lines in the above) and use $\sum_{q_{i}%
}C_{s,q_{i}}e^{i\vec{q}_{i}\cdot\vec{r}_{\nu}}=NC_{s}^{\nu}$ then we have%

\begin{align}
\dot{C}_{s,q_{i}}  & =g_{s}^{\ast}(\epsilon_{_{s}}^{\ast}\cdot\hat{d}_{s}%
)\sum_{\mu}e^{-i(\vec{k}_{s}+\vec{q}_{i})\cdot\vec{r}_{\mu}}e^{i(\omega
_{ks}-\omega_{23}-\Delta_{2})t}B_{\mu}-\frac{3}{8\pi}\oint d\Omega_{i}%
[1-(\hat{k}_{i}\cdot\hat{d}_{i})^{2}]\frac{\Gamma_{3}}{2}\times\nonumber\\
& \sum_{\mu}e^{i(\vec{k}_{i}-\vec{q}_{i})\cdot\vec{r}_{\mu}}\sum_{\nu
}e^{-i\vec{k}_{i}\cdot\vec{r}_{\nu}}\frac{1}{N}\sum_{q_{i}^{\prime}}%
e^{i\vec{q}_{i}^{\prime}\cdot\vec{r}_{\nu}}C_{s,q_{i}^{\prime}}\Big|_{|\vec
{k}_{i}|=k_{3}}+i\frac{3}{8\pi}\oint d\Omega_{i}[1-(\hat{k}_{i}\cdot\hat
{d}_{i})^{2}]\times\nonumber\\
& \int_{0}^{\infty}d\omega_{i}\frac{\Gamma_{i}}{2\pi}\Big[\text{P.V.}%
(\omega_{i}-\omega_{3})^{-1}+\text{P.V.}(\omega_{i}+\omega_{3})^{-1}%
\Big]\Big[\sum_{\mu}e^{i(\vec{k}_{i}-\vec{q}_{i})\cdot\vec{r}_{\mu}}\sum
_{\nu\neq\mu}e^{-i\vec{k}_{i}\cdot\vec{r}_{\nu}}C_{s}^{\nu}\Big]\nonumber\\
& =g_{s}^{\ast}(\epsilon_{_{s}}^{\ast}\cdot\hat{d}_{s})\sum_{\mu}e^{-i(\vec
{k}_{s}+\vec{q}_{i})\cdot\vec{r}_{\mu}}e^{i(\omega_{ks}-\omega_{23}-\Delta
_{2})t}B_{\mu}-\frac{3}{8\pi}\oint d\Omega_{i}[1-(\hat{k}_{i}\cdot\hat{d}%
_{i})^{2}]\frac{\Gamma_{3}}{2}\nonumber\\
& \times\frac{1}{N}\sum_{\mu}e^{i(\vec{k}_{i}-\vec{q}_{i})\cdot\vec{r}_{\mu}%
}\sum_{q_{i}^{\prime}}\sum_{\nu}e^{i(\vec{q}_{i}^{\prime}-\vec{k}_{i}%
)\cdot\vec{r}_{\nu}}C_{s,q_{i}^{\prime}}\Big|_{|\vec{k}_{i}|=k_{3}}+i\frac
{3}{8\pi}\oint d\Omega_{i}[1-(\hat{k}_{i}\cdot\hat{d}_{i})^{2}]\nonumber\\
& \times\int_{0}^{\infty}d\omega_{i}\frac{\Gamma_{i}}{2\pi}\Big[\text{P.V.}%
(\omega_{i}-\omega_{3})^{-1}+\text{P.V.}(\omega_{i}+\omega_{3})^{-1}%
\Big]\frac{1}{N}\sum_{\mu}e^{i(\vec{k}_{i}-\vec{q}_{i})\cdot\vec{r}_{\mu}}%
\sum_{q_{i}^{\prime}}\nonumber\\
& \Big[\sum_{\nu}e^{i(\vec{q}_{i}^{\prime}-\vec{k}_{i})\cdot\vec{r}_{\nu}%
}-e^{i(\vec{q}_{i}^{\prime}-\vec{k}_{i})\cdot\vec{r}_{\mu}}\Big]C_{s,q_{i}%
^{\prime}}.
\end{align}

Due to the summation of exponential factors from the above, the coupling from
the other modes $q_{i}^{\prime}$ is significant only when $q_{i}^{\prime
}=k_{i}=q_{i}$, so finally we have%

\begin{align}
\dot{C}_{s,q_{i}}  & =g_{s}^{\ast}(\epsilon_{_{s}}^{\ast}\cdot\hat{d}_{s}%
)\sum_{\mu}e^{-i(\vec{k}_{s}+\vec{q}_{i})\cdot\vec{r}_{\mu}}e^{i(\omega
_{ks}-\omega_{23}-\Delta_{2})t}B_{\mu}-\frac{\Gamma_{3}}{2}(N\bar{\mu
}+1)C_{s,q_{i}}\nonumber\\
& +i\delta\omega_{i}C_{s,q_{i}}%
\end{align}
where the collective decay rate is \cite{mu}%

\begin{equation}
\frac{\Gamma_{3}}{2}(N\bar{\mu}+1)\equiv\frac{\Gamma_{3}}{2}\frac{3}{8\pi
}\oint d\Omega_{i}[1-(\hat{k}_{i}\cdot\hat{d}_{i})^{2}]\frac{1}{N}\sum
_{\mu,\nu}e^{i(\vec{k}_{i}-\vec{q}_{i})\cdot(\vec{r}_{\mu}-\vec{r}_{\nu})},
\end{equation}
\bigskip and the collective frequency shift expressed in terms of the
continuous integral over a frequency space is%

\begin{align}
\delta\omega_{i}  & \equiv\int_{0}^{\infty}d\omega_{i}\frac{\Gamma_{i}}{2\pi
}\Big[\text{P.V.}(\omega_{i}-\omega_{3})^{-1}+\text{P.V.}(\omega_{i}%
+\omega_{3})^{-1}\Big]N\bar{\mu}(k_{i}),\nonumber\\
& =\int_{0}^{\infty}d\omega_{i}\frac{\Gamma_{i}}{2\pi}\Big[\text{P.V.}%
(\omega_{i}-\omega_{3})^{-1}+\text{P.V.}(\omega_{i}+\omega_{3})^{-1}%
\Big]\frac{1}{N}\sum_{\mu,\nu\neq\mu}e^{i(\vec{k}_{i}-\vec{q}_{i})\cdot
(\vec{r}_{\mu}-\vec{r}_{\nu})}.\nonumber\\
&
\end{align}

The geometrical constant $\bar{\mu}$ for a cylindrical ensemble (of height $h
$ and radius $a$) is
\begin{equation}
\bar{\mu}(k_{3})=\frac{6(N-1)}{NA^{2}H^{2}}\int_{-1}^{1}\frac{dx(1+x^{2}%
)}{(1-x)^{2}(1-x^{2})}\text{sin}^{2}[\frac{1}{2}H(1-x)]J_{1}^{2}%
[A(1-x^{2})^{1/2}]\label{mu}%
\end{equation}
where $H=k_{3}h$ and $A=k_{3}a$ are dimensionless length scales, and circular
polarizations are considered \cite{mu}. $J_{1}$ is the Bessel function of the
first kind.

The alternative way to express the collective decay rate and shift is
\cite{Lehm}%

\begin{align}
\frac{\Gamma_{3}^{N}}{2}  & =\frac{\Gamma_{3}}{2}(N\bar{\mu}+1)\equiv
\frac{\Gamma_{3}}{2}\frac{1}{N}\sum_{\mu,\nu}F_{\mu\nu}(k_{3}r_{\mu\nu
})e^{-i\vec{q}_{i}\cdot(\vec{r}_{\mu}-\vec{r}_{\nu})},\\
\delta\omega_{i}  & =-\frac{\Gamma_{3}}{2}\frac{2}{N}\sum_{\mu,\nu\neq\mu
}G_{\mu\nu}(k_{3}r_{\mu\nu})e^{-i\vec{q}_{i}\cdot(\vec{r}_{\mu}-\vec{r}_{\nu
})}\nonumber\\
& =\frac{\Gamma_{3}}{Nk_{3}^{3}}\text{P.V.}\int_{-\infty}^{\infty}\frac
{dk}{2\pi}\frac{k^{3}}{k-k_{3}}\sum_{\mu,\nu\neq\mu}F_{\mu\nu}(kr_{\mu\nu
})e^{-i\vec{q}_{i}\cdot(\vec{r}_{\mu}-\vec{r}_{\nu})}.
\end{align}
where%
\begin{align}
F_{\alpha\beta}(\xi)  & =\frac{3}{2}\{[1-(\hat{p}\cdot\hat{r}_{\alpha\beta
})^{2}]\frac{\sin\xi}{\xi}+[1-3(\hat{p}\cdot\hat{r}_{\alpha\beta})^{2}%
](\frac{\cos\xi}{\xi^{2}}-\frac{\sin\xi}{\xi^{3}})\},\nonumber\\
G_{\alpha\beta}(\xi)  & =\frac{3}{4}\{-[1-(\hat{p}\cdot\hat{r}_{\alpha\beta
})^{2}]\frac{\cos\xi}{\xi}+[1-3(\hat{p}\cdot\hat{r}_{\alpha\beta})^{2}%
](\frac{\sin\xi}{\xi^{2}}+\frac{\cos\xi}{\xi^{3}})\},\label{dd}%
\end{align}
and note that $\xi=k_{3}r_{\alpha\beta}$.

\section{Adiabatic Approximation}

Under the conditions of large detuned laser excitations, we may use the
adiabatic approximation to eliminate the laser-excited states and solve for
the signal-idler probability amplitude. \ Before proceeding to the adiabatic
approximation, we solve $C_{s,q_{i}}$ first and substitute it to solve
$B_{\mu}$.%

\begin{equation}
C_{s,q_{i}}(t)=g_{s}^{\ast}(\epsilon_{_{s}}^{\ast}\cdot\hat{d}_{s})\sum_{\mu
}e^{-i(\vec{k}_{s}+\vec{q}_{i})\cdot\vec{r}_{\mu}}\int_{0}^{t}dt^{\prime
}e^{i(\omega_{s}-\omega_{23}-\Delta_{2})t^{\prime}}e^{(-\frac{\Gamma_{3}^{N}%
}{2}+i\delta\omega_{i})(t-t^{\prime})}B_{\mu}(t^{\prime}).
\end{equation}

Let $B_{k_{a}+k_{b}}\equiv\sum_{\mu}e^{-i(\vec{k}_{a}+\vec{k}_{b})\cdot\vec
{r}_{\mu}}B_{\mu}$ and $A_{k_{a}}\equiv\sum_{\mu}e^{-i\vec{k}_{a}\cdot\vec
{r}_{\mu}}A_{\mu}$, we have%

\begin{align}
\dot{B}_{k_{a}+k_{b}}  & =i\Delta_{2}B_{k_{a}+k_{b}}+i\frac{\Omega_{b}}%
{2}A_{k_{a}}-\sum_{ks,\lambda_{s}}|g_{s}|^{2}|\epsilon_{k_{s},\lambda_{s}%
}^{\ast}\cdot\hat{d}_{s}|^{2}\int_{0}^{t}dt^{\prime}e^{i(\omega_{ks}%
-\omega_{23}-\Delta_{2})(t^{\prime}-t)}\nonumber\\
& e^{(-\frac{\Gamma_{3}^{N}}{2}+i\delta\omega_{i})(t-t^{\prime})}%
B_{k_{a}+k_{b}}(t^{\prime})\nonumber\\
& =i\Delta_{2}B_{k_{a}+k_{b}}+i\frac{\Omega_{b}}{2}A_{k_{a}}-\frac{\Gamma_{2}%
}{2}B_{k_{a}+k_{b}}+iB_{k_{a}+k_{b}}\int_{0}^{\infty}d\omega_{s}\frac
{\Gamma_{s}}{2\pi}\times\nonumber\\
& \text{P.V.}(\omega_{s}-\omega_{23}-\Delta_{2})^{-1}%
\end{align}
where the Weisskopf-Wigner approach is used to derive the decay rate for the
signal transition, and in conjunction with the result of $\dot{C}_{\mu},$ the
Lamb shift is also derived as the difference of level shifts that $\int
_{0}^{\infty}d\omega_{s}\frac{\Gamma_{s}}{2\pi}[$P.V.$(\omega_{s}-\omega
_{23}-\Delta_{2})^{-1}-$P.V.$(\omega_{s}+\omega_{23}+\Delta_{2})^{-1}]$.
\ \ We then renormalize it and apply the adiabatic approximation.

When the detunings are large enough that%
\[
|\Delta_{1}|,|\Delta_{2}|\gg\frac{|\Omega_{a}|}{2},\frac{|\Omega_{b}|}%
{2},\frac{\Gamma_{2}}{2}.
\]

We can solve the coupled equations of motion by adiabatically eliminating the
intermediate and upper excited states in the excitation process. \ The
adiabatic approximation requires that the driving pulses are smoothly turned
on, and we will show under what condition of the pulses that the approximation
is valid.

First we use integration by parts to solve the probability amplitudes in the
adiabatic approximation (zeroth order) and their first-order correction.
\ Note that we allow time-varying Rabi frequencies.%

\begin{align}
A_{k_{a}}(t)  & =e^{i\Delta_{1}t}\Big[\frac{i}{2}\int_{-\infty}^{t}%
e^{-i\Delta_{1}t^{\prime}}\Omega_{a}(t^{\prime})\mathcal{E}(t^{\prime
})dt^{\prime}+\frac{i}{2}\int_{-\infty}^{t}e^{-i\Delta_{1}t^{\prime}}%
\Omega_{b}^{\ast}(t^{\prime})B_{k_{a}+k_{b}}(t^{\prime})dt^{\prime
}\Big]\nonumber\\
& \approx-\frac{N\Omega_{a}(t)\mathcal{E}(t)}{2\Delta_{1}}-\frac{\Omega
_{b}^{\ast}(t)B_{k_{a}+k_{b}}(t)}{2\Delta_{1}}+\frac{i}{2\Delta_{1}^{2}}%
\frac{d}{dt}\Big(\Omega_{a}(t)\mathcal{E}(t)\Big)\nonumber\\
& +\frac{i}{2\Delta_{1}^{2}}\frac{d}{dt}\Big(\Omega_{b}^{\ast}(t)B_{k_{a}%
+k_{b}}(t)\Big),\\
B_{k_{a}+k_{b}}(t)  & =e^{i(\Delta_{2}+i\Gamma_{2}/2)t}\Big[\frac{i}{2}%
\int_{-\infty}^{t}e^{-i(\Delta_{2}+i\Gamma_{2}/2)t^{\prime}}\Omega
_{b}(t^{\prime})A_{k_{a}}(t^{\prime})dt^{\prime}\Big]\nonumber\\
& \approx-\frac{\Omega_{b}(t)A_{k_{a}}(t)}{2(\Delta_{2}+i\Gamma_{2}/2)}%
+\frac{i\frac{d}{dt}\Big(\Omega_{b}(t)A_{k_{a}}(t)\Big)}{2(\Delta_{2}%
+i\Gamma_{2}/2)^{2}}%
\end{align}
where higher order terms involving a second derivative of the fields are
neglected due to their feature of slow variation. \ The initial conditions are
used in the below,
\begin{align}
& B_{k_{a}+k_{b}}({-\infty})=A_{k_{a}}({-\infty})=0,~\frac{d}{dt^{\prime}%
}\Big(\Omega_{a}(t^{\prime})\mathcal{E}(t^{\prime})\Big)\Big|_{-\infty
}=0,\nonumber\\
& \frac{d}{dt^{\prime}}\Big(\Omega_{b}^{\ast}(t^{\prime})B_{k_{a}+k_{b}%
}(t^{\prime})\Big)\Big|_{-\infty}=\frac{d}{dt^{\prime}}\Big(\Omega
_{b}(t^{\prime})A_{k_{a}}(t^{\prime})\Big)\Big|_{-\infty}=0.\nonumber
\end{align}

With conditions in the following (i) to (iii),%

\begin{align}
(\text{i})  & \left\vert \frac{\frac{d}{dt}\Big(\Omega_{a}(t)\mathcal{E}%
(t)\Big)}{\Delta_{1}\Omega_{a}(t)\mathcal{E}(t)}\right\vert \ll1,\\
(\text{ii})  & \left\vert \frac{\frac{d}{dt}\Big(\Omega_{b}^{\ast}%
(t)B_{k_{a}+k_{b}}(t)\Big)}{\Delta_{1}\Omega_{b}^{\ast}(t)B_{k_{a}+k_{b}}%
(t)}\right\vert \ll1,\\
(\text{iii})  & \left\vert \frac{\frac{d}{dt}\Big(\Omega_{b}(t)A_{k_{a}%
}(t)\Big)}{(\Delta_{2}+i\Gamma_{2}/2)\Omega_{b}(t)A_{k_{a}}(t)}\right\vert
\ll1,
\end{align}
we can derive $A_{k_{a}}(t)$, $B_{k_{a}+k_{b}}(t),$ and $\mathcal{E}(t)$ in
the adiabatic approximation,%
\begin{align}
A_{k_{a}}(t)  & =\frac{-\frac{N\Omega_{a}(t)\mathcal{E}(t)}{2\Delta_{1}}%
}{1-\frac{|\Omega_{b}(t)|^{2}}{4\Delta_{1}(\Delta_{2}+i\Gamma_{2}/2)}}%
\approx-\frac{N\Omega_{a}(t)}{2\Delta_{1}}\mathcal{E}(t),\\
\mathcal{E}(t)  & =e^{-\frac{iN}{4\Delta_{1}}\int_{-\infty}^{t}|\Omega
_{a}(t^{\prime})|^{2}dt^{\prime}}\approx1-\frac{iN}{4\Delta_{1}}\int_{-\infty
}^{t}|\Omega_{a}(t^{\prime})|^{2}dt^{\prime},\\
B_{k_{a}+k_{b}}(t)  & =\frac{\frac{N\Omega_{a}(t)\Omega_{b}(t)}{4\Delta
_{1}\Delta_{2}}\mathcal{E}(t)}{1-\frac{|\Omega_{b}(t)|^{2}}{4\Delta_{1}%
(\Delta_{2}+i\Gamma_{2}/2)}}\approx\frac{N\Omega_{a}(t)\Omega_{b}(t)}%
{4\Delta_{1}\Delta_{2}}\equiv Nb(t),
\end{align}
where the probability amplitude of the first excited state follows the first
laser field, and the upper excited state follows the products of two laser
fields. \ The AC Stark shift is present in the ground state that can be
ignored if $\Delta_{1}\gg N\int_{-\infty}^{t}|\Omega_{a}(t^{\prime}%
)|^{2}dt^{\prime}/4$. \ This condition is also required for the assumption of
single excitations states we consider.

Finally, we have the probability amplitudes associated with the signal
$C_{s,k_{i}}(t)$ and signal-idler photons $D_{s,i}(t)$,%

\begin{align}
& C_{s,k_{i}}(t)\nonumber\\
& =g_{s}^{\ast}(\epsilon_{_{s}}^{\ast}\cdot\hat{d}_{s})\int_{0}^{t}dt^{\prime
}e^{i(\omega_{s}-\omega_{23}-\Delta_{2})t^{\prime}}e^{(-\frac{\Gamma_{3}^{N}%
}{2}+i\delta\omega_{i})(t-t^{\prime})}\sum_{\mu}e^{i\Delta\vec{k}\cdot\vec
{r}_{\mu}}e^{-i(\vec{k}_{a}+\vec{k}_{b})\cdot\vec{r}_{\mu}}B_{\mu}(t^{\prime
})\nonumber\\
& =g_{s}^{\ast}(\epsilon_{_{s}}^{\ast}\cdot\hat{d}_{s})\frac{1}{N}\sum_{\mu
}e^{i\Delta\vec{k}\cdot\vec{r}_{\mu}}\int_{0}^{t}dt^{\prime}e^{i(\omega
_{s}-\omega_{23}-\Delta_{2})t^{\prime}}e^{(-\frac{\Gamma_{3}^{N}}{2}%
+i\delta\omega_{i})(t-t^{\prime})}B_{k_{a}+k_{b}}(t^{\prime}),
\end{align}%
\begin{align}
D_{s,i}(t)  & =g_{i}^{\ast}g_{s}^{\ast}(\epsilon_{k_{i},\lambda_{i}}^{\ast
}\cdot\hat{d}_{i})(\epsilon_{k_{s},\lambda_{s}}^{\ast}\cdot\hat{d}_{s}%
)\sum_{\mu}e^{i\Delta\vec{k}\cdot\vec{r}_{\mu}}\int_{0}^{t}\int_{0}%
^{t^{\prime}}dt^{\prime\prime}dt^{\prime}e^{(-\frac{\Gamma_{3}^{N}}{2}%
+i\delta\omega_{i})(t^{\prime}-t^{\prime\prime})}\nonumber\\
& e^{i(\omega_{i}-\omega_{3})t^{\prime}}e^{i(\omega_{s}-\omega_{23}-\Delta
_{2})t^{\prime\prime}}b(t^{\prime\prime}).\label{two}%
\end{align}

Note that $e^{-i(\vec{k}_{a}+\vec{k}_{b})\cdot\vec{r}_{\mu}}B_{\mu}(t^{\prime
})$ does not depend on the atomic index $\mu$ under the adiabatic
approximation, and $\Delta\vec{k}=\vec{k}_{a}+\vec{k}_{b}-\vec{k}_{s}-\vec
{k}_{i}$ is the phase mismatch.

The above expressions are the main results of this Appendix and we proceed to
investigate their properties when Gaussian pump pulses are used in Chapter 3.



\chapter{Derivation of a c-number Langevin equation for the cascade emission}

In this appendix, we show the details in the derivations of c-number Langevin
equations that are the foundation for numerical approaches of the cascade
emission in Chapter 4. \ First we describe how to quantize the free
electromagnetic field \cite{quantization}, and we formulate the Fokker-Planck
equation for our system using the positive P-representation. \ We derive the
Fokker-Planck equations by characteristic functions \cite{LT:Haken}, and the
corresponding c-number Langevin equations are derived. \ The noise
correlations are found from the diffusion coefficients in Fokker-Planck
equations. \ 

\section{Quantized Electromagnetic Field}

To describe the propagating quantum fields in one dimension, we take the
approach of the reference \cite{quantization}. \ Before proceeding, we specify
the positive frequency of a free propagating field operator in the discrete space,%

\begin{equation}
\hat{E}^{+}(\vec{x})=i\sum_{k,\lambda}\sqrt{\frac{\hbar\omega_{k}}%
{2\epsilon_{0}V}}\hat{a}_{k,\lambda}\vec{\epsilon}_{k,\lambda}e^{i\vec{k}%
\cdot\vec{x}}%
\end{equation}
where $\vec{\epsilon}_{k,\lambda}$ and $\vec{\epsilon}_{\lambda}(\vec{k})$
specify polarizations of the field, and the interchange of discrete and
continuous space has relation,
\[
\sum_{k}\rightarrow\frac{V}{(2\pi)^{3}}\int d^{3}k~,~~\hat{a}_{k,\lambda
}\rightarrow\sqrt{\frac{V}{(2\pi)^{3}}}\int d^{3}k\hat{a}_{\lambda}(\vec{k}),
\]
where creation and annihilation operators satisfy commutation relations,
\begin{equation}
\lbrack\hat{a}_{k,\lambda},\hat{a}_{k^{\prime},\lambda^{\prime}}^{\dag
}]=\delta_{\lambda,\lambda^{\prime}}\delta_{k,k^{\prime}}.
\end{equation}

For the purpose of describing one-dimensional propagating field (paraxial
approximation), we discretize the space along the propagation ($\hat{z}$) and
denote $\vec{r}$ as the vectors on the cross section. We then have%
\begin{align}
\hat{E}^{+}(z,\vec{r})  & =i\sum_{n=-M}^{M}\sqrt{\frac{\hbar\omega_{s,n}%
}{2\epsilon_{0}}}e^{i(k_{s}+k_{n})z}\frac{1}{\sqrt{V}}\sum_{\lambda}%
\sum_{k_{n\perp}}e^{i\vec{k}_{n\perp}\cdot\vec{r}}\vec{\epsilon}%
_{k_{n},\lambda}\hat{a}_{k,\lambda}\\
k_{n}  & =\frac{2\pi n}{L}~,~~\omega_{s,n}=\omega_{s}+k_{n}c~,~~\omega
_{s}=k_{s}c~,~~n=-M,...,M\nonumber
\end{align}
where $L$ is the length of propagation that is equally split into $2M+1$
elements, and the center of the interval is $z=z_{m}=\frac{mL}{2M+1}$
with$~m=-M,...,M$. $\ k_{s}$ is the central longitudinal mode of the field.
\ Note that the polarization $\vec{\epsilon}_{k,\lambda}$\ with paraxial
approximation has $k\approx k_{n}$.

The next step is to characterize the transverse mode of propagating field, and
we introduce a set of orthonormal transverse mode functions ($f_{i,k_{n\perp}%
}$) that $\sum_{k_{n\perp}}f_{i,k_{n\perp}}^{\ast}f_{j,k_{n\perp}}=\delta
_{ij}$. \ A longitudinal annihilation operator is defined as
\[
\hat{c}_{n,i,\lambda}\equiv\sum_{k_{n\perp}}f_{i,k_{n\perp}}^{\ast}\hat
{a}_{k,\lambda},
\]
which also satisfies commutation relations $[\hat{c}_{n,i,\lambda},\hat
{c}_{n^{\prime},j,\lambda^{\prime}}^{\dag}]=\delta_{nn^{\prime}}\delta
_{ij}\delta_{\lambda\lambda^{\prime}}$. \ We can substitute $\hat
{a}_{k,\lambda}=\sum_{i}\hat{c}_{n,i,\lambda}f_{i,k_{n\perp}}$ that
\begin{equation}
\hat{E}^{+}(z,\vec{r})=i\sum_{n=-M}^{M}\sqrt{\frac{\hbar\omega_{s,n}%
}{2\epsilon_{0}}}e^{i(k_{s}+k_{n})z}\frac{1}{\sqrt{V}}\sum_{\lambda}%
\sum_{k_{n\perp}}e^{i\vec{k}_{n\perp}\cdot\vec{r}}\vec{\epsilon}%
_{k_{n},\lambda}f_{i,k_{n\perp}}\hat{c}_{n,i,\lambda}.
\end{equation}

Let the spatial transverse mode function
\[
u_{i}(\vec{r})\equiv\frac{i}{\sqrt{V}}\sum_{k_{n\perp}}e^{i\vec{k}_{n\perp
}\cdot\vec{r}}f_{i,k_{n\perp}}~,~~\text{where}~~\int d^{2}rdzu_{i}^{\ast}%
u_{i}=1
\]
and we have
\begin{equation}
\hat{E}^{+}(z,\vec{r})=\sum_{n=-M}^{M}\sqrt{\frac{\hbar\omega_{s,n}}%
{2\epsilon_{0}}}e^{i(k_{s}+k_{n})z}\sum_{i,\lambda}\vec{\epsilon}%
_{k_{n},\lambda}\hat{c}_{n,i,\lambda}u_{i}(\vec{r}).
\end{equation}

An approximation of a single transverse mode can be applied if only single
mode is collected for the experiment, and a flat transverse mode can be
assumed ($u_{i}=\frac{1}{\sqrt{V}}$) if the collected mode has a narrower
spatial bandwidth than the mode function. \ Finally, we have
\begin{equation}
\hat{E}^{+}(z,\vec{r})=\sum_{n=-M}^{M}\sqrt{\frac{\hbar w_{s,n}}{2\epsilon
_{0}V}}e^{i(k_{s}+k_{n})z}\sum_{\lambda}\vec{\epsilon}_{k_{n},\lambda}\hat
{c}_{n,\lambda}.
\end{equation}

For a demonstration of deriving an interaction Hamiltonian and Maxwell-Bloch
equations, we use a two-state system ($|0\rangle$ and $|1\rangle$), and the
polarization is not concerned here. \ The free field and interaction
Hamiltonian (interacting with atomic ensemble with $N$ atoms) is
\begin{align}
H_{0}  & =\sum_{n}\hbar\omega_{s,n}\hat{c}_{n}^{\dag}\hat{c}_{n}+\hbar
\omega_{1}|1\rangle\langle1|,\\
V  & =-\sum_{\mu}^{N}\vec{d}^{\mu}\cdot\vec{E}(r_{\mu})=\left[  -\hbar
g\sum_{\mu}\sum_{n,m=-M}^{M}\hat{\sigma}^{\mu,m\dag}\hat{c}_{n}e^{i(k_{s}%
+k_{n})z_{m}}+h.c.\right]  _{\text{RWA}},\\
g  & \equiv\frac{d}{\hbar}\sqrt{\frac{\hbar\omega_{s}}{2\epsilon_{0}V}}%
,~\vec{d}^{\mu}\equiv\hat{\sigma}^{\mu}+\hat{\sigma}^{\mu,\dag},~\vec{E}%
\equiv\hat{E}^{+}+\hat{E}^{-}%
\end{align}
where $\sum_{\mu}$ sums over $\frac{N}{2M+1}$ atoms in the cross sections and
the index $m$ on raising and lowering atomic operators $\hat{\sigma}$
characterizes the position of the atoms. \ The rotating wave approximation
(RWA) is made in the interaction Hamiltonian and slowly varying coupling
constant $g$ is taken out of the discrete mode sum and is assigned a central
frequency, which is the narrow band assumption for the field.

Now we introduce a new operator
\begin{equation}
\hat{b}_{l}=\frac{1}{\sqrt{2M+1}}\sum_{n=-M}^{M}\hat{c}_{n}e^{ik_{n}z_{l}%
}~,~~l=-M,...,M,
\end{equation}
which satisfies commutation relations $[\hat{b}_{l},\hat{b}_{l^{\prime}}%
^{\dag}]=\delta_{ll^{\prime}},$ and the Hamiltonian can be re-expressed as
\begin{align}
H_{0}  & =\hbar\omega_{s}\sum_{l}\hat{b}_{l}^{\dag}\hat{b}_{l}+\hbar
\sum_{ll^{\prime}}\omega_{ll^{\prime}}\hat{b}_{l}^{\dag}\hat{b}_{l^{\prime}%
}+\hbar\omega_{1}|1\rangle\langle1|,\\
V  & =-\hbar g\sum_{\mu,l}\sqrt{2M+1}\hat{\sigma}^{\mu,l\dag}\hat{b}%
_{l}e^{ik_{s}z_{l}}+h.c..
\end{align}

The Heisenberg equation of slowly varying field operators ($\tilde{b}%
_{l}\equiv\hat{b}_{l}e^{i\omega_{s}t}$) is
\begin{equation}
\dot{\tilde{b}}_{l}=-i\sum_{l^{\prime}}\omega_{ll^{\prime}}\tilde
{b}_{l^{\prime}}+ig^{\ast}\sum_{\mu}\sqrt{2M+1}\hat{\sigma}^{\mu,l}%
e^{-ik_{s}z_{l}+i\omega_{s}t}~,~~\omega_{ll^{\prime}}\equiv\sum_{n}\frac
{k_{n}c}{2M+1}e^{ik_{n}(z_{l}-z_{l^{\prime}})},
\end{equation}
and we may use the limit of $M\rightarrow\infty$ that
\begin{equation}
z_{m}=\frac{mL}{2M+1}\rightarrow z~,~~\sqrt{2M+1}~\tilde{b}_{l}\rightarrow
\tilde{E}_{s}^{+}(z,t)~,~~-i\sum_{l^{\prime}}\omega_{ll^{\prime}}\tilde
{b}_{l^{\prime}}\sqrt{2M+1}\rightarrow-c\frac{d}{dz}\tilde{E}_{s}^{+}(z,t)
\end{equation}
where the derivative can be shown from
\begin{align}
-i\sum_{l^{\prime}}\omega_{ll^{\prime}}\tilde{b}_{l^{\prime}}  &
=-i\sum_{l^{\prime}}\sum_{n}\frac{k_{n}c}{2M+1}e^{ik_{n}(z_{l}-z_{l^{\prime}%
})}\tilde{b}_{l^{\prime}}=-\frac{c}{2M+1}\sum_{l^{\prime}}\sum_{n}\frac
{d}{dz_{l}}[e^{ik_{n}(z_{l}-z_{l^{\prime}})}]\tilde{b}_{l^{\prime}}\nonumber\\
& =-c\sum_{l^{\prime}}\frac{d}{dz_{l}}\delta_{ll^{\prime}}\tilde{b}%
_{l^{\prime}}=-c\frac{d}{dz_{l}}\tilde{b}_{l}.
\end{align}

In the end, we have
\begin{equation}
(\frac{\partial}{\partial t}+c\frac{\partial}{\partial z})\tilde{E}_{s}%
^{+}(z,t)=ig^{\ast}\text{lim}_{M\rightarrow\infty}(2M+1)\sum_{\mu}\hat{\sigma
}^{\mu,l}e^{-ik_{s}z_{l}+i\omega_{s}t}\Big|_{z_{l}\rightarrow z},
\end{equation}
and use the limit,
\[
\text{lim}_{M\rightarrow\infty}\frac{2M+1}{L}\delta_{ll^{\prime}}%
\rightarrow\delta(z-z^{\prime}),
\]
then we have (define slowly varying atomic operators $\tilde{\sigma}^{\mu
,l}=\hat{\sigma}^{\mu,l}e^{-ik_{s}z_{l}+i\omega_{s}t}$)
\begin{equation}
\text{lim}_{M\rightarrow\infty}\frac{2M+1}{L}\sum_{\mu=1}^{N}\tilde{\sigma
}^{\mu,l}\delta_{z_{\mu},z_{l}}L\Big|_{z_{l}\rightarrow z}\rightarrow\sum
_{\mu=1}^{N}\tilde{\sigma}^{\mu}\delta(z_{\mu}-z)L=\frac{N}{N_{z}}\sum_{\mu
}^{N_{z}}\tilde{\sigma}^{\mu}.
\end{equation}

The field propagation equation in Maxwell-Bloch equations becomes%
\begin{equation}
(\frac{\partial}{\partial t}+c\frac{\partial}{\partial z})\tilde{E}_{s}%
^{+}(z,t)=ig^{\ast}\sum_{\mu=1}^{N}\tilde{\sigma}^{\mu}\delta(z_{\mu
}-z)L=ig^{\ast}\frac{N}{N_{z}}\sum_{\mu=1}^{N_{z}}\tilde{\sigma}^{\mu}.
\end{equation}

\section{Positive P-representation}

The phase space methods \cite{QN:Gardiner} that mainly include P-, Q-, and
Wigner (W) representations are techniques of using classical analogues to
study quantum systems, especially harmonic oscillators. \ The eigenstate of
harmonic oscillator is a coherent state that provides the basis expansion to
construct various representations. \ P and Q-representation are associated
respectively with evaluations of normal and anti-normal order correlations of
creation and destruction operators. \ W-representation is invented for the
purpose of describing symmetrically ordered creation and destruction
operators. \ Since P-representation describes normally ordered quantities that
are relevant in experiments, we are interested in investigating one class of
generalized P-representations, the positive P-representation that has
semi-definite property in the diffusion process, which is important in
describing quantum noise systems.

Postive-P representation \cite{QO:Walls, drummond80} is an extension to
Glauber-Sudarshan P-representation that uses coherent state ($|\alpha\rangle$)
as a basis expansion of density operator $\rho$. \ In terms of diagonal
coherent states with a quasi-probability distribution, $P(\alpha,\alpha^{\ast
})$, a density operator in P-representation is
\begin{equation}
\rho=\int_{D}|\alpha\rangle\langle\alpha|P(\alpha,\alpha^{\ast})d^{2}\alpha,
\end{equation}
where $D$ represents the integration domain. \ The normalization condition of
$\rho,$ which is Tr\{$\rho$\}$=1,$ indicates the normalization for $P$ as
well, $\int_{D}P(\alpha,\alpha^{\ast})d^{2}\alpha=1$. \ 

Positive P-representation uses a non-diagonal coherent state expansion and the
density operator can be expressed as%
\begin{equation}
\rho=\int_{D}\Lambda(\alpha,\beta)P(\alpha,\beta)d\mu(\alpha,\beta),
\end{equation}
where%

\begin{equation}
d\mu(\alpha,\beta)=d^{2}\alpha d^{2}\beta\text{ and }\Lambda(\alpha
,\beta)=\frac{|\alpha\rangle\langle\beta^{\ast}|}{\langle\beta^{\ast}%
|\alpha\rangle},
\end{equation}
and $\langle\beta^{\ast}|\alpha\rangle$ in non-diagonal projection operators,
$\Lambda(\alpha,\beta),$ makes sure of the normalization condition in
distribution function, $P(\alpha,\beta).$

Any normally ordered observable can be deduced from the distribution function
$P(\alpha,\beta)$ that%

\begin{equation}
\langle(a^{\dag})^{m}a^{n}\rangle=\int_{D}\beta^{m}\alpha^{n}P(\alpha
,\beta)d\mu(\alpha,\beta).
\end{equation}

A characteristic function $\chi_{p}(\lambda_{\alpha},\lambda_{\beta})$
(Fourier-transformed distribution function in Glauber-Sudarshan
P-representation but now is extended into a larger dimension) can help
formulate distribution function, which is%

\begin{equation}
\chi_{p}(\lambda_{\alpha},\lambda_{\beta})=\int_{D}e^{i\lambda_{\alpha}%
\alpha+i\lambda_{\beta}\beta}P(\alpha,\beta)d\mu(\alpha,\beta).
\end{equation}
It is calculated from a normally ordered exponential operator $E(\lambda),$%

\begin{equation}
\chi_{p}(\lambda_{\alpha},\lambda_{\beta})=\text{Tr\{}\rho E(\lambda)\text{\},
}E(\lambda)=e^{i\lambda_{\beta}a^{\dagger}}e^{i\lambda_{\alpha}a}.
\end{equation}

Then a Fokker-Planck equation can be derived from the time derivative of
characteristic function,%

\begin{equation}
\frac{\partial\chi_{p}}{\partial t}=\frac{\partial}{\partial t}\text{Tr\{}\rho
E(\lambda)\text{\}=Tr\{}\frac{\partial\rho}{\partial t}E(\lambda)\text{\}}%
\end{equation}
by Liouville equations,%
\begin{equation}
\frac{\partial\rho}{\partial t}=\frac{1}{i\hbar}[H,\rho].
\end{equation}

In laser theory \cite{LT:Haken}, a P-representation method is extended to
describe atomic and atom-field interaction systems. \ When a large number of
atoms is considered, which is indeed the case of the actual laser, a
macroscopic variable can be defined. \ Then a generalized Fokker-Planck
equation can be derived from characteristic functions by neglecting higher
order terms that are proportional to the inverse of number of atoms. \ It is
the similar to our case when we solve light-matter interactions in an atomic
ensemble that the large number cuts off the higher order terms in
characteristic functions, which we will demonstrate in the next subsection.

\subsection{Hamiltonian}

The Hamiltonian is in Schr\"{o}dinger picture, and we separate it into two
parts where $H_{0}$ is the free Hamiltonian of the atomic ensemble and one
dimensional counter-propagating signal and idler fields, and $H_{I}$ is the
interaction Hamiltonian of atoms interacting with two classical fields and two
quantum fields (signal and idler). \ Dipole approximation of $-\vec{d}%
\cdot\vec{E}$ and rotating wave approximation (RWA) have been made to these
interactions. \ Similar to the previous Appendix, we have%

\begin{align}
H  & =H_{0}+H_{I}\text{ ,}\nonumber\\
H_{0}  & =\sum_{i=1}^{3}\sum_{l=-M}^{M}\hbar\omega_{i}\tilde{\sigma}_{ii}%
^{l}+\hbar\omega_{s}\sum_{l=-M}^{M}\hat{a}_{s,l}^{\dag}\hat{a}_{s,l}+\hbar
\sum_{l,l^{\prime}}\omega_{l^{\prime}l}\hat{a}_{s,l}^{\dag}\hat{a}%
_{s,l^{\prime}}\nonumber\\
& +\hbar\omega_{i}\sum_{l=-M}^{M}\hat{a}_{i,l}^{\dag}\hat{a}_{i,l}+\hbar
\sum_{l,l^{\prime}}\omega_{ll^{\prime}}\hat{a}_{i,l}^{\dag}\hat{a}%
_{i,l^{\prime}}\text{ ,}\\
H_{I}  & =-\hbar\sum_{l=-M}^{M}\Big[\Omega_{a}(t)\tilde{\sigma}_{01}%
^{l\dagger}e^{ik_{a}z_{l}-i\omega_{a}t}+\Omega_{b}(t)\tilde{\sigma}%
_{12}^{l\dagger}e^{-ik_{b}z_{l}-i\omega_{b}t}+h.c.\Big]\\
& -\hbar\sum_{l=-M}^{M}\left[  g_{s}\sqrt{2M+1}\tilde{\sigma}_{32}^{l\dagger
}\hat{a}_{s,l}e^{-ik_{s}z_{l}}+g_{i}\sqrt{2M+1}\tilde{\sigma}_{03}^{l\dagger
}\hat{a}_{i,l}e^{ik_{i}z_{l}}+h.c.\right] \nonumber\\
&
\end{align}
where $\tilde{\sigma}_{mn}^{l}\equiv\sum_{\mu}^{N_{z}}\hat{\sigma}_{mn}%
^{\mu,l}=\sum_{\mu}^{N_{z}}|m\rangle_{\mu}\langle n|\Big|_{r_{\mu}=z_{l}%
},~\Omega_{a}(t)\equiv f_{a}(t)d_{10}\mathcal{E}(k_{a})/(2\hbar),$ and $f_{a}$
is slow varying temporal profile without spatial dependence (ensemble scale
much less than pulse length). $g_{s}\equiv d_{23}\mathcal{E}(k_{s}%
)/\hbar,~\mathcal{E}(k)=\sqrt{\hbar\omega/2\epsilon_{0}V}$ and $z_{m}%
=\frac{mL}{2M+1},~m=-M,...,M,$ and $L$ is the length of propagation. \ Note
that the Rabi frequency is half of the standard definition.

The normally ordered exponential operator is chosen as%

\begin{align}
E(\lambda)  & =\prod_{l}E^{l}(\lambda),\nonumber\\
E^{l}(\lambda)  & =e^{i\lambda_{19}^{l}\tilde{\sigma}_{01}^{l\dagger}%
}e^{i\lambda_{18}^{l}\tilde{\sigma}_{12}^{l\dagger}}e^{i\lambda_{17}^{l}%
\tilde{\sigma}_{02}^{l\dagger}}e^{i\lambda_{16}^{l}\tilde{\sigma}%
_{13}^{l\dagger}}e^{i\lambda_{15}^{l}\tilde{\sigma}_{03}^{l\dagger}%
}e^{i\lambda_{14}^{l}\tilde{\sigma}_{32}^{l\dagger}}e^{i\lambda_{13}^{l}%
\tilde{\sigma}_{11}^{l}}e^{i\lambda_{12}^{l}\tilde{\sigma}_{22}^{l}%
}e^{i\lambda_{11}^{l}\tilde{\sigma}_{33}^{l}}e^{i\lambda_{10}^{l}\tilde
{\sigma}_{32}^{l}}\nonumber\\
& e^{i\lambda_{9}^{l}\tilde{\sigma}_{03}^{l}}e^{i\lambda_{8}^{l}\tilde{\sigma
}_{13}^{l}}e^{i\lambda_{7}^{l}\tilde{\sigma}_{02}^{l}}e^{i\lambda_{6}%
^{l}\tilde{\sigma}_{12}^{l}}e^{i\lambda_{5}^{l}\tilde{\sigma}_{01}^{l}%
}e^{i\lambda_{4}^{l}\hat{a}_{s,l}^{\dagger}}e^{i\lambda_{3}^{l}\hat{a}_{s,l}%
}e^{i\lambda_{2}^{l}\hat{a}_{i,l}^{\dagger}}e^{i\lambda_{1}^{l}\hat{a}_{i,l}%
}.\label{order}%
\end{align}

Aside from the atom-field interaction $\frac{\partial\rho}{\partial t}%
=\frac{1}{i\hbar}[H,\rho],$\ when dissipation from vacuum is considered
(single atomic decay), we can express them in terms of a Lindblad form where
we have for the four-level atomic system,
\begin{align}
\big(\frac{\partial\rho}{\partial t}\big)_{sp}  & =\sum_{l=-M}^{M}\sum_{\mu
}^{N_{z}}\Big\{\frac{\gamma_{01}}{2}[2\hat{\sigma}_{01}^{\mu,l}\rho\hat
{\sigma}_{01}^{\mu,l\dagger}-\hat{\sigma}_{01}^{\mu,l\dagger}\hat{\sigma}%
_{01}^{\mu,l}\rho-\rho\hat{\sigma}_{01}^{\mu,l\dagger}\hat{\sigma}_{01}%
^{\mu,l}]\nonumber\\
& +\frac{\gamma_{12}}{2}[2\hat{\sigma}_{12}^{\mu,l}\rho\hat{\sigma}_{12}%
^{\mu,l\dagger}-\hat{\sigma}_{12}^{\mu,l\dagger}\hat{\sigma}_{12}^{\mu,l}%
\rho-\rho\hat{\sigma}_{12}^{\mu,l\dagger}\hat{\sigma}_{12}^{\mu,l}]\nonumber\\
& +\frac{\gamma_{32}}{2}[2\hat{\sigma}_{_{32}}^{\mu,l}\rho\hat{\sigma}_{_{32}%
}^{\mu,l\dagger}-\hat{\sigma}_{_{32}}^{\mu,l\dagger}\hat{\sigma}_{_{32}}%
^{\mu,l}\rho-\rho\hat{\sigma}_{_{32}}^{\mu,l\dagger}\hat{\sigma}_{_{32}}%
^{\mu,l}]\nonumber\\
& +\frac{\gamma_{03}}{2}[2\hat{\sigma}_{03}^{\mu,l}\rho\hat{\sigma}_{03}%
^{\mu,l\dagger}-\hat{\sigma}_{03}^{\mu,l\dagger}\hat{\sigma}_{03}^{\mu,l}%
\rho-\rho\hat{\sigma}_{03}^{\mu,l\dagger}\hat{\sigma}_{03}^{\mu,l}]\Big\}.
\end{align}

The characteristic functions can be calculated,%

\begin{align}
\chi & =\text{Tr\{}E(\lambda)\rho\text{\},}\\
\frac{\partial\chi}{\partial t}  & =\text{Tr\{}E(\lambda)\frac{\partial\rho
}{\partial t}\text{\}}=\big(\frac{\partial\chi}{\partial t}\big)_{A}%
+\big(\frac{\partial\chi}{\partial t}\big)_{L}\nonumber\\
& +\big(\frac{\partial\chi}{\partial t}\big)_{A-L}+\big(\frac{\partial\chi
}{\partial t}\big)_{sp},\\
\big(\frac{\partial\chi}{\partial t}\big)_{A}  & =\text{Tr\{}E(\lambda
)\frac{1}{i\hbar}[H_{A},\rho]\text{\}, }\big(\frac{\partial\chi}{\partial
t}\big)_{L}=\text{Tr\{}E(\lambda)\frac{1}{i\hbar}[H_{L},\rho]\text{\},}%
\nonumber\\
\big(\frac{\partial\chi}{\partial t}\big)_{A-L}  & =\text{Tr\{}E(\lambda
)\frac{1}{i\hbar}[H_{A-L},\rho]\text{\}, }\big(\frac{\partial\chi}{\partial
t}\big)_{sp}=\text{Tr\{}E(\lambda)\big(\frac{\partial\rho}{\partial
t}\big)_{sp}\text{\}}%
\end{align}
where $H_{0}=H_{A}+H_{L}$, $H_{A}$ is the atomic free evolution Hamiltonian,
$H_{L}$ is the Hamiltonian for laser fields, and $H_{A-L}=H_{I}.$ \ Now we
continue to derive the time derivative in each part of characteristic functions.

\subsection{Characteristic function - atomic part}

The atomic part in characteristic function is deduced from $\big(\frac
{\partial\rho}{\partial t}\big)_{A}$%

\begin{align}
& \big(\frac{\partial\chi}{\partial t}\big)_{A}=\text{Tr\{}E(\lambda
)\big(\frac{\partial\rho}{\partial t}\big)_{A}\text{\},}\nonumber\\
& \big(\frac{\partial\rho}{\partial t}\big)_{A}=\frac{1}{i\hbar}[H_{A}%
,\rho]\nonumber\\
& =\sum_{l}\Big[-i\omega_{1}(\tilde{\sigma}_{11}^{l}\rho-\rho\tilde{\sigma
}_{11}^{l})-i\omega_{2}(\tilde{\sigma}_{22}^{l}\rho-\rho\tilde{\sigma}%
_{22}^{l})-i\omega_{3}(\tilde{\sigma}_{33}^{l}\rho-\rho\tilde{\sigma}_{33}%
^{l})\Big]
\end{align}

so various components in $\big(\frac{\partial\chi}{\partial t}\big)_{A}$ are%

\begin{align}
\text{Tr\{}E(\lambda)\sum_{l}\tilde{\sigma}_{11}^{l}\rho\text{\}}  & =\sum
_{l}\text{Tr\{}E(\lambda)\tilde{\sigma}_{11}^{l}\rho\text{\}}\nonumber\\
& =\sum_{l}[i\lambda_{5}\frac{\partial}{\partial(i\lambda_{5})}-i\lambda
_{6}\frac{\partial}{\partial(i\lambda_{6})}-i\lambda_{8}\frac{\partial
}{\partial(i\lambda_{8})}+\frac{\partial}{\partial(i\lambda_{13})}]_{l}%
\chi,\nonumber\\
\text{Tr\{}E(\lambda)\sum_{l}\rho\tilde{\sigma}_{11}^{l}\text{\}}  & =\sum
_{l}[i\lambda_{19}\frac{\partial}{\partial(i\lambda_{19})}-i\lambda_{18}%
\frac{\partial}{\partial(i\lambda_{18})}-i\lambda_{16}\frac{\partial}%
{\partial(i\lambda_{16})}+\frac{\partial}{\partial(i\lambda_{13})}]_{l}%
\chi,\nonumber\\
\text{Tr\{}E(\lambda)\sum_{l}\tilde{\sigma}_{22}^{l}\rho\text{\}}  & =\sum
_{l}[i\lambda_{6}\frac{\partial}{\partial(i\lambda_{6})}+i\lambda_{7}%
\frac{\partial}{\partial(i\lambda_{7})}+i\lambda_{10}\frac{\partial}%
{\partial(i\lambda_{10})}+\frac{\partial}{\partial(i\lambda_{12})}]_{l}%
\chi,\nonumber\\
\text{Tr\{}E(\lambda)\sum_{l}\rho\tilde{\sigma}_{22}^{l}\text{\}}  & =\sum
_{l}[i\lambda_{18}\frac{\partial}{\partial(i\lambda_{18})}+i\lambda_{17}%
\frac{\partial}{\partial(i\lambda_{17})}+i\lambda_{14}\frac{\partial}%
{\partial(i\lambda_{14})}+\frac{\partial}{\partial(i\lambda_{12})}]_{l}%
\chi,\nonumber\\
\text{Tr\{}E(\lambda)\sum_{l}\tilde{\sigma}_{33}^{l}\rho\text{\}}  & =\sum
_{l}[i\lambda_{8}\frac{\partial}{\partial(i\lambda_{8})}+i\lambda_{9}%
\frac{\partial}{\partial(i\lambda_{9})}-i\lambda_{10}\frac{\partial}%
{\partial(i\lambda_{10})}+\frac{\partial}{\partial(i\lambda_{11})}]_{l}%
\chi,\nonumber\\
\text{Tr\{}E(\lambda)\sum_{l}\rho\tilde{\sigma}_{33}^{l}\text{\}}  & =\sum
_{l}[i\lambda_{16}\frac{\partial}{\partial(i\lambda_{16})}+i\lambda_{15}%
\frac{\partial}{\partial(i\lambda_{15})}-i\lambda_{14}\frac{\partial}%
{\partial(i\lambda_{14})}+\frac{\partial}{\partial(i\lambda_{11})}]_{l}%
\chi\nonumber\\
&
\end{align}
where the subscript $l$ on the bracket reminds us the derivatives inside the
bracket operate on $l$th component of the characteristic functions.

\subsection{Characteristic function - field part}

The field part in characteristic function is deduced from $\big(\frac
{\partial\rho}{\partial t}\big)_{L}$%

\begin{align}
\big(\frac{\partial\chi}{\partial t}\big)_{L}  & =\text{Tr\{}E(\lambda
)\big(\frac{\partial\rho}{\partial t}\big)_{L}\text{\},}\nonumber\\
\big(\frac{\partial\rho}{\partial t}\big)_{L}  & =\frac{1}{i\hbar}[H_{L}%
,\rho]=\sum_{l}\Big[-i\omega_{s}(\hat{a}_{s,l}^{\dagger}\hat{a}_{s,l}\rho
-\rho\hat{a}_{s,l}^{\dagger}\hat{a}_{s,l})-i\omega_{i}(\hat{a}_{i,l}^{\dagger
}\hat{a}_{i,l}\rho-\rho\hat{a}_{i,l}^{\dagger}\hat{a}_{i,l})\Big]+\nonumber\\
& \sum_{l,l^{\prime}}\Big[-i\omega_{l^{\prime}l}(\hat{a}_{s,l}^{\dagger}%
\hat{a}_{s,l^{\prime}}\rho-\rho\hat{a}_{s,l}^{\dagger}\hat{a}_{s,l^{\prime}%
})-i\omega_{ll^{\prime}}(\hat{a}_{i,l}^{\dagger}\hat{a}_{i,l^{\prime}}%
\rho-\rho\hat{a}_{i,l}^{\dagger}\hat{a}_{i,l^{\prime}})\Big],
\end{align}
and various components in $\big(\frac{\partial\chi}{\partial t}\big)_{L}$ are%

\begin{align}
\text{Tr\{}E(\lambda)\sum_{l}\hat{a}_{s,l}^{\dagger}\hat{a}_{s,l}%
\rho\text{\}}  & =\sum_{l}[\frac{\partial^{2}}{\partial(i\lambda_{4}%
)\partial(i\lambda_{3})}+i\lambda_{3}\frac{\partial}{\partial(i\lambda_{3}%
)}]_{l}\chi,\nonumber\\
\text{Tr\{}E(\lambda)\sum_{l}\rho\hat{a}_{s,l}^{\dagger}\hat{a}_{s,l}%
\text{\}}  & =\sum_{l}[\frac{\partial^{2}}{\partial(i\lambda_{4}%
)\partial(i\lambda_{3})}+i\lambda_{4}\frac{\partial}{\partial(i\lambda_{4}%
)}]_{l}\chi,\nonumber\\
\text{Tr\{}E(\lambda)\sum_{l}\hat{a}_{i,l}^{\dagger}\hat{a}_{i,l}%
\rho\text{\}}  & =\sum_{l}[\frac{\partial^{2}}{\partial(i\lambda_{2}%
)\partial(i\lambda_{1})}+i\lambda_{1}\frac{\partial}{\partial(i\lambda_{1}%
)}]_{l}\chi,\nonumber\\
\text{Tr\{}E(\lambda)\sum_{l}\rho\hat{a}_{i,l}^{\dagger}\hat{a}_{i,l}%
\text{\}}  & =\sum_{l}[\frac{\partial^{2}}{\partial(i\lambda_{2}%
)\partial(i\lambda_{1})}+i\lambda_{2}\frac{\partial}{\partial(i\lambda_{2}%
)}]_{l}\chi,
\end{align}
and%

\begin{align}
\text{Tr\{}E(\lambda)\sum_{l,l^{\prime}}\omega_{l^{\prime}l}\hat{a}%
_{s,l}^{\dagger}\hat{a}_{s,l^{\prime}}\rho\text{\}}  & =\sum_{l,l^{\prime}%
}\omega_{l^{\prime}l}[\frac{\partial^{2}}{\partial(i\lambda_{4}^{l}%
)\partial(i\lambda_{3}^{l^{\prime}})}+i\lambda_{3}^{l^{\prime}}\frac{\partial
}{\partial(i\lambda_{3}^{l^{\prime}})}]\chi,\nonumber\\
\text{Tr\{}E(\lambda)\sum_{l,l^{\prime}}\omega_{l^{\prime}l}\rho\hat{a}%
_{s,l}^{\dagger}\hat{a}_{s,l^{\prime}}\text{\}}  & =\sum_{l,l^{\prime}}%
\omega_{l^{\prime}l}[\frac{\partial^{2}}{\partial(i\lambda_{4}^{l}%
)\partial(i\lambda_{3}^{l^{\prime}})}+i\lambda_{4}^{l}\frac{\partial}%
{\partial(i\lambda_{4}^{l})}]\chi,\nonumber\\
\text{Tr\{}E(\lambda)\sum_{l,l^{\prime}}\omega_{ll^{\prime}}\hat{a}%
_{i,l}^{\dagger}\hat{a}_{i,l^{\prime}}\rho\text{\}}  & =\sum_{l,l^{\prime}%
}\omega_{ll^{\prime}}[\frac{\partial^{2}}{\partial(i\lambda_{2}^{l}%
)\partial(i\lambda_{1}^{l^{\prime}})}+i\lambda_{1}^{l^{\prime}}\frac{\partial
}{\partial(i\lambda_{1}^{l^{\prime}})}]\chi,\nonumber\\
\text{Tr\{}E(\lambda)\sum_{l,l^{\prime}}\omega_{ll^{\prime}}\rho\hat{a}%
_{i,l}^{\dagger}\hat{a}_{i,l^{\prime}}\text{\}}  & =\sum_{l,l^{\prime}}%
\omega_{ll^{\prime}}[\frac{\partial^{2}}{\partial(i\lambda_{2}^{l}%
)\partial(i\lambda_{1}^{l^{\prime}})}+i\lambda_{2}^{l}\frac{\partial}%
{\partial(i\lambda_{2}^{l})}]\chi.
\end{align}

\subsection{Characteristic function - atom-field part}

The atom-field interaction part in characteristic function is deduced from
$\big(\frac{\partial\rho}{\partial t}\big)_{A-L},$ and we denote part (a) for
the classical field interaction.%

\begin{align}
\big(\frac{\partial\chi}{\partial t}\big)_{A-L}^{(a)}  & =\text{Tr\{}%
E(\lambda)\big(\frac{\partial\rho}{\partial t}\big)_{A-L}^{(a)}\text{\},}%
\nonumber\\
\big(\frac{\partial\rho}{\partial t}\big)_{A-L}^{(a)}  & =\frac{1}{i\hbar
}\Big[-\hbar\sum_{l=-M}^{M}\big[\Omega_{a}(t)\tilde{\sigma}_{01}^{l\dagger
}e^{ik_{a}z_{l}-i\omega_{a}t}+\Omega_{b}(t)\tilde{\sigma}_{12}^{l\dagger
}e^{-ik_{b}z_{l}-i\omega_{b}t}+h.c.\big],\rho\Big],\nonumber\\
&
\end{align}
and various components in $\big(\frac{\partial\chi}{\partial t}\big)_{A-L}%
^{(a)}$ are%

\begin{align}
& \text{Tr\{}E(\lambda)\sum_{l}e^{ik_{a}z_{l}}\tilde{\sigma}_{01}^{l\dagger
}\rho\text{\}}=\nonumber\\
& \sum_{l}e^{ik_{a}z_{l}}\Big[-(i\lambda_{5})^{2}\frac{\partial}%
{\partial(i\lambda_{5})}+(i\lambda_{5})(i\lambda_{6})\frac{\partial}%
{\partial(i\lambda_{6})}-(i\lambda_{5})(i\lambda_{7})\frac{\partial}%
{\partial(i\lambda_{7})}+(i\lambda_{5})(i\lambda_{8})\frac{\partial}%
{\partial(i\lambda_{8})}\nonumber\\
& -(i\lambda_{5})(i\lambda_{9})\frac{\partial}{\partial(i\lambda_{9}%
)}-i\lambda_{5}\frac{\partial}{\partial(i\lambda_{11})}-i\lambda_{5}%
\frac{\partial}{\partial(i\lambda_{12})}-2i\lambda_{5}\frac{\partial}%
{\partial(i\lambda_{13})}+i\lambda_{5}N_{z}-i\lambda_{7}\frac{\partial
}{\partial(i\lambda_{6})}\nonumber\\
& -i\lambda_{9}\frac{\partial}{\partial(i\lambda_{8})}+i\lambda_{16}%
e^{i\lambda_{13}}\frac{\partial}{\partial(i\lambda_{15})}+i\lambda
_{18}e^{i\lambda_{13}}\frac{\partial}{\partial(i\lambda_{17})}+e^{i\lambda
_{13}}\frac{\partial}{\partial(i\lambda_{19})}\Big]_{l}\chi,\nonumber\\
& \text{Tr\{}E(\lambda)\sum_{l}e^{-ik_{a}z_{l}}\tilde{\sigma}_{01}^{l}%
\rho\text{\}}=\sum_{l}e^{-ik_{a}z_{l}}\Big[\frac{\partial}{\partial
(i\lambda_{5})}\Big]_{l}\chi,\nonumber\\
& \text{Tr\{}E(\lambda)\sum_{l}\rho e^{ik_{a}z_{l}}\tilde{\sigma}%
_{01}^{l\dagger}\text{\}}=\sum_{l}e^{ik_{a}z_{l}}\Big[\frac{\partial}%
{\partial(i\lambda_{19})}\Big]_{l}\chi,\nonumber\\
& \text{Tr\{}E(\lambda)\sum_{l}\rho e^{-ik_{a}z_{l}}\tilde{\sigma}_{01}%
^{l}\text{\}}=\text{Tr\{}E(\lambda)\sum_{l}e^{ik_{a}z_{l}}\tilde{\sigma}%
_{01}^{l\dagger}\rho\text{\}}_{\substack{\lambda_{5}^{\ast}\leftrightarrow
-\lambda_{19,}\lambda_{6}^{\ast}\leftrightarrow-\lambda_{18,}\lambda_{7}%
^{\ast}\leftrightarrow-\lambda_{17,}\lambda_{8}^{\ast}\leftrightarrow
-\lambda_{16,} \\\lambda_{9}^{\ast}\leftrightarrow-\lambda_{15,}\lambda
_{11}^{\ast}\leftrightarrow-\lambda_{11,}\lambda_{12}^{\ast}\leftrightarrow
-\lambda_{12,}\lambda_{13}^{\ast}\leftrightarrow-\lambda_{13}}}^{\ast
}\nonumber\\
&
\end{align}
where a correspondence that we denote as $C$ later is $\lambda_{5}^{\ast
}\leftrightarrow-\lambda_{19,\text{ }}\lambda_{6}^{\ast}\leftrightarrow
-\lambda_{18,}$ $\lambda_{7}^{\ast}\leftrightarrow-\lambda_{17,}$ $\lambda
_{8}^{\ast}\leftrightarrow-\lambda_{16,}$ $\lambda_{9}^{\ast}\leftrightarrow
-\lambda_{15,}$ $\lambda_{11}^{\ast}\leftrightarrow-\lambda_{11,}$
$\lambda_{12}^{\ast}\leftrightarrow-\lambda_{12,}$ $\lambda_{13}^{\ast
}\leftrightarrow-\lambda_{13}$, can be observed to help calculate the
characteristic function. \ Also%

\begin{align}
& \text{Tr\{}E(\lambda)\sum_{l}e^{-ik_{b}z_{l}}\tilde{\sigma}_{12}^{l\dagger
}\rho\text{\}}=\nonumber\\
& \sum_{l}e^{-ik_{b}z_{l}}\Big[-(i\lambda_{6})^{2}\frac{\partial}%
{\partial(i\lambda_{6})}-(i\lambda_{6})(i\lambda_{8})\frac{\partial}%
{\partial(i\lambda_{8})}-(i\lambda_{6})(i\lambda_{7})\frac{\partial}%
{\partial(i\lambda_{7})}\\
& -(i\lambda_{6})(i\lambda_{10})\frac{\partial}{\partial(i\lambda_{10}%
)}+(i\lambda_{7})(i\lambda_{6})\frac{\partial}{\partial(i\lambda_{7}%
)}+i\lambda_{6}\frac{\partial}{\partial(i\lambda_{13})}-i\lambda_{6}%
\frac{\partial}{\partial(i\lambda_{12})}+i\lambda_{7}\frac{\partial}%
{\partial(i\lambda_{5})}\nonumber\\
& +(i\lambda_{8})(i\lambda_{10})(\frac{\partial}{\partial(i\lambda_{12}%
)}-\frac{\partial}{\partial(i\lambda_{11})})-i\lambda_{8}(i\lambda_{10}%
)^{2}\frac{\partial}{\partial(i\lambda_{10})}-i\lambda_{8}e^{i\lambda
_{12}-i\lambda_{11}}\frac{\partial}{\partial(i\lambda_{14})}\nonumber\\
& +i\lambda_{10}e^{i\lambda_{11}-i\lambda_{13}}\frac{\partial}{\partial
(i\lambda_{16})}+((i\lambda_{10})(i\lambda_{14})e^{i\lambda_{11}-i\lambda
_{13}}+e^{i\lambda_{12}-i\lambda_{13}})\frac{\partial}{\partial(i\lambda
_{18})}\Big]_{l}\chi,\nonumber\\
& \text{Tr\{}E(\lambda)\sum_{l}e^{ik_{b}z_{l}}\tilde{\sigma}_{12}^{l}%
\rho\text{\}}=\sum_{l}e^{ik_{b}z_{l}}\Big[\frac{\partial}{\partial
(i\lambda_{6})}+i\lambda_{5}\frac{\partial}{\partial(i\lambda_{7})}%
\Big]_{l}\chi,\nonumber\\
& \text{Tr\{}E(\lambda)\sum_{l}\rho e^{-ik_{b}z_{l}}\tilde{\sigma}%
_{12}^{l\dagger}\text{\}}=\text{Tr\{}E(\lambda)\sum_{l}e^{ik_{b}z_{l}}%
\tilde{\sigma}_{12}^{l}\rho\text{\}}_{C}^{\ast},\nonumber\\
& \text{Tr\{}E(\lambda)\sum_{l}\rho e^{ik_{b}z_{l}}\tilde{\sigma}_{12}%
^{l}\text{\}}=\text{Tr\{}E(\lambda)\sum_{l}e^{-ik_{b}z_{l}}\tilde{\sigma}%
_{12}^{l\dagger}\rho\text{\}}_{C}^{\ast},
\end{align}
and the atom-field interaction characteristic function for quantum fields,
which we denote as part (b), is%

\begin{align}
\big(\frac{\partial\chi}{\partial t}\big)_{A-L}^{(b)}  & =\text{Tr\{}%
E(\lambda)\big(\frac{\partial\rho}{\partial t}\big)_{A-L}^{(b)}\text{\},}%
\nonumber\\
\big(\frac{\partial\rho}{\partial t}\big)_{A-L}^{(b)}  & =\frac{1}{i\hbar
}\Big[-\hbar\sum_{l=-M}^{M}\big[g_{s}\sqrt{2M+1}\tilde{\sigma}_{32}^{l\dagger
}\hat{a}_{s,l}e^{-ik_{s}z_{l}}\nonumber\\
& +g_{i}\sqrt{2M+1}\tilde{\sigma}_{03}^{l\dagger}\hat{a}_{i,l}e^{ik_{i}z_{l}%
}+h.c.\big],\rho\Big].
\end{align}

For the part of fields only,%

\begin{align}
\text{Tr\{}E(\lambda)\hat{a}_{s,l}\rho\text{\}}  & =[\frac{\partial}%
{\partial(i\lambda_{3}^{l})}]\chi\text{, \ Tr\{}E(\lambda)\rho\hat{a}%
_{s,l}\text{\}}=[\frac{\partial}{\partial(i\lambda_{3}^{l})}+i\lambda_{4}%
^{l}]\chi,\nonumber\\
\text{Tr\{}E(\lambda)\hat{a}_{s,l}^{\dagger}\rho\text{\}}  & =[\frac{\partial
}{\partial(i\lambda_{4}^{l})}+i\lambda_{3}^{l}]\chi\text{, \ Tr\{}%
E(\lambda)\rho\hat{a}_{s,l}^{\dagger}\text{\}}=[\frac{\partial}{\partial
(i\lambda_{4}^{l})}]\chi,\nonumber\\
\text{Tr\{}E(\lambda)\hat{a}_{i,l}\rho\text{\}}  & =[\frac{\partial}%
{\partial(i\lambda_{1}^{l})}]\chi\text{, \ Tr\{}E(\lambda)\rho\hat{a}%
_{i,l}\text{\}}=[\frac{\partial}{\partial(i\lambda_{1}^{l})}+i\lambda_{2}%
^{l}]\chi,\nonumber\\
\text{Tr\{}E(\lambda)\hat{a}_{i,l}^{\dagger}\rho\text{\}}  & =[\frac{\partial
}{\partial(i\lambda_{2}^{l})}+i\lambda_{1}^{l}]\chi\text{, \ Tr\{}%
E(\lambda)\rho\hat{a}_{i,l}^{\dagger}\text{\}}=[\frac{\partial}{\partial
(i\lambda_{2}^{l})}]\chi,
\end{align}
and for the part of atomic operators associated with signal field,%
\begin{align}
\text{Tr\{}E(\lambda)\tilde{\sigma}_{32}^{l\dagger}\rho\text{\}}  &
=\Big[i\lambda_{6}\frac{\partial}{\partial(i\lambda_{8})}+i\lambda_{7}%
\frac{\partial}{\partial(i\lambda_{9})}-(i\lambda_{10})^{2}\frac{\partial
}{\partial(i\lambda_{10})}\nonumber\\
& +i\lambda_{10}(\frac{\partial}{\partial(i\lambda_{11})}-\frac{\partial
}{\partial(i\lambda_{12})})+e^{i\lambda_{12}-i\lambda_{11}}\frac{\partial
}{\partial(i\lambda_{14})}\Big]_{l}\chi,\nonumber\\
\text{Tr\{}E(\lambda)\tilde{\sigma}_{32}^{l}\rho\text{\}}  & =\Big[i\lambda
_{8}\frac{\partial}{\partial(i\lambda_{6})}+\frac{\partial}{\partial
(i\lambda_{10})}+i\lambda_{9}\frac{\partial}{\partial(i\lambda_{7})}%
\Big]_{l}\chi,\nonumber\\
\text{Tr\{}E(\lambda)\rho\tilde{\sigma}_{32}^{l\dagger}\text{\}}  &
=\text{Tr\{}E(\lambda)\tilde{\sigma}_{32}^{l}\rho\text{\}}_{C}^{\ast
},\nonumber\\
\text{Tr\{}E(\lambda)\rho\tilde{\sigma}_{32}^{l}\text{\}}  & =\text{Tr\{}%
E(\lambda)\tilde{\sigma}_{32}^{l\dagger}\rho\text{\}}_{C}^{\ast},
\end{align}
and for the part of atomic operators associated with idler field,%

\begin{align}
& \text{Tr\{}E(\lambda)\tilde{\sigma}_{03}^{l\dagger}\rho\text{\}}\nonumber\\
& =\Big[(i\lambda_{6})(i\lambda_{8})^{2}\frac{\partial}{\partial(i\lambda
_{8})}+(i\lambda_{5})(i\lambda_{6})(i\lambda_{8})\frac{\partial}%
{\partial(i\lambda_{6})}-(i\lambda_{5})(i\lambda_{8})(\frac{\partial}%
{\partial(i\lambda_{13})}-i\lambda_{9}\frac{\partial}{\partial(i\lambda_{9}%
)}\nonumber\\
& +(i\lambda_{10})\frac{\partial}{\partial(i\lambda_{10})}-\frac{\partial
}{\partial(i\lambda_{11})})+(i\lambda_{5})(i\lambda_{6})\frac{\partial
}{\partial(i\lambda_{10})}-i\lambda_{5}e^{i\lambda_{11}-i\lambda_{13}}%
\frac{\partial}{\partial(i\lambda_{16})}\nonumber\\
& -(i\lambda_{5})(i\lambda_{14})e^{i\lambda_{11}-i\lambda_{13}}\frac{\partial
}{\partial(i\lambda_{18})}-(i\lambda_{5})(i\lambda_{9})\frac{\partial
}{\partial(i\lambda_{5})}-(i\lambda_{7})(i\lambda_{8})\frac{\partial}%
{\partial(i\lambda_{6})}-i\lambda_{7}\frac{\partial}{\partial(i\lambda_{10}%
)}\nonumber\\
& -(i\lambda_{7})(i\lambda_{9})\frac{\partial}{\partial(i\lambda_{7}%
)}+i\lambda_{9}N_{z}-(i\lambda_{9})^{2}\frac{\partial}{\partial(i\lambda_{9}%
)}+i\lambda_{8}[-(i\lambda_{9})\frac{\partial}{\partial(i\lambda_{8}%
)}+i\lambda_{16}e^{i\lambda_{13}}\frac{\partial}{\partial(i\lambda_{15}%
)}\nonumber\\
& +i\lambda_{18}e^{i\lambda_{13}}\frac{\partial}{\partial(i\lambda_{17}%
)}+e^{i\lambda_{13}}\frac{\partial}{\partial(i\lambda_{19})}]+e^{i\lambda
_{11}}\frac{\partial}{\partial(i\lambda_{15})}+i\lambda_{9}[-\frac{\partial
}{\partial(i\lambda_{13})}-i\lambda_{10}\frac{\partial}{\partial(i\lambda
_{10})}\nonumber\\
& -\frac{\partial}{\partial(i\lambda_{12})}+2i\lambda_{10}\frac{\partial
}{\partial(i\lambda_{10})}-2\frac{\partial}{\partial(i\lambda_{11})}%
]+i\lambda_{14}e^{i\lambda_{11}}\frac{\partial}{\partial(i\lambda_{17}%
)}\Big]_{l}\chi,\nonumber\\
& \text{Tr\{}E(\lambda)\tilde{\sigma}_{03}^{l}\rho\text{\}}=\Big[\frac
{\partial}{\partial(i\lambda_{9})}\Big]_{l}\chi,\nonumber\\
& \text{Tr\{}E(\lambda)\rho\tilde{\sigma}_{03}^{l\dagger}\text{\}}%
=\text{Tr\{}E(\lambda)\tilde{\sigma}_{03}^{l}\rho\text{\}}_{C}^{\ast
},\nonumber\\
& \text{Tr\{}E(\lambda)\rho\tilde{\sigma}_{03}^{l}\text{\}}=\text{Tr\{}%
E(\lambda)\tilde{\sigma}_{03}^{l\dagger}\rho\text{\}}_{C}^{\ast}.
\end{align}

\subsection{Characteristic function - dissipation part}

We calculate the characteristic function from $\big(\frac{\partial\rho
}{\partial t}\big)_{sp}$ up to the second order of various $\lambda$'s (where
we denote $(2)$) that account for drift and diffusion terms in Fokker-Planck
equation. \ Below we drop the summation over spatial slices $l,$ which we will
retrieve later,%
\begin{align}
& \text{Tr\{}E(\lambda)\hat{\sigma}_{01}\rho\hat{\sigma}_{01}^{\dagger
}\text{\}}^{(2)}\nonumber\\
& =\text{Tr\{}[(i\lambda_{8})(i\lambda_{10})\hat{\sigma}_{12}+(i\lambda
_{6})(i\lambda_{14})\hat{\sigma}_{13}+(i\lambda_{8})(i\lambda_{15})\hat
{\sigma}_{10}+\nonumber\\
& (i\lambda_{8})(i\lambda_{16})\hat{\sigma}_{11}+(i\lambda_{6})(i\lambda
_{17})\hat{\sigma}_{10}+e^{-i\lambda_{13}}\hat{\sigma}_{11}-i\lambda
_{19}e^{-i\lambda_{13}}\hat{\sigma}_{10}\nonumber\\
& +(i\lambda_{6})(i\lambda_{18})\hat{\sigma}_{11}-i\lambda_{8}e^{-i\lambda
_{11}}\hat{\sigma}_{13}-i\lambda_{6}e^{-i\lambda_{12}}\hat{\sigma}%
_{12}]E(\lambda)\rho\text{\}}^{(2)}%
\end{align}
where various properties of tracing can be found in previous sections, and the
one we did not have before is (up to first order)%

\begin{align}
& \text{Tr\{}\hat{\sigma}_{13}E(\lambda)\rho\text{\}}^{(1)}\nonumber\\
& =[i\lambda_{15}\frac{\partial}{\partial(i\lambda_{19})}-i\lambda_{16}%
(\frac{\partial}{\partial(i\lambda_{11})}-\frac{\partial}{\partial
(i\lambda_{13})})-i\lambda_{18}\frac{\partial}{\partial(i\lambda_{14}%
)}\nonumber\\
& +e^{i\lambda_{11}-i\lambda_{13}}\frac{\partial}{\partial(i\lambda_{8}%
)}+i\lambda_{10}\frac{\partial}{\partial(i\lambda_{6})}]\chi.
\end{align}

Put everything together, and for the dissipation of first laser transition we have%

\begin{align}
& \gamma_{01}\text{Tr\{}E(\lambda)[\hat{\sigma}_{01}\rho\hat{\sigma}%
_{01}^{\dagger}-\frac{1}{2}\hat{\sigma}_{11}\rho-\frac{1}{2}\rho\hat{\sigma
}_{11}]\text{\}}^{(2)}=\nonumber\\
& \gamma_{01}[-\frac{i\lambda_{5}}{2}\frac{\partial}{\partial(i\lambda_{5}%
)}-\frac{i\lambda_{19}}{2}\frac{\partial}{\partial(i\lambda_{19})}%
-\frac{i\lambda_{6}}{2}\frac{\partial}{\partial(i\lambda_{6})}-\frac
{i\lambda_{18}}{2}\frac{\partial}{\partial(i\lambda_{18})}-\frac{i\lambda_{8}%
}{2}\frac{\partial}{\partial(i\lambda_{8})}\nonumber\\
& -\frac{i\lambda_{16}}{2}\frac{\partial}{\partial(i\lambda_{16})}%
-i\lambda_{13}\frac{\partial}{\partial(i\lambda_{13})}+(i\lambda
_{13})(i\lambda_{18})\frac{\partial}{\partial(i\lambda_{18})}+(i\lambda
_{13})(i\lambda_{16})\frac{\partial}{\partial(i\lambda_{6})}\nonumber\\
& +(i\lambda_{13})(i\lambda_{16})\frac{\partial}{\partial(i\lambda_{16}%
)}+(i\lambda_{13})(i\lambda_{8})\frac{\partial}{\partial(i\lambda_{8}%
)}+(i\lambda_{8})(i\lambda_{18})\frac{\partial}{\partial(i\lambda_{14}%
)}+(i\lambda_{6})(i\lambda_{16})\frac{\partial}{\partial(i\lambda_{10}%
)}\nonumber\\
& +(i\lambda_{8})(i\lambda_{16})\frac{\partial}{\partial(i\lambda_{11}%
)}+(i\lambda_{6})(i\lambda_{18})\frac{\partial}{\partial(i\lambda_{12})}%
+\frac{(i\lambda_{13})^{2}}{2}\frac{\partial}{\partial(i\lambda_{13})}]\chi.
\end{align}

And for the second laser,%

\begin{align}
\text{Tr\{}E(\lambda)\hat{\sigma}_{12}\rho\hat{\sigma}_{12}^{\dagger}%
\text{\}}^{(2)}  & =\text{Tr\{}\Big[(i\lambda_{14})(i\lambda_{15})\hat{\sigma
}_{20}+(i\lambda_{16})(i\lambda_{14})\hat{\sigma}_{21}+e^{i\lambda
_{13}-i\lambda_{12}}\Big(\hat{\sigma}_{22}-i\lambda_{17}\hat{\sigma}%
_{20}\nonumber\\
& -i\lambda_{14}\hat{\sigma}_{23}-i\lambda_{18}\hat{\sigma}_{21}%
+(i\lambda_{19})(i\lambda_{18})\hat{\sigma}_{20}\Big)\Big]E(\lambda
)\rho\text{\}}^{(2)}.
\end{align}

The above requires%

\[
\text{Tr}\{\hat{\sigma}_{20}E(\lambda)\rho\}=[\frac{\partial}{\partial
(i\lambda_{17})}]\chi.
\]

Then we have%

\begin{align}
& \gamma_{12}\text{Tr\{}E(\lambda)[\hat{\sigma}_{12}\rho\hat{\sigma}%
_{12}^{\dagger}-\frac{1}{2}\hat{\sigma}_{22}\rho-\frac{1}{2}\rho\hat{\sigma
}_{22}]\text{\}}^{(2)}=\nonumber\\
& \gamma_{12}[-\frac{i\lambda_{6}}{2}\frac{\partial}{\partial(i\lambda_{6}%
)}-\frac{i\lambda_{18}}{2}\frac{\partial}{\partial(i\lambda_{18})}%
-\frac{i\lambda_{7}}{2}\frac{\partial}{\partial(i\lambda_{7})}-\frac
{i\lambda_{17}}{2}\frac{\partial}{\partial(i\lambda_{17})}-\frac{i\lambda
_{10}}{2}\frac{\partial}{\partial(i\lambda_{10})}\nonumber\\
& -\frac{i\lambda_{14}}{2}\frac{\partial}{\partial(i\lambda_{14})}%
+(i\lambda_{13}-i\lambda_{12})\frac{\partial}{\partial(i\lambda_{12}%
)}+(i\lambda_{5})(i\lambda_{19})\frac{\partial}{\partial(i\lambda_{12})}%
+\frac{(i\lambda_{13}-i\lambda_{12})^{2}}{2}\frac{\partial}{\partial
(i\lambda_{12})}]\chi.\nonumber\\
&
\end{align}

And the dissipation for the signal transition,%

\begin{align}
& \text{Tr\{}E(\lambda)\hat{\sigma}_{32}\rho\hat{\sigma}_{32}^{\dagger
}\text{\}}^{(2)}=\nonumber\\
& \text{Tr\{}[(i\lambda_{8})(i\lambda_{16})\hat{\sigma}_{22}+(i\lambda
_{9})(i\lambda_{15})\hat{\sigma}_{22}+(i\lambda_{14})(i\lambda_{15}%
)\hat{\sigma}_{20}+(i\lambda_{14})(i\lambda_{16})\hat{\sigma}_{21}\nonumber\\
& +e^{i\lambda_{11}-i\lambda_{12}}(-i\lambda_{17}\hat{\sigma}_{20}%
-i\lambda_{14}\hat{\sigma}_{23}+\hat{\sigma}_{22}-i\lambda_{18}\hat{\sigma
}_{21}+(i\lambda_{19})(i\lambda_{18})\hat{\sigma}_{20})]E(\lambda
)\rho\text{\},}\nonumber\\
&
\end{align}
so we have%

\begin{align}
& \gamma_{32}\text{Tr\{}E(\lambda)[\hat{\sigma}_{32}\rho\hat{\sigma}%
_{32}^{\dagger}-\frac{1}{2}\hat{\sigma}_{22}\rho-\frac{1}{2}\rho\hat{\sigma
}_{22}]\text{\}}^{(2)}=\nonumber\\
& \gamma_{32}[-\frac{i\lambda_{6}}{2}\frac{\partial}{\partial(i\lambda_{6}%
)}-\frac{i\lambda_{18}}{2}\frac{\partial}{\partial(i\lambda_{18})}%
-\frac{i\lambda_{7}}{2}\frac{\partial}{\partial(i\lambda_{7})}-\frac
{i\lambda_{17}}{2}\frac{\partial}{\partial(i\lambda_{17})}-\frac{i\lambda
_{10}}{2}\frac{\partial}{\partial(i\lambda_{10})}\nonumber\\
& -\frac{i\lambda_{14}}{2}\frac{\partial}{\partial(i\lambda_{14})}%
+(i\lambda_{11}-i\lambda_{12})\frac{\partial}{\partial(i\lambda_{12}%
)}+(i\lambda_{5})(i\lambda_{19})\frac{\partial}{\partial(i\lambda_{12}%
)}+(i\lambda_{8})(i\lambda_{16})\frac{\partial}{\partial(i\lambda_{12}%
)}\nonumber\\
& +\frac{(i\lambda_{11}-i\lambda_{12})^{2}}{2}\frac{\partial}{\partial
(i\lambda_{12})}]\chi.
\end{align}

And for idler transition,%

\begin{align}
& \text{Tr\{}E(\lambda)\hat{\sigma}_{03}\rho\hat{\sigma}_{03}^{\dagger
}\text{\}}^{(2)}=\nonumber\\
& \text{Tr\{}[(i\lambda_{10})(i\lambda_{18})\hat{\sigma}_{31}+(i\lambda
_{10})(i\lambda_{14})\hat{\sigma}_{33}+(i\lambda_{10})(i\lambda_{17}%
)\hat{\sigma}_{30}-i\lambda_{10}e^{-i\lambda_{12}}\hat{\sigma}_{32}\nonumber\\
& e^{-i\lambda_{11}}(\hat{\sigma}_{33}-i\lambda_{16}\hat{\sigma}%
_{31}+(i\lambda_{16})(i\lambda_{19})\hat{\sigma}_{30}-i\lambda_{15}\hat
{\sigma}_{30}]E(\lambda)\rho\text{\}}^{(2)}.
\end{align}

The above needs%

\[
\text{Tr}\{\hat{\sigma}_{31}E(\lambda)\rho\}=[\frac{\partial}{\partial
(i\lambda_{16})}+i\lambda_{19}\frac{\partial}{\partial(i\lambda_{15})}]\chi,
\]
then we have%

\begin{align}
& \gamma_{03}\text{Tr\{}E(\lambda)[\hat{\sigma}_{03}\rho\hat{\sigma}%
_{03}^{\dagger}-\frac{1}{2}\hat{\sigma}_{33}\rho-\frac{1}{2}\rho\hat{\sigma
}_{33}]\text{\}}^{(2)}\nonumber\\
& =\gamma_{03}[-\frac{i\lambda_{8}}{2}\frac{\partial}{\partial(i\lambda_{8}%
)}-\frac{i\lambda_{16}}{2}\frac{\partial}{\partial(i\lambda_{16})}%
-\frac{i\lambda_{9}}{2}\frac{\partial}{\partial(i\lambda_{9})}-\frac
{i\lambda_{15}}{2}\frac{\partial}{\partial(i\lambda_{15})}-\frac{i\lambda
_{10}}{2}\frac{\partial}{\partial(i\lambda_{10})}\nonumber\\
& -\frac{i\lambda_{14}}{2}\frac{\partial}{\partial(i\lambda_{14})}%
-i\lambda_{11}\frac{\partial}{\partial(i\lambda_{11})}+(i\lambda
_{11})(i\lambda_{14})\frac{\partial}{\partial(i\lambda_{14})}+(i\lambda
_{10})(i\lambda_{11})\frac{\partial}{\partial(i\lambda_{10})}\nonumber\\
& +(i\lambda_{10})(i\lambda_{14})\frac{\partial}{\partial(i\lambda_{12}%
)}+\frac{(i\lambda_{11})^{2}}{2}\frac{\partial}{\partial(i\lambda_{11})}]\chi.
\end{align}

\section{Stochastic Differential Equation}

A distribution function can be found by Fourier transforming the
characteristic functions,%

\begin{equation}
f(\vec{\alpha})=\frac{1}{(2\pi)^{n}}\int...\int e^{-i\vec{\alpha}\cdot
\vec{\lambda}}\chi(\vec{\lambda})d\lambda_{1}...d\lambda_{n},
\end{equation}
then%

\begin{equation}
\frac{\partial f}{\partial t}=\frac{1}{(2\pi)^{n}}\int...\int e^{-i\vec
{\alpha}\cdot\vec{\lambda}}\frac{\partial\chi}{\partial t}d\lambda
_{1}...d\lambda_{n}.
\end{equation}

If $\frac{\partial\chi}{\partial t}=i\lambda_{\beta}\frac{\partial\chi
}{\partial(i\lambda_{\gamma})}$, use integration by parts and neglect the
boundary terms, we have $\frac{\partial f}{\partial t}=-\frac{\partial
}{\partial(\alpha_{\beta})}\alpha_{\gamma}f$ where a minus sign is from
$i\lambda_{\beta}$. \ Correspondingly, if $\frac{\partial\chi}{\partial
t}=e^{i\lambda_{\beta}}$, we have \bigskip$\frac{\partial f}{\partial
t}=e^{-\frac{\partial}{\partial(\alpha_{\beta})}}$.

\subsection{Fokker-Planck equation}

Let
\begin{equation}
\frac{\partial f}{\partial t}=\mathcal{L}f=\sum_{l,l^{\prime}}[\mathcal{L}%
_{A}\delta_{ll^{\prime}}+\mathcal{L}_{L}+\mathcal{L}_{A-L}^{(a)}%
\delta_{ll^{\prime}}+\mathcal{L}_{A-L}^{(b)}\delta_{ll^{\prime}}%
+\mathcal{L}_{sp}\delta_{ll^{\prime}}]f,
\end{equation}
then we have for the atomic part,%

\begin{align}
\mathcal{L}_{A}  & =-i\omega_{1}[\frac{\partial}{\partial\alpha_{5}^{l}%
}(-\alpha_{5}^{l})-\frac{\partial}{\partial\alpha_{6}^{l}}(-\alpha_{6}%
^{l})-\frac{\partial}{\partial\alpha_{8}^{l}}(-\alpha_{8}^{l})]\nonumber\\
& -i\omega_{2}[\frac{\partial}{\partial\alpha_{6}^{l}}(-\alpha_{6}^{l}%
)+\frac{\partial}{\partial\alpha_{7}^{l}}(-\alpha_{7}^{l})+\frac{\partial
}{\partial\alpha_{10}^{l}}(-\alpha_{10}^{l})]\nonumber\\
& -i\omega_{3}[\frac{\partial}{\partial\alpha_{8}^{l}}(-\alpha_{8}^{l}%
)+\frac{\partial}{\partial\alpha_{9}^{l}}(-\alpha_{9}^{l})-\frac{\partial
}{\partial\alpha_{10}^{l}}(-\alpha_{10}^{l})]+(c.c.\text{ with }C^{\prime})
\end{align}
where $C^{\prime}$ is $\alpha_{5}^{\ast}\leftrightarrow\alpha_{19,}$
$\alpha_{6}^{\ast}\leftrightarrow\alpha_{18,}$ $\alpha_{7}^{\ast
}\leftrightarrow\alpha_{17,}$ $\alpha_{8}^{\ast}\leftrightarrow\alpha_{16,}$
$\alpha_{9}^{\ast}\leftrightarrow\alpha_{15,}$ $\alpha_{10}^{\ast
}\leftrightarrow\alpha_{14,}$ $\alpha_{11}^{\ast}\leftrightarrow\alpha_{11,}$
$\alpha_{12}^{\ast}\leftrightarrow\alpha_{12,}$ $\alpha_{13}^{\ast
}\leftrightarrow\alpha_{13,}$ $\alpha_{1}^{\ast}\leftrightarrow\alpha_{2,}$
$\alpha_{3}^{\ast}\leftrightarrow\alpha_{4},$ and $c.c.$\ is complex
conjugation. \ Also for the field part,%

\begin{align}
\mathcal{L}_{L}  & =[i\omega_{s}\frac{\partial}{\partial\alpha_{3}^{l}}%
\alpha_{3}^{l}-i\omega_{s}\frac{\partial}{\partial\alpha_{4}^{l}}\alpha
_{4}^{l}+i\omega_{i}\frac{\partial}{\partial\alpha_{1}^{l}}\alpha_{1}%
^{l}-i\omega_{i}\frac{\partial}{\partial\alpha_{2}^{l}}\alpha_{2}^{l}%
]\delta_{ll^{\prime}}\nonumber\\
& +i\omega_{l^{\prime}l}\frac{\partial}{\partial\alpha_{3}^{l}}\alpha
_{3}^{l^{\prime}}-i\omega_{l^{\prime}l}\frac{\partial}{\partial\alpha
_{4}^{l^{\prime}}}\alpha_{4}^{l}+i\omega_{ll^{\prime}}\frac{\partial}%
{\partial\alpha_{1}^{l}}\alpha_{1}^{l^{\prime}}-i\omega_{ll^{\prime}}%
\frac{\partial}{\partial\alpha_{2}^{l^{\prime}}}\alpha_{2}^{l}.
\end{align}

The atom-field interaction part (a) is%
\begin{align}
& \mathcal{L}_{A-L}^{(a)}=\nonumber\\
& i\Omega_{a}e^{ik_{a}z_{l}-i\omega_{a}t}[-\frac{\partial^{2}}{\partial
\alpha_{5}^{l}\partial\alpha_{5}^{l}}(\alpha_{5}^{l})+\frac{\partial^{2}%
}{\partial\alpha_{5}^{l}\partial\alpha_{6}^{l}}(\alpha_{6}^{l})-\frac
{\partial^{2}}{\partial\alpha_{5}^{l}\partial\alpha_{7}^{l}}(\alpha_{7}%
^{l})+\frac{\partial^{2}}{\partial\alpha_{5}^{l}\partial\alpha_{8}^{l}}%
(\alpha_{8}^{l})\nonumber\\
& -\frac{\partial^{2}}{\partial\alpha_{5}^{l}\partial\alpha_{9}^{l}}%
(\alpha_{9}^{l})-\frac{\partial}{\partial\alpha_{5}^{l}}(-\alpha_{11}%
^{l}-\alpha_{12}^{l}-2\alpha_{13}^{l}+N_{z})-\frac{\partial}{\partial
\alpha_{7}^{l}}(-\alpha_{6}^{l})-\frac{\partial}{\partial\alpha_{9}^{l}%
}(-\alpha_{8}^{l})\nonumber\\
& +\frac{\partial}{\partial\alpha_{16}^{l}}e^{-\frac{\partial}{\partial
\alpha_{13}^{l}}}(-\alpha_{15}^{l})+\frac{\partial}{\partial\alpha_{18}^{l}%
}e^{-\frac{\partial}{\partial\alpha_{13}^{l}}}(-\alpha_{17}^{l})+e^{-\frac
{\partial}{\partial\alpha_{13}^{l}}}(\alpha_{19}^{l})]-i\Omega_{a}%
e^{ik_{a}z_{l}-i\omega_{a}t}(\alpha_{19}^{l})\nonumber\\
& +i\Omega_{b}e^{-ik_{b}z_{l}-i\omega_{b}t}[-\frac{\partial^{2}}%
{\partial\alpha_{6}^{l}\partial\alpha_{6}^{l}}(\alpha_{6}^{l})-\frac
{\partial^{2}}{\partial\alpha_{6}^{l}\partial\alpha_{8}^{l}}(\alpha_{8}%
^{l})-\frac{\partial^{2}}{\partial\alpha_{6}^{l}\partial\alpha_{10}^{l}%
}(\alpha_{10}^{l})\nonumber\\
& +\frac{\partial}{\partial\alpha_{6}^{l}}(-\alpha_{13}^{l}+\alpha_{12}%
^{l})+\frac{\partial}{\partial\alpha_{7}^{l}}(-\alpha_{5}^{l})+\frac
{\partial^{2}}{\partial\alpha_{8}^{l}\partial\alpha_{10}^{l}}(\alpha_{12}%
^{l}-\alpha_{11}^{l})-\frac{\partial^{3}}{\partial\alpha_{8}^{l}\partial
\alpha_{10}^{l}\partial\alpha_{10}^{l}}(\alpha_{10}^{l})\nonumber\\
& -\frac{\partial}{\partial\alpha_{8}^{l}}e^{-\frac{\partial}{\partial
\alpha_{12}^{l}}+\frac{\partial}{\partial\alpha_{11}^{l}}}(-\alpha_{14}%
^{l})+\frac{\partial}{\partial\alpha_{10}^{l}}e^{-\frac{\partial}%
{\partial\alpha_{11}^{l}}+\frac{\partial}{\partial\alpha_{13}^{l}}}%
(-\alpha_{16}^{l})+(\frac{\partial^{2}}{\partial\alpha_{10}^{l}\partial
\alpha_{14}^{l}}e^{-\frac{\partial}{\partial\alpha_{11}^{l}}+\frac{\partial
}{\partial\alpha_{13}^{l}}}\nonumber\\
& +e^{-\frac{\partial}{\partial\alpha_{12}^{l}}+\frac{\partial}{\partial
\alpha_{13}^{l}}})(\alpha_{18}^{l})]-i\Omega_{b}e^{-ik_{b}z_{l}-i\omega_{b}%
t}[\alpha_{18}^{l}+\frac{\partial}{\partial\alpha_{19}^{l}}(-\alpha_{17}%
^{l})]+(c.c.\text{ with }C^{\prime}),
\end{align}
and let $\mathcal{L}_{A-L}^{(b)}=\mathcal{L}_{A-L,S}^{(b)}+\mathcal{L}%
_{A-L,I}^{(b)},$ which are the terms for signal (S) and idler (I) parts,%

\begin{align}
\mathcal{L}_{A-L,S}^{(b)}  & =ig_{s}\sqrt{2M+1}e^{-ik_{s}z_{l}}[\frac
{\partial}{\partial\alpha_{6}^{l}}(-\alpha_{8}^{l})+\frac{\partial}%
{\partial\alpha_{7}^{l}}(-\alpha_{9}^{l})-\frac{\partial^{2}}{\partial
\alpha_{10}^{l}\partial\alpha_{10}^{l}}(\alpha_{10}^{l})\nonumber\\
& +\frac{\partial}{\partial\alpha_{10}^{l}}(-\alpha_{11}^{l}+\alpha_{12}%
^{l})+e^{-\frac{\partial}{\partial\alpha_{12}^{l}}+\frac{\partial}%
{\partial\alpha_{11}^{l}}}(\alpha_{14}^{l})]\alpha_{3}^{l}+ig_{s}^{\ast}%
\sqrt{2M+1}e^{ik_{s}z_{l}}\nonumber\\
& \lbrack\frac{\partial}{\partial\alpha_{8}^{l}}(-\alpha_{6}^{l})+\alpha
_{10}^{l}+\frac{\partial}{\partial\alpha_{9}^{l}}(-\alpha_{7}^{l})](\alpha
_{4}^{l}-\frac{\partial}{\partial\alpha_{3}^{l}})+(c.c.\text{ with }C^{\prime
})
\end{align}
\bigskip

and%

\begin{align}
\mathcal{L}_{A-L,I}^{(b)}  & =ig_{i}\sqrt{2M+1}e^{ik_{i}z_{l}}[\frac
{\partial^{3}}{\partial\alpha_{5}^{l}\partial\alpha_{8}^{l}\partial\alpha
_{8}^{l}}(-\alpha_{8}^{l})+\frac{\partial^{3}}{\partial\alpha_{5}^{l}%
\partial\alpha_{6}^{l}\partial\alpha_{8}^{l}}(-\alpha_{6}^{l})+\frac
{\partial^{2}}{\partial\alpha_{5}^{l}\partial\alpha_{6}^{l}}(\alpha_{10}%
^{l})\nonumber\\
& -\frac{\partial^{2}}{\partial\alpha_{5}^{l}\partial\alpha_{9}^{l}}%
(\alpha_{5}^{l})-\frac{\partial^{2}}{\partial\alpha_{5}^{l}\partial\alpha
_{8}^{l}}(\alpha_{13}^{l}+\frac{\partial}{\partial\alpha_{9}^{l}}\alpha
_{9}^{l}-\frac{\partial}{\partial\alpha_{10}^{l}}\alpha_{10}^{l}-\alpha
_{11}^{l})\nonumber\\
& -\frac{\partial}{\partial\alpha_{5}^{l}}e^{-\frac{\partial}{\partial
\alpha_{11}^{l}}+\frac{\partial}{\partial\alpha_{13}^{l}}}(-\alpha_{16}%
^{l})-\frac{\partial^{2}}{\partial\alpha_{5}^{l}\partial\alpha_{14}^{l}%
}e^{-\frac{\partial}{\partial\alpha_{11}^{l}}+\frac{\partial}{\partial
\alpha_{13}^{l}}}(\alpha_{18}^{l})-\frac{\partial^{2}}{\partial\alpha_{7}%
^{l}\partial\alpha_{8}^{l}}(\alpha_{6}^{l})\nonumber\\
& -\frac{\partial}{\partial\alpha_{7}^{l}}(-\alpha_{10}^{l})-\frac
{\partial^{2}}{\partial\alpha_{7}^{l}\partial\alpha_{9}^{l}}(\alpha_{7}%
^{l})-\frac{\partial^{2}}{\partial\alpha_{9}^{l}\partial\alpha_{9}^{l}}%
(\alpha_{9}^{l})-N_{z}\frac{\partial}{\partial\alpha_{9}^{l}}-\frac{\partial
}{\partial\alpha_{8}^{l}}\big(\nonumber\\
& -\frac{\partial}{\partial\alpha_{9}^{l}}(-\alpha_{8}^{l})+\frac{\partial
}{\partial\alpha_{16}^{l}}e^{-\frac{\partial}{\partial\alpha_{13}^{l}}%
}(-\alpha_{15}^{l})+\frac{\partial}{\partial\alpha_{18}^{l}}e^{-\frac
{\partial}{\partial\alpha_{13}^{l}}}(-\alpha_{17}^{l})+e^{-\frac{\partial
}{\partial\alpha_{13}^{l}}}\alpha_{19}^{l}\big)\nonumber\\
& +e^{-\frac{\partial}{\partial\alpha_{11}^{l}}}\alpha_{15}^{l}-\frac
{\partial}{\partial\alpha_{9}^{l}}\big(-\alpha_{13}^{l}+\frac{\partial
}{\partial\alpha_{10}^{l}}(\alpha_{10}^{l})-\alpha_{12}^{l}+2\frac{\partial
}{\partial\alpha_{10}^{l}}(-\alpha_{10}^{l})\nonumber\\
& -2\alpha_{11}^{l}\big)+\frac{\partial}{\partial\alpha_{14}^{l}}%
e^{-\frac{\partial}{\partial\alpha_{11}^{l}}}(-\alpha_{17}^{l})]\alpha_{1}%
^{l}\nonumber\\
& +ig_{i}^{\ast}\sqrt{2M+1}e^{-ik_{i}z_{l}}(\alpha_{9}^{l})(\alpha_{2}%
^{l}-\frac{\partial}{\partial\alpha_{1}^{l}})+(c.c.\text{ with }C^{\prime}).
\end{align}

The dissipation part $\mathcal{L}_{sp}$ can be derived accordingly and the
above equation, which involves higher order derivatives (third order and
higher), is neglected. \ The validity of truncation to second order is due to
the expansion in the small parameter $1/N_{z}$. \ 

If the Fokker-Planck equation is
\begin{equation}
\frac{\partial f}{\partial t}=-\frac{\partial}{\partial\alpha}A_{\alpha
}f-\frac{\partial}{\partial\beta}A_{\beta}f+\frac{1}{2}(\frac{\partial^{2}%
}{\partial\alpha\partial\beta}+\frac{\partial^{2}}{\partial\beta\partial
\alpha})D_{\alpha\beta}f
\end{equation}
where $A$ and $D$ are drift and diffusion terms then we have a corresponding
classical Langevin equation%

\begin{equation}
\frac{\partial\alpha}{\partial t}=A_{\alpha}+\Gamma_{\alpha}\text{, }%
\frac{\partial\beta}{\partial t}=A_{\beta}+\Gamma_{\beta}%
\end{equation}
with a correlation function $\langle\Gamma_{\alpha}\Gamma_{\beta}%
\rangle=\delta(t-t^{\prime})D_{\alpha\beta}$. \ So we have according to
various $\mathcal{L}$'s,%
\begin{align}
\dot{\alpha}_{5}^{l}  & =(-i\omega_{1}-\frac{\gamma_{01}}{2})\alpha_{5}%
^{l}+i\Omega_{a}e^{ik_{a}z_{l}-i\omega_{a}t}(\alpha_{0}^{l}-\alpha_{13}%
^{l})+i\Omega_{b}^{\ast}e^{ik_{b}z_{l}+i\omega_{b}t}\alpha_{7}^{l}\nonumber\\
& -ig_{i}\sqrt{2M+1}e^{ik_{i}z_{l}}\alpha_{16}^{l}\alpha_{1}^{l}+\Gamma
_{5}^{l},\nonumber\\
\dot{\alpha}_{6}^{l}  & =i(\omega_{1}-\omega_{2}+i\frac{\gamma_{01}+\gamma
_{2}}{2})\alpha_{6}^{l}-i\Omega_{a}^{\ast}e^{-ik_{a}z_{l}+i\omega_{a}t}%
\alpha_{7}^{l}+i\Omega_{b}e^{-ik_{b}z_{l}-i\omega_{b}t}(\alpha_{13}^{l}%
-\alpha_{12}^{l})\nonumber\\
& +ig_{s}\sqrt{2M+1}e^{-ik_{s}z_{l}}\alpha_{8}^{l}\alpha_{3}^{l}+\Gamma
_{6}^{l},\nonumber\\
\dot{\alpha}_{7}^{l}  & =(-i\omega_{2}-\frac{\gamma_{2}}{2})\alpha_{7}%
^{l}-i\Omega_{a}e^{ik_{a}z_{l}-i\omega_{a}t}\alpha_{6}^{l}+i\Omega
_{b}e^{-ik_{b}z_{l}-i\omega_{b}t}\alpha_{5}^{l}\nonumber\\
& +ig_{s}\sqrt{2M+1}e^{-ik_{s}z_{l}}\alpha_{9}^{l}\alpha_{3}^{l}-ig_{i}%
\sqrt{2M+1}e^{ik_{i}z_{l}}\alpha_{10}^{l}\alpha_{1}^{l}+\Gamma_{7}%
^{l},\nonumber\\
\dot{\alpha}_{13}^{l}  & =-\gamma_{01}\alpha_{13}^{l}+\gamma_{12}\alpha
_{12}^{l}+i\Omega_{a}e^{ik_{a}z_{l}-i\omega_{a}t}\alpha_{19}^{l}-i\Omega
_{a}^{\ast}e^{-ik_{a}z_{l}+i\omega_{a}t}\alpha_{5}^{l}\nonumber\\
& -i\Omega_{b}e^{-ik_{b}z_{l}-i\omega_{b}t}\alpha_{18}^{l}+i\Omega_{b}^{\ast
}e^{ik_{b}z_{l}+i\omega_{b}t}\alpha_{6}^{l}+\Gamma_{13}^{l},\nonumber\\
\dot{\alpha}_{12}^{l}  & =-\gamma_{2}\alpha_{12}^{l}+i\Omega_{b}%
e^{-ik_{b}z_{l}-i\omega_{b}t}\alpha_{18}^{l}-i\Omega_{b}^{\ast}e^{ik_{b}%
z_{l}+i\omega_{b}t}\alpha_{6}^{l}+ig_{s}\sqrt{2M+1}e^{-ik_{s}z_{l}}\alpha
_{14}^{l}\alpha_{3}^{l}\nonumber\\
& -ig_{s}^{\ast}\sqrt{2M+1}e^{ik_{s}z_{l}}\alpha_{10}^{l}\alpha_{4}^{l}%
+\Gamma_{12}^{l},\nonumber\\
\dot{\alpha}_{11}^{l}  & =-\gamma_{03}\alpha_{11}^{l}+\gamma_{32}\alpha
_{12}^{l}-ig_{s}\sqrt{2M+1}e^{-ik_{s}z_{l}}\alpha_{14}^{l}\alpha_{3}%
^{l}+ig_{s}\sqrt{2M+1}e^{ik_{s}z_{l}}\alpha_{10}^{l}\alpha_{4}^{l}\nonumber\\
& +ig_{i}\sqrt{2M+1}e^{ik_{i}z_{l}}\alpha_{15}^{l}\alpha_{1}^{l}-ig_{i}^{\ast
}\sqrt{2M+1}e^{-ik_{i}z_{l}}\alpha_{9}^{l}\alpha_{2}^{l}+\Gamma_{11}%
^{l},\nonumber\\
\dot{\alpha}_{8}^{l}  & =i(\omega_{1}-\omega_{3}+i\frac{\gamma_{01}%
+\gamma_{03}}{2})\alpha_{8}^{l}-i\Omega_{a}^{\ast}e^{-ik_{a}z_{l}+i\omega
_{a}t}\alpha_{9}^{l}-i\Omega_{b}e^{-ik_{b}z_{l}-i\omega_{b}t}\alpha_{14}%
^{l}\nonumber\\
& +ig_{s}^{\ast}\sqrt{2M+1}e^{ik_{s}z_{l}}\alpha_{6}^{l}\alpha_{4}^{l}%
+ig_{i}\sqrt{2M+1}e^{ik_{i}z_{l}}\alpha_{19}^{l}\alpha_{1}^{l}+\Gamma_{8}%
^{l},\nonumber\\
\dot{\alpha}_{9}^{l}  & =(-i\omega_{3}-\frac{\gamma_{03}}{2})\alpha_{9}%
^{l}-i\Omega_{a}e^{ik_{a}z_{l}-i\omega_{a}t}\alpha_{8}^{l}+ig_{s}^{\ast}%
\sqrt{2M+1}e^{ik_{s}z_{l}}\alpha_{7}^{l}\alpha_{4}^{l}\nonumber\\
& +ig_{i}\sqrt{2M+1}e^{ik_{i}z_{l}}(\alpha_{0}^{l}-\alpha_{11}^{l})\alpha
_{1}^{l}+\Gamma_{9}^{l},\nonumber\\
\dot{\alpha}_{14}^{l}  & =i(\omega_{2}-\omega_{3}+i\frac{\gamma_{03}%
+\gamma_{2}}{2})\alpha_{14}^{l}-i\Omega_{b}^{\ast}e^{ik_{b}z_{l}+i\omega_{b}%
t}\alpha_{8}^{l}\nonumber\\
& +ig_{s}^{\ast}\sqrt{2M+1}e^{ik_{s}z_{l}}(\alpha_{12}^{l}-\alpha_{11}%
^{l})\alpha_{4}^{l}+ig_{i}\sqrt{2M+1}e^{ik_{i}z_{l}}\alpha_{17}^{l}\alpha
_{1}^{l}+\Gamma_{14}^{l},\nonumber\\
\dot{\alpha}_{4}^{l}  & =i\omega_{s}\alpha_{4}^{l}+i\sum_{l^{\prime}}%
\omega_{ll^{\prime}}\alpha_{4}^{l^{\prime}}-ig_{s}\sqrt{2M+1}e^{-ik_{s}z_{l}%
}\alpha_{14}^{l}+\Gamma_{4}^{l},\nonumber\\
\dot{\alpha}_{1}^{l}  & =-i\omega_{i}\alpha_{1}^{l}-i\sum_{l^{\prime}}%
\omega_{ll^{\prime}}\alpha_{1}^{l^{\prime}}+ig_{i}^{\ast}\sqrt{2M+1}%
e^{-ik_{i}z_{l}}\alpha_{9}^{l}+\Gamma_{1}^{l},
\end{align}
where $\gamma_{2}=\gamma_{12}+\gamma_{32}.$ \ We postpone the derivations of
diffusion coefficients after the scaling is made in the next section, and note
that the complete equations of motion are found by making complex conjugate of
the above with correspondence $C^{\prime}$ and changing Langevin noises
correspondingly, say $\Gamma_{5}^{\ast}\rightarrow\Gamma_{19}.$

\subsection{Slowly varying envelopes and scaled equations of motion}

Here we introduce the slowly varying envelopes and define our cross-grained
collective atomic and field observables, then finally transform the equations
in a dimensionless form for later numerical simulations. \ We note that%

\begin{equation}
i\sum_{l^{\prime}}\omega_{ll^{\prime}}\alpha_{4}^{l^{\prime}}=c\frac{d}%
{dz_{l}}\alpha_{4}^{l}\text{, }-i\sum_{l^{\prime}}\omega_{ll^{\prime}}%
\alpha_{1}^{l^{\prime}}=-c\frac{\partial}{\partial z_{l}}\alpha_{1}^{l},
\end{equation}

and $\alpha_{0}^{l}=N_{z}-\alpha_{13}^{l}-\alpha_{12}^{l}-\alpha_{11}^{l} $.
\ Define slow varying observables that
\begin{align}
& \widetilde{\alpha}_{5}(z,t)\equiv\frac{1}{N_{z}}\alpha_{5}^{l}%
e^{-ik_{a}z_{l}+i\omega_{a}t},~\widetilde{\alpha}_{6}(z,t)\equiv\frac{1}%
{N_{z}}\alpha_{6}^{l}e^{ik_{b}z_{l}+i\omega_{b}t},\nonumber\\
& \widetilde{\alpha}_{7}(z,t)\equiv\frac{1}{N_{z}}\alpha_{7}^{l}%
e^{-ik_{a}z_{l}+ik_{b}z_{l}+i\omega_{b}t+i\omega_{a}t},~\widetilde{\alpha}%
_{8}(z,t)\equiv\frac{1}{N_{z}}\alpha_{8}^{l}e^{-i\omega_{a}t+i\omega
_{3}t+ik_{a}z_{l}-ik_{i}z_{l}},\nonumber\\
\text{ }  & \widetilde{\alpha}_{9}(z,t)\equiv\frac{1}{N_{z}}\alpha_{9}%
^{l}e^{-ik_{i}z_{l}+i\omega_{3}t},\widetilde{\alpha}_{11}(z,t)\equiv\frac
{1}{N_{z}}\alpha_{11}^{l},\nonumber\\
& \widetilde{\alpha}_{12}(z,t)\equiv\frac{1}{N_{z}}\alpha_{12}^{l}%
,\widetilde{\alpha}_{13}(z,t)\equiv\frac{1}{N_{z}}\alpha_{13}^{l},\nonumber\\
& \text{ }\widetilde{\alpha}_{14}(z,t)\equiv\frac{1}{N_{z}}\alpha_{14}%
^{l}e^{-i(\omega_{23}+\Delta_{2})t}e^{ik_{a}z_{l}-ik_{b}z_{l}-ik_{i}z_{l}%
}\text{ }%
\end{align}
where $e^{i\Delta kz}=e^{ik_{a}z_{l}-ik_{b}z_{l}-ik_{i}z_{l}+ik_{s}z_{l}}$.
\ Also for the field variables,%

\begin{equation}
E_{s}^{-}(z,t)\equiv\frac{g_{s}^{\ast}}{d_{i}/\hbar}\sqrt{2M+1}\alpha_{4}%
^{l}e^{-i\omega_{s}t},\text{ }E_{i}^{+}(z,t)\equiv\frac{g_{i}}{d_{i}/\hbar
}\sqrt{2M+1}\alpha_{1}^{l}e^{i\omega_{i}t},
\end{equation}
where we use the idler dipole moment in signal field scaling for the purpose
of scale-free atomic equation of motions, so we need to keep in mind that in
calculating signal intensity or correlation function, an extra factor of
$(d_{i}/d_{s})^{2}$ needs to be taken care of.

We choose the central frequency of signal and idler as $\omega_{s}=\omega
_{23}+\Delta_{2},\omega_{i}=\omega_{3}$ where $\Delta_{1}=\omega_{a}%
-\omega_{1}$ and $\Delta_{2}=\omega_{a}+\omega_{b}-\omega_{2}$. \ With a
scaling of Arecchi-Courtens cooperation length \cite{scale}, we set up the
units of field strength, time, and length in the following,%

\begin{equation}
\frac{E_{c}}{T_{c}}=\frac{N|g_{i}|^{2}}{d_{i}/\hbar},\text{ }L_{c}%
=cT_{c},\text{ }\frac{1}{T_{c}}=\sqrt{\frac{d_{i}^{2}n\omega_{i}}%
{2\hbar\epsilon_{0}}},\text{ }E_{c}=\sqrt{\frac{n\hbar\omega_{i}}%
{2\epsilon_{0}}}=\frac{1}{T_{c}}\frac{1}{d_{i}/\hbar}.
\end{equation}

Compared with optical density and superradiant time scale, we have (in terms
of single atomic decay rate $\gamma$)%

\begin{equation}
N|g_{i}|^{2}=\frac{\gamma_{N}}{L/c},\text{ }\gamma_{N}=N\frac{3}{8\pi}%
\frac{\lambda^{2}}{A}\gamma,n=\frac{N}{V}.
\end{equation}

Now the slowly varying and dimensionless equations of motion with Langevin
noises in Ito's form are%

\begin{align}
\frac{\partial}{\partial t}\widetilde{\alpha}_{5}  & =(i\Delta_{1}%
-\frac{\gamma_{01}}{2})\widetilde{\alpha}_{5}+i\Omega_{a}(\widetilde{\alpha
}_{0}-\widetilde{\alpha}_{13})+i\Omega_{b}^{\ast}\widetilde{\alpha}%
_{7}-i\widetilde{\alpha}_{16}E_{i}^{+}+\mathcal{F}_{5},\nonumber\\
\frac{\partial}{\partial t}\widetilde{\alpha}_{6}  & =i(\Delta_{2}-\Delta
_{1}+i\frac{\gamma_{01}+\gamma_{2}}{2})\widetilde{\alpha}_{6}-i\Omega
_{a}^{\ast}\widetilde{\alpha}_{7}+i\Omega_{b}(\widetilde{\alpha}%
_{13}-\widetilde{\alpha}_{12})+i\widetilde{\alpha}_{8}E_{s}^{+}e^{-i\Delta
kz}+\mathcal{F}_{6},\nonumber\\
\frac{\partial}{\partial t}\widetilde{\alpha}_{7}  & =(i\Delta_{2}%
-\frac{\gamma_{2}}{2})\widetilde{\alpha}_{7}-i\Omega_{a}\widetilde{\alpha}%
_{6}+i\Omega_{b}\widetilde{\alpha}_{5}+i\widetilde{\alpha}_{9}E_{s}%
^{+}e^{-i\Delta kz}-i\widetilde{\alpha}_{10}E_{i}^{+}+\mathcal{F}%
_{7},\nonumber\\
\frac{\partial}{\partial t}\widetilde{\alpha}_{13}  & =-\gamma_{01}%
\widetilde{\alpha}_{13}+\gamma_{12}\widetilde{\alpha}_{12}+i\Omega
_{a}\widetilde{\alpha}_{19}-i\Omega_{a}^{\ast}\widetilde{\alpha}_{5}%
-i\Omega_{b}\widetilde{\alpha}_{18}+i\Omega_{b}^{\ast}\widetilde{\alpha}%
_{6}+\mathcal{F}_{13},\nonumber\\
\frac{\partial}{\partial t}\widetilde{\alpha}_{12}  & =-\gamma_{2}%
\widetilde{\alpha}_{12}+i\Omega_{b}\widetilde{\alpha}_{18}-i\Omega_{b}^{\ast
}\widetilde{\alpha}_{6}+i\widetilde{\alpha}_{14}E_{s}^{+}e^{-i\Delta
kz}-i\widetilde{\alpha}_{10}E_{s}^{-}e^{i\Delta kz}+\mathcal{F}_{12}%
,\nonumber\\
\frac{\partial}{\partial t}\widetilde{\alpha}_{11}  & =-\gamma_{03}%
\widetilde{\alpha}_{11}+\gamma_{32}\widetilde{\alpha}_{12}-i\widetilde{\alpha
}_{14}E_{s}^{+}e^{-i\Delta kz}+i\widetilde{\alpha}_{10}E_{s}^{-}e^{i\Delta
kz}+i\widetilde{\alpha}_{15}E_{i}^{+}-i\widetilde{\alpha}_{9}E_{i}%
^{-}+\mathcal{F}_{11},\nonumber\\
\frac{\partial}{\partial t}\widetilde{\alpha}_{8}  & =-(i\Delta_{1}%
+\frac{\gamma_{01}+\gamma_{03}}{2})\widetilde{\alpha}_{8}-i\Omega_{a}^{\ast
}\widetilde{\alpha}_{9}-i\Omega_{b}\widetilde{\alpha}_{14}+i\widetilde{\alpha
}_{6}E_{s}^{-}e^{i\Delta kz}+i\widetilde{\alpha}_{19}E_{i}^{+}+\mathcal{F}%
_{8},\nonumber\\
\frac{\partial}{\partial t}\widetilde{\alpha}_{9}  & =-\frac{\gamma_{03}}%
{2}\widetilde{\alpha}_{9}-i\Omega_{a}\widetilde{\alpha}_{8}+i\widetilde
{\alpha}_{7}E_{s}^{-}e^{i\Delta kz}+i(\widetilde{\alpha}_{0}-\widetilde
{\alpha}_{11})E_{i}^{+}+\mathcal{F}_{9},\nonumber\\
\frac{\partial}{\partial t}\widetilde{\alpha}_{14}  & =-(i\Delta_{2}%
+\frac{\gamma_{03}+\gamma_{2}}{2})\widetilde{\alpha}_{14}-i\Omega_{b}^{\ast
}\widetilde{\alpha}_{8}+i(\widetilde{\alpha}_{12}-\widetilde{\alpha}%
_{11})E_{s}^{-}e^{i\Delta kz}+i\widetilde{\alpha}_{17}E_{i}^{+}+\mathcal{F}%
_{14},\nonumber\\
& \label{bloch2}%
\end{align}
and field propagation equations are%
\begin{align}
(\frac{\partial}{\partial t}-\frac{\partial}{\partial z})E_{s}^{-}  &
=-i\widetilde{\alpha}_{14}e^{-i\Delta kz}\frac{|g_{s}|^{2}}{|g_{i}|^{2}%
}+\mathcal{F}_{4},\nonumber\\
(\frac{\partial}{\partial t}+\frac{\partial}{\partial z})E_{i}^{+}  &
=i\widetilde{\alpha}_{9}+\mathcal{F}_{1},
\end{align}
where $\frac{|g_{s}|^{2}}{|g_{i}|^{2}}$ is a unit transformation factor from
the signal field strength to the idler one. \ For a recognizable format of the
above equations used in the text of Chapter 4, we change the labels in the below,%

\begin{align}
& \widetilde{\alpha}_{5}\leftrightarrow\pi_{01},\text{ }\widetilde{\alpha}%
_{6}\leftrightarrow\pi_{12},\text{ }\widetilde{\alpha}_{7}\leftrightarrow
\pi_{02},\text{ }\widetilde{\alpha}_{8}\leftrightarrow\pi_{13},\text{
}\widetilde{\alpha}_{9}\leftrightarrow\pi_{03},\text{ }\widetilde{\alpha}%
_{10}\leftrightarrow\pi_{32},\text{ }\widetilde{\alpha}_{11}\leftrightarrow
\pi_{33},\nonumber\\
& \widetilde{\alpha}_{12}\leftrightarrow\pi_{22},\text{ }\widetilde{\alpha
}_{13}\leftrightarrow\pi_{11},\text{ }\widetilde{\alpha}_{14}\leftrightarrow
\pi_{32}^{\dag},\text{ }\widetilde{\alpha}_{15}\leftrightarrow\pi_{03}^{\dag
},\text{ }\widetilde{\alpha}_{16}\leftrightarrow\pi_{13}^{\dag},\text{
}\widetilde{\alpha}_{17}\leftrightarrow\pi_{02}^{\dag},\text{ }\nonumber\\
& \widetilde{\alpha}_{18}\leftrightarrow\pi_{12}^{\dag},\text{ }%
\widetilde{\alpha}_{19}\leftrightarrow\pi_{01}^{\dag},
\end{align}
where $\pi_{ij}$ is the stochastic variable that corresponds to the atomic
populations of state $|i\rangle$ when $i=j$ and to atomic coherence when
$i\neq j$. \ Note that the associated c-number Langevin noises are changed accordingly.

The Langevin noises are defined as%

\begin{align}
\mathcal{F}_{5}(z,t)  & =\frac{1}{N_{z}}\Gamma_{5}^{l}e^{-ik_{a}z_{l}%
+i\omega_{a}t},\mathcal{F}_{6}(z,t)=\frac{1}{N_{z}}\Gamma_{6}^{l}%
e^{ik_{b}z_{l}+i\omega_{b}t},\nonumber\\
\mathcal{F}_{7}(z,t)  & =\frac{1}{N_{z}}\Gamma_{7}^{l}e^{-ik_{a}z_{l}%
+ik_{b}z_{l}+i\omega_{b}t+i\omega_{a}t},\mathcal{F}_{13}(z,t)=\frac{1}{N_{z}%
}\Gamma_{13}^{l},\mathcal{F}_{12}(z,t)=\frac{1}{N_{z}}\Gamma_{12}%
^{l},\nonumber\\
\mathcal{F}_{11}(z,t)  & =\frac{1}{N_{z}}\Gamma_{11}^{l},\mathcal{F}%
_{8}(z,t)=\frac{1}{N_{z}}\Gamma_{8}^{l}e^{-i\omega_{a}t+i\omega_{3}%
t+ik_{a}z_{l}-ik_{i}z_{l}},\nonumber\\
\mathcal{F}_{9}(z,t)  & =\frac{1}{N_{z}}\Gamma_{9}^{l}e^{-ik_{i}z_{l}%
+i\omega_{3}t},\mathcal{F}_{14}(z,t)=\frac{1}{N_{z}}\Gamma_{14}^{l}%
e^{-i(\omega_{23}+\Delta_{2})t}e^{ik_{a}z_{l}-ik_{b}z_{l}-ik_{i}z_{l}%
},\nonumber\\
\mathcal{F}_{4}(z,t)  & =\frac{g_{s}^{\ast}}{d_{i}/\hbar}\sqrt{2M+1}%
e^{-i\omega_{s}t}\Gamma_{4}^{l},\mathcal{F}_{1}(z,t)=\frac{g_{i}}{d_{i}/\hbar
}\sqrt{2M+1}e^{i\omega_{i}t}\Gamma_{1}^{l}%
\end{align}
where other Langevin noises can be found by using the correspondence similar
to $C^{\prime}$, for example, $\mathcal{F}_{5}^{\ast}\leftrightarrow
\mathcal{F}_{19}$.

Before we proceed to formulate the diffusion coefficients, we need to be
careful about the scaling factor for the transformation to continuous
variables when numerical simulation is applied. \ Take $\left\langle
\mathcal{F}_{6}\mathcal{F}_{5}\right\rangle $ for example,%
\begin{align}
& \left\langle \mathcal{F}_{6}(z,t)\mathcal{F}_{5}(z^{\prime},t^{\prime
})\right\rangle \nonumber\\
& =\frac{1}{N_{z}^{2}}e^{ik_{b}z_{l}+i\omega_{b}t}e^{-ik_{a}z_{l^{\prime}%
}+i\omega_{a}t^{\prime}}\left\langle \Gamma_{6}^{l}\Gamma_{5}^{l^{\prime}%
}\right\rangle \nonumber\\
& =\frac{1}{N_{z}^{2}}e^{ik_{b}z_{l}+i\omega_{b}t}e^{-ik_{a}z_{l}+i\omega
_{a}t}[i\Omega_{a}e^{ik_{a}z_{l}-i\omega_{a}t}\alpha_{6}^{l}+ig_{i}\sqrt
{2M+1}e^{ik_{i}z_{l}}\alpha_{10}^{l}\alpha_{1}^{l}]\delta(t-t^{\prime}%
)\delta_{ll^{\prime}}\nonumber\\
& =\frac{1}{N_{z}}\left[  i\frac{\Omega_{a}}{T_{c}}T_{c}\widetilde{\alpha}%
_{6}+i\frac{d_{i}}{\hbar}E_{c}\widetilde{\alpha}_{10}\frac{E_{i}^{+}}{E_{c}%
}\right]  \delta(t-t^{\prime})\delta(z-z^{\prime})\frac{L}{2M+1}\nonumber\\
& =\left[  i(\Omega_{a}T_{c})\widetilde{\alpha}_{6}+i\widetilde{\alpha}%
_{10}(E_{i}^{+}/E_{c})\right]  \frac{1}{T_{c}^{2}}\delta(t-t^{\prime}%
)T_{c}\delta(z-z^{\prime})L_{c}\frac{L}{L_{c}}\frac{N_{z}}{N}\frac{1}{N_{z}%
}\nonumber\\
& =\frac{1}{N_{c}}\left[  i(\Omega_{a}T_{c})\widetilde{\alpha}_{6}%
+i\widetilde{\alpha}_{10}(E_{i}^{+}/E_{c})\right]  \frac{1}{T_{c}^{2}}%
\delta(t-t^{\prime})T_{c}\delta(z-z^{\prime})L_{c}%
\end{align}
where we have used $\lim_{M\rightarrow\infty}\frac{2M+1}{L}\delta_{ll^{\prime
}}=\delta(z-z^{\prime})$, $2M+1=\frac{N}{N_{z}},$ and $N_{c}=\frac{NL_{c}}{L}$
is the cooperation number. \ Then we have the dimensionless form of diffusion coefficients.%

\begin{align}
T_{c}^{2}\left\langle \mathcal{F}_{6}(z,t)\mathcal{F}_{5}(z^{\prime}%
,t^{\prime})\right\rangle  & =\frac{1}{N_{c}}D_{6,5}\delta(t-t^{\prime}%
)\delta(z-z^{\prime})\\
D_{6,5}  & =\left[  i\Omega_{a}\widetilde{\alpha}_{6}+i\widetilde{\alpha}%
_{10}E_{i}^{+}\right]  .
\end{align}

The dimensionless diffusion coefficients $D_{ij}$ are%

\begin{align}
(\text{i})D_{5,5}  & =-i2\Omega_{a}\widetilde{\alpha}_{5};\text{ }%
D_{5,6}=i(\Omega_{a}\widetilde{\alpha}_{6}+\widetilde{\alpha}_{10}E_{i}%
^{+});\text{ }D_{5,7}=-i\Omega_{a}\widetilde{\alpha}_{7};\text{ }\nonumber\\
D_{5,8}  & =i(\Omega_{a}\widetilde{\alpha}_{8}+(\widetilde{\alpha}%
_{11}-\widetilde{\alpha}_{13})E_{i}^{+});\text{ }D_{5,9}=-i(\Omega
_{a}\widetilde{\alpha}_{9}+\widetilde{\alpha}_{5}E_{i}^{+});\text{
}\nonumber\\
D_{5,11}  & =-i\widetilde{\alpha}_{16}E_{i}^{+};\text{ }D_{5,13}%
=i\widetilde{\alpha}_{16}E_{i}^{+};\text{ }D_{5,14}=-i\widetilde{\alpha}%
_{18}E_{i}^{+};\text{ }D_{5,19}=\gamma_{12}\widetilde{\alpha}_{12};\nonumber\\
(\text{ii})D_{6,6}  & =-i2\Omega_{b}\widetilde{\alpha}_{6};\text{ }%
D_{6,8}=-i\Omega_{b}\widetilde{\alpha}_{8};\text{ }D_{6,10}=-i\Omega
_{b}\widetilde{\alpha}_{10};\text{ }\nonumber\\
D_{6,13}  & =-i\Omega_{a}^{\ast}\widetilde{\alpha}_{7}+\gamma_{01}%
\widetilde{\alpha}_{6};\text{ }D_{6,16}=-i\widetilde{\alpha}_{7}E_{i}%
^{-}+\gamma_{01}\widetilde{\alpha}_{10};\text{ }D_{6,18}=\gamma_{01}%
\widetilde{\alpha}_{12};\nonumber\\
(\text{iii})D_{7,8}  & =-i\widetilde{\alpha}_{6}E_{i}^{+};\text{ }%
D_{7,9}=-i\widetilde{\alpha}_{7}E_{i}^{+};\nonumber\\
(\text{iv})D_{8,9}  & =-i\widetilde{\alpha}_{8}E_{i}^{+};\text{ }%
D_{8,10}=i\Omega_{b}(\widetilde{\alpha}_{12}-\widetilde{\alpha}_{11});\text{
}D_{8,11}=i\Omega_{b}\widetilde{\alpha}_{14};\text{ }\nonumber\\
D_{8,12}  & =-i\Omega_{b}\widetilde{\alpha}_{14};D_{8,13}=-i\Omega_{a}^{\ast
}\widetilde{\alpha}_{9}+i\widetilde{\alpha}_{19}E_{i}^{+}+\gamma
_{01}\widetilde{\alpha}_{8};\text{ }\nonumber\\
D_{8,16}  & =i\widetilde{\alpha}_{15}E_{i}^{+}-i\widetilde{\alpha}_{9}%
E_{i}^{-}+\gamma_{01}\widetilde{\alpha}_{11}+\gamma_{32}\widetilde{\alpha
}_{12};\text{ }D_{8,18}=i\widetilde{\alpha}_{17}E_{i}^{+}+\gamma
_{01}\widetilde{\alpha}_{14};\nonumber\\
(\text{v})D_{9,9}  & =-i2\widetilde{\alpha}_{9}E_{i}^{+};\text{ }%
D_{9,10}=i\widetilde{\alpha}_{10}E_{i}^{+};\text{ }D_{9,15}=\gamma
_{32}\widetilde{\alpha}_{12};\nonumber\\
(\text{vi})D_{10,10}  & =-i2\widetilde{\alpha}_{10}E_{s}^{+}e^{-i\Delta
kz};\text{ }D_{10,11}=i(\Omega_{b}\widetilde{\alpha}_{16}-\widetilde{\alpha
}_{7}E_{i}^{-})+\gamma_{03}\widetilde{\alpha}_{10};\text{ }\nonumber\\
D_{10,13}  & =-i\Omega_{b}\widetilde{\alpha}_{16};D_{10,14}=i\Omega
_{b}\widetilde{\alpha}_{18}-i\Omega_{b}^{\ast}\widetilde{\alpha}_{6}%
+\gamma_{03}\widetilde{\alpha}_{12};\text{ }D_{10,19}=i\widetilde{\alpha}%
_{6}E_{i}^{-};\nonumber\\
(\text{vii})D_{11,11}  & =i\widetilde{\alpha}_{14}E_{s}^{+}e^{-i\Delta
kz}-i\widetilde{\alpha}_{10}E_{s}^{-}e^{i\Delta kz}+i\widetilde{\alpha}%
_{15}E_{i}^{+}-i\widetilde{\alpha}_{9}E_{i}^{-}+\gamma_{32}\widetilde{\alpha
}_{12}+\gamma_{03}\widetilde{\alpha}_{11};\nonumber\\
D_{11,12}  & =i\widetilde{\alpha}_{10}E_{s}^{-}e^{i\Delta kz}-i\widetilde
{\alpha}_{14}E_{s}^{+}e^{-i\Delta kz}-\gamma_{32}\widetilde{\alpha}%
_{12};\nonumber\\
(\text{viii})D_{12,12}  & =i\Omega_{b}\widetilde{\alpha}_{18}-i\Omega
_{b}^{\ast}\widetilde{\alpha}_{6}-i\widetilde{\alpha}_{10}E_{s}^{-}e^{i\Delta
kz}+i\widetilde{\alpha}_{14}E_{s}^{+}e^{-i\Delta kz}+\gamma_{2}\widetilde
{\alpha}_{12};\nonumber\\
D_{12,13}  & =-i\Omega_{b}\widetilde{\alpha}_{18}+i\Omega_{b}^{\ast}%
\widetilde{\alpha}_{6}-\gamma_{12}\widetilde{\alpha}_{12};\nonumber\\
(\text{ix})D_{13,13}  & =i\Omega_{a}\widetilde{\alpha}_{19}-i\Omega_{a}^{\ast
}\widetilde{\alpha}_{5}+i\Omega_{b}\widetilde{\alpha}_{18}-i\Omega_{b}^{\ast
}\widetilde{\alpha}_{6}+\gamma_{01}\widetilde{\alpha}_{13}+\gamma
_{12}\widetilde{\alpha}_{12};\nonumber\\
(\text{x})D_{3,8}  & =\frac{|g_{s}|^{2}}{|g_{i}|^{2}}i\widetilde{\alpha}%
_{6}e^{i\Delta kz};\text{ }D_{3,9}=\frac{|g_{s}|^{2}}{|g_{i}|^{2}}%
i\widetilde{\alpha}_{7}e^{i\Delta kz}.
\end{align}

\subsection{Alternative method to derive diffusion coefficients by Einstein
relations}

Before going further to set up the stochastic differential equation, we show
here how we derive the diffusion coefficients from the Heisenberg-Langevin
approach with Einstein relations, and it provides the important check for
Fokker-Planck equations.\ \ We note here that a symmetric property of the
diffusion coefficients is within Fokker-Planck equation, whereas the quantum
diffusion coefficients in quantum Langevin equation do not have symmetric
property simply because the quantum operators do not necessarily commute with
each other. \ 

The approach involves a quantum-classical correspondence in deriving c-number
Langevin equations and requires a chosen normal ordering of quantum operators.
\ We use the same ordering as we use for deriving Fokker-Planck equations in
Eq. (\ref{order}),
\[
\tilde{\sigma}_{01}^{\dagger},\tilde{\sigma}_{12}^{\dagger},\tilde{\sigma
}_{02}^{\dagger},\tilde{\sigma}_{13}^{\dagger},\tilde{\sigma}_{03}^{\dagger
},\tilde{\sigma}_{32}^{\dagger},\tilde{\sigma}_{11},\tilde{\sigma}_{22}%
,\tilde{\sigma}_{33},\tilde{\sigma}_{32},\tilde{\sigma}_{03},\tilde{\sigma
}_{13},\tilde{\sigma}_{02},\tilde{\sigma}_{12},\tilde{\sigma}_{01},\hat{a}%
_{s}^{\dagger},\hat{a}_{s},\hat{a}_{i}^{\dagger},\hat{a}_{i}%
\]
and its classical correspondence is $\widetilde{\alpha}_{19,}\widetilde
{\alpha}_{18,}...\widetilde{\alpha}_{1.}$

We take $D_{8,13}=D_{13,8}$ for a demonstration.\ \ We first calculate the
quantum diffusion coefficient, $\mathcal{\hat{D}}_{13,8},$ using Einstein
relations where we attach the hat to it, and then we can find $\mathcal{\bar
{D}}_{13,8},$ a classical diffusion coefficient, which is reviewed in Chapter
2. \ Note that in calculating the quantum coefficients, we take advantage of
Eq. (\ref{bloch2}) where the drift terms are directly corresponded to quantum
Langevin equations. \ For clarity, $\mathcal{\hat{D}}_{13,8}=\mathcal{\hat{D}%
}_{\tilde{\sigma}_{11},\tilde{\sigma}_{13}}$ with $\widetilde{\alpha}%
_{13}\rightarrow\tilde{\sigma}_{11},\widetilde{\alpha}_{8}\rightarrow
\tilde{\sigma}_{13}$ representing a correspondence to quantum Langevin
equations. \ The index in the classical variables $\widetilde{\alpha}$
represents the ordering we choose as defined above, and in various quantum
operators $\tilde{\sigma}$, the index represents the atomic levels for atomic
coherences or populations. \ We should find $\mathcal{\bar{D}}_{13,8}%
=D_{13,8}=D_{8,13},$ and the proof is illustrated below by the Einstein's
relation, Eq. (\ref{Einstein_1}),%

\begin{align}
&  \left\langle \mathcal{\hat{D}}_{13,8}\right\rangle =\nonumber\\
&  -\left\langle \left[  -\gamma_{01}\tilde{\sigma}_{11}+\gamma_{12}%
\tilde{\sigma}_{22}+i\Omega_{a}\tilde{\sigma}_{01}^{\dagger}-i\Omega_{a}%
^{\ast}\tilde{\sigma}_{01}-i\Omega_{b}\tilde{\sigma}_{12}^{\dagger}%
+i\Omega_{b}^{\ast}\tilde{\sigma}_{12}\right]  \tilde{\sigma}_{13}%
\right\rangle \nonumber\\
&  -\left\langle \tilde{\sigma}_{11}\left[  -(i\Delta_{1}+\frac{\gamma
_{01}+\gamma_{03}}{2})\tilde{\sigma}_{13}-i\Omega_{a}^{\ast}\tilde{\sigma
}_{03}-i\Omega_{b}\tilde{\sigma}_{32}^{\dagger}+i\tilde{\sigma}_{12}E_{s}%
^{-}e^{i\Delta kz}+i\tilde{\sigma}_{01}^{\dagger}E_{i}^{+}\right]
\right\rangle \nonumber\\
&  +\frac{\partial}{\partial t}\left\langle \tilde{\sigma}_{11}\tilde{\sigma
}_{13}\right\rangle \nonumber\\
&  =\left\langle \gamma_{01}\tilde{\sigma}_{13}\right\rangle ,
\end{align}
where the term $\frac{\partial}{\partial t}\left\langle \tilde{\sigma}%
_{11}\tilde{\sigma}_{13}\right\rangle =$ $\frac{\partial}{\partial
t}\left\langle \tilde{\sigma}_{13}\right\rangle $ is the drift term of the
quantum Langevin equation that can be found from Eq. (\ref{bloch2}),%

\begin{equation}
\frac{\partial}{\partial t}\tilde{\sigma}_{13}=-(i\Delta_{1}+\frac{\gamma
_{01}+\gamma_{03}}{2})\tilde{\sigma}_{13}-i\Omega_{a}^{\ast}\tilde{\sigma
}_{03}-i\Omega_{b}\tilde{\sigma}_{32}^{\dagger}+i\tilde{\sigma}_{12}E_{s}%
^{-}e^{i\Delta kz}+i\tilde{\sigma}_{01}^{\dagger}E_{i}^{+}.
\end{equation}

From Eq. (\ref{Einstein_2}), we have%
\begin{align}
&  \left\langle \mathcal{\bar{D}}_{13,8}\right\rangle =\nonumber\\
&  \left\langle \mathcal{\hat{D}}_{13,8}\right\rangle +\Big\{\left\langle
\left[  -\gamma_{01}\tilde{\sigma}_{11}+\gamma_{12}\tilde{\sigma}_{22}%
+i\Omega_{a}\tilde{\sigma}_{01}^{\dagger}-i\Omega_{a}^{\ast}\tilde{\sigma
}_{01}-i\Omega_{b}\tilde{\sigma}_{12}^{\dagger}+i\Omega_{b}^{\ast}%
\tilde{\sigma}_{12}\right]  \tilde{\sigma}_{13}\right\rangle +\nonumber\\
&  \left\langle \tilde{\sigma}_{11}\left[  -(i\Delta_{1}+\frac{\gamma
_{01}+\gamma_{03}}{2})\tilde{\sigma}_{13}-i\Omega_{a}^{\ast}\tilde{\sigma
}_{03}-i\Omega_{b}\tilde{\sigma}_{32}^{\dagger}+i\tilde{\sigma}_{12}E_{s}%
^{-}e^{i\Delta kz}+i\tilde{\sigma}_{01}^{\dagger}E_{i}^{+}\right]
\right\rangle \nonumber\\
&  -\text{classical counterpart}\Big\}\nonumber\\
&  =\left\langle \gamma_{01}\widetilde{\alpha}_{8}-i\Omega_{a}^{\ast
}\widetilde{\alpha}_{9}+i\widetilde{\alpha}_{19}E_{i}^{+}\right\rangle ,
\end{align}
where classical counterpart represents the last two terms of Eq.
(\ref{Einstein_2}). \ We have used the commutation relations for non-normal
correlation functions that%

\begin{align}
\left[  \tilde{\sigma}_{01},\tilde{\sigma}_{13}\right]   & =\tilde{\sigma
}_{03},\text{ }\nonumber\\
\left[  \tilde{\sigma}_{12},\tilde{\sigma}_{13}\right]   & =0,\nonumber\\
\left[  \tilde{\sigma}_{11},\tilde{\sigma}_{32}^{\dagger}\right]   &
=0,\nonumber\\
\left[  \tilde{\sigma}_{11},\tilde{\sigma}_{01}^{\dagger}\right]   &
=\tilde{\sigma}_{01}^{\dagger},
\end{align}
and use the correspondence $\tilde{\sigma}_{03}\rightarrow\widetilde{\alpha
}_{9}$ and $\tilde{\sigma}_{01}^{\dagger}\rightarrow\widetilde{\alpha}_{19}$
in $\mathcal{\bar{D}}_{13,8}.$ \ The rest of the diffusion coefficients are
confirmed by the method of Einstein relations illustrated above.

\subsection{Ito and Stratonovich stochastic differential equations}

The c-number Langevin equations derived from Fokker-Planck equations have a
direct correspondence to Ito-type stochastic differential equations. \ In
stochastic simulations, it is important to find the expressions of Langevin
noises\ from diffusion coefficients.

For any symmetric diffusion matrix $D(\alpha)$, it can always be factorized into%

\begin{equation}
D(\alpha)=B(\alpha)B^{T}(\alpha)
\end{equation}

where $B$ $\rightarrow$ $BS$ (an orthogonal matrix$S$ that $SS^{T}=I$)
preserves the diffusion matrix so $B$ is not unique. \ The matrix $B$ is in
terms of the Langevin noises where $\xi_{i}dt=dW_{t}^{i}$ (Wiener process) and
$\left\langle \xi_{i}(t)\xi_{j}(t^{\prime})\right\rangle =\delta_{ij}%
\delta(t-t^{\prime})$ and the $\xi_{i}$ below is just a random number in
Gaussian distribution with zero mean and unit variance.

In numerical simulation, we use the semi-implicit algorithm that guarantees
the stability and convergence in the integration of stochastic differential
equations. \ So a transformation from Ito to Stratonovich-type stochastic
differential equation is necessary,%

\begin{align}
dx_{t}^{i}  & =A_{i}(t,\overrightarrow{x_{t}})dt+\sum\limits_{j}%
B_{ij}(t,\overrightarrow{x_{t}})dW_{t}^{j}\text{ \ (Ito)}\\
dx_{t}^{i}  & =[A_{i}(t,\overrightarrow{x_{t}})-\frac{1}{2}\sum\limits_{j}%
\sum\limits_{k}B_{jk}(t,\overrightarrow{x_{t}})\frac{\partial}{\partial x^{j}%
}B_{ik}(t,\overrightarrow{x_{t}})]dt\nonumber\\
& +\sum\limits_{j}B_{ij}(t,\overrightarrow{x_{t}})dW_{t}^{j}\text{
\ (Stratonovich)}%
\end{align}
where a correction in drift term appears due to the transformation.

Here we have the full equations with 19 variables in the positive-P
representation, 64 diffusion matrix elements, and 117 noise terms (random
number generators). \ A correction in drift term is underlined and we have (S
for Stratonovich)%
\begin{align*}
\frac{\partial}{\partial\tau}\widetilde{\alpha}_{5}  & =\underbrace
{(\frac{i\Omega_{a}}{2})}+(i\Delta_{1}-\frac{\gamma_{01}}{2})\widetilde
{\alpha}_{5}+i\Omega_{a}(\widetilde{\alpha}_{0}-\widetilde{\alpha}%
_{13})+i\Omega_{b}^{\ast}\widetilde{\alpha}_{7}-i\widetilde{\alpha}_{16}%
E_{i}^{+}+\mathcal{F}_{5},\text{ \ (S)}\\
\frac{\partial}{\partial\tau}\widetilde{\alpha}_{19}  & =\underbrace
{(\frac{-i\Omega_{a}^{\ast}}{2})}+(-i\Delta_{1}-\frac{\gamma_{01}}%
{2})\widetilde{\alpha}_{19}-i\Omega_{a}^{\ast}(\widetilde{\alpha}%
_{0}-\widetilde{\alpha}_{13})-i\Omega_{b}\widetilde{\alpha}_{17}%
+i\widetilde{\alpha}_{8}E_{i}^{-}+\mathcal{F}_{19},\\
\frac{\partial}{\partial\tau}\widetilde{\alpha}_{6}  & =\underbrace
{(i\Omega_{b})+}i(\Delta_{2}-\Delta_{1}+i\frac{\gamma_{01}+\gamma_{2}}%
{2})\widetilde{\alpha}_{6}-i\Omega_{a}^{\ast}\widetilde{\alpha}_{7}%
+i\Omega_{b}(\widetilde{\alpha}_{13}-\widetilde{\alpha}_{12})\\
& +i\widetilde{\alpha}_{8}E_{s}^{+}e^{-i\Delta kz}+\mathcal{F}_{6},\\
\frac{\partial}{\partial\tau}\widetilde{\alpha}_{18}  & =\underbrace
{(-i\Omega_{b}^{\ast})}-i(\Delta_{2}-\Delta_{1}-i\frac{\gamma_{01}+\gamma_{2}%
}{2})\widetilde{\alpha}_{18}+i\Omega_{a}\widetilde{\alpha}_{17}-i\Omega
_{b}^{\ast}(\widetilde{\alpha}_{13}-\widetilde{\alpha}_{12})\\
& -i\widetilde{\alpha}_{16}E_{s}^{-}e^{i\Delta kz}+\mathcal{F}_{18},\\
\frac{\partial}{\partial\tau}\widetilde{\alpha}_{7}  & =(i\Delta_{2}%
-\frac{\gamma_{2}}{2})\widetilde{\alpha}_{7}-i\Omega_{a}\widetilde{\alpha}%
_{6}+i\Omega_{b}\widetilde{\alpha}_{5}+i\widetilde{\alpha}_{9}E_{s}%
^{+}e^{-i\Delta kz}-i\widetilde{\alpha}_{10}E_{i}^{+}+\mathcal{F}_{7},\\
\frac{\partial}{\partial\tau}\widetilde{\alpha}_{17}  & =(-i\Delta_{2}%
-\frac{\gamma_{2}}{2})\widetilde{\alpha}_{17}+i\Omega_{a}^{\ast}%
\widetilde{\alpha}_{18}-i\Omega_{b}^{\ast}\widetilde{\alpha}_{19}%
-i\widetilde{\alpha}_{15}E_{s}^{-}e^{i\Delta kz}+i\widetilde{\alpha}_{14}%
E_{i}^{-}+\mathcal{F}_{17},\\
\frac{\partial}{\partial\tau}\widetilde{\alpha}_{8}  & =(-i\Delta_{1}%
-\frac{\gamma_{01}+\gamma_{03}}{2})\widetilde{\alpha}_{8}-i\Omega_{a}^{\ast
}\widetilde{\alpha}_{9}-i\Omega_{b}\widetilde{\alpha}_{14}+i\widetilde{\alpha
}_{6}E_{s}^{-}e^{i\Delta kz}+i\widetilde{\alpha}_{19}E_{i}^{+}+\mathcal{F}%
_{8},\\
\frac{\partial}{\partial\tau}\widetilde{\alpha}_{16}  & =(i\Delta_{1}%
-\frac{\gamma_{01}+\gamma_{03}}{2})\widetilde{\alpha}_{16}+i\Omega
_{a}\widetilde{\alpha}_{15}+i\Omega_{b}^{\ast}\widetilde{\alpha}%
_{10}-i\widetilde{\alpha}_{18}E_{s}^{+}e^{-i\Delta kz}-i\widetilde{\alpha}%
_{5}E_{i}^{-}+\mathcal{F}_{16},\\
\frac{\partial}{\partial\tau}\widetilde{\alpha}_{9}  & =\underbrace
{(iE_{i}^{+})}-\frac{\gamma_{03}}{2}\widetilde{\alpha}_{9}-i\Omega
_{a}\widetilde{\alpha}_{8}+i\widetilde{\alpha}_{7}E_{s}^{-}+i(\widetilde
{\alpha}_{0}-\widetilde{\alpha}_{11})E_{i}^{+}+\mathcal{F}_{9},\\
\frac{\partial}{\partial\tau}\widetilde{\alpha}_{15}  & =\underbrace
{(-iE_{i}^{-})}-\frac{\gamma_{03}}{2}\widetilde{\alpha}_{15}+i\Omega_{a}%
^{\ast}\widetilde{\alpha}_{16}-i\widetilde{\alpha}_{17}E_{s}^{+}%
-i(\widetilde{\alpha}_{0}-\widetilde{\alpha}_{11})E_{i}^{-}+\mathcal{F}%
_{15},\\
\frac{\partial}{\partial\tau}\widetilde{\alpha}_{10}  & =\underbrace{(\frac
{i}{2}E_{s}^{+})}+(i\Delta_{2}-\frac{\gamma_{03}+\gamma_{2}}{2})\widetilde
{\alpha}_{10}+i\Omega_{b}\widetilde{\alpha}_{16}-i(\widetilde{\alpha}%
_{12}-\widetilde{\alpha}_{11})E_{s}^{+}e^{-i\Delta kz}\\
& -i\widetilde{\alpha}_{7}E_{i}^{-}+\mathcal{F}_{10},\\
\frac{\partial}{\partial\tau}\widetilde{\alpha}_{14}  & =\underbrace
{(-\frac{i}{2}E_{s}^{-})}+(-i\Delta_{2}-\frac{\gamma_{03}+\gamma_{2}}%
{2})\widetilde{\alpha}_{14}-i\Omega_{b}^{\ast}\widetilde{\alpha}%
_{8}+i(\widetilde{\alpha}_{12}-\widetilde{\alpha}_{11})E_{s}^{-}e^{i\Delta
kz}\\
& +i\widetilde{\alpha}_{17}E_{i}^{+}+\mathcal{F}_{14},\\
\frac{\partial}{\partial\tau}\widetilde{\alpha}_{13}  & =\underbrace
{\frac{-5\gamma_{01}+\gamma_{12}}{4}}-\gamma_{01}\widetilde{\alpha}%
_{13}+\gamma_{12}\widetilde{\alpha}_{12}+i\Omega_{a}\widetilde{\alpha}%
_{19}-i\Omega_{a}^{\ast}\widetilde{\alpha}_{5}-i\Omega_{b}\widetilde{\alpha
}_{18}\\
& +i\Omega_{b}^{\ast}\widetilde{\alpha}_{6}+\mathcal{F}_{13},\\
\frac{\partial}{\partial\tau}\widetilde{\alpha}_{12}  & =\underbrace
{-\frac{\gamma_{2}}{4}}-\gamma_{2}\widetilde{\alpha}_{12}+i\Omega
_{b}\widetilde{\alpha}_{18}-i\Omega_{b}^{\ast}\widetilde{\alpha}%
_{6}+i\widetilde{\alpha}_{14}E_{s}^{+}e^{-i\Delta kz}-i\widetilde{\alpha}%
_{10}E_{s}^{-}e^{i\Delta kz}+\mathcal{F}_{12},\\
\frac{\partial}{\partial\tau}\widetilde{\alpha}_{11}  & =\underbrace
{\frac{-3\gamma_{03}+\gamma_{32}}{4}}-\gamma_{03}\widetilde{\alpha}%
_{11}+\gamma_{32}\widetilde{\alpha}_{12}-i\widetilde{\alpha}_{14}E_{s}%
^{+}e^{-i\Delta kz}+i\widetilde{\alpha}_{10}E_{s}^{-}e^{i\Delta kz}%
+i\widetilde{\alpha}_{15}E_{i}^{+}\\
& -i\widetilde{\alpha}_{9}E_{i}^{-}+\mathcal{F}_{11},
\end{align*}

\begin{align}
-\frac{\partial}{\partial z}E_{s}^{+}  & =i\widetilde{\alpha}_{10}\frac
{|g_{s}|^{2}}{|g_{i}|^{2}}+\mathcal{F}_{3},\text{ }\nonumber\\
-\frac{\partial}{\partial z}E_{s}^{-}  & =-i\widetilde{\alpha}_{14}%
\frac{|g_{s}|^{2}}{|g_{i}|^{2}}+\mathcal{F}_{4},\nonumber\\
\frac{\partial}{\partial z}E_{i}^{+}  & =i\widetilde{\alpha}_{9}%
+\mathcal{F}_{1},\text{ }\nonumber\\
\frac{\partial}{\partial z}E_{i}^{-}  & =-i\widetilde{\alpha}_{15}%
+\mathcal{F}_{2}.
\end{align}

The Langevin noises are formulated as a non-square form \cite{QO:Walls,Smith1}%
\begin{align*}
\mathcal{F}_{1}  & =\mathcal{F}_{2}=0;\text{ }\\
\mathcal{F}_{5}  & =\sqrt{D_{5,5}}\xi_{1}+\sqrt{\frac{D_{5,19}}{2}}(\xi
_{12}+i\xi_{13})+\sqrt{\frac{D_{5,6}}{2}}(\xi_{14}+i\xi_{15})+\sqrt
{\frac{D_{5,7}}{2}}(\xi_{16}+i\xi_{17})\\
& +\sqrt{\frac{D_{5,8}}{2}}(\xi_{18}+i\xi_{19})+\sqrt{\frac{D_{5,9}}{2}}%
(\xi_{20}+i\xi_{21})+\sqrt{\frac{D_{5,14}}{2}}(\xi_{22}+i\xi_{23})\\
& +\sqrt{\frac{D_{5,13}}{2}}(\xi_{24}+i\xi_{25})+\sqrt{\frac{D_{5,11}}{2}}%
(\xi_{26}+i\xi_{27});\text{ }\\
\mathcal{F}_{19}  & =\sqrt{\frac{D_{5,19}}{2}}(\xi_{12}-i\xi_{13}%
)+\sqrt{D_{19,19}}\xi_{2}+\sqrt{\frac{D_{19,18}}{2}}(\xi_{28}+i\xi_{29}%
)+\sqrt{\frac{D_{19,17}}{2}}(\xi_{30}+i\xi_{31})\\
& +\sqrt{\frac{D_{19,16}}{2}}(\xi_{32}+i\xi_{33})+\sqrt{\frac{D_{19,15}}{2}%
}(\xi_{34}+i\xi_{35})+\sqrt{\frac{D_{19,10}}{2}}(\xi_{36}+i\xi_{37})\\
& +\sqrt{\frac{D_{19,13}}{2}}(\xi_{38}+i\xi_{39})+\sqrt{\frac{D_{19,11}}{2}%
}(\xi_{40}+i\xi_{41});
\end{align*}%
\begin{align*}
\mathcal{F}_{6}  & =\sqrt{\frac{D_{5,6}}{2}}(\xi_{14}-i\xi_{15})+\sqrt
{D_{6,6}}\xi_{3}+\sqrt{\frac{D_{6,18}}{2}}(\xi_{42}+i\xi_{43})+\sqrt
{\frac{D_{6,8}}{2}}(\xi_{44}+i\xi_{45})\\
& +\sqrt{\frac{D_{6,16}}{2}}(\xi_{46}+i\xi_{47})+\sqrt{\frac{D_{6,10}}{2}}%
(\xi_{48}+i\xi_{49})+\sqrt{\frac{D_{6,13}}{2}}(\xi_{50}+i\xi_{51});\\
\mathcal{F}_{18}  & =\sqrt{\frac{D_{19,18}}{2}}(\xi_{28}-i\xi_{29}%
)+\sqrt{\frac{D_{6,18}}{2}}(\xi_{42}-i\xi_{43})+\sqrt{D_{18,18}}\xi_{4}%
+\sqrt{\frac{D_{18,8}}{2}}(\xi_{52}+i\xi_{53})\\
& +\sqrt{\frac{D_{18,16}}{2}}(\xi_{54}+i\xi_{55})+\sqrt{\frac{D_{18,14}}{2}%
}(\xi_{56}+i\xi_{57})+\sqrt{\frac{D_{18,13}}{2}}(\xi_{58}+i\xi_{59});\\
\mathcal{F}_{7}  & =\sqrt{\frac{D_{5,7}}{2}}(\xi_{16}-i\xi_{17})+\sqrt
{\frac{D_{7,8}}{2}}(\xi_{60}+i\xi_{61})+\sqrt{\frac{D_{7,9}}{2}}(\xi_{62}%
+i\xi_{63});\\
\mathcal{F}_{17}  & =\sqrt{\frac{D_{19,17}}{2}}(\xi_{30}-i\xi_{31}%
)+\sqrt{\frac{D_{17,16}}{2}}(\xi_{64}+i\xi_{65})+\sqrt{\frac{D_{17,15}}{2}%
}(\xi_{66}+i\xi_{67});
\end{align*}

\begin{align*}
\mathcal{F}_{8}  & =\sqrt{\frac{D_{5,8}}{2}}(\xi_{18}-i\xi_{19})+\sqrt
{\frac{D_{6,8}}{2}}(\xi_{44}-i\xi_{45})+\sqrt{\frac{D_{18,8}}{2}}(\xi
_{52}-i\xi_{53})\\
& +\sqrt{\frac{D_{7,8}}{2}}(\xi_{60}-i\xi_{61})+\sqrt{\frac{D_{8,16}}{2}}%
(\xi_{68}+i\xi_{69})+\sqrt{\frac{D_{8,9}}{2}}(\xi_{70}+i\xi_{71})\\
& +\sqrt{\frac{D_{8,10}}{2}}(\xi_{72}+i\xi_{73})+\sqrt{\frac{D_{8,13}}{2}}%
(\xi_{74}+i\xi_{75})+\sqrt{\frac{D_{8,12}}{2}}(\xi_{76}+i\xi_{77})\\
& +\sqrt{\frac{D_{8,11}}{2}}(\xi_{78}+i\xi_{79})+\sqrt{\frac{D_{8,3}}{2}}%
(\xi_{80}+i\xi_{81});\\
\mathcal{F}_{16}  & =\sqrt{\frac{D_{19,16}}{2}}(\xi_{32}-i\xi_{33}%
)+\sqrt{\frac{D_{6,16}}{2}}(\xi_{46}-i\xi_{47})+\sqrt{\frac{D_{18,16}}{2}}%
(\xi_{54}-i\xi_{55})\\
& +\sqrt{\frac{D_{17,16}}{2}}(\xi_{64}-i\xi_{65})+\sqrt{\frac{D_{8,16}}{2}%
}(\xi_{68}-i\xi_{69})+\sqrt{\frac{D_{16,15}}{2}}(\xi_{82}+i\xi_{83})\\
& +\sqrt{\frac{D_{16,14}}{2}}(\xi_{84}+i\xi_{85})+\sqrt{\frac{D_{16,13}}{2}%
}(\xi_{86}+i\xi_{87})+\sqrt{\frac{D_{16,12}}{2}}(\xi_{88}+i\xi_{89})\\
& +\sqrt{\frac{D_{16,11}}{2}}(\xi_{90}+i\xi_{91})+\sqrt{\frac{D_{16,4}}{2}%
}(\xi_{92}+i\xi_{93});\\
\mathcal{F}_{9}  & =\sqrt{\frac{D_{5,9}}{2}}(\xi_{20}-i\xi_{21})+\sqrt
{\frac{D_{7,9}}{2}}(\xi_{62}-i\xi_{63})+\sqrt{\frac{D_{8,9}}{2}}(\xi_{70}%
-i\xi_{71})\\
& +\sqrt{D_{9,9}}\xi_{5}+\sqrt{\frac{D_{9,15}}{2}}(\xi_{94}+i\xi_{95}%
)+\sqrt{\frac{D_{9,10}}{2}}(\xi_{96}+i\xi_{97})+\sqrt{\frac{D_{9,3}}{2}}%
(\xi_{98}+i\xi_{99});\text{ }\\
\mathcal{F}_{15}  & =\sqrt{\frac{D_{19,15}}{2}}(\xi_{34}-i\xi_{35}%
)+\sqrt{\frac{D_{17,15}}{2}}(\xi_{66}-i\xi_{67})+\sqrt{\frac{D_{16,15}}{2}%
}(\xi_{82}-i\xi_{83})\\
& +\sqrt{\frac{D_{9,15}}{2}}(\xi_{94}-i\xi_{95})+\sqrt{D_{15,15}}\xi_{6}%
+\sqrt{\frac{D_{15,14}}{2}}(\xi_{100}+i\xi_{101})\\
& +\sqrt{\frac{D_{15,4}}{2}}(\xi_{102}+i\xi_{103});
\end{align*}

\begin{align}
\mathcal{F}_{10}  & =\sqrt{\frac{D_{19,10}}{2}}(\xi_{36}-i\xi_{37}%
)+\sqrt{\frac{D_{6,10}}{2}}(\xi_{48}-i\xi_{49})+\sqrt{\frac{D_{8,10}}{2}}%
(\xi_{72}-i\xi_{73})\nonumber\\
& +\sqrt{\frac{D_{9,10}}{2}}(\xi_{96}-i\xi_{97})+\sqrt{D_{10,10}}\xi_{7}%
+\sqrt{\frac{D_{10,14}}{2}}(\xi_{104}+i\xi_{105})\nonumber\\
& +\sqrt{\frac{D_{10,13}}{2}}(\xi_{106}+i\xi_{107})+\sqrt{\frac{D_{10,11}}{2}%
}(\xi_{108}+i\xi_{109});\nonumber\\
\text{ }\mathcal{F}_{14}  & =\sqrt{\frac{D_{5,14}}{2}}(\xi_{22}-i\xi
_{23})+\sqrt{\frac{D_{18,14}}{2}}(\xi_{56}-i\xi_{57})+\sqrt{\frac{D_{16,14}%
}{2}}(\xi_{84}-i\xi_{85})\nonumber\\
& +\sqrt{\frac{D_{15,14}}{2}}(\xi_{100}-i\xi_{101})+\sqrt{\frac{D_{10,14}}{2}%
}(\xi_{104}-i\xi_{105})+\sqrt{D_{14,14}}\xi_{8}\nonumber\\
& +\sqrt{\frac{D_{14,13}}{2}}(\xi_{110}+i\xi_{111})+\sqrt{\frac{D_{14,11}}{2}%
}(\xi_{112}+i\xi_{113});\nonumber\\
\mathcal{F}_{13}  & =\sqrt{\frac{D_{5,13}}{2}}(\xi_{24}-i\xi_{25})+\sqrt
{\frac{D_{19,13}}{2}}(\xi_{38}-i\xi_{39})+\sqrt{\frac{D_{6,13}}{2}}(\xi
_{50}-i\xi_{51})\nonumber\\
& +\sqrt{\frac{D_{18,13}}{2}}(\xi_{58}-i\xi_{59})+\sqrt{\frac{D_{8,13}}{2}%
}(\xi_{74}-i\xi_{75})+\sqrt{\frac{D_{16,13}}{2}}(\xi_{86}-i\xi_{87}%
)\nonumber\\
& +\sqrt{\frac{D_{10,13}}{2}}(\xi_{106}-i\xi_{107})+\sqrt{\frac{D_{14,13}}{2}%
}(\xi_{110}-i\xi_{111})+\sqrt{D_{13,13}}\xi_{9}\nonumber\\
& +\sqrt{\frac{D_{12,13}}{2}}(\xi_{114}+i\xi_{115});\nonumber\\
\mathcal{F}_{12}  & =\sqrt{\frac{D_{8,12}}{2}}(\xi_{76}-i\xi_{77})+\sqrt
{\frac{D_{16,12}}{2}}(\xi_{88}-i\xi_{89})+\sqrt{\frac{D_{12,13}}{2}}(\xi
_{114}-i\xi_{115})\nonumber\\
& +\sqrt{D_{12,12}}\xi_{10}+\sqrt{\frac{D_{11,12}}{2}}(\xi_{116}+i\xi
_{117});\text{ }\nonumber\\
\mathcal{F}_{11}  & =\sqrt{\frac{D_{5,11}}{2}}(\xi_{26}-i\xi_{27})+\sqrt
{\frac{D_{19,11}}{2}}(\xi_{40}-i\xi_{41})+\sqrt{\frac{D_{8,11}}{2}}(\xi
_{78}-i\xi_{79})\nonumber\\
& +\sqrt{\frac{D_{16,11}}{2}}(\xi_{90}-i\xi_{91})+\sqrt{\frac{D_{10,11}}{2}%
}(\xi_{108}-i\xi_{109})+\sqrt{\frac{D_{14,11}}{2}}(\xi_{112}-i\xi
_{113})\nonumber\\
& +\sqrt{\frac{D_{11,12}}{2}}(\xi_{116}-i\xi_{117})+\sqrt{D_{11,11}}\xi
_{11};\nonumber\\
\mathcal{F}_{3}  & =\sqrt{\frac{D_{8,3}}{2}}(\xi_{80}-i\xi_{81})+\sqrt
{\frac{D_{9,3}}{2}}(\xi_{98}-i\xi_{99});\text{ }\nonumber\\
\mathcal{F}_{4}  & =\sqrt{\frac{D_{16,4}}{2}}(\xi_{92}-i\xi_{93})+\sqrt
{\frac{D_{15,4}}{2}}(\xi_{102}-i\xi_{103}).
\end{align}

In numerical simulations, we have a factor $\frac{1}{\sqrt{N_{c}\Delta t\Delta
z}}$ for Langevin noises $\mathcal{F}$ and $\frac{1}{N_{c}\Delta t\Delta z}$
for correction terms.



\chapter{Multimode description of correlated two-photon state}

In this Appendix, we introduce a general model for quantum detection
efficiency for multimode analysis in various quantum communication scheme.
\ Based on this detection model with the spectral description of correlated
two-photon state, we derive the effective density matrix conditioning on the
detection events of entanglement swapping, polarization maximally entangled
(PME) state projection, and quantum teleportation.

\section{Quantum Efficiency of Detector}

\begin{figure}
[ptb]
\begin{center}
\includegraphics[
natheight=7.499600in,
natwidth=9.999800in,
height=2.9447in,
width=3.9167in
]%
{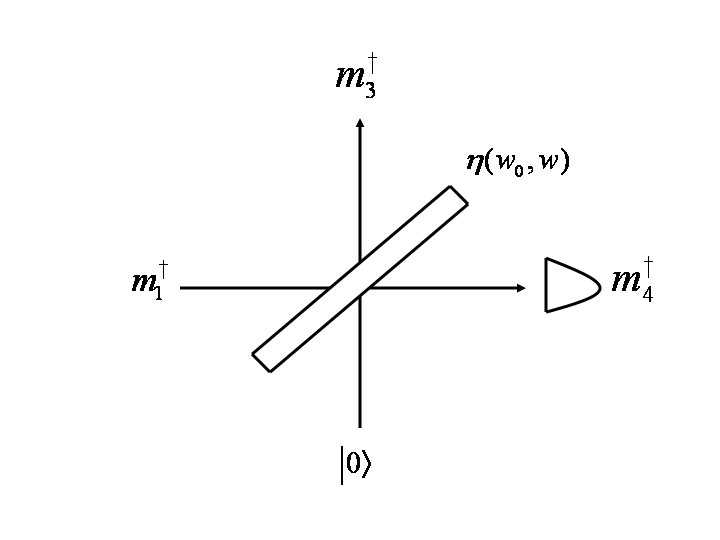}%
\caption{Model of quantum efficiency of detector.}%
\label{detect}%
\end{center}
\end{figure}

To account for quantum efficiency of detector and the affect of its own
spectrum filtering, we introduce an extra beam splitter (B.S.) with a
transmissivity $\eta(\omega,\omega_{0})$ \cite{det} before the detection
event. $\ \eta$ models the quantum efficiency of the detectors in the
microscopic level (response at frequency $\omega_{0}$) and the macroscopic
level (time-integrated detection). \ One example of conditioning on the single
click of the detector, the output density operator becomes%

\begin{align}
\hat{\rho}_{out}  & =\int_{-\infty}^{\infty}d\omega_{0}\hat{\Pi}_{1}%
\text{Tr}_{ref}\big[\hat{U}_{BS}\hat{\rho}_{in}\hat{U}_{BS}^{\dag}%
\big]\hat{\Pi}_{1}\label{model}\\
\hat{\Pi}_{1}  & \equiv\int_{-\infty}^{\infty}d\omega|\omega\rangle
\langle\omega|\\
\hat{U}_{BS}  & \equiv\left(
\begin{array}
[c]{cc}%
\sqrt{1-\eta} & \sqrt{\eta}\\
\sqrt{\eta} & -\sqrt{1-\eta}%
\end{array}
\right)
\end{align}
where $\text{Tr}_{ref}$ is the trace over the reflected modes $m_{3}^{\dag}, $
and the flat spectrum projection operator $\hat{\Pi}_{1}$ (only photon number
is projected and no frequency resolution) is considered in the measurement
process \cite{spectral}. \ In Figure \ref{detect}, $m_{1}^{\dag} $ is the
incoming photon operator before the detection, $m_{3}^{\dag}$ is the reflected
mode, and $m_{4}^{\dag}$ is now the detection mode with a modelling of
spectral quantum efficiency and an effective quantum efficiency is defined as%

\begin{equation}
\int_{-\infty}^{\infty}\eta(\omega,\omega_{0})d\omega_{0}=\eta_{eff}(\omega).
\end{equation}
%

\section{Multimode Description of Entanglement Swapping}

From Eq. (\ref{mode}), we use single mode $\Phi(\omega)$ for Raman photon and
a multimode description $f(\omega_{s},\omega_{i})$ for cascade photons and
rewrite the effective state. \ Note that a symmetric setup is considered so
the mode description is the same for both sides A and B in the scheme of
entanglement swapping.%

\begin{align}
& |\Psi\rangle_{eff}=\eta_{1}(1-\eta_{2})\times\nonumber\\
& \int f(\omega_{s},\omega_{i})\hat{a}_{s}^{\dag,A}(\omega_{s})\hat{a}%
_{i}^{\dag,A}(\omega_{i})d\omega_{s}d\omega_{i}\int f(\omega_{s}^{\prime
},\omega_{i}^{\prime})\hat{a}_{s}^{\dag,B}(\omega_{s}^{\prime})\hat{a}%
_{i}^{\dag,B}(\omega_{i}^{\prime})d\omega_{s}^{\prime}d\omega_{i}^{\prime
}|0\rangle+\nonumber\\
& \eta_{2}(1-\eta_{1})\int\Phi(\omega)d\omega\hat{a}_{r}^{\dag,A}(\omega
)\hat{S}_{A}^{\dag}\int\Phi(\omega^{\prime})d\omega^{\prime}\hat{a}_{r}%
^{\dag,B}(\omega^{\prime})\hat{S}_{B}^{\dag}|0\rangle+\sqrt{\eta_{1}%
(1-\eta_{1})}\times\nonumber\\
& \sqrt{\eta_{2}(1-\eta_{2})}\int f(\omega_{s},\omega_{i})d\omega_{s}%
d\omega_{i}\times\hat{a}_{s}^{\dag,A}(\omega_{s})\hat{a}_{i}^{\dag,A}%
(\omega_{i})\int\Phi(\omega^{\prime})d\omega^{\prime}\hat{a}_{r}^{\dag
,B}(\omega^{\prime})\hat{S}_{B}^{\dag}|0\rangle+\nonumber\\
& \sqrt{\eta_{1}\eta_{2}(1-\eta_{1})(1-\eta_{2})}\int\Phi(\omega)d\omega
\hat{a}_{r}^{\dag,A}(\omega)\hat{S}_{A}^{\dag}\int f(\omega_{s}^{\prime
},\omega_{i}^{\prime})\hat{a}_{s}^{\dag,B}(\omega_{s}^{\prime})\hat{a}%
_{i}^{\dag,B}(\omega_{i}^{\prime})d\omega_{s}^{\prime}d\omega_{i}^{\prime
}|0\rangle.\nonumber\\
&
\end{align}

With the B.S., we have $\hat{a}_{i}^{\dag,A}=\frac{\hat{m}_{1}^{\dag}+\hat
{m}_{2}^{\dag}}{\sqrt{2}}$, $\hat{a}_{i}^{\dag,B}=\frac{\hat{n}_{1}^{\dag
}+\hat{n}_{2}^{\dag}}{\sqrt{2}}$, $\hat{a}_{r}^{\dag,A}=\frac{\hat{m}%
_{1}^{\dag}-\hat{m}_{2}^{\dag}}{\sqrt{2}}$, $\hat{a}_{r}^{\dag,B}=\frac
{\hat{n}_{1}^{\dag}-\hat{n}_{2}^{\dag}}{\sqrt{2}}$, where $\hat{a}%
_{i}^{\dagger}$ is the creation operator for idler photon and $\hat{a}%
_{r}^{\dagger}$ is for Raman photon. \ The input density operator is
$\hat{\rho}_{in}=|\Psi\rangle_{eff}\langle\Psi|$ and conditioning on the pair
of single click ($\hat{m}_{1,2}^{\dag},\hat{n}_{1,2}^{\dag}$), we are able to
generate maximally entangled singlet or triplet state $|\Psi\rangle
_{DLCZ}=\frac{S_{A}^{\dag}\pm S_{B}^{\dag}}{\sqrt{2}}|0\rangle_{A,B}$. Without
loss of generality, we consider a triplet state along with a pair of clicks
($\hat{m}_{1}^{\dag},\hat{n}_{1}^{\dag}$) and use the model of quantum
efficiency in Eq. (\ref{model}) with tracing over the detection modes
($\hat{m}_{4}^{\dag},\hat{n}_{4}^{\dag}$). \ Note that $\hat{m}_{1}^{\dag
}=\sqrt{1-\eta}\hat{m}_{3}^{\dag}+\sqrt{\eta}\hat{m}_{4}^{\dag}$ and $\hat
{n}_{1}^{\dag}=\sqrt{1-\eta}\hat{n}_{3}^{\dag}+\sqrt{\eta}\hat{n}_{4}^{\dag}$
as we model the quantum efficiency in the previous Section.%

\begin{align}
\hat{\rho}_{out}  & =\int_{-\infty}^{\infty}d\omega_{0}\text{Tr}%
_{m4,n4}\big\{\text{Tr}_{m3,n3}\big[\hat{U}_{BS}^{B}\hat{U}_{BS}^{A}\hat{\rho
}_{in}\hat{U}_{BS}^{\dag,A}\hat{U}_{BS}^{\dag,B}\big]\hat{M}_{4,4}\big\}\\
\hat{M}_{4,4}  & \equiv(\hat{I}_{m4}^{\dag}-|0\rangle_{m4}\langle
0|)\otimes|0\rangle_{m2}\langle0|\otimes(\hat{I}_{n4}^{\dag}-|0\rangle
_{n4}\langle0|)\otimes|0\rangle_{n2}\langle0|
\end{align}
where the unitary B.S. operator is denoted by both sides (A and B) and NRPD
projection operators are used \cite{shapiro}. \ These operators project the
state with single click of the detected mode without resolving the number of
photons. $\ \hat{I}$ is identity operator. \ The un-normalized output density
operator after tracing out these modes becomes%

\begin{align}
& \hat{\rho}_{out}=\frac{\eta_{1}^{2}(1-\eta_{2})^{2}}{4}\times\nonumber\\
& \int d\omega_{i}d\omega_{i}^{\prime}\eta_{eff}(\omega_{i})\eta_{eff}%
(\omega_{i}^{\prime})\big[\int f(\omega_{s},\omega_{i})\hat{a}_{s}^{\dag
,A}(\omega_{s})d\omega_{s}\int f(\omega_{s}^{\prime},\omega_{i}^{\prime}%
)\hat{a}_{s}^{\dag,B}(\omega_{s}^{\prime})d\omega_{s}^{\prime}\big]\nonumber\\
& |0\rangle\langle0|\big[\int f^{\ast}(\omega_{s}^{\prime\prime},\omega
_{i})\hat{a}_{s}^{A}(\omega_{s}^{\prime\prime})d\omega_{s}^{\prime\prime}\int
f^{\ast}(\omega_{s}^{\prime\prime\prime},\omega_{i}^{\prime})\hat{a}_{s}%
^{B}(\omega_{s}^{\prime\prime\prime})d\omega_{s}^{\prime\prime\prime
}\big]\nonumber\\
& +\frac{\eta_{1}\eta_{2}(1-\eta_{1})(1-\eta_{2})}{4}\bigg\{\int d\omega
_{i}\eta_{eff}(\omega_{i})\Big[\int f(\omega_{s},\omega_{i})dw_{s}\int
f^{\ast}(\omega_{s}^{\prime},\omega_{i})d\omega_{s}^{\prime}\nonumber\\
& \int|\Phi(\omega)|^{2}\eta_{eff}(\omega)d\omega\Big]\Big(\hat{a}_{s}%
^{\dag,A}(\omega_{s})\hat{S}_{B}^{\dag}|0\rangle\langle0|\hat{S}_{B}\hat
{a}_{s}^{A}(\omega_{s}^{\prime})+\nonumber\\
& \hat{a}_{s}^{\dag,B}(\omega_{s})\hat{S}_{A}^{\dag}|0\rangle\langle0|\hat
{S}_{A}\hat{a}_{s}^{B}(\omega_{s}^{\prime})\Big)+\int\int f(\omega_{s}%
,\omega_{i})d\omega_{s}\Phi^{\ast}(\omega_{i})\eta_{eff}(\omega_{i}%
)d\omega_{i}\times\nonumber\\
& \int\int f^{\ast}(\omega_{s}^{\prime},\omega_{i}^{\prime})d\omega
_{s}^{\prime}\Phi(\omega_{i}^{\prime})\eta_{eff}(\omega_{i}^{\prime}%
)d\omega_{i}^{\prime}\Big(\hat{a}_{s}^{\dag,A}(\omega_{s})\hat{S}_{B}^{\dag
}|0\rangle\langle0|\hat{S}_{A}\hat{a}_{s}^{B}(\omega_{s}^{\prime})+\nonumber\\
& \hat{a}_{s}^{\dag,B}(\omega_{s})\hat{S}_{A}^{\dag}|0\rangle\langle0|\hat
{S}_{B}\hat{a}_{s}^{A}(\omega_{s}^{\prime})\Big)\bigg\}+\hat{\rho}%
_{out}^{\prime}%
\end{align}
where $\eta_{eff}(\omega)$ is introduced after integration of $\omega_{0}, $
and we denote it as an effective quantum efficiency for idler field
$\omega_{i}$ or Raman photon at frequency $\omega$ (wavelength $780$ nm for D2
line of Rb atom). $\ \hat{\rho}_{out}^{\prime}$ includes the terms that won't
survive after the interference of telecom photons in the middle B.S.
(conditioning on a single click of detector). \ They involve operators like
$\hat{a}_{s}^{\dag,A}\hat{a}_{s}^{\dag,B}|0\rangle\langle0|\hat{a}_{s}^{A}%
\hat{S}_{B}$, $\hat{a}_{s}^{\dag,A}\hat{a}_{s}^{\dag,B}|0\rangle\langle
0|\hat{S}_{A}\hat{S}_{B}$ and $\hat{S}^{\dag,A}\hat{S}^{\dag,B}|0\rangle
\langle0|\hat{S}_{A}\hat{S}_{B}$.

The normalization factor is derived by tracing over the atomic degree of
freedom.
\begin{align}
& \text{Tr}(\hat{\rho}_{out})\equiv\mathcal{N}=\nonumber\\
& \frac{\eta_{1}^{2}(1-\eta_{2})^{2}}{4}\int d\omega_{s}d\omega_{i}\eta
_{eff}(\omega_{i})|f(\omega_{s},\omega_{i})|^{2}\int d\omega_{s}^{\prime
}d\omega_{i}^{\prime}\eta_{eff}(\omega_{i}^{\prime})|f(\omega_{s}^{\prime
},\omega_{i}^{\prime})|^{2}+\nonumber\\
& \frac{\eta_{1}\eta_{2}(1-\eta_{1})(1-\eta_{2})}{2}\int d\omega_{s}%
d\omega_{i}\eta_{eff}(\omega_{i})|f(\omega_{s},\omega_{i})|^{2}\int|\Phi
|^{2}(\omega)\eta_{eff}(\omega)d\omega+\nonumber\\
& \frac{\eta_{2}^{2}(1-\eta_{1})^{2}}{4}\int|\Phi|^{2}(\omega)\eta
_{eff}(\omega)d\omega\int|\Phi|^{2}(\omega^{\prime})\eta_{eff}(\omega^{\prime
})d\omega^{\prime}\label{normalization}%
\end{align}
which will be put back when we calculate the heralding and success probabilities.

Next we interfere telecom photons with B.S. that $\hat{a}_{s}^{\dag,A}%
=\frac{\hat{c}_{1}^{\dag}+\hat{c}_{2}^{\dag}}{\sqrt{2}}$, $\hat{a}_{s}%
^{\dag,B}=\frac{\hat{c}_{1}^{\dag}-\hat{c}_{2}^{\dag}}{\sqrt{2}},$ and again a
quantum efficiency $\eta(\omega,\omega_{0})$ for telecom photon is introduced.
\ Use $\hat{c}_{1}^{\dag}=\sqrt{1-\eta}\hat{c}_{3}^{\dag}+\sqrt{\eta}\hat
{c}_{4}^{\dag}$ and trace over the reflected mode $\hat{c}_{3}^{\dag}$
conditioning on the click of $\hat{c}_{4}^{\dag}$ from NRPD. \ The effective
density matrix becomes
\begin{align}
\hat{\rho}_{out}^{(2)}  & =\int_{-\infty}^{\infty}d\omega_{0}\text{Tr}%
_{c4}\big\{\text{Tr}_{c3}\big[\hat{U}_{BS}^{C}\hat{\rho}_{in}\hat{U}%
_{BS}^{\dag,C}\big]\hat{M}_{4}\big\}\nonumber\\
& \equiv\int_{-\infty}^{\infty}d\omega_{0}\hat{\rho}_{out}^{(2)}(\omega
_{0}),\\
\hat{\rho}_{out}^{(2)}(\omega_{0})  & \equiv\text{Tr}_{c4}\big\{\hat{\rho
}_{in}^{(2)}(\omega_{0})\big\}\\
\hat{M}_{4}  & \equiv(\hat{I}_{c4}^{\dag}-|0\rangle_{c4}\langle0|)\otimes
|0\rangle_{c2}\langle0|,
\end{align}

\begin{align}
& \hat{\rho}_{in}^{(2)}(\omega_{0})=\frac{\eta_{1}^{2}(1-\eta_{2})^{2}}%
{16}\int d\omega_{i}d\omega_{i}^{\prime}\eta_{eff}(\omega_{i})\eta
_{eff}(\omega_{i}^{\prime})\bigg\{\nonumber\\
& \int d\omega_{s}(1-\eta(\omega_{s}))f(s,i)f^{\ast}(s,i^{\prime})\int
d\omega_{s}^{\prime}f(s^{\prime},i^{\prime})\sqrt{\eta(\omega_{s}^{\prime}%
)}\hat{c}_{4}^{\dag}(\omega_{s}^{\prime})|0\rangle\langle0|\times\nonumber\\
& \int d\omega_{s}^{\prime\prime}\hat{c}_{4}(\omega_{s}^{\prime\prime}%
)\sqrt{\eta(\omega_{s}^{\prime\prime})}f^{\ast}(s^{\prime\prime},i)+\int
d\omega_{s}(1-\eta(\omega_{s}))f(s,i)f^{\ast}(s,i)\times\nonumber\\
& \int d\omega_{s}^{\prime}f(s^{\prime},i^{\prime})\sqrt{\eta(\omega
_{s}^{\prime})}\hat{c}_{4}^{\dag}(\omega_{s}^{\prime})|0\rangle\langle0|\int
d\omega_{s}^{\prime\prime\prime}\hat{c}_{4}(\omega_{s}^{\prime\prime\prime
})\sqrt{\eta(\omega_{s}^{\prime\prime\prime})}f^{\ast}(s^{\prime\prime\prime
},i^{\prime})+\nonumber\\
& \int d\omega_{s}^{\prime}(1-\eta(\omega_{s}^{\prime}))f(s^{\prime}%
,i^{\prime})f^{\ast}(s^{\prime},i^{\prime})\int d\omega_{s}f(s,i)\sqrt
{\eta(\omega_{s})}\hat{c}_{4}^{\dag}(\omega_{s})|0\rangle\langle
0|\times\nonumber\\
& \int d\omega_{s}^{\prime\prime}\hat{c}_{4}(\omega_{s}^{\prime\prime}%
)\sqrt{\eta(\omega_{s}^{\prime\prime})}f^{\ast}(s^{\prime\prime},i)+\int
d\omega_{s}^{\prime}(1-\eta(\omega_{s}^{\prime}))f(s^{\prime},i^{\prime
})f^{\ast}(s^{\prime},i)\times\nonumber\\
& \int d\omega_{s}f(s,i)\sqrt{\eta(\omega_{s})}\hat{c}_{4}^{\dag}(\omega
_{s})|0\rangle\langle0|\int d\omega_{s}^{\prime\prime\prime}\hat{c}_{4}%
(\omega_{s}^{\prime\prime\prime})\sqrt{\eta(\omega_{s}^{\prime\prime\prime}%
)}f^{\ast}(s^{\prime\prime\prime},i^{\prime})+\nonumber\\
& \int d\omega_{s}^{\prime}\sqrt{\eta(\omega_{s}^{\prime})}f(s^{\prime
},i^{\prime})\int d\omega_{s}\sqrt{\eta(\omega_{s})}f(s,i)\hat{c}_{4}^{\dag
}(\omega_{s})\hat{c}_{4}^{\dag}(\omega_{s}^{\prime})|0\rangle\langle
0|\times\nonumber\\
& \int d\omega_{s}^{\prime\prime}\sqrt{\eta(\omega_{s}^{\prime\prime})}%
f^{\ast}(s^{\prime\prime},i)\int d\omega_{s}^{\prime\prime\prime}\sqrt
{\eta(\omega_{s}^{\prime\prime\prime})}f^{\ast}(s^{\prime\prime\prime
},i^{\prime})\hat{c}_{4}(\omega_{s}^{\prime\prime})\hat{c}_{4}(\omega
_{s}^{\prime\prime\prime})\bigg\}+\nonumber\\
& \frac{\eta_{1}\eta_{2}(1-\eta_{1})(1-\eta_{2})}{8}\bigg\{\int d\omega
_{i}\eta_{eff}(\omega_{i})\int f(s,i)d\omega_{s}\int f^{\ast}(s^{\prime
},i)d\omega_{s}^{\prime}\times\nonumber\\
& \int d\omega|\Phi(\omega)|^{2}\eta_{eff}(\omega)\sqrt{\eta(\omega_{s})}%
\hat{c}_{4}^{\dag}(\omega_{s})\Big(\hat{S}_{B}^{\dag}|0\rangle\langle0|\hat
{S}_{B}+\hat{S}_{A}^{\dag}|0\rangle\langle0|\hat{S}_{A}\Big)\times\nonumber\\
& \hat{c}_{4}(\omega_{s}^{\prime})\sqrt{\eta(\omega_{s}^{\prime})}\int\int
f(s,i)d\omega_{s}\Phi^{\ast}(\omega_{i})\eta_{eff}(\omega_{i})d\omega
_{i}\times\nonumber\\
& \int\int f^{\ast}(s^{\prime},i^{\prime})d\omega_{s}^{\prime}\Phi(\omega
_{i}^{\prime})\eta_{eff}(\omega_{i}^{\prime})d\omega_{i}^{\prime}\sqrt
{\eta(\omega_{s})}\hat{c}_{4}^{\dag}(\omega_{s})\times\nonumber\\
& \Big(\hat{S}_{B}^{\dag}|0\rangle\langle0|\hat{S}_{A}+\hat{S}_{A}^{\dag
}|0\rangle\langle0|\hat{S}_{B}\Big)\hat{c}_{4}(\omega_{s}^{\prime})\sqrt
{\eta(\omega_{s}^{\prime})}\bigg\}
\end{align}
where a brief notation for spectrum $f(s,i)\equiv f(\omega_{s},\omega_{i})$
and quantum efficiency $\eta(\omega)\equiv\eta(\omega,\omega_{0})$. \ This
quantum efficiency refers to the telecom photon. \ We proceed to trace over
the detected modes and the density matrix can be simplified by interchange of
variables in integration.%

\begin{align}
& \hat{\rho}_{out}^{(2)}(\omega_{0})=\frac{\eta_{1}^{2}(1-\eta_{2})^{2}}%
{8}\int d\omega_{i}d\omega_{i}^{\prime}\eta_{eff}(\omega_{i})\eta_{eff}%
(\omega_{i}^{\prime})\bigg\{\nonumber\\
& \int d\omega_{s}(1-\eta(\omega_{s},\omega_{0}))f(\omega_{s},\omega
_{i})f^{\ast}(\omega_{s},\omega_{i}^{\prime})\int d\omega_{s}^{\prime}%
f(\omega_{s}^{\prime},\omega_{i}^{\prime})f^{\ast}(\omega_{s}^{\prime}%
,\omega_{i})\eta(\omega_{s}^{\prime},\omega_{0})+\nonumber\\
& \int d\omega_{s}(1-\eta(\omega_{s},\omega_{0}))|f(\omega_{s},\omega
_{i})|^{2}\int d\omega_{s}^{\prime}|f(\omega_{s}^{\prime},\omega_{i}^{\prime
})|^{2}\eta(\omega_{s}^{\prime},\omega_{0})+\nonumber\\
& \frac{1}{2}\int d\omega_{s}^{\prime}\eta(\omega_{s}^{\prime},\omega
_{0})|f(\omega_{s}^{\prime},\omega_{i}^{\prime})|^{2}\int d\omega_{s}%
\eta(\omega_{s},\omega_{0})|f(\omega_{s},\omega_{i})|^{2}+\frac{1}{2}%
\times\nonumber\\
& \int d\omega_{s}^{\prime}\eta(\omega_{s}^{\prime},\omega_{0})f(\omega
_{s}^{\prime},\omega_{i}^{\prime})f^{\ast}(\omega_{s}^{\prime},\omega_{i})\int
d\omega_{s}\eta(\omega_{s},\omega_{0})f(\omega_{s},\omega_{i})f^{\ast}%
(\omega_{s},\omega_{i}^{\prime})\bigg\}|0\rangle\langle0|\nonumber\\
& +\frac{\eta_{1}\eta_{2}(1-\eta_{1})(1-\eta_{2})}{8}\bigg\{\int d\omega
_{i}\eta_{eff}(\omega_{i})\int\eta(\omega_{s},\omega_{0})|f(\omega_{s}%
,\omega_{i})|^{2}d\omega_{s}\times\nonumber\\
& \int d\omega|\Phi(\omega)|^{2}\eta_{eff}(\omega)\Big(\hat{S}_{B}^{\dag
}|0\rangle\langle0|\hat{S}_{B}+\hat{S}_{A}^{\dag}|0\rangle\langle0|\hat{S}%
_{A}\Big)+\nonumber\\
& \int\int\eta(\omega_{s},\omega_{0})f(\omega_{s},\omega_{i})d\omega_{s}%
\Phi^{\ast}(\omega_{i})\eta_{eff}(\omega_{i})d\omega_{i}\int f^{\ast}%
(\omega_{s},\omega_{i}^{\prime})\Phi(\omega_{i}^{\prime})\eta_{eff}(\omega
_{i}^{\prime})d\omega_{i}^{\prime}\times\nonumber\\
& \Big(\hat{S}_{B}^{\dag}|0\rangle\langle0|\hat{S}_{A}+\hat{S}_{A}^{\dag
}|0\rangle\langle0|\hat{S}_{B}\Big)\bigg\}
\end{align}
where the trace over two photon states requires the commutation relation of
photon operators.%

\begin{align}
& \text{Tr}[\hat{m}_{4}^{\dag}(\omega_{s})\hat{m}_{4}^{\dag}(\omega
_{s}^{\prime})|0\rangle\langle0|\hat{m}_{4}(\omega_{s}^{\prime\prime})\hat
{m}_{4}(\omega_{s}^{\prime\prime\prime})]\nonumber\\
& =\langle0|\hat{m}_{4}(\omega_{s}^{\prime\prime})[\delta(\omega_{s}%
,\omega_{s}^{\prime\prime\prime})+\hat{m}_{4}^{\dag}(\omega_{s})\hat{m}%
_{4}(\omega_{s}^{\prime\prime\prime})]\hat{m}_{4}^{\dag}(\omega_{s}^{\prime
})|0\rangle\nonumber\\
& =\delta(\omega_{s},\omega_{s}^{\prime\prime\prime})\delta(\omega_{s}%
^{\prime\prime},\omega_{s}^{\prime})+\delta(\omega_{s},\omega_{s}%
^{\prime\prime})\delta(\omega_{s}^{\prime},\omega_{s}^{\prime\prime\prime}).
\end{align}

The above is the general formulation for the un-normalized density matrix
conditioning on three clicks of NRPD's. \ We've included spectral quantum
efficiency of the detector either for near-infrared ($\eta_{eff}$) or telecom
wavelength ($\eta_{t}\equiv\int_{-\infty}^{\infty}\eta(\omega,\omega
_{0})d\omega_{0}$)

To proceed, we assume a flat and finite spectrum response ($\eta_{eff}%
(\omega)=\eta_{eff}$, $\eta_{t}(\omega)=\eta_{t}$) with the range $\omega
_{0}\in\lbrack\Omega-\Delta,\Omega+\Delta]$ centered at $\Omega$
(near-infrared or telecom) and $\omega\in\lbrack\omega_{0}-\delta,\omega
_{0}+\delta]$. \ The widths $2\Delta$ and $2\delta$ are large enough compared
to our source bandwidth so these detection events do not give us any
information of spectrum for our source. \ A perfect efficiency also means no
photon loss during detection. \ Note that the integral involves multiplication
of two telecom photon efficiency $\int_{-\infty}^{\infty}\eta(\omega
,\omega_{0})\eta(\omega^{\prime},\omega_{0})d\omega_{0}=\eta_{t}^{2}(\omega)$
that is valid if the source bandwidth is smaller than detector's.

After the integration of $\omega_{0}$, we have%

\begin{align}
& \hat{\rho}_{out}^{(2)}=\frac{\eta_{1}^{2}(1-\eta_{2})^{2}}{8}\eta_{eff}%
^{2}\int d\omega_{i}d\omega_{i}^{\prime}\bigg\{(1-\eta_{t})\eta_{t}\int
d\omega_{s}f(\omega_{s},\omega_{i})f^{\ast}(\omega_{s},\omega_{i}^{\prime
})\times\nonumber\\
& \int d\omega_{s}^{\prime}f(\omega_{s}^{\prime},\omega_{i}^{\prime})f^{\ast
}(\omega_{s}^{\prime},\omega_{i})+(1-\eta_{t})\eta_{t}\int d\omega
_{s}|f(\omega_{s},\omega_{i})|^{2}\int d\omega_{s}^{\prime}|f(\omega
_{s}^{\prime},\omega_{i}^{\prime})|^{2}+\nonumber\\
& \frac{\eta_{t}^{2}}{2}\int d\omega_{s}^{\prime}|f(\omega_{s}^{\prime}%
,\omega_{i}^{\prime})|^{2}\int d\omega_{s}|f(\omega_{s},\omega_{i})|^{2}%
+\frac{\eta_{t}^{2}}{2}\int d\omega_{s}^{\prime}f(\omega_{s}^{\prime}%
,\omega_{i}^{\prime})f^{\ast}(\omega_{s}^{\prime},\omega_{i})\times\nonumber\\
& \int d\omega_{s}f(\omega_{s},\omega_{i})f^{\ast}(\omega_{s},\omega
_{i}^{\prime})\bigg\}|0\rangle\langle0|+\frac{\eta_{1}\eta_{2}(1-\eta
_{1})(1-\eta_{2})}{8}\eta_{t}\eta_{eff}^{2}\times\nonumber\\
& \bigg\{\int d\omega_{i}\int|f(\omega_{s},\omega_{i})|^{2}d\omega_{s}\int
d\omega|\Phi(\omega)|^{2}\Big(\hat{S}_{B}^{\dag}|0\rangle\langle0|\hat{S}%
_{B}+\hat{S}_{A}^{\dag}|0\rangle\langle0|\hat{S}_{A}\Big)+\nonumber\\
& \int\int f(\omega_{s},\omega_{i})d\omega_{s}\Phi^{\ast}(\omega_{i}%
)d\omega_{i}\int f^{\ast}(\omega_{s},\omega_{i}^{\prime})\Phi(\omega
_{i}^{\prime})d\omega_{i}^{\prime}\nonumber\\
& \Big(\hat{S}_{B}^{\dag}|0\rangle\langle0|\hat{S}_{A}+\hat{S}_{A}^{\dag
}|0\rangle\langle0|\hat{S}_{B}\Big)\bigg\}.\label{out}%
\end{align}

\section{Density Matrix of PME Projection and Quantum Teleportation}

In Chapter 5.4, we have the normalized density operator $\hat{\rho}%
_{out,n}^{(2),AB}$\ of the DLCZ entangled state through entanglement swapping.
\ With another pair of DLCZ entangled state, $\hat{\rho}_{out,n}^{(2),CD}$,
the joint density operator for these two pairs constructs the polarization
maximally entangled state (PME) projection and is interpreted as%

\begin{align}
& \hat{\rho}_{out,n}^{(2),AB}\otimes\hat{\rho}_{out,n}^{(2),CD}=\nonumber\\
& \frac{1}{(a+b)^{2}}\bigg\{a^{2}|0\rangle\langle0|+\frac{ab}{2}%
\Big[|0\rangle_{AB}\langle0|\Big(\hat{S}_{C}^{\dag}|0\rangle\langle0|\hat
{S}_{C}+\hat{S}_{D}^{\dag}|0\rangle\langle0|\hat{S}_{D}\nonumber\\
& +\lambda_{1}\hat{S}_{C}^{\dag}|0\rangle\langle0|\hat{S}_{D}+\lambda_{1}%
\hat{S}_{D}^{\dag}|0\rangle\langle0|\hat{S}_{C}\Big)+|0\rangle_{CD}%
\langle0|\Big(\hat{S}_{B}^{\dag}|0\rangle\langle0|\hat{S}_{B}+\hat{S}%
_{A}^{\dag}|0\rangle\langle0|\hat{S}_{A}\nonumber\\
& +\lambda_{1}\hat{S}_{B}^{\dag}|0\rangle\langle0|\hat{S}_{A}+\lambda_{1}%
\hat{S}_{A}^{\dag}|0\rangle\langle0|\hat{S}_{B}\Big)\Big]+\frac{b^{2}}%
{4}\Big(\hat{S}_{C}^{\dag}|0\rangle\langle0|\hat{S}_{C}+\hat{S}_{D}^{\dag
}|0\rangle\langle0|\hat{S}_{D}\nonumber\\
& +\lambda_{1}\hat{S}_{C}^{\dag}|0\rangle\langle0|\hat{S}_{D}+\lambda_{1}%
\hat{S}_{D}^{\dag}|0\rangle\langle0|\hat{S}_{C}\Big)\otimes\Big(\hat{S}%
_{B}^{\dag}|0\rangle\langle0|\hat{S}_{B}+\hat{S}_{A}^{\dag}|0\rangle
\langle0|\hat{S}_{A}\nonumber\\
& +\lambda_{1}\hat{S}_{B}^{\dag}|0\rangle\langle0|\hat{S}_{A}+\lambda_{1}%
\hat{S}_{A}^{\dag}|0\rangle\langle0|\hat{S}_{B}\Big)\bigg\},
\end{align}
which is used to calculate the success probability after post measurement [a
click from each side, the side of (A or C) and (B or D)]. \ $a=\eta_{r}%
(2-\eta)\Big(1+\sum_{j}\lambda_{j}^{2}\Big),b=4$, and $\eta_{r}=\eta_{1}%
/\eta_{2}$, $\eta=\eta_{t}$, $\lambda_{j}$ is Schmidt number that is used to
decompose the two-photon source from the cascade transition. \ 

In DLCZ protocol, quantum teleportation uses the similar setup in PME
projection and combines with the desired teleported state, $|\Phi
\rangle=(d_{0}\hat{S}_{I_{1}}^{\dag}+d_{1}\hat{S}_{I_{2}}^{\dag})|0\rangle$,
which is represented by two other atomic ensembles $I_{1}$ and $I_{2}$.
$\ $The requirement of normalization of the state is $d_{0}|^{2}+|d_{1}%
|^{2}=1$, and the density operator of quantum teleportation is $\hat{\rho
}_{QT}=|\Phi\rangle\langle\Phi|\otimes\hat{\rho}_{out,n}^{(2),AB}\otimes
\hat{\rho}_{out,n}^{(2),CD}$. \ Conditioning on clicks of $\hat{D}_{I_{1}}$
and $\hat{D}_{I_{2}}$, the effective density matrix for quantum teleportation
is (using $\hat{S}_{I_{1}}^{\dag}=(\hat{D}_{I_{1}}+\hat{D}_{A})/\sqrt{2},$
$\hat{S}_{I_{2}}^{\dag}=(\hat{D}_{I_{2}}+\hat{D}_{C})/\sqrt{2}$ for the effect
of beam splitter)%

\begin{align}
& \hat{\rho}_{QT,eff}=\Big[\frac{|d_{0}|^{2}}{2}(\hat{D}_{I_{1}}^{\dag
}|0\rangle\langle0|\hat{D}_{I_{1}})+\frac{|d_{1}|^{2}}{2}(\hat{D}_{I_{2}%
}^{\dag}|0\rangle\langle0|\hat{D}_{I_{2}})+\frac{d_{0}d_{1}^{\ast}}{2}(\hat
{D}_{I_{1}}^{\dag}|0\rangle\langle0|\hat{D}_{I_{2}})\nonumber\\
& +\frac{d_{0}^{\ast}d_{1}}{2}(\hat{D}_{I_{2}}^{\dag}|0\rangle\langle0|\hat
{D}_{I_{1}})\Big]\otimes\frac{1}{(a+b)^{2}}\bigg\{a^{2}|0\rangle
\langle0|+\frac{ab}{2}\Big[|0\rangle_{AB}\langle0|\nonumber\\
& \Big(\frac{\hat{D}_{I_{2}}^{\dag}|0\rangle\langle0|\hat{D}_{I_{2}}}{2}%
+\hat{S}_{D}^{\dag}|0\rangle\langle0|\hat{S}_{D}+\lambda_{1}\frac{\hat
{D}_{I_{2}}^{\dag}}{\sqrt{2}}|0\rangle\langle0|\hat{S}_{D}+\lambda_{1}\hat
{S}_{D}^{\dag}|0\rangle\langle0|\frac{\hat{D}_{I_{2}}}{\sqrt{2}}%
\Big)\nonumber\\
& +|0\rangle_{CD}\langle0|\Big(\hat{S}_{B}^{\dag}|0\rangle\langle0|\hat{S}%
_{B}+\frac{\hat{D}_{I_{1}}^{\dag}|0\rangle\langle0|\hat{D}_{I_{1}}}{2}%
+\lambda_{1}\hat{S}_{B}^{\dag}|0\rangle\langle0|\frac{\hat{D}_{I_{1}}}%
{\sqrt{2}}+\lambda_{1}\frac{\hat{D}_{I_{2}}^{\dag}}{\sqrt{2}}|0\rangle
\langle0|\hat{S}_{B}\Big)\Big]\nonumber\\
& +\frac{b^{2}}{4}\Big(\frac{\hat{D}_{I_{2}}^{\dag}|0\rangle\langle0|\hat
{D}_{I_{2}}}{2}+\hat{S}_{D}^{\dag}|0\rangle\langle0|\hat{S}_{D}+\lambda
_{1}\frac{\hat{D}_{I_{2}}^{\dag}}{\sqrt{2}}|0\rangle\langle0|\hat{S}%
_{D}+\lambda_{1}\hat{S}_{D}^{\dag}|0\rangle\langle0|\frac{\hat{D}_{I_{2}}%
}{\sqrt{2}}\Big)\otimes\nonumber\\
& \Big(\hat{S}_{B}^{\dag}|0\rangle\langle0|\hat{S}_{B}+\frac{\hat{D}_{I_{1}%
}^{\dag}|0\rangle\langle0|\hat{D}_{I_{1}}}{2}+\lambda_{1}\hat{S}_{B}^{\dag
}|0\rangle\langle0|\frac{\hat{D}_{I_{1}}}{\sqrt{2}}+\lambda_{1}\frac{\hat
{D}_{I_{2}}^{\dag}}{\sqrt{2}}|0\rangle\langle0|\hat{S}_{B}%
\Big)\bigg\},\label{QT}%
\end{align}
which is used to calculate the success probability for teleported state.



\chapter{Hamiltonian and equation of motion for frequency conversion in a
diamond type atomic ensemble}

In this appendix, we derive the Hamiltonian and the Maxwell-Bloch equation for
frequency conversion in ladder-type transition. \ The steady state solutions
for atoms are solved, and the solution to the field equations are discussed in
Chapter 6. \ Similar to the derivation in Appendix B where the cascade
emissions are investigated, the conversion scheme here also involves four-wave
mixing with two classical driving lasers and two quantum fields, signal and
idler. \ The driving lasers are applied in a way that signal or idler is
converted only when an idler or signal is put into interaction with the atoms
(see Figure \ref{conversion}). \ We will use the same quantization procedure
for electromagnetic fields as discussed in Appendix B.2.1.

\section{Hamiltonian and Maxwell-Bloch Equation}

To derive the coupled Maxwell-Bloch equations it is convenient to employ a
quantized description of the electromagnetic field \cite{quantization} and use
Heisenberg-Langevin equation methods, and then invoke a standard semiclassical
factorization assumption. The propagation length $L$ is discretized into
$2M+1$ elements. \ The positive frequency component of the electric field
operator is given by $\hat{E}^{+}(z)=\sum_{n=-M}^{M}\sqrt{\frac{\hbar
\omega_{s,n}}{2\epsilon_{0}V}}e^{i(k_{s}+k_{n})z}\hat{c}_{n}$ where $[\hat
{c}_{n},\hat{c}_{n^{\prime}}^{\dag}]=\delta_{nn^{\prime}}$, $k_{n}=\frac{2\pi
n}{L}~,$ $\omega_{s,n}=\omega_{s}+k_{n}c~,$ $n=-M,...,M$ and $\omega_{s}%
=k_{s}c~$is the central frequency. \ Define the local boson operators $\hat
{a}_{l}=\frac{1}{\sqrt{2M+1}}\sum_{n=-M}^{M}\hat{c}_{n}e^{ik_{n}z_{l}}$ where
$[\hat{a}_{l},\hat{a}_{l^{\prime}}^{\dag}]=\delta_{ll^{\prime}}$. Similar
definitions hold for the signal, s, and idler field, $i$, which carry an
additional index in the following.\ 

The Hamiltonian for the interacting system, $\hat{H}_{I}=-\vec{d}\cdot\vec{E}%
$, depicted in Figure \ref{conversion} is given by, (we ignore the
interactions responsible for atomic spontaneous emission for the moment)%

\begin{equation}
\hat{H}=\hat{H}_{0}+\hat{H}_{I},
\end{equation}
where%

\begin{align}
\hat{H}_{0}  & =\sum_{i=1}^{3}\sum_{l=-M}^{M}\hbar\omega_{i}\hat{\sigma}%
_{ii}^{l}+\hbar\omega_{s}\sum_{l=-M}^{M}\hat{a}_{s,l}^{\dag}\hat{a}%
_{s,l}+\hbar\sum_{l,l^{\prime}}\omega_{ll^{\prime}}\hat{a}_{s,l}^{\dag}\hat
{a}_{s,l^{\prime}}\nonumber\\
& +\hbar\omega_{i}\sum_{l=-M}^{M}\hat{a}_{i,l}^{\dag}\hat{a}_{i,l}+\hbar
\sum_{l,l^{\prime}}\omega_{ll^{\prime}}\hat{a}_{i,l}^{\dag}\hat{a}%
_{i,l^{\prime}},
\end{align}
and%

\begin{align}
\hat{H}_{I} &  =-\hbar\sum_{l=-M}^{M}\Big\{\Omega_{a}(t)\hat{\sigma}%
_{01}^{l\dagger}e^{ik_{a}z_{l}-i\omega_{a}t}+\Omega_{b}(t)\hat{\sigma}%
_{32}^{l\dagger}e^{-ik_{b}z_{l}-i\omega_{b}t}\nonumber\\
&  +g_{s}\sqrt{2M+1}\hat{\sigma}_{12}^{l\dagger}\hat{a}_{s,l}e^{-ik_{s}z_{l}%
}+g_{i}\sqrt{2M+1}\hat{\sigma}_{03}^{l\dagger}\hat{a}_{i,l}e^{ik_{i}z_{l}%
}+h.c.\Big\}
\end{align}
where $\hat{\sigma}_{mn}^{l}\equiv\sum_{\mu}^{N_{z}}\hat{\sigma}_{mn}^{\mu
,l}=\sum_{\mu}^{N_{z}}|m\rangle_{\mu}\langle n|\Big|_{r_{\mu}=z_{l}},~$the
Rabi frequencies $\ \ \Omega_{a,(b)}(t)=f_{a,(b)}(t)d_{10,(23)}\mathcal{E}%
(k_{a,(b)})/(2\hbar)$ is half the standard definition, and $f_{a,(b)}$ is a
slowly varying temporal profile without spatial dependence (ensemble scale
much less than pulse length). \ The dipole matrix element $d_{mn}\equiv$
$\langle m|\hat{d}|n\rangle$, coupling strength $g_{s,(i)}\equiv
d_{21,(30)}\mathcal{E}(k_{s,(i)})/\hbar,~\mathcal{E}(k)=\sqrt{\hbar
\omega/2\epsilon_{0}V},$ and $z_{p}=\frac{pL}{2M+1},$ $p=-M,...,M$. \ The
matrix $\omega_{ll^{\prime}}\equiv\sum_{n=-M}^{M}k_{n}e^{ik_{n}(z_{l}%
-z_{l^{\prime}})}/(2M+1)$ accounts for field propagation by coupling the local
mode operators.

The dynamical equations including dissipation due to spontaneous emission may
be treated by standard Langevin-Heisenberg equation methods \cite{QO:Scully},
and we define $\gamma_{ij}$ as the natural transition rate from $|j\rangle
\rightarrow|i\rangle.$ \ Since we are interested in a semiclassical
description, we replace the field operators by c-numbers in the Langevin
equations, and drop the zero-mean Langevin noise sources. All atomic spin
operators are also replaced by their expectation values. \ Finally, in the
co-moving frame coordinates $z$ and $\tau=t-z/c$ the atomic equations are%

\begin{align}
\frac{\partial}{\partial\tau}\tilde{\sigma}_{01}  & =(i\Delta_{1}-\frac
{\gamma_{01}}{2})\tilde{\sigma}_{01}+i\Omega_{a}(\tilde{\sigma}_{00}%
-\tilde{\sigma}_{11})+ig_{s}^{\ast}\tilde{\sigma}_{02}E_{s}^{-}-ig_{i}%
\tilde{\sigma}_{13}^{\dag}E_{i}^{+},\nonumber\\
\frac{\partial}{\partial\tau}\tilde{\sigma}_{12}  & =(i\Delta\omega_{s}%
-\frac{\gamma_{01}+\gamma_{2}}{2})\tilde{\sigma}_{12}-i\Omega_{a}^{\ast}%
\tilde{\sigma}_{02}+ig_{s}(\tilde{\sigma}_{11}-\tilde{\sigma}_{22})E_{s}%
^{+}+iP^{\ast}\Omega_{b}\tilde{\sigma}_{13},\nonumber\\
\frac{\partial}{\partial\tau}\tilde{\sigma}_{02}  & =(i\Delta_{2}-\frac
{\gamma_{2}}{2})\tilde{\sigma}_{02}-i\tilde{\sigma}_{12}\Omega_{a}%
+ig_{s}\tilde{\sigma}_{01}E_{s}^{+}+iP^{\ast}\tilde{\sigma}_{03}\Omega
_{b}-iP^{\ast}g_{i}\tilde{\sigma}_{32}E_{i}^{+},\nonumber\\
\frac{\partial}{\partial\tau}\tilde{\sigma}_{11}  & =-\gamma_{01}\tilde
{\sigma}_{11}+\gamma_{12}\tilde{\sigma}_{22}+i\Omega_{a}\tilde{\sigma}%
_{01}^{\dag}-i\Omega_{a}^{\ast}\tilde{\sigma}_{01}-ig_{s}\tilde{\sigma}%
_{12}^{\dag}E_{s}^{+}+ig_{s}^{\ast}\tilde{\sigma}_{12}E_{s}^{-},\nonumber\\
\frac{\partial}{\partial\tau}\tilde{\sigma}_{22}  & =-\gamma_{2}\tilde{\sigma
}_{22}+ig_{s}\tilde{\sigma}_{12}^{\dag}E_{s}^{+}-ig_{s}^{\ast}\tilde{\sigma
}_{12}E_{s}^{-}+i\Omega_{b}\tilde{\sigma}_{32}^{\dag}-i\Omega_{b}^{\ast}%
\tilde{\sigma}_{32},\nonumber\\
\frac{\partial}{\partial\tau}\tilde{\sigma}_{33}  & =-\gamma_{03}\tilde
{\sigma}_{33}+\gamma_{32}\tilde{\sigma}_{22}-i\Omega_{b}\tilde{\sigma}%
_{32}^{\dag}+i\Omega_{b}^{\ast}\tilde{\sigma}_{32}+ig_{i}\tilde{\sigma}%
_{03}^{\dag}E_{i}^{+}-ig_{i}^{\ast}\tilde{\sigma}_{03}E_{i}^{-},\nonumber\\
\frac{\partial}{\partial\tau}\tilde{\sigma}_{13}  & =(i\Delta\omega
_{i}-i\Delta_{1}-\frac{\gamma_{01}+\gamma_{03}}{2})\tilde{\sigma}_{13}%
-i\Omega_{a}^{\ast}\tilde{\sigma}_{03}-iPg_{s}\tilde{\sigma}_{32}^{\dag}%
E_{s}^{+}+iP\Omega_{b}^{\ast}\tilde{\sigma}_{12}\nonumber\\
& +ig_{i}\tilde{\sigma}_{01}^{\dag}E_{i}^{+},\nonumber\\
\frac{\partial}{\partial\tau}\tilde{\sigma}_{03}  & =(i\Delta\omega_{i}%
-\frac{\gamma_{03}}{2})\tilde{\sigma}_{03}-i\Omega_{a}\tilde{\sigma}%
_{13}+iP\Omega_{b}^{\ast}\tilde{\sigma}_{02}+ig_{i}(\tilde{\sigma}_{00}%
-\tilde{\sigma}_{33})E_{i}^{+},\nonumber\\
\frac{\partial}{\partial\tau}\tilde{\sigma}_{32}^{\dag}  & =(-i\Delta
_{b}-\frac{\gamma_{03}+\gamma_{2}}{2})\tilde{\sigma}_{32}^{\dag}-iP^{\ast
}g_{s}^{\ast}\tilde{\sigma}_{13}E_{s}^{-}+i\Omega_{b}^{\ast}(\tilde{\sigma
}_{22}-\tilde{\sigma}_{33})+iP^{\ast}g_{i}\tilde{\sigma}_{02}^{\dag}E_{i}%
^{+}\nonumber\\
& \label{bloch_conv}%
\end{align}
where\ $\gamma_{2}=\gamma_{12}+\gamma_{32},$ $P\equiv e^{i\Delta
kz-i\Delta\omega t},$ the four-wave mixing mismatch wavevector $\Delta
k=k_{a}-k_{s}+k_{b}-k_{i},$ the frequency mismatch $\Delta\omega=\omega
_{a}+\omega_{s}-\omega_{b}-\omega_{i}=\Delta_{1}-\Delta_{b}+\Delta\omega
_{s}-\Delta\omega_{i},$\ and various detunings are defined as $\Delta
\omega_{i}=\omega_{i}-\omega_{3}$, $\Delta\omega_{s}=\omega_{s}-\omega_{12},$
$\Delta_{1}=\omega_{a}-\omega_{1},$ $\Delta_{2}=\omega_{a}+\omega_{s}%
-\omega_{2}=\Delta_{1}+\Delta\omega_{s}$ , $\Delta_{b}=\omega_{b}-\omega
_{23}.$ The slow-varying atomic operators are defined%

\begin{align}
\tilde{\sigma}_{01}  & \equiv\frac{1}{N_{z}}\sigma_{01}^{l}e^{-ik_{a}%
z_{l}+i\omega_{a}t},\text{ }\tilde{\sigma}_{12}\equiv\frac{1}{N_{z}}%
\sigma_{12}^{l}e^{ik_{s}z_{l}+i\omega_{s}t},\text{ }\tilde{\sigma}_{02}%
\equiv\frac{1}{N_{z}}\sigma_{02}^{l}e^{-ik_{a}z_{l}+ik_{s}z_{l}+i\omega
_{s}t+i\omega_{a}t},\text{ }\nonumber\\
\tilde{\sigma}_{13}  & \equiv\frac{1}{N_{z}}\sigma_{13}^{l}e^{-i\omega
_{a}t+i\omega_{i}t+ik_{a}z_{l}-ik_{i}z_{l}},\tilde{\sigma}_{03}\equiv\frac
{1}{N_{z}}\sigma_{03}^{l}e^{-ik_{i}z_{l}+i\omega_{i}t},\tilde{\sigma}%
_{32}^{\dag}\equiv\frac{1}{N_{z}}\sigma_{32}^{l\dag}e^{-i\omega_{b}%
t}e^{-ik_{b}z_{l}},\nonumber\\
\tilde{\sigma}_{22}  & \equiv\frac{1}{N_{z}}\tilde{\sigma}_{22}^{l}%
,\tilde{\sigma}_{33}\equiv\frac{1}{N_{z}}\tilde{\sigma}_{33}^{l},\tilde
{\sigma}_{11}\equiv\frac{1}{N_{z}}\tilde{\sigma}_{11}^{l}%
\end{align}
where $N_{z}(2M+1)=N$.

The field equations are%
\begin{align}
\frac{\partial}{\partial z}E_{s}^{+}  & =\frac{iNg_{s}^{\ast}}{c}\tilde
{\sigma}_{12},\label{maxwell1}\\
\frac{\partial}{\partial z}E_{i}^{+}  & =\frac{iNg_{i}^{\ast}}{c}\tilde
{\sigma}_{03}\label{maxwell2}%
\end{align}
where the field operators are defined as%

\begin{equation}
E_{s}^{-}(z,t)\equiv\sqrt{2M+1}\hat{a}_{s,l}^{\dag}e^{-i\omega_{s}t},\text{
}E_{i}^{+}(z,t)\equiv\sqrt{2M+1}\hat{a}_{i,l}e^{i\omega_{i}t}.
\end{equation}

\ Langevin noises are not concerned here for we are interested in the
normally-ordered quantity, frequency conversion efficiency, of input field and
additional quantum noise corrections vanish as the $|2\rangle\rightarrow
|3\rangle$ transition driven by pump laser b has vanishing populations and
atomic coherence. \ For energy and momentum conservation ($P=1$), and in
the\ weak field limit, we solve atomic operators in steady state after
linearizing with respect to the probe fields%

\begin{align}
T_{01}\tilde{\sigma}_{01}  & =i\Omega_{a}(1-2\tilde{\sigma}_{11}-\tilde
{\sigma}_{22}-\tilde{\sigma}_{33}),\nonumber\\
T_{32}^{\ast}\tilde{\sigma}_{32}^{\dag}  & =i\Omega_{b}^{\ast}(\tilde{\sigma
}_{22}-\tilde{\sigma}_{33}),\nonumber\\
T_{02}\tilde{\sigma}_{02}  & =-i\Omega_{a}\tilde{\sigma}_{12}+ig_{s}%
\tilde{\sigma}_{01}E_{s}^{+}+i\Omega_{b}\tilde{\sigma}_{03}-ig_{i}%
\tilde{\sigma}_{32}E_{i}^{+},\nonumber\\
T_{13}\tilde{\sigma}_{13}  & =-i\Omega_{a}^{\ast}\tilde{\sigma}_{03}%
-ig_{s}\tilde{\sigma}_{32}^{\dag}E_{s}^{+}+i\Omega_{b}^{\ast}\tilde{\sigma
}_{12}+ig_{i}\tilde{\sigma}_{01}^{\dag}E_{i}^{+},\nonumber\\
T_{12}\tilde{\sigma}_{12}  & =-i\Omega_{a}^{\ast}\tilde{\sigma}_{02}%
+ig_{s}(\tilde{\sigma}_{11}-\tilde{\sigma}_{22})E_{s}^{+}+i\tilde{\sigma}%
_{13}\Omega_{b},\nonumber\\
T_{03}\tilde{\sigma}_{03}  & =-i\Omega_{a}\tilde{\sigma}_{13}+i\tilde{\sigma
}_{02}\Omega_{b}^{\ast}+ig_{i}(\tilde{\sigma}_{00}-\tilde{\sigma}_{33}%
)E_{i}^{+}\label{steady}%
\end{align}
where $T_{01}=\frac{\gamma_{01}}{2}-i\Delta_{1},$ $T_{32}^{\ast}=\frac
{\gamma_{03}+\gamma_{2}}{2}+i\Delta_{b},$ $T_{02}=\frac{\gamma_{2}}{2}%
-i\Delta_{2},$ $T_{13}=\frac{\gamma_{01}+\gamma_{03}}{2}+i\Delta_{1}%
-i\Delta\omega_{i},$ $T_{12}=\frac{\gamma_{01}+\gamma_{2}}{2}-i\Delta
\omega_{s},$ $T_{03}=\frac{\gamma_{03}}{2}-i\Delta\omega_{i},$ and note that
$\tilde{\sigma}_{02},$ $\tilde{\sigma}_{13},$ $\tilde{\sigma}_{12},$
$\tilde{\sigma}_{03}$ are expressed in first order of fields and
$\tilde{\sigma}_{01},$ $\tilde{\sigma}_{32}^{\dag}$ in zeroth order. \ For
population operators, we solve them in the zeroth order of fields and the
nonzero steady states of population and coherence operator are (s denotes
steady state solution)%

\begin{equation}
\tilde{\sigma}_{11,s}=\frac{|\Omega_{a}|^{2}}{\Delta_{1}^{2}+\frac{\gamma
_{01}^{2}}{4}+2|\Omega_{a}|^{2}},\text{ }\tilde{\sigma}_{00,s}=1-\tilde
{\sigma}_{11,s},\text{ }\tilde{\sigma}_{01,s}=\frac{i\Omega_{a}}{\frac
{\gamma_{01}}{2}-i\Delta_{1}}(1-2\tilde{\sigma}_{11,s}).
\end{equation}

Substitute the above back into Eq. (\ref{steady}) and solve for $\tilde
{\sigma}_{12}$ and $\tilde{\sigma}_{03}.$ The parametric coupling equations
for the signal and idler fields become%

\begin{align}
\frac{\partial}{\partial z}E_{s}^{+}  & =\beta_{s}E_{s}^{+}+\kappa_{s}%
E_{i}^{+}\nonumber\\
\frac{\partial}{\partial z}E_{i}^{+}  & =\kappa_{i}E_{s}^{+}+\alpha_{i}%
E_{i}^{+}\label{field}%
\end{align}
where%

\begin{align}
\beta_{s}  & =\frac{-N|g_{s}|^{2}}{cD}[\tilde{\sigma}_{11,s}(T_{03}%
+\frac{|\Omega_{a}|^{2}}{T_{13}}+\frac{|\Omega_{b}|^{2}}{T_{02}}%
)-\frac{i\Omega_{a}^{\ast}\tilde{\sigma}_{01,s}}{T_{02}}(T_{03}+\frac
{|\Omega_{a}|^{2}-|\Omega_{b}|^{2}}{T_{13}})],\nonumber\\
& \\
\kappa_{s}  & =\frac{-Ng_{i}g_{s}^{\ast}}{cD}[\tilde{\sigma}_{00,s}%
(\frac{\Omega_{a}^{\ast}\Omega_{b}}{T_{02}}+\frac{\Omega_{a}^{\ast}\Omega_{b}%
}{T_{13}})+\frac{i\Omega_{b}\tilde{\sigma}_{01,s}^{\dag}}{T_{13}}(T_{03}%
+\frac{|\Omega_{b}|^{2}-|\Omega_{a}|^{2}}{T_{02}})],\\
\kappa_{i}  & =\frac{-Ng_{s}g_{i}^{\ast}}{cD}[\tilde{\sigma}_{11,s}%
(\frac{\Omega_{a}\Omega_{b}^{\ast}}{T_{02}}+\frac{\Omega_{a}\Omega_{b}^{\ast}%
}{T_{13}})+\frac{i\Omega_{b}^{\ast}\tilde{\sigma}_{01,s}}{T_{02}}(T_{12}%
+\frac{|\Omega_{b}|^{2}-|\Omega_{a}|^{2}}{T_{13}})],\\
\alpha_{i}  & =\frac{-N|g_{i}|^{2}}{cD}[\tilde{\sigma}_{00,s}(T_{12}%
+\frac{|\Omega_{a}|^{2}}{T_{02}}+\frac{|\Omega_{b}|^{2}}{T_{13}}%
)-\frac{i\Omega_{a}\tilde{\sigma}_{01,s}^{\dag}}{T_{13}}(T_{12}+\frac
{|\Omega_{a}|^{2}-|\Omega_{b}|^{2}}{T_{02}})],\nonumber\\
& \\
D  & \equiv T_{12}T_{03}+T_{12}(\frac{|\Omega_{a}|^{2}}{T_{13}}+\frac
{|\Omega_{b}|^{2}}{T_{02}})+T_{03}(\frac{|\Omega_{a}|^{2}}{T_{02}}%
+\frac{|\Omega_{b}|^{2}}{T_{13}})+\frac{(|\Omega_{a}|^{2}-|\Omega_{b}%
|^{2})^{2}}{T_{02}T_{13}}.\nonumber\\
&
\end{align}
%

\begin{figure}
[ptb]
\begin{center}
\includegraphics[
natheight=17.333600in,
natwidth=24.239400in,
height=3.2941in,
width=6.02in
]%
{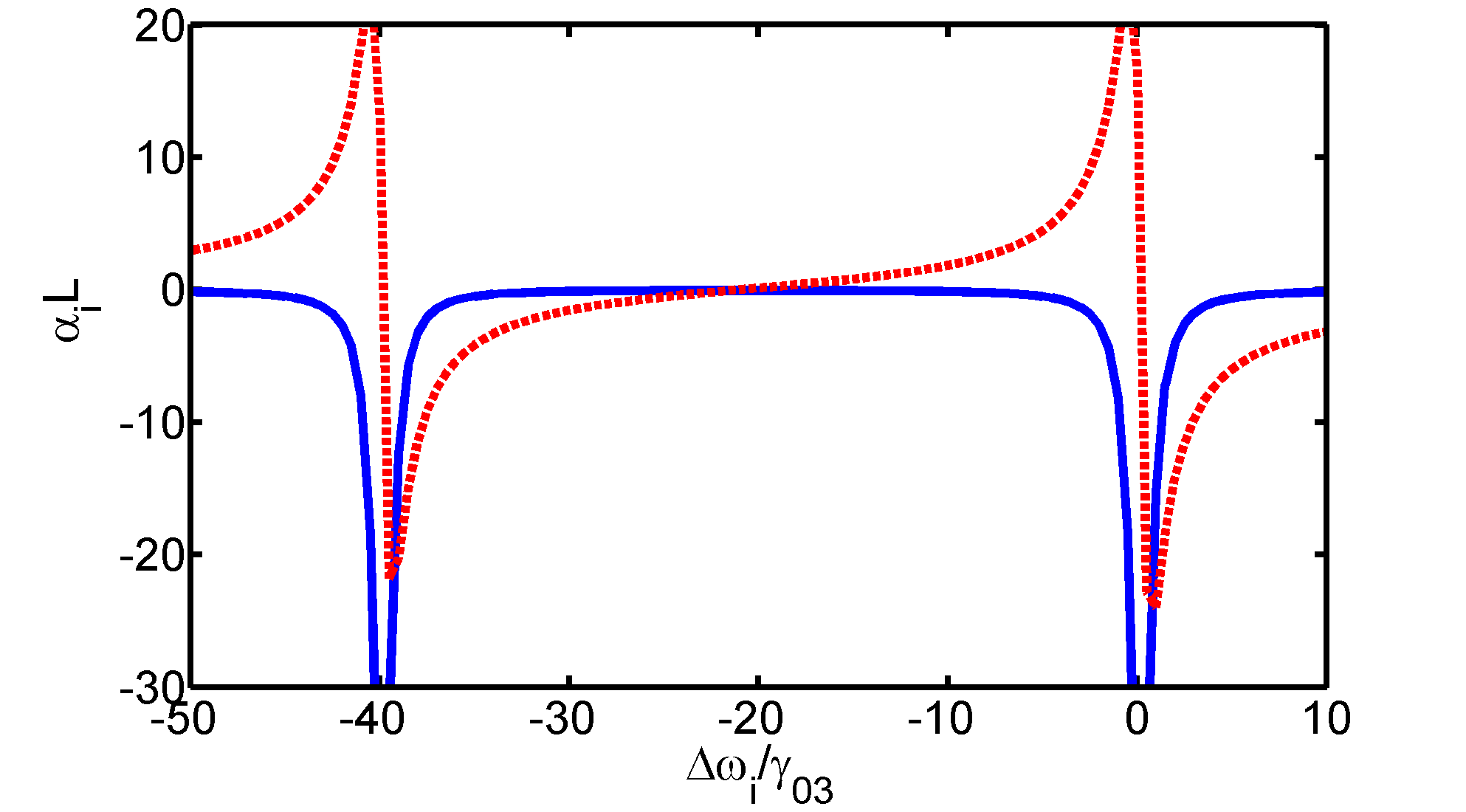}%
\caption{Self-coupling coefficient $\alpha_{i}.$ $\ $A dimensionless quantity
$\alpha_{i}L$ is plotted with real (solid blue) and imaginery (dashed red)
parts as a dependence of idler detuning $\Delta\omega_{i}$ showing a normal
dispersion inside the EIT window. \ }%
\label{alpha_i}%
\end{center}
\end{figure}

The absorption coefficient for idler field is the real part of $(-\alpha_{i})$
and phase velocity $v_{p}=\omega/k(\omega)=c/n(\omega)$ that $k(\omega
)=n(\omega)\omega/c.$ \ The wavevector is related to coefficient $\alpha_{i}$
that $k(\omega)=$ Im$(\alpha_{i})+\omega/c$ so $n(\omega)=1+ $ Im$(\alpha
_{i})/(\omega/c)$. \ The group velocity is $v_{g}=d\omega/dk(\omega
)=c/(n+\omega dn/d\omega)$ where $n\approx1,$ and it is this steep slope of
refractive index that makes a large group delay ($dn/d\omega>0$ inside EIT
window). \ As an example in Figure \ref{alpha_i}, we demonstrate the real and
imaginary parts of self-coupling coefficient $\alpha_{i}$ with the optical
depth (opd) $\rho\sigma L=150$ (see Chapter 6 for more details on other
parameters)$.$ \ The dispersion curve (Im($\alpha_{i}L$)) inside the left
parametric coupling window bounded by two absorption peaks (Re($\alpha_{i}L$))
shows a normal dispersion indicating a group delay at the center of the window
(see Figure \ref{coefficient} for complete parametric coupling windows).
\ Note that we plot out unitless $\alpha_{i}L$ where $L$ is in the order of
millimeter for regular cold atomic ensemble, and see Sec. II and III for
detail discussion of various coupling coefficients in Eq. (\ref{field}) and
efficiency dependence on optical depth.\

\begin{singlespaced}\begin{postliminary}
\ \ \cleardoublepage\addcontentsline{toc}{chapter}{BIBLIOGRAPHY}

%

\end{postliminary}\end{singlespaced}%

\end{document}